\documentclass[PhD, 11pt]{msu-thesis}
\usepackage[raggedright]{titlesec}
\usepackage[T1]{fontenc}
\usepackage{newtxtext,newtxmath} 
\usepackage{graphicx}
\usepackage{ragged2e}
\usepackage[numbers,sort&compress]{natbib}
\renewcommand{\bibname}{\centering \large BIBLIOGRAPHY}

\usepackage{color}   
\usepackage{hyperref}
\hypersetup{
    colorlinks=false, 
    linktoc=all,     
    linkcolor=black,  
}
\usepackage{etaremune}
\DeclareUnicodeCharacter{2009}{\,}
\title{\large Nuclear Data to Quantify Urca Cooling in Accreting Neutron Stars}
\author{\large Rahul Jain}
\dualmajor{\large Physics}{\large Computational Mathematics, Science, and Engineering}
\dedication{\large To Maa --\\ Who sacrificed her dreams, \\ so that she could fuel mine.}
\date{\large 2024}

\begin{document}

\frontmatter
\maketitlepage

\chapter*{\centering \large ABSTRACT}
\DoubleSpacing
\large
Neutron stars in Low Mass X-ray Binaries (LMXBs) can accrete matter onto their surface from the companion star. Transiently accreting neutron stars go through alternating phases of active accretion outbursts and quiescence. X-ray observations during the quiescence phase show a drop in X-ray luminosity with the time in quiescence. This is also inferred as the drop in surface temperature or the cooling of accreting neutron stars in quiescence. Analyzing these cooling curves reveals a great deal of information about the structure and composition of neutron stars. However, model-observation comparisons of such cooling curves are challenging - partly due to observational uncertainties, and partly due to incomplete knowledge of heating mechanisms during accretion outbursts. This situation is further exacerbated by the recent discovery of Urca cooling in the neutron star crust. These are cycles that alternate between electron-capture and $\beta$-decay to produce a large flux of neutrinos and anti-neutrinos. These freely stream out of the star and carry energy with them, essentially cooling the neutron star crust without changing the composition. As a result, it is necessary to accurately quantify the strength of Urca cooling to constrain the heat sources in neutron star crusts and facilitate better model-observation comparisons of the cooling curves.  

Urca cooling is effective only for a certain subset of nuclei with specific properties. One of the required conditions is a strong ground-state to ground-state $\beta$-decay transition strength for a nucleus. Previous studies have shown $^{33}$Mg to be a strong Urca cooling agent for neutron star crusts composed of X-ray burst ashes. This is attributed to a 37\% strong ground-state branch in the $\beta$-decay of $^{33}$Mg inferred from high-resolution $\beta$-delayed $\gamma$ spectroscopy. This is, however, a first-forbidden transition and the strong ground-state branch seems anomalously high compared to theoretical calculations. A goal of this dissertation is to remeasure this transition strength using Total Absorption Spectroscopy with the SuN detector. $\beta$-delayed neutron branching ratio is also measured with the NERO detector. A combination of SuN and NERO helps mitigate the Pandemonium effect, which is shown to systematically overestimate low-energy branchings in high-resolution $\gamma$ spectroscopy. 

The ground-state branch for the $\beta$-decay of $^{33}$Mg $\rightarrow$ $^{33}$Al was measured to be 0.7(24)\% corresponding to a log-ft value of 7.0$_{-0.7}^{+\infty}$. This is significantly lower than the previous measurement and is consistent with the first-forbidden nature of the transition arising from the recently confirmed negative parity ground state of $^{33}$Mg. It further translates into a substantially reduced intrinsic Urca cooling luminosity of L$_\text{34}$ = 60.0. This highlights the importance of Total Absorption Spectroscopy and motivates future experiments with this technique to refine calculations of Urca cooling. 

The rate of Urca cooling (L$_{34}$) is extremely sensitive to electron-capture thresholds (Q$_\text{EC}$) and is proportional to |(Q$_\text{EC}$)|$^5$. These electron-capture thresholds depend on nuclear masses. Several of the potential Urca cooling candidates are neutron-rich exotic nuclei whose masses have not been measured experimentally and Urca cooling calculations have to rely on theoretical mass predictions. However, theoretical model predictions diverge as they move away from the stable nuclei and do not have uncertainties. A global nuclear mass model with quantified uncertainties is also developed as a part of this dissertation using Bayesian Gaussian Process Regression and Bayesian Model Averaging (BMA). Updated neutron star crust calculations with the BMA mass model change not only the magnitude of Urca cooling but also the depth at which it happens. This has important implications for the overall thermal profile of the accreting neutron star crust.

\SingleSpacing
\clearpage

\makecopyrightpage 

%
\makededicationpage
\clearpage
\chapter*{\centering \large ACKNOWLEDGEMENTS}
\DoubleSpacing 
\large 

The challenging endeavor of getting a Ph.D. would certainly not have been possible without the help and support of many that I received along the way. I am extremely grateful to everyone who helped me in whatever way possible, no matter how small. While I try my best to recognize as many people as I can, I apologize in advance if I inadvertently miss someone. Please know that I sincerely appreciate your time and help. 

First and foremost, I would like to thank my advisor, Prof. Hendrik Schatz, for his excellent mentorship and guidance throughout these years. I could not have asked for a better graduate school experience than under his advising. A true expert in the field of Nuclear Astrophysics, every single interaction with you is filled with tremendous learning. I just wish I `walked' into your office more often. Thank you for always being open to and appreciative of my ideas and providing resources to try them out. Along with the academic opportunities, the extra-curricular opportunities I received as a graduate student helped me gain a big-picture perspective of the field and ultimately shaped me into a better scientist.

I also want to thank Prof. Witek Nazarewicz, who gracefully agreed to advise me on the CMSE part of my dissertation and welcomed me into the `Nuclear Theory' group at FRIB. I am fascinated by your passion for nuclear physics and hope that it rubbed off on me too. I would also like to thank other members of my committee - Prof. Artemis Spyrou for collaborating with us on my thesis experiment along with her group, supporting me in her role as the Associate Director for Education at NSCL, and eventually writing recommendation letters during job searching process; Prof. Ed Brown for all his help with astrophysics and introducing me to his seemingly simple yet powerful back-of-the-envelope calculations for neutron stars; and Prof. Michael Murillo for introducing me to Gaussian Process and Machine Learning. I am also grateful to all my committee members for taking time out of their busy schedules for annual meetings and ensuring I was on track and making progress. Additionally, I would also like to thank Prof. Sean Liddick for teaching me the basics of detector setup and data acquisition systems in his BCS lab. 

A big thanks to Wei-Jia Ong and Kirby Hermansen for helping me through all the stages of my thesis experiment. I would never have been able to do this successfully without you both. Your help on multiple fronts prevented me from reinventing the wheel on so many occasions. Thanks also to Andrea Richard, Mallory Smith, and Caley Harris for all your help in the S2 vault with SuN during the experiment in the middle of the COVID-19 pandemic, and Nabin Rijal for helping with the vacuum systems. I am also thankful to the entire NSCL operations group for successfully delivering the desired rare isotope beams for the experiment. In addition, I would like to thank Hannah Berg and Adriana Sweet for helping with GEANT4 simulations, Artemis Tsantiri and Eleanor Ronning for helping with RAINIER simulations, and Sudhanva Lalit for helping with crust model calculations. I would also like to thank Leo Neufcourt for helping me with the formalism of Bayesian Machine Learning and letting me use his code for the nuclear masses project and Samuel Giuliani for helpful discussions.

I am extremely grateful for the additional mentoring I received as a graduate student. I want to thank Fernando Montes for his valuable guidance on the professional development front and for always being open to talk about anything. I also want to thank Ana Becerril-Reyes for supporting me in all my student leadership roles within JINA and IReNA. Special thanks to Jaspreet Singh Randhawa who has been an incredible mentor and is always looking out for me. Thanks also to Steven Thomas and the entire AGEP community for providing a valuable space to rewind and relax from the hustle of graduate school. Finally, I would also like to thank all members of the Nuclear Astrophysics Group as well as the Nuclear Theory Group at FRIB who made it fun to work at the laboratory. 

I have also benefited from the support of the wonderful community in East Lansing. Thank you Larry Lee for being a great mentor and helping me navigate the American way of life. Thank you also for many of my `first-ever' traditions and for giving me some truly unimaginable life experiences. I am also thankful to Shay Manawar and Anum Mughal for being dear friends and helping me in times of crisis. Finally, thanks to all members of the squash community at the Michigan Athletic Club for helping me continue playing squash and keep improving my game. This has been a huge factor in preserving my sanity through all these years. 

I am exceptionally lucky to have the kind of love and support I get from my family. Mom and Dad, you are amazing role models and with you by my side, I can confidently chase any goal I set for myself. Thank you Chimu for your tough love and humbling me whenever needed. Finally and most importantly, thank you Karishma for always being there for me. You inspire me to keep pushing my limits and be the better version of myself, every single day. I hope I make you proud. 

\clearpage
\SingleSpacing
\tableofcontents* 
\clearpage
\listoftables 
\clearpage
\listoffigures 
%
%
%
\mainmatter
%

    
\huge \chapter{Introduction}\label{chp:intro}
\large 
Neutron stars are the densest stellar objects in the universe. They were discovered in 1967 as pulsars with radio telescopes \cite{Bell1969}. They weigh as much as about 1 {--} 2 times the mass of our sun (M$_\odot$) but measure only a few kilometers (10 {--} 12 km) in their radii \cite{ozel2016}. As a result, they have densities of more than $10^{17}$ kg/m$^3$. When massive stars (M $\geq$ 8 {--} 10 M$_\odot$) are nearing the end of their evolution, they form an inert iron core at their center that cannot undergo nuclear fusion any further to form heavier elements. This is because iron is the most stable element with the highest binding energy per nucleon. After a certain point, the iron core becomes so massive that is no longer able to support itself against its gravitational pull, resulting in a core collapse. The core is compressed to nuclear saturation densities until it begins to feel the repulsive nature of the strong nuclear force. This creates a shockwave that propagates outward and the infalling matter bounces off this rigid core, resulting in a core-collapse supernova \cite{Burrows2021}. The remnant of these explosions is a proto-neutron star. At this point, if the neutron degeneracy pressure is enough to support the compact object remnant, it becomes a neutron star. If not, the proto-neutron star collapses into a black hole.

Neutron stars are interesting sites to study the properties of matter under extreme conditions. Unlike a black hole where gravity turns everything into a singularity, all four fundamental forces {---} the strong force, the weak force, the electromagnetic force, and the gravitational force {---} play a role in the interaction of matter in neutron stars \cite{Lattimer2016,Baym2018,Oertel2017}. They are also prime candidates for multi-messenger astronomy \cite{Bailes2021}. Astrophysical sites habitating a neutron star are sources for gravitational waves \cite{Abbott2017}, neutrino signals \cite{Lattimer1989}, X-ray observations, radio wave signals, etc. Recently, neutron star mergers have also been identified as prominent sites for the production of heavy elements in the universe via the rapid neutron capture process, or the r-process \cite{Pian2017}. UV, visible, and infrared spectroscopy of such sites also help determine the abundance of each heavy element formed in the r-process \cite{Watson2019}. All these characteristics make neutron stars a prime object of interest to a wide range of physicists and astronomers \cite{Schatz2022horizons}. 
\bigskip

\LARGE \section{Accreting Neutron Stars}\label{sec:accretion}
\bigskip
\large
Up to 85\% of the stars in our universe are born in binary systems, with some in triple or even higher multiple systems \cite{Duchene2013-ez}. For binary systems with stars having unequal masses, each star evolves on a different timescale. Massive stars evolve much faster than lighter stars. If the lighter companion star can withstand the supernova explosion of the heavier star that is nearing the end of its evolution and if the binary system remains bound, a compact object binary is formed. It is a binary system with a regular low mass star and a compact object. If the compact object is a neutron star, it is called a neutron star binary. 

\begin{figure}
    \centering
    \includegraphics[width=300 pt,keepaspectratio]{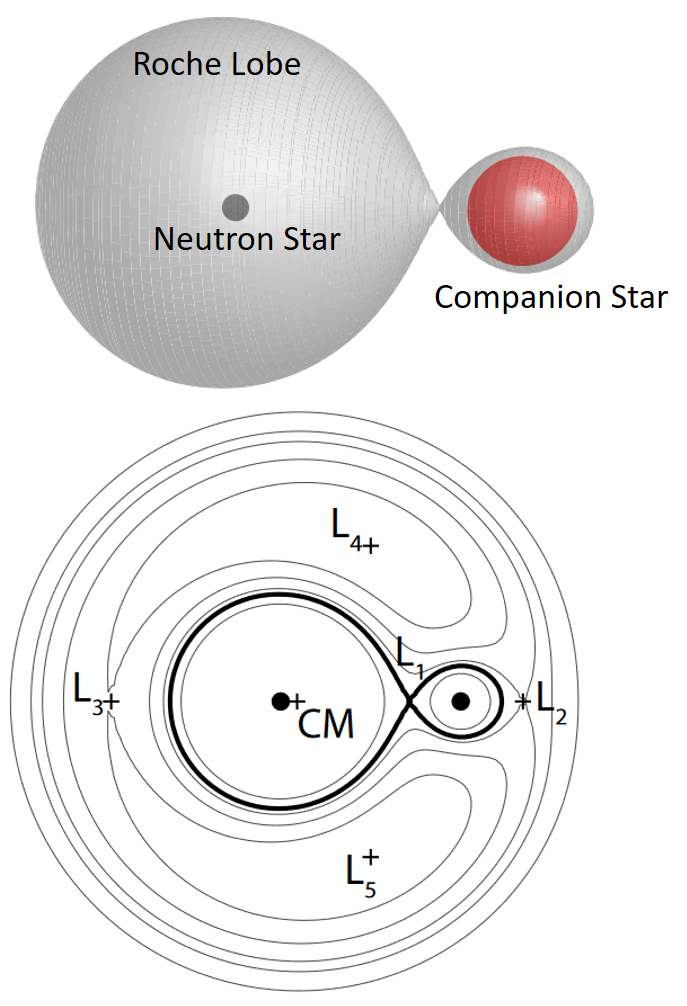}
    \caption{\large Accretion onto a neutron star via the mechanism of Roche lobe overflow. Matter is funneled into an accretion disk around the neutron star through the L1 Lagrange point.Figure modified from Ed Brown (private communication).}
    \label{fig:roche}
\end{figure}

Each star in a compact object binary has an imaginary envelope, called the Roche lobe, where matter present within the envelope will be gravitationally bound to that star. The Roche lobes of two stars in a binary system meet at a point called the L1 Lagrange point (Refer to Figure \ref{fig:roche}). Due to the internal evolutionary changes in the companion star or changes in it's orbit, it can expand and fill up its Roche lobe. At this point, matter is funneled through the L1 Lagrange point \cite{Charles2011} into an accretion disk around the compact star. Figure \ref{fig:roche} explains the process of accretion through the mechanism of Roche lobe overflow. An accretion disk forms around the neutron star because angular momentum still has to be conserved. The angular momentum, however, is eventually lost due to friction in the disk, and the matter gets deposited onto the surface of the neutron star.

The accreted matter releases about 200 MeV/u of gravitational potential energy while falling onto the surface of a neutron star. All this energy is radiated away in the form of X-rays. As a result, an accreting neutron star binary that is accreting matter from a sun-like companion star is also called a Low Mass X-ray Binary (LMXB). Such systems have been observed and monitored with modern space-based X-ray telescopes for a few decades now \cite{wijnands:ks1731,wijnands.ea:xmm_1731,Cackett2006Cooling-of-the-,Cackett2010Continued-Cooli}. LMXBs are significantly important to understand the properties of the crust of neutron stars as we will see in the following sections. 
\bigskip

\Large \subsection{Quasipersistent X-ray Transients}\label{sec:transients}
\bigskip
\large

Accretion onto a neutron star can be unstable and it can pause after several months or years. Such transient systems undergo periodic episodes of accretion, i.e., they alternate between periods of active accretion and quiescence. Accretion outburst periods are characterized by persistently high X-ray luminosity ($10^{35} - 10^{37}$ erg/s), whereas the quiescence phase is characterized by a 2 -- 3 orders of magnitude drop in persistent X-ray flux. While it is agreed that this drop in luminosity is due to the halting of accretion, the exact mechanism for pausing and resuming accretion on timescales of just a few years is unknown. Modern sensitive X-ray telescopes allow tracking of such quasipersistent transients even in their quiescence phase. So far, about a dozen such systems have been discovered and monitored \cite{Wijnands2017}. Some of them have also been monitored for multiple outbursts and quiescent periods \cite{Parikh2019,Degenaar2019}. 

\begin{figure}
    \centering
\includegraphics[width=450 pt,keepaspectratio]{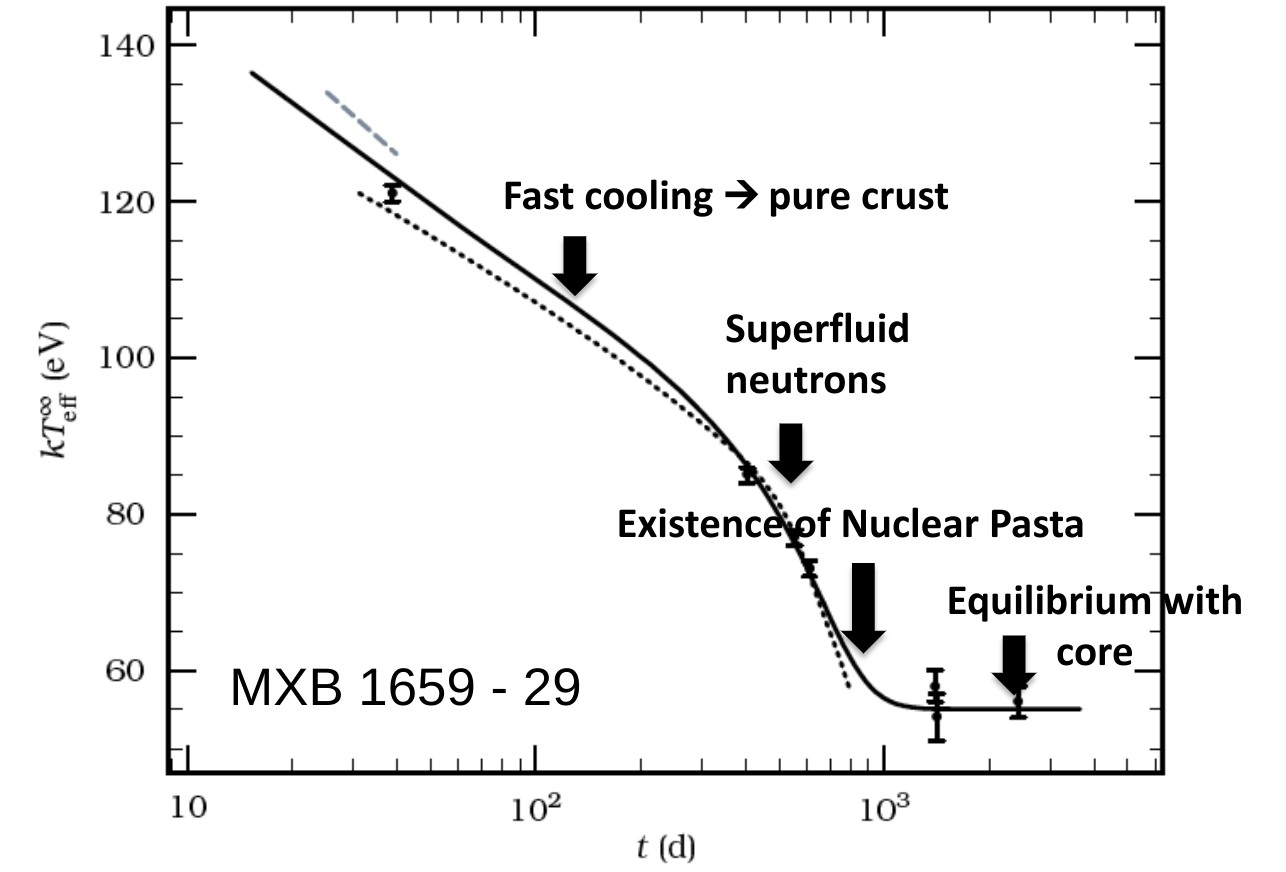}
    \caption{\large X-ray observations of the surface temperature of the neutron star in system MXB 1659--29 as a function of the number of days spent in quiescence. Different lines correspond to different models of the cooling curves. Solid arrows show examples of insights drawn from modeling the cooling curves. Figure reproduced and modified from \cite{Brown2009Mapping-Crustal}.}
    \label{fig:coolingmxb}
\end{figure}

Fitting the spectrum of X-ray photon energies to a black body spectrum, we can deduce the surface temperatures of accreting neutron stars. Tracking transient X-ray binaries in quiescence reveals a drop in surface temperature with the number of days spent in quiescence. This cooling of the crust over timescales of a few years is very different from the cooling of the entire neutron star that happens over timescales of thousands of years. The observed cooling curves, as they are called, for transiently accreting neutron stars provide insights into the structure and composition of the crusts of neutron stars \cite{Rutledge2002,Cackett2006,Shternin2007,Brown2009Mapping-Crustal}. As the neutron star spends more time in quiescence, the surface temperature probes successively deeper layers of the crust. Figure \ref{fig:coolingmxb} shows the observed cooling curve for the system MXB 1659--29 after 2.5 years of active accretion and possible insights drawn about the properties of its crust from cooling curve models \cite{Brown2009Mapping-Crustal}. The relatively fast cooling at earlier times reveals a well-ordered lattice structure of the outer crust with high thermal conductivity \cite{Cackett2006,Shternin2007,Brown2009Mapping-Crustal}. The bend in the cooling curve after a few hundred days in quiescence is interpreted as a phase transition marked by the onset of neutron superfluidity in the inner crust \cite{Shternin2007,Brown2009Mapping-Crustal}. The cooling curves may also carry evidence for the existence of nuclear pasta at the crust-core transition \cite{Horowitz2015,Deibel2017}. Molecular dynamics simulations of nuclear pasta predict high impurity scattering which impacts the thermal conductivity \cite{Horowitz2015}. 

During accretion outbursts, hydrogen and helium-rich matter from the outer layers of the companion star is deposited on the surface of neutron stars. This matter undergoes either stable or explosive nuclear burning in the envelope, depending on the composition of accreted matter, the rate of accretion, and the properties of the neutron star. \cite{Cyburt2016}. The ashes of this burning are submerged into the crust as the infalling matter from continued mass accretion pushes them deeper. For the typical accretion rates in Low Mass X-ray Binaries (LMXBs), it is estimated that the entire crust of the neutron star will be replaced by accreted matter in about $\sim$10,000 years. Since most LMXBs are expected to be older than 1 {--} 10 Myr \cite{Naylor1993}, their crusts are completely replaced by accreted matter.  

As the ashes are pushed deeper into the crust, the rising pressure and density induce nuclear reactions (Section \ref{sec:intro-reactions}) \cite{Haensel1990,Haensel2008,Lau2018}. The energy generated by nuclear reactions heats the crust and brings it out of thermal equilibrium with the core. Once accretion stops and the system enters the quiescent phase, the crust relaxes until it reaches thermal equilibrium with the core again. This is the origin of observed decrease in surface temperature during the quiescence phase for accreting neutron stars. The fraction of heat that flows to the surface vs. the fraction that flows to the core depends on the distribution of heat sources within the crust, the composition of the crust, and the thermal conductivity of the different layers of the crust. The composition of the crust, in turn, depends on the nuclear reactions that process the accreted matter. As a result, a detailed understanding of all the different nuclear reactions that can occur throughout the neutron star is important in the modeling of the cooling curves of accreting neutron stars. 
\bigskip

\Large \subsection{Nuclear Reactions during Accretion}\label{sec:intro-reactions}
\bigskip
\large 

\begin{figure}
    \centering
\includegraphics[width=350 pt,keepaspectratio]{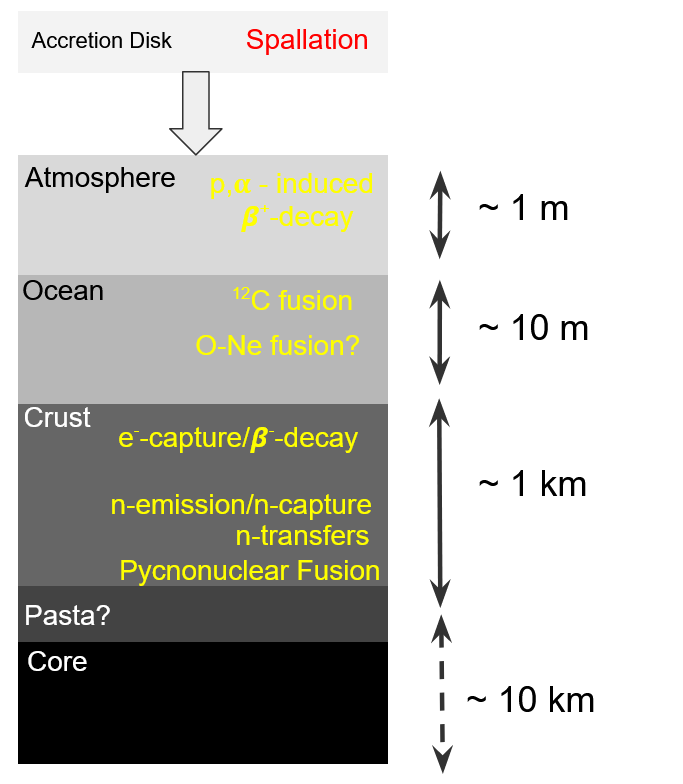}
    \caption{\large Nuclear reactions that take place in the different layers of an accreting neutron star. H/He-rich accreted matter from a companion star is processed by these reactions as it is buried in the star. Note that the thicknesses of different layers are not drawn to scale.}
    \label{fig:nuclearreactions}
\end{figure}

A neutron star has a structure, composed of different layers, analogous to that of the Earth (Refer to Figure \ref{fig:nuclearreactions}). It has a thin atmosphere of gaseous atomic plasma, followed by an ocean consisting of liquid plasma with free electrons and ionized nuclei. Beneath this is a solid crust, composed of a crystallized lattice of nuclei and degenerate electrons. The inner crust also consists of free and superfluid neutrons. The mantle is composed of nuclear pasta which are structures of pure nucleonic matter. Finally, we have the core which is thought to be composed of exotic nuclear matter. All these layers cover a very wide range in temperature, pressure, and density conditions. For example, density runs from around 10$^{-4}$ g/cm$^{3}$ in the atmosphere to 10$^{15}$ g/cm$^{3}$ and beyond in the core. As a result, many different types of nuclear reactions are enabled in different layers of the neutron star that change the composition of accreted matter as it is buried in the neutron star. Figure \ref{fig:nuclearreactions} shows the different layers and possible nuclear reactions at those depths. 

Nuclear reactions begin altering the composition of accreted material even before it is deposited on the surface of the neutron star. As the accreted material slows down in the atmosphere, it is exposed to a flux of high energy protons, resulting in spallation reactions. These reactions reduce the CNO abundance, and result in hydrogen being present at ignition. The spallation of heavy elements also helps replenish some of the CNO elements \cite{Randhawa2019}. This alters the rate of hydrogen burning in the atmosphere which primarily proceeds through the CNO cycle, thus requiring heavy elements as catalysts. Stable hydrogen burning produces helium, whereas stable helium burning through the 3-$\alpha$ reaction produces carbon. Once the temperature rises beyond $\sim$ 10$^{8}$ K, the accelerated rate of nuclear reactions results in a thermonuclear runaway, triggering a type-I X-ray burst powered by the rapid proton capture or the rp-process \cite{Schatz1998}. The thermonuclear runaway causes a breakout from the CNO cycle and allows for the formation of heavy elements through proton captures and $\beta^{+}$-decay. (p, $\alpha$) reactions create cycles in the rp-process. The ashes of nuclear burning in the atmosphere are deposited in the ocean, and may contain significant amounts of carbon. If certain conditions are met, the carbon layer can be ignited and undergo explosive burning triggered by carbon fusion, resulting in a superburst \cite{Meisel2022Constraining-Ac}. These are similar to the type-I X-ray bursts but are a thousand times more energetic and last much longer. Recently, a hyperburst powered by explosive oxygen and neon fusion in the ocean was also proposed to explain the anomalously hot crust in MAXI J0556-332 \cite{Page2022A-Hyperburst-in}.  

In the crust, the densities are greater than 10$^{6-8}$ g/cm$^{3}$. At these densities, the Fermi energies of electrons are sufficiently high to overcome the electron-capture threshold. As a result, a series of electron-captures occur along isobaric chains. Electron-captures often proceed in pairs of two because of the odd-even staggering of nuclear masses, and hence, electron capture thresholds. The reverse $\beta$-decay is usually blocked but can happen under certain conditions as described in Section \ref{sec:urca-cooling} \cite{Schatz2013}. When nuclei hit the neutron dripline, they emit neutrons that can be captured by light elements. Recently, another channel for exchanging neutrons, i.e. the n-transfer reactions was proposed \cite{Chugunov2018,Schatz2022}. At the high densities in the neutron star crust, the wave functions of neighboring nuclei overlap. As a result, a weakly bound neutron can quantum mechanically tunnel from one nucleus to another. This changes the abundances in different isobaric mass chains. Even deeper in the crust, the zero-point energy vibrations overlap and result in density-induced fusion, also called pycnonuclear fusion \cite{Beard2010,Yakovlev2006,Jain2023}. 

All these reactions deposit heat at different locations in the crust during active accretion. \texttt{xnet} is a nuclear reaction network that accounts for all possible reactions in the crust and calculates the composition and nuclear energy deposition profile in the crust. It accepts an initial composition as input which depends on nuclear burning stages outside the crust. It then calculates the steady-state composition of the crust in a planar geometry, where the crust is approximated to be very thin which leads to a constant surface gravity across the crust. It tracks the changes in the composition of an accreted fluid element due to increasing pressure, $P = \dot{m}gt$, where $\dot{m}$ is the local mass accretion rate, and $g$ is the constant local surface gravity. The density, $\rho$, is calculated using an equation of state as prescribed in \cite{Gupta2007}. Increasing time is a proxy for increasing pressure, density, and depth in the neutron star crust. Reference \cite{Lau2018} lists more details about the nuclear physics properties used in \texttt{xnet}. 

Recently, an alternative approach has been suggested in the literature \cite{Gusakov2020,Gusakov2021,Shchechilin2021} to calculate the heat deposited during accretion in the neutron star crust. Availability of free neutrons in the inner crust leads to the diffusion of neutrons, violating the condition of hydrostatic equilibrium and the assumption of a constant red-shifted neutron chemical potential ($\mu_N^\infty$). This leads to a very different Equation of State (EOS) for the accreted crust that is close to catalyzed EOS in the inner crust \cite{Gusakov2020}. There are no pycnonuclear fusion reactions in the crust as discussed above. As a result, the total heat deposited in the crust is only 10--30\% of previous calculations with pycnonuclear fusion reactions \cite{Shchechilin2021}, although the complete reaction pathways are yet to be identified. The heat deposition only depends on the pressure $P_{oi}$ at the outer-inner crust interface that is different from the traditional neutron-drip pressure \cite{Gusakov2021}. 

\bigskip

\Large \subsection{Shallow Heat Source}\label{sec:shallow-heating}
\bigskip
\large 

\begin{figure}
    \centering
    \includegraphics[width=400 pt,keepaspectratio]{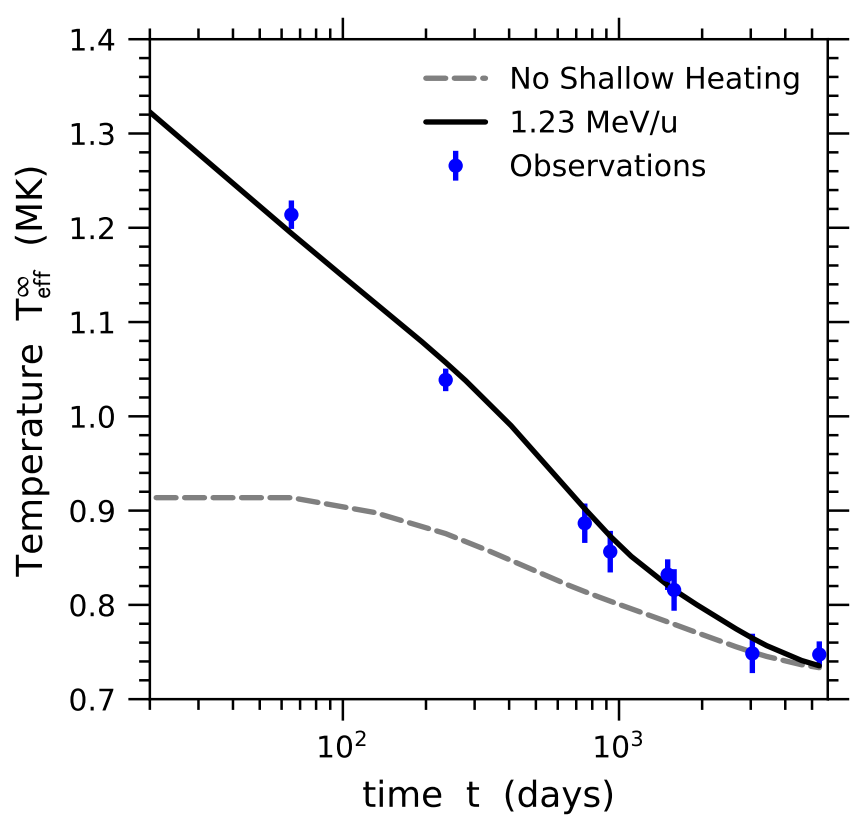}
    \caption{\large Modeled cooling curves in quiescence for the quasipersistent transient KS 1731--260 with and without an artificial shallow heat source. Shallow heating of 1.23 MeV/u is required by the models to match quiescent X-ray observations.}
    \label{fig:shallowheating}
\end{figure}

The nuclear energy output from the above network calculations is not sufficient to explain the quiescent X-ray observations. Almost all models of accreting neutron star systems need to account for an additional artificial heat source at relatively shallow depths to match the observed cooling curves at early times \cite{Brown2009Mapping-Crustal,Waterhouse2016,Turlione2015,Degenaar2014Probing-the-Cru,Potekhin2021}. The physical origins of this mysterious heat source is still unknown and is one of the most challenging open questions in neutron star crust physics. While most systems require a few MeV/u of additional shallow heating to match the observations, some require no shallow heating \cite{Degenaar2015}, whereas some require $\sim$ 10 MeV/u \cite{Deibel2015A-Strong-Shallo}. Even the same systems over different cycles of outburst and quiescence can require either the same \cite{Parikh2019} or different \cite{Degenaar2019} levels of shallow heating. Further, the inferred properties of the shallow heat source also depend on the composition of the envelope and crust of the neutron star. Figure \ref{fig:shallowheating} shows how accounting for additional shallow heating of 1.23 MeV/u in KS 1731-260 helps match the cooling curves models with observations. Several mechanisms ranging from nuclear reactions to astrophysical processes for the origin of shallow heating have been proposed. They include residual accretion during quiescence, viscous heating caused by accretion-induced shear \cite{Piro2007Turbulent-Mixin}, convective mixing, nuclear fusion of light neutron-rich elements \cite{Horowitz2008}, electron captures in the crust \cite{Gupta2007,Chamel2020}, pion-induced heating \cite{Fattoyev2018}, uncertainties in pycnonuclear fusion \cite{Jain2023}, gravity-wave transport of angular momentum \cite{sunyaev:transient}. However, none of these mechanisms can reproduce observations and the origin of shallow heating remains unknown. 
\bigskip

\LARGE \section{Urca cooling}\label{sec:urca-cooling}
\bigskip
\large 

Nuclear reactions not only heat the crust but can also cool the crust. This is possible through a process known as Urca cooling, where successive electron captures and beta-decays alternating between the same nuclei emit a large flux of neutrinos and anti-neutrinos, respectively, that carry energy away with them. An Urca type process was first theorized by George Gamow and M\'{a}rio Schenberg while visiting Cassino da Urca in Urca, Rio de Janeiro. Even though they named the process Urca cooling to commemorate their visit, they later backronymed it to UnRecordable Cooling Agent. Urca cooling is the dominant mechanism for the cooling of very hot, newly born neutron stars \cite{Negreiros2012}. This happens in the neutron star core via the direct-Urca process where free protons and neutrons cycle as:
\begin{align*}
     \large p + e^- &\rightarrow n + \color{blue}{\nu_e} \\
    \large n &\rightarrow p + e^- + \color{blue}{\bar{\nu_e}}
\end{align*}
Depending on the proton and neutron densities in the core that set their Fermi momenta, bystander nucleons may also participate in the process to enable momentum conservation. This leads to a modified Urca process which is less effective. Urca cooling in the crust happens via a new type of crust Urca process \cite{Schatz2013} where neutron-rich radioactive nuclei undergo this cycle as:
\begin{align*}
     \large ^{A}_{Z}X + e^- &\rightarrow \hspace{0pt} ^{A}_{Z-1}Y + \color{blue}{\nu_e} \\
    \large ^{A}_{Z-1}Y &\rightarrow \hspace{0pt} ^{A}_{Z}X + e^- + \color{blue}{\bar{\nu_e}}
\end{align*}
Urca cooling was initially expected to not play a role in the cooling of accreting neutron star crusts since the properties of the crust were calculated using zero-temperature limit approximation. This is valid since the crust has high density and the Boltzmann temperature is negligible compared to the electron Fermi energy. However, it was later shown that heat deposited by nuclear reactions during accretion is enough to raise the temperature to $T \sim 10^9$ K where Urca cooling can affect X-ray observations \cite{Deibel2015}. Hence, it is important in the modeling of quiescent cooling curves. 

\begin{figure}
    \centering
    \includegraphics[width=450 pt,keepaspectratio]{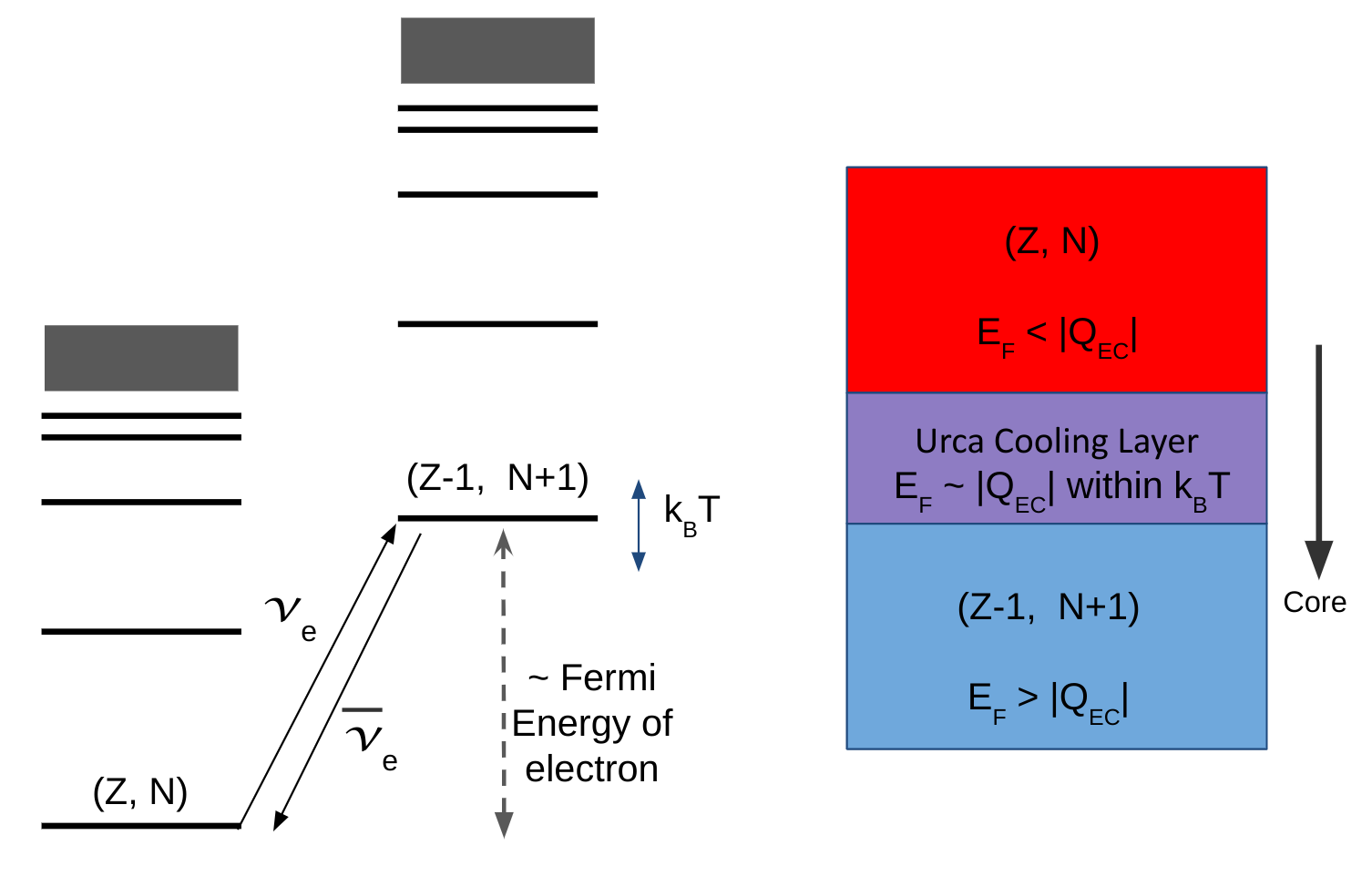}
    \caption{\large Formation of an Urca cooling layer in the crust of an accreting neutron star at a depth where the electron Fermi energy is equivalent to the electron-capture threshold for a neutron-rich isotope. e$^{-}$-capture/$\beta^{-}$-decay cycles generate a flux of neutrinos and anti-neutrinos that stream freely out of the star since the crust is not dense enough to trap the neutrinos. This effectively cools the star, which is also known as Urca cooling. }
    \label{fig:urcacooling}
\end{figure}

Electrons in the crust form a degenerate electron gas and the electron population follows Fermi-Dirac statistics. The probability distribution for a given energy state (E) to be filled is given by: 
\begin{equation}\label{eqn:Fermi}
    \large f(E) = \frac{1}{e^{(E-E_F)/k_BT}+1}
\end{equation}
where $E_F$ is the Fermi energy of the electron gas, $k_B$ is the Boltzmann constant, and T is the temperature. At zero temperatures, the probability distribution is a step-function, where all the energy states below the Fermi energy are filled and none of the energy states above the Fermi energy are occupied. This zero-temperature limit approximation was used in earlier reaction network calculations that find sharp composition changes when Fermi energy reaches the electron capture threshold. The reverse $\beta^-$-decay is blocked since there is no phase space available for the emitted electron as all the states below Fermi energy are filled. 

However, at finite temperatures, there is a diffusion of occupied states at the Fermi energy. A few electrons jump from states just below to Fermi level to states just above the Fermi level. The higher the temperature, the higher the diffusion. Now the electrons from the states above Fermi energy can be captured before the Fermi energy reaches the electron capture threshold. There are also unoccupied states below Fermi energy for the electron from subsequent $\beta^-$-decay. Essentially, this is opening up the phase space for reverse $\beta^-$-decays in a very thin shell at the boundary between two composition layers. This Urca shell contains both the nuclides that rapidly undergo e$^{-}$-capture/$\beta^{-}$-decay cycles and enable Urca cooling. 

Urca cooling pairs are mostly formed by odd-mass nuclei. Even mass nuclei have either both even or odd numbers of neutrons and protons. The odd-even staggering in the nuclear binding energy leads to a lower electron capture threshold on odd-odd nuclei than on even-even nuclei. As a result, the electron captures mostly proceed in steps of two in even mass nuclei, and Urca cooling is inhibited. If a ground state capture is forbidden and the capture occurs into excited states, the captures could proceed in single steps, but when the electron Fermi energy exceeds the ground state threshold, beta decay is also blocked. Odd mass nuclei, on the other hand, can either be odd-even or even-odd nuclei and allow for e$^{-}$-capture/$\beta^{-}$-decay cycles at each step. 

The rate of Urca cooling is extremely sensitive to temperature. The higher temperature opens a wider phase space for all participating particles which leads to a $\sim T^5$ dependence on temperature \cite{Shapiro1983}. Additionally, higher temperatures allow for a wider diffusion of the electron occupancy probability function given by Fermi-Dirac statistics (Eqn. \ref{eqn:Fermi}) leading to a thicker Urca shell. Several odd mass open-shell nuclei are also deformed which leads to many low-lying excitations. Any excited state within $\sim k_BT$ can also contribute to Urca cooling. This leads to a thermal decoupling of the layers below and above the Urca shell. If a heat source is present just below the Urca shell, most of the energy it deposits will just be carried away by neutrinos from enhanced Urca cooling. This puts a very strong constraint on the location of the shallow heat source in the crust. By extension, it is very important to account for accurate Urca cooling while modeling the cooling curves of accreting neutron stars in quiescence. 
\bigskip

\LARGE \section{Nuclear Data for Modeling}\label{sec:nuclear-data}
\bigskip
\large 

The total neutrino luminosity from Urca cooling for a given Urca pair can be approximated analytically as:

\begin{equation}
    \large L_\nu \approx L_{34} \times 10^{34}  \hspace{1pt}\text{ergs/s} \times X(A) \hspace{1pt} T_9^{5} \left ( \frac{g_{14}}{2} \right)^{-1}R_{10}^{2}
\end{equation}
where $T_9$ is the temperature in GK, $g_{14}$ is the surface gravity in $10^{14}$ g/cm$^2$, $R_{10}$ is the neutron star radius in 10 km, X(A) is the mass fraction of the nuclear isotope of mass A undergoing Urca process, and $L_{34}$ is a constant depending on the nuclear properties of that isotope as:
\begin{equation}\label{eqn:urca}
    \large L_{34} = 0.87 \left( \frac{10^6 s}{ft} \right) \left( \frac{56}{A} \right) \left( \frac{|Q_{EC}|}{4  \hspace{1pt}\text{MeV}} \right)^5 \left( \frac{\langle F \rangle ^*}{0.5} \right)
\end{equation}
where $ft$ is the nuclear $ft$ value representing the comparative half-life of the weak transition (refer Section \ref{sec:weak-interactions}), A is the nuclear mass number, $Q_{EC}$ is the electron-capture Q-value, and $\langle F \rangle ^*$ is the Coulomb factor given by 
\begin{equation}
    \large \langle F \rangle ^* = \frac{F^+ F^-}{F^+ + F^-}
\end{equation}
with
\begin{equation}
     \large F ^{\pm} = \frac{2 \pi \alpha Z}{1 - e^{\mp 2 \pi \alpha Z}}
\end{equation}
where $\alpha = \frac{1}{137}$ is the electrodynamic fine structure constant, and Z is the atomic charge number. 

The rate of Urca cooling depends on the equation of state of dense matter that sets the mass and radius of the neutron star, which also sets the surface gravity. In addition, it is also very sensitive to the temperature as discussed in the previous section. The nuclear physics parameters of importance are the mass fraction, comparative half-life, and the absolute Q-value of the weak transition. Mass fraction mainly depends on the dynamics of nuclear burning of accreted matter outside the crust and the composition of ashes formed that get buried in the crust. As a result, the comparative half-life that depends on the weak transition strength, and the Q-value that depends on nuclear masses are the key pieces of nuclear data that are required to calculate the rate of Urca cooling in accreting neutron star crusts.  
\bigskip

\Large \subsection{Weak Interactions}\label{sec:weak-interactions}
\bigskip
\large 

The transition strengths between nuclear levels in the parent and the daughter nuclei are required to calculate the strength of Urca cooling. These are extracted from the laboratory $\beta$-decay process. This section explains how the laboratory $\beta$-decay half-life and the feeding intensities relate to the transition strengths. 

$\beta$-decays are mediated by the weak nuclear force. The transition rate, in general, treating the decay-causing interaction as a weak perturbation is given by Fermi's Golden Rule \cite{Krane1989} as:
\begin{equation}\label{eqn:gold}
    \large \lambda = \frac{2\pi}{\hbar}|V_{fi}|^2\rho(E_f)
\end{equation}
where the matrix element $V_{fi}$ is the integral of interaction V between the initial and final states given by 
\begin{equation}
    \large V_{fi} = \int \psi_f^* \hspace{1pt} V \hspace{1pt} \psi_i dv
\end{equation}
and $\rho(E_f)$ is the density of states with final energy $E_f$. Since $\beta$-decay is a three-body process with the nucleus, electron, and antineutrino in the final state, the matrix element can be written as:
\begin{equation}\label{eqn:vfi}
    \large V_{fi} = g \int [\psi_f^*\phi_e^*\phi_\nu^* ]\hspace{1pt} \mathcal{O}_\chi \hspace{1pt} \psi_i dv
\end{equation}
where $\psi_f^*, \hspace{1pt} \phi_e^*, \hspace{1pt} \phi_\nu^*$ are the final states of the nucleus, electron, and the antineutrino, respectively, $\mathcal{O}_\chi$ is the mathematical operator for the interaction, and $g$ is the $\beta$-decay strength constant with
\begin{equation}
    \large g = 0.88 \times 10^{-4} MeV.fm^3
\end{equation}

The kinetic energy released in the decay will be shared by the recoiling nucleus, the electron, and the antineutrino. Let $p$ be the final momentum of the electron and $q$ be the final momentum of the antineutrino. $p$ and $q$ will be related as:
\begin{equation}\label{eqn:q}
    \large q = \frac{Q - T_e}{c} = \frac{Q-\sqrt{p^2c^2+m_e^2c^4}+m_ec^2}{c}
\end{equation}
where $Q$ is the Q-value of the decay and $T_e$ is the kinetic energy of the electron which depends on $p$. The number of final states with the momentum of electron between $p$ and $p+dp$ and the momentum of antineutrino between $q$ and $q+dq$ is given by:
\begin{equation}\label{eqn:states}
    \large d^2n = dn_edn_\nu = \frac{(4\pi)^2V^2p^2dpq^2dq}{h^6}
\end{equation}
Substituting Eqn. \ref{eqn:states} and Eqn. \ref{eqn:vfi} in Eqn. \ref{eqn:gold}, we can write the partial transition rate as:
\begin{equation}\label{eqn:partial}
    \large d\lambda = \frac{2\pi}{\hbar}g^2|M_{fi}|^2(4\pi)^2\frac{p^2dp \hspace{1pt} q^2}{h^6}\frac{dq}{dE_f}
\end{equation}\label{eqn:mfi}
where 
\begin{equation}
    \large M_{fi} = \int \psi_f^* \hspace{1pt} \mathcal O \hspace{1pt} \psi_i \hspace{1pt} dv
\end{equation}
is the nuclear matrix element. Integrating Eqn. \ref{eqn:partial} over all the possible values of electron momenta, and substituting for $q$ from Eqn. \ref{eqn:q}, we can write the total transition rate as:
\begin{equation}\label{eqn:total}
    \large \lambda = \frac{g^2 |M_{fi}|^2}{2\pi^3\hbar^7c^3} \int_{0}^{p_{max}} F(Z',p)\hspace{1pt} p^2 \hspace{1pt} (Q-T_e)^2 \hspace{1pt} dp
\end{equation}
where $F(Z',p)$ is the Fermi function which accounts for the influence of the nuclear Coulomb field. Eqn. \ref{eqn:total} can be rewritten as:
\begin{equation}\label{eqn:final}
    \large \lambda = \frac{f(Z',E_0)g^2m_e^5c^4|M_{fi}|^2}{2\pi^3\hbar^7}
\end{equation}
where 
\begin{equation}
    \large f(Z',E_0) = \frac{1}{(m_ec)^3(m_ec^2)^2} \int_{0}^{p_{max}} F(Z',p) \hspace{1pt} p^2 \hspace{1pt} (E_0 - E_e)^2 \hspace{1pt} dp
\end{equation}
is the Fermi Integral. The half-life of $\beta$-decay is related to the transition rate as:
\begin{equation}
   \large  t_{1/2} = \frac{ln(2)}{\lambda}
\end{equation}
Substituting for $\lambda$ and rearranging the Fermi Integral, we get
\begin{equation}\label{eqn:ft}
    \large f \hspace{1pt} t_{1/2} = ln(2)\frac{2\pi^3\hbar^7}{g^2m_e^5c^4|M_{fi}|^2}
\end{equation}
This is also known as the comparative half-life for the weak transition. This $ft$ value depends only on the nuclear matrix element, whereas all other effects are embedded in the Fermi Integral. This helps in comparing the overlap between different sets of initial and final nuclear wave functions. Since the $ft$ values range in many orders of magnitude, they are often quoted as log-ft values for a transition. The square of the nuclear matrix element $|M_{fi}|^2$ is referred to as the transition strength. 

\begin{table}[]
    \centering
    \begin{tabular}{cccc}
    \toprule
       \large Transition  & \large L & \large $\Delta J$ & \large $\Delta \pi$ \\
    \midrule
      \large Fermi (``Superallowed'')   & \large 0 & \large 0 & \large 0 \\
      \large Gamow-Teller (``Allowed'')   & \large 0 & \large 0,1 & \large 0 \\
      \large First-forbidden   & \large 1 & \large 0,1,2 & \large 1 \\
      \large Second-forbidden   & \large 2 & \large 1,2,3 & \large 0\\
      \large Third-forbidden   & \large 3 & \large 2,3,4 & \large 1 \\
      \large \vdots & \vdots & \vdots & \vdots \\
    \bottomrule
    \end{tabular}
    \caption{\large Table listing the quantum mechanical selection rules for different degrees of allowed/forbidden weak transitions. L denotes the angular momentum carried by the decay particle (in units of $\hbar$), whereas $(\Delta J, \Delta \pi)$ denotes the change in spin-parity of the initial and final nuclear states. Note that $\Delta J$ is dependent on L.}
    \label{tab:selection}
\end{table}

In the $\beta$-decay process, along with the energy and momentum, angular momentum should also be conserved. Since quantum mechanical particles participate in the transition, the total angular momentum is quantized and is given by the vector sum of orbital angular momentum and spin angular momentum. Naturally, there occur some selection rules between the spin-parity of initial and final nuclear states $(\Delta J, \Delta \pi)$, that must be followed as listed in Table \ref{tab:selection}. The difference in angular momentum between nuclear states is carried by the decay particles. This difference, along with the change in parity characterizes the `allowedness' of the weak transition. Allowed decays are where the decay particles carry no orbital angular momentum $(L = 0)$ whereas forbidden decays are where the decay particles carry the difference in angular momentum $(L > 0)$. It also sets the degree of `forbiddenness' of a transition. Forbidden transitions do occur but each degree of forbiddenness slows down the rate of transition by a few orders of magnitude. For the electron to carry away the difference in angular momentum, it should be emitted off-center of the nucleus where the wavefunction drops off exponentially.
\bigskip

\Large \subsection{Nuclear Masses}\label{sec:intro-masses}
\bigskip
\large 

The amount of energy released in a nuclear reaction or decay (also called as Q-value) is related to the total mass loss by Einstein's mass-energy equivalence principle,
\begin{equation}
    \large E = mc^2.
\end{equation}
As a result, the mass of a nucleus is an important quantity in any calculation involving nuclear reactions. The Q-value is given by the difference in the mass of reactants and the mass of products. For nuclear decays mediated by the weak force, there are three possible types of $\beta$-decays in the laboratory:
\begin{align*}
    \large \beta^-&: \hspace{2pt} ^A_ZX \rightarrow ^A_{Z+1}Y + e^- + \bar{\nu_e} \\ 
    \large \beta^+&: \hspace{2pt} ^A_ZX  \rightarrow ^A_{Z-1}Y + e^+ +\nu_e \\
    \large EC&: \hspace{2pt} ^A_ZX + e^- \rightarrow ^A_{Z-1}Y + \nu_e.
\end{align*}
The Q-values for each of these reactions are given by:
\begin{align*}
    \large Q(\beta^-)&: [m(^A_ZX) - m(^A_{Z+1}Y)]c^2 \\ 
    \large Q(\beta^+)&: [m(^A_ZX) - m(^A_{Z-1}Y) - 2m_e]c^2  \\
    \large \large Q(EC)&: [m(^A_ZX) - m(^A_{Z-1}Y)]c^2 - B_n.
\end{align*}

\begin{figure}
    \centering
    \includegraphics[width=450 pt,keepaspectratio]{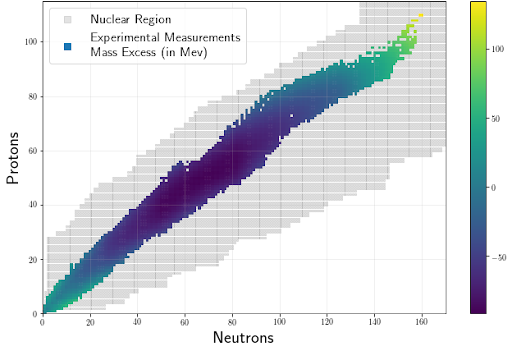}
    \caption{\large The nuclear masses landscape on the chart of nuclides. The colored region is for nuclei whose masses have been experimentally measured as of AME2020 database \cite{Wang2021}. The grey region is the extent of nuclear existence, i.e. the region between the proton and the neutron driplines. Points in the grey area that aren't colored are nuclei whose masses remain to be measured experimentally.}
    \label{fig:masses}
\end{figure}

Note that these are atomic masses (that in turn depend on the nuclear mass), so the electron mass cancels for $\beta^-$-decay but adds up for $\beta^+$-decay. The neutrino mass is considered to be negligible here. $B_n$ is the binding energy of the captured n-shell electron. To calculate Urca cooling strength as per Equation (\ref{eqn:urca}), the $|Q_{EC}|$ term is the Q-value for $e^-$-capture. However, since the electron is captured from the degenerate electron gas, $B_n$ can be neglected. Thus $Q'(EC) = Q(\beta^-)$ is used to estimate the Urca cooling luminosity. Consequently, it depends on the nuclear masses.   

In crusts of accreting neutron stars, electron captures can happen all the way up to the neutron dripline, where nuclei in their ground state are no longer neutron bound. As a result, we need to know accurate nuclear masses for all the neutron-rich nuclei to calculate the total Urca cooling luminosity throughout the crust. Figure \ref{fig:masses} shows the status of current experimental mass measurements and theoretical approximates of the extent of driplines on the nuclear chart. It is immediately clear that many of the nuclear masses for neutron-rich nuclei still need to be measured. Even after considerable experimental efforts, many of these nuclear mass measurements remain out of our reach for the foreseeable future and we need to rely on theoretical mass model predictions. There is, however, a variety of mass models to choose from \cite{Mller2016,Goriely2013}. Each theoretical model has different assumptions and approximations and is calibrated on different subsets of experimental data. Moreover, most of the predictions lack uncertainties, and extrapolating nuclear masses in the very neutron-rich region yields completely different trends. This becomes a problem for crust cooling calculations since we do not know apriori which mass model is correct and should be relied on. As a result, there is a dire need for a quantified global nuclear mass model that can accurately predict nuclear masses throughout the nuclear chart with quantified uncertainties. Chapter \ref{chp:massmodel} outlines the procedure to create such a mass model for use in astrophysical models by averaging the prediction of multiple theoretical mass models using Bayesian Model Averaging (BMA).
\bigskip

\LARGE \section{$^{33}$Mg - $^{33}$Al Transition for Urca Cooling}\label{sec:intro-transition}
\bigskip
\large 

\begin{figure}
    \centering
    \includegraphics[width=450 pt,keepaspectratio]{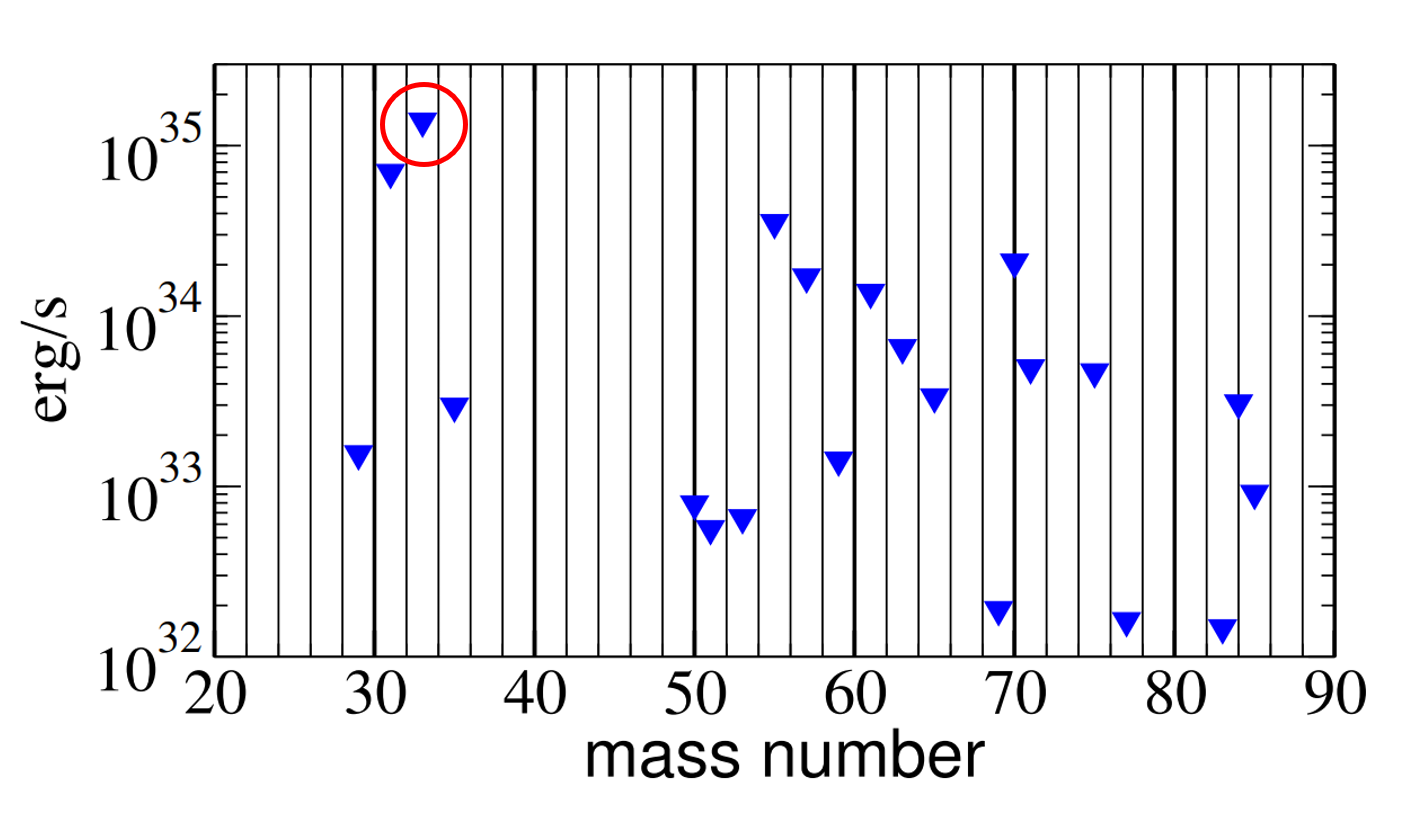}
    \caption{\large Urca cooling luminosity in a particular isobaric mass chain for a neutron star crust composed of typical Type-I X-ray burst ashes powered by rp-process as a function of mass number. These calculations are performed with the \texttt{Xnet} nuclear reaction network for a crust with a constant temperature of 0.5 GK. Mass \emph{A} = 33 chain, circled in red, has the highest Urca cooling luminosity.}
    \label{fig:mass33}
\end{figure}

Effective Urca cooling requires a large abundance of nuclei in the Urca pair and a strong ground-state to ground-state $\beta$-decay/$e^-$-capture transition strength. \texttt{Xnet} calculations starting with ashes of typical Type-I X-ray bursts powered by rp-process show significant Urca cooling. Figure \ref{fig:mass33} shows neutrino cooling luminosity in such calculations as a function of the mass number. Transitions in the isobaric mass chain with \emph{A} = 33 provide the strongest Urca cooling. The dominance of mass \emph{A} = 33 cooling stems primarily from two Urca pairs with strong ground-state to ground-state transitions: $^{33}$Si $\leftrightarrow$ $^{33}$Al \cite{Morton2002}, $^{33}$Al $\leftrightarrow$ $^{33}$Mg \cite{Tripathi2008}. However, there are questions in the literature about the strong ground-state to ground-state transition strength of $^{33}$Mg $\leftrightarrow$ $^{33}$Al \cite{Yordanov2010}. The measured nuclear log-ft value of 5.2 \cite{Tripathi2008} is anomalously low for the first-forbidden nature of this decay transition. As a result, addressing the nuclear physics uncertainties in this transition is critical for a quantitative prediction of Urca cooling strength. 

\bigskip

\Large \subsection{Anomaly in Nuclear Data}\label{sec:intro-anomaly}
\bigskip
\large 

The strong ground-state to ground-state transition strength in $^{33}$Al $\leftrightarrow$ $^{33}$Mg has been inferred in the $\beta$-decay direction with high-resolution $\beta$-delayed $\gamma$-ray spectroscopy. A strong ground-state branch of 37(8)\% was determined by Tripathi et al. \cite{Tripathi2008}. This translates into a comparative half-life of log-ft = 5.2. They did this by subtracting the feeding to all excited states inferred from $\gamma$-ray intensities from the various de-excitation cascades. Since the spin-parity of the ground state of $^{33}$Al is $5/2^+$, they conclude that the strong ground-state branch is only consistent with a positive parity of the ground state in $^{33}$Mg. They assigned a ground-state spin-parity of $3/2^+$ to $^{33}$Mg, so that the $^{33}$Mg $\leftrightarrow$ $^{33}$Al ground-state to ground-state transition is an allowed transition as per selection rules outlined in Table \ref{tab:selection}. However, Yordanov et al. measured a negative magnetic moment for the ground-state of $^{33}$Mg \cite{Yordanov2007} and assigned a $3/2^-$ ground-state spin-parity for $^{33}$Mg to be consistent with their observations. This is further supported by Coulomb dissociation measurements \cite{Datta2016} and spectroscopy of $^{33}$Mg with knockout reactions \cite{Bazin2021} and is typically adopted in the literature \cite{Richard2017}. This makes the $^{33}$Mg $\leftrightarrow$ $^{33}$Al ground-state to ground-state transition a first-forbidden transition. Even the Evaluated Nuclear Structure Data Files (ENSDF) states that log ft = 5.2 is small for first-forbidden decay. Yordanov et al., in their comment on Tripathi et al. \cite{Yordanov2010} provide a possible explanation stating \emph{``the large Q-value of 13.4 MeV opens a wide window for $\beta$-decay up to the 5.5 MeV neutron-separation energy. With the highest reported level
at 4.73 MeV, a gap of nearly 1 MeV is left for non-observed states. In such cases, much of the
$\beta$-intensity remains undetected, a problem known as Pandemonium.''} Furthermore, there are also questions about the $\beta$-delayed neutron branching ratio, or the $(P_n)$ value of $^{33}Mg$ decay. The eXperimental Unevaluated Nuclear Data List (XUNDL), unlike ENSDF presents a reanalysis of the $\beta$-delayed $\gamma$-rays from the neutron emission daughter $^{32}$Al measured by Tripathi et al. Along with the likely incomplete neutron spectra from an experiment at the GANIL facility in France published only in a short conference proceeding \cite{Angelique2006}, they conclude that P$_n$ = 14\% as reported by Langevin et al. \cite{Langevin1984} seems too low by almost a factor of 2. This leads to the following possibilities or a combination of these possibilities as an explanation for the anomalously high ground-state to ground-state transition strength in the decay of $^{33}$Mg:
\begin{enumerate}
    \large \item The high transition strength with a low log ft value measured by Tripathi et al. in the $\beta$-decay of $^{33}$Mg is due to nuclear structure effects from the deformed $^{33}$Mg nucleus being in the island of inversion near the N = 20 closed neutron shell.
    \large \item The level scheme of $^{33}$Mg is likely incomplete  and there are unobserved weak $\gamma$-transitions from higher energy levels that are currently unaccounted for.
    \large \item The P$_n$ value is higher than currently reported, which will reduce the inferred $\beta$-decay ground-state branching. 
\end{enumerate}
\bigskip

\LARGE \section{Dissertation Goals and Outline}\label{sec:outline}
\bigskip
\large 

This dissertation aims to improve the quality of nuclear data to help accurately quantify the strength of Urca cooling in accreting neutron star crusts. It tackles this problem on two fronts: experimental and computational. For my dissertation, I have performed nuclear physics experiments in the laboratory and also used state-of-the-art statistical analysis methods in nuclear theory to refine the quality of nuclear data that goes into nuclear reaction network calculations of accreting neutron stars. More specifically, the goals of this dissertation are two-fold.
\begin{enumerate}
    \large \item To infer the correct ground-state to ground-state $\beta$-decay transition strength in the decay of $^{33}$Mg by remeasuring the complete $\beta$-delayed $\gamma$-intensity distribution using Total Absorption Spectroscopy (TAS) that is free of the Pandemonium effect \cite{Hardy1977} (Refer Section \ref{sec:pandemoniumtas}). In addition to $\beta$-feeding intensities in $^{33}$Al, this will also require the measurement of $\beta$-delayed neutron-branching ratio (P$_n$ value). 
    \large \item To create a global nuclear mass model with quantified uncertainties that agree with current experimental mass measurements and reliably extrapolates nuclear masses to the neutron-rich regions on the nuclear chart. 
\end{enumerate}

Chapter \ref{chp:expt} presents the experimental technique and the methods used for the $\beta$-decay measurement of $^{33}$Mg. The experimental data analysis procedure is outlined in Chapter \ref{chp:analysis} and the results from this experiment are presented in Chapter \ref{chp:results}. Chapter \ref{chp:massmodel} then switches gears and presents the statistical framework for building a quantified nuclear mass model using Bayesian Machine Learning techniques. The astrophysical impact of experimental measurement as well as the refined nuclear mass model is discussed in Chapter \ref{chp:impact}. Finally, the dissertation ends with conclusions and future outlook in Chapter \ref{chp:conclusions}.

\huge \chapter{Experimental Measurement}\label{chp:expt}
\large 

This chapter focuses on the experimental measurement technique for the ground-state to ground-state $\beta$-decay transition strength from $^{33}$Mg -- $^{33}$Al. This quantity is being remeasured to resolve the conflict in the literature arising from an anomalously high ground-state branch \cite{Yordanov2010,Tripathi2010}. This measurement is performed using Total Absorption Spectroscopy (TAS) to avoid any systematic errors, especially those arising from the Pandemonium effect \cite{Hardy1977}. Subsequent sections discuss the experimental methodology and $\beta$-delayed spectroscopy, production and delivery of $^{33}$Mg beam, details about the BCS, NERO, and SuN detectors used during the experiment, and aspects of the data acquisition system to record and store the data from the experiment. 

\bigskip
\LARGE \section{Experimental Technique}
\bigskip
\large 

\begin{figure}
    \centering
    \includegraphics[width=450pt,keepaspectratio]{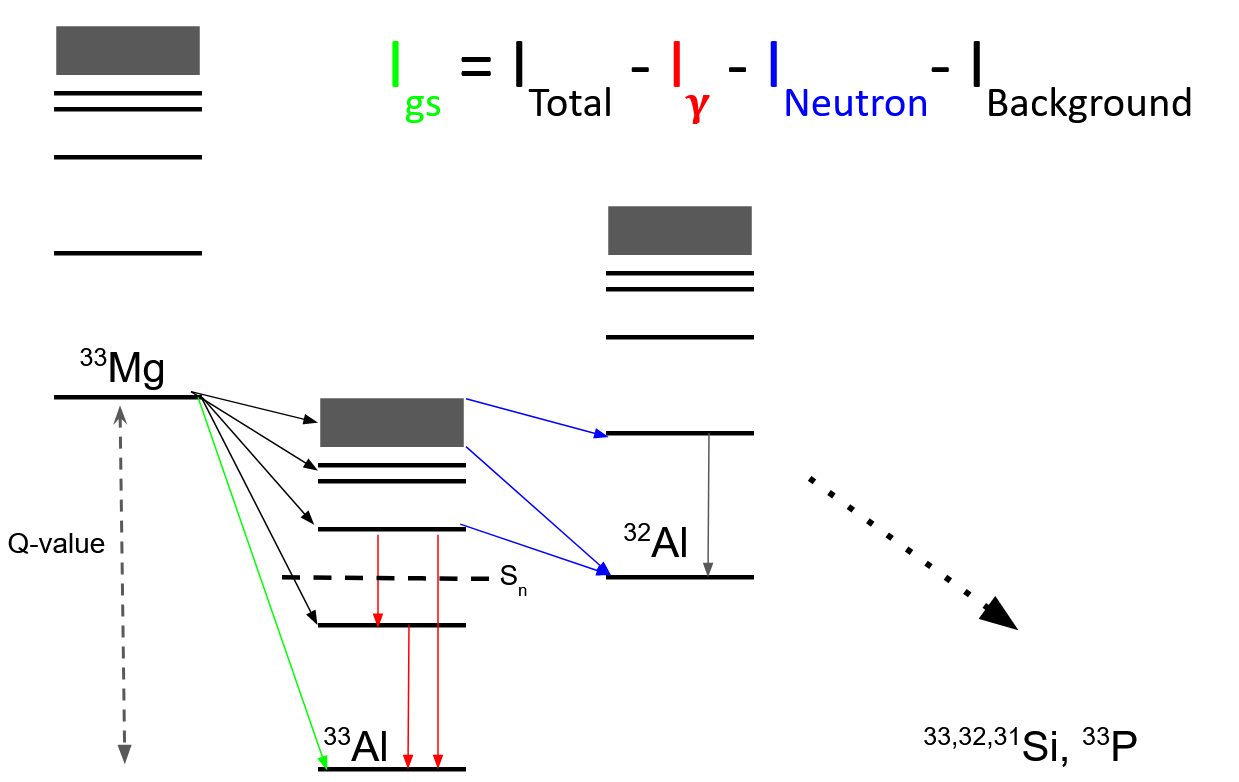}
    \caption{\large Schematic of the experimental technique to measure the ground-state to ground-state $\beta$-decay transition strength of $^{33}$Mg. Feeding intensities to excited states in $^{33}$Al are measured by detecting $\beta$-delayed $\gamma$ rays with SuN detector whereas $\beta$-delayed neutron branching ratio to $^{32}$Al is measured by counting neutrons with NERO detector. Subtracting both from the total transition strength yields the ground-state branch. }
    \label{fig:expt}
\end{figure}

$\beta$-decay is a three body process involving the nucleus undergoing $\beta$-decay, the emitted electron, and the emitted neutrino. Energy from the Q-value of the decay is predominantly shared by the electron and the neutrino. A very small fraction also goes to the recoiling nucleus to conserve momentum. Due to the three-body nature of $\beta$-decay, the energy of the electron from $\beta$-decay forms a continuous spectrum. Transitions to each excited state in the daughter nucleus would contribute to the electron energy spectrum and deconvoluting it to get the contribution from individual transitions would be very difficult. It is, therefore, not very useful to measure the electron energy from the $\beta$-decay. Instead, the contribution to excited states is estimated by detecting the $\beta$-delayed $\gamma$ rays, as the $\gamma$ rays have precise energies depending on the level scheme of the daughter nucleus. $\beta$-decay transitions to the ground state in the daughter nucleus will not have any $\beta$-delayed $\gamma$ radiation. The ground-state branch can then be inferred by subtracting the contributions of all the excited states from the total transition strength. In addition, since we are dealing with very neutron-rich nuclei, the neutron separation energies are lower than the $\beta$-decay Q-value. If the $\beta$-decay transition happens to an excited level above the neutron separation energy in the daughter nucleus, it can deexcite either via $\gamma$-decay or neutron emission. If there is a $\beta$-delayed neutron emission, the $\beta$-delayed neutron branch also needs to be measured and subtracted. In summary, the experimental technique to measure the ground-state to ground-state $\beta$-decay transition strength is to subtract the contributions from excited states by counting $\beta$-delayed $\gamma$ rays and $\beta$-delayed neutrons. Figure \ref{fig:expt} shows the schematic for this experimental technique. 

\bigskip
\Large \subsection{$\beta$-decay Kinetics}\label{sec:bateman}
\bigskip
\large 

Nuclear decay is kinetically a first-order transition, where the number of decays at time $t$ is directly proportional to the total number of nuclei in a sample at time $t$. This can be represented as 
\begin{equation}\label{eqn:decay}
    \large \frac{dN(t)}{dt} = -\lambda N(t)
\end{equation}
The proportionality constant, $\lambda$ is the rate of $\beta$-decay. Solving the differential Equation \ref{eqn:decay} yields the solution for the number of parent nuclei in the sample at time $t$ as 
\begin{equation}\label{eqn:nuclear}
    \large N(t) = N_0e^{-\lambda t}
\end{equation}
where $N_0$ is the number of parent nuclei in the sample at time $t = 0$. The rate $\lambda$ is related to the half-life $T_{1/2}$ as
\begin{equation}\label{eqn:halflife}
    \large \lambda = \frac{ln(2)}{T_{1/2}}
\end{equation}
The number of daughter nuclei in a sample at time $t$ is equal to the number of decays of parent nuclei and is given by
\begin{equation}\label{eqn:daughter1}
    \large N'(t) = N_0(1 - e^{-\lambda t})
\end{equation}
Equation \ref{eqn:daughter1} is only valid when the daughter nucleus is stable against further nuclear decay. However, for experimental measurements on exotic nuclei in the laboratory, the daughter nuclei are often unstable against $\beta$-decay, and Equation \ref{eqn:daughter1} has to be appropriately modified. For example, in the $\beta$-decay of $^{33}$Mg, the daughter $^{33}$Al has an even shorter half-life than the parent and starts undergoing $\beta$-decay as soon as it is produced. Considering that the daughter nucleus decays with rate $\lambda'$, the differential equation can be rewritten as 
\begin{equation}
    \large \frac{dN'(t)}{dt} = \lambda N - \lambda'N'
\end{equation}
and the solution for the number of daughter nuclei in a sample is given by 
\begin{equation}
    \large N'(t) = N_0'e^{-\lambda't} - N_0\frac{\lambda}{\lambda'-\lambda} \big [ e^{-\lambda't} - e^{-\lambda t}\big]
\end{equation}
Note that $N'$ depends on both $\lambda$ and $\lambda'$. 
For a nuclear decay chain involving multiple transitions, the amount of any $k^{th}$ nucleus in the decay chain at time $t$ is described by Bateman equations \cite{Bateman1910}, and the analytical solutions \cite{Cetnar2006} are given by 
\begin{equation}\label{eqn:bateman}
    \large N_k(t) = \sum_{j=1}^{k}A_je^{-\lambda_jt} \prod_{l=j}^{l=k-2}\frac{-\lambda_{l+1}}{\lambda_{l+1} - \lambda_j}
\end{equation}
where $A_j$ are constants relating to the initial quantity of the $j_{th}$ species, and $\lambda_j$ are the corresponding decay constants.

For an experimental decay curve (histogram of decay times for a nucleus) based on the time of electron detection that does not have 100\% efficiency, even the first registered electron could have been produced from the $\beta$-decay of any nucleus in the decay chain. As a result, to extract the half-life of the parent nucleus from the experimental decay curve, it has to be fit to the sum of contributions from all decaying species in the decay chain. The half-lives of all daughter nuclei have to be either known or estimated from a multi-parameter fit to the decay curve. 

\bigskip
\Large \subsection{$\beta$-delayed Spectroscopy}

\bigskip
\Large \subsubsection{$\beta$-delayed $\gamma$ decay}
\bigskip
\large

An excited nucleus predominantly deexcites by emitting $\gamma$ rays, unless other competing channels are present (for example, light particle emission). The lifetimes of these excited states are usually so short that the decay can be considered almost instantaneous for all practical purposes (except when an isomer is present). The energy of the emitted $\gamma$-ray photon is equal to the difference between the energies of the initial and the final level in a nucleus. Since the photon also carries some angular momentum depending on its multipolarity (L), the total angular momentum should also be conserved. Let $(I_i,\pi_i)$ be the total angular momentum and parity of the initial excited state and $(I_f,\pi_f)$ be the total angular momentum and parity of the final deexcited state. For a multipole radiation carrying angular momentum L, 
\begin{equation}
    \large \overrightarrow{I_i} = \overrightarrow{L} + \overrightarrow{I_f}
\end{equation}
L = 0 is not allowed since there is no monopole radiation. L = 1 corresponds to dipole radiation, L = 2 is the quadrupole radiation, and so on. Whether the emitted radiation is of the electric or magnetic type is determined by the change in parity $(\Delta \pi)$ between the initial and final levels. ($\Delta \pi$ = no) corresponds to even parity for the radiation field. This is valid for electric transition where L = even and magnetic transition where L = odd. On the other hand, ($\Delta \pi$ = yes) corresponds to odd parity for the radiation field which is given by electric transition where L = odd and magnetic transitions where L = even. These selection rules can be summarized as follows:
\begin{equation*}
    \large |I_i - I_f| \leq L \leq |I_i + I_f| \hspace{33pt} (L \neq 0)
\end{equation*}
\begin{equation*}
    \large \Delta \pi = \text{no: \hspace{18pt} even electric, odd magnetic}
\end{equation*}
\begin{equation*}
    \large \Delta \pi = \text{yes: \hspace{18pt} odd electric, even magnetic}
\end{equation*}
Consider, for example, a nucleus deexciting from a 5/2$^{+}$ state to a 3/2$^{-}$ state. The possible values of L would range from |5/2 - 3/2| = 1 to |5/2 + 3/2| = 4. And since the parity is changing, there will be electric transitions with odd L and magnetic transitions with even L. As a result, a $\gamma$-deexcitation from 5/2$^{+} \rightarrow$ 3/2$^{-}$ will happen via E1, M2, E3, and M4 radiation. According to Weisskopf estimates for the transition rates for each electromagnetic radiation type, the lower order multipolarities are preferred and the transition rate drops down by about five orders of magnitude for one unit increase in polarity. For the same order of multipolarity, electric radiation is two orders of magnitude more likely than magnetic radiation. An excited state often has multiple choices for deexcitation, and the fastest transitions are always preferred in nature. As a result, most practical purposes consider only E1, M1, and E2 transitions. 

\bigskip
\Large \subsubsection{$\beta$-delayed neutron emission}
\bigskip
\large

For exotic neutron-rich nuclei, the neutron separation energies can be lower than the $\beta$-decay Q-value. In that case, $\beta$-decay can populate excited levels in the daughter nucleus above the neutron separation threshold. This opens an additional channel for the excited daughter nucleus to deexcite via $\beta$-delayed neutron emission. This neutron emission can happen either into the ground state or to excited states in $\beta$-delayed neutron daughter, if energetically possible. The energy of the emitted neutron, $E_n$, is given by:
\begin{equation}
    \large E_n = E_{x,i} - E_{x,f} - S_n
\end{equation}
where $E_{x,i}$ is the excitation energy in the $\beta$-decay daughter nucleus, $E_{x,f}$ is the excitation energy of the $\beta$-delayed neutron daughter ($E_{x,f}$ = 0 for neutron emission into the ground state), and S$_n$ is the neutron separation energy of the $\beta$-decay daughter. The sum of neutron emission probabilities from all excited states is referred to as the total $\beta$-delayed neutron emission branching ratio, or the P$_n$ value.

\bigskip
\Large \subsection{Pandemonium Effect and Total Absorption Spectroscopy}\label{sec:pandemoniumtas}
\bigskip
\large 

\begin{figure}
    \centering
    \includegraphics[width=400pt,keepaspectratio]{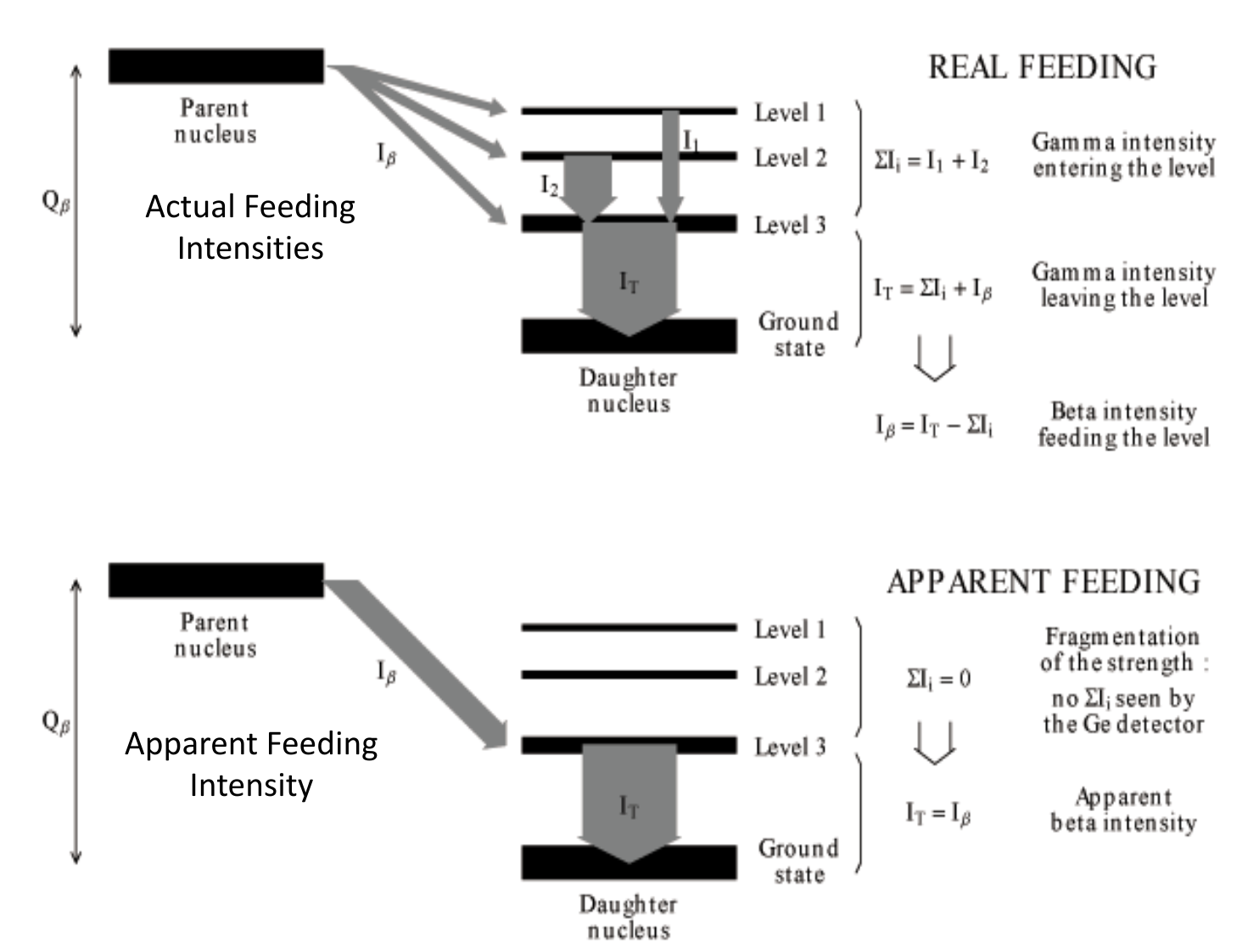}
    \caption{\large Schematic explanation of the Pandemonium effect \cite{Hardy1977}. Weak $\beta$-delayed $\gamma$-transitions from higher excitation energies are hard to detect with high-resolution $\gamma$-ray detectors. The inferred $\beta$-feeding intensities are then wrongly attributed to lower excitation energies. This is a known source of systematic error in data analysis. Figure reproduced from \cite{Aguado2012}.}
    \label{fig:pan}
\end{figure}

\begin{figure}
    \centering
    \includegraphics[width=450pt,keepaspectratio]{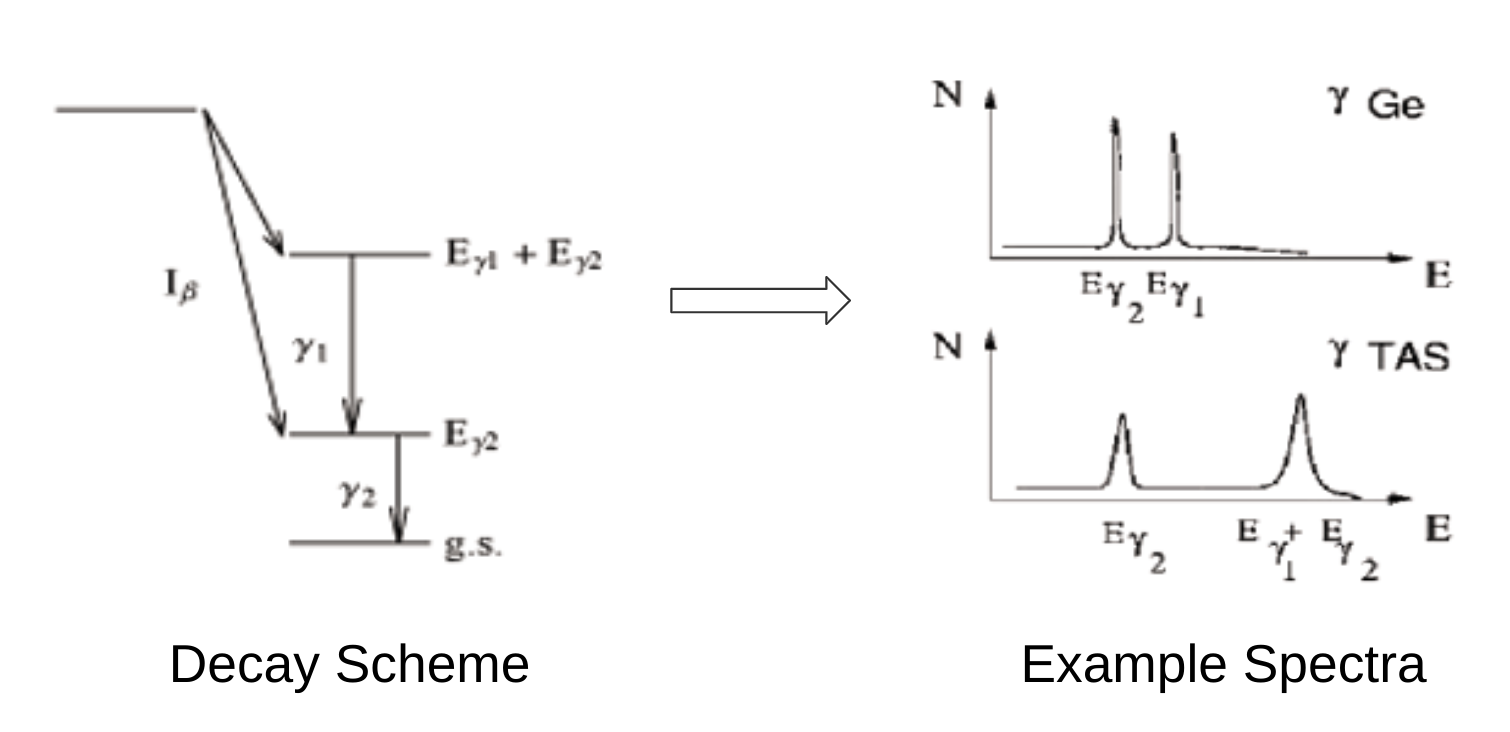}
    \caption{\large Comparison of $\gamma$-ray spectra with high-resolution Ge detectors and high-efficiency TAS detectors for a hypothetical decay to two feeding states in daughter nucleus. While Ge detectors show peaks at individual transitions, TAS detectors sum the energies in each event, such that the peaks correspond to excitation energies. The high efficiency of TAS detectors then helps eliminate the Pandemonium effect shown in Figure \ref{fig:pan}. Figure reproduced from \cite{Aguado2012}.}
    \label{fig:tas}
\end{figure}

The $\beta$ feeding intensity to an excited state in the daughter nucleus is measured by detecting the $\gamma$ ray intensity from that level. However, deexcitation of higher energy levels can also occur via that intermediate state. The total $\gamma$ intensity ($I_T$) from that level will then be the sum of $\beta$ feeding intensity ($I_\beta$)and the feeding intensities from higher energy levels ($\Sigma I_i$). 
\begin{equation}
    \large I_T = I_\beta + \Sigma I_i \hspace{11pt} \Rightarrow \hspace{11pt} I_\beta = I_T - \Sigma I_i
\end{equation}
To apply this formalism in practice, a complete level scheme of the daughter nucleus should be known to accurately assign observed $\gamma$ rays to corresponding transitions between excited states. To this end, a high-resolution germanium detector is used to isolate individual $\gamma$-rays. However, this high resolution comes at the cost of low efficiency where weaker transitions go undetected. This is true, especially for $\beta$-decays with high Q-value and daughter nuclei having high level densities. Each additional $\gamma$ ray in a high multiplicity cascade diminishes the likelihood of detecting all the transitions. Due to this, $\Sigma I_i$ is underestimated and $I_\beta$ is overestimated. This causes systematic errors in the measurement of $\beta$ feeding intensities, a problem known as the Pandemonium effect. Figure \ref{fig:pan} gives a schematic explanation of this problem. This was first observed in the Monte Carlo simulation of $\beta$-decay in a fictional nucleus called `Pandemonium' \cite{Hardy1977}. 

This problem is addressed in this work by a method called Total Absorption Spectroscopy (TAS). In the trade-off between resolution and efficiency of $\gamma$-ray detection, TAS employs very high efficiency ($\sim$100\%) scintillator detectors with close to 4$\pi$ solid angle coverage that are on the opposite end of the trade-off spectrum from Ge detectors. High efficiency allows for the detection of even the weakest branch and the energies of all transitions in a single cascade are summed on an event-by-event basis. Unless there is an isomer present in the daughter nucleus, a peak in the TAS spectrum will directly correspond to an excited level in the daughter nucleus. Figure \ref{fig:tas} shows the difference between a $\gamma$ spectrum from Ge detector v/s TAS spectrum for a hypothetical decay scheme. In the TAS method, instead of worrying about systematic errors in $\Sigma I_i$, the $\beta$ feeding intensity is simply $I_\beta$ = $I_T$. This is why TAS measurements are free of the Pandemonium effect. 

\bigskip
\LARGE \section{Beam Delivery}\label{sec:beamdelivery}
\bigskip
\large 

\begin{figure}
    \centering
    \includegraphics[width=450 pt,keepaspectratio]{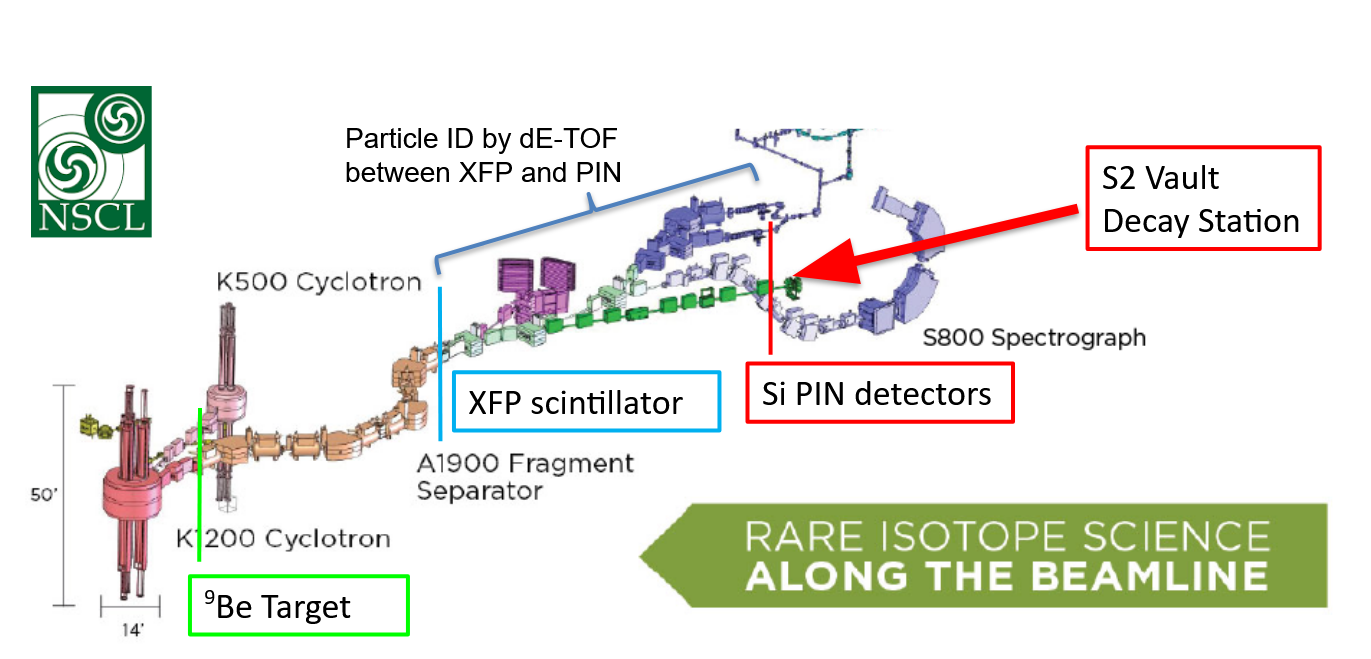}
    \caption{\large Beamline at National Superconducting Cyclotron Laboratory (NSCL) for producing and delivering $^{33}$Mg to the experimental end-station in S2 vault. Shown here are the coupled cyclotrons - K500 and K1200, $^{9}$Be fragmentation target, A1900 fragment separator, and detectors for particle identification. Figure adapted from the NSCL website.}
    \label{fig:nscl}
\end{figure}

\begin{figure}
    \centering
    \includegraphics[width=350pt,keepaspectratio]{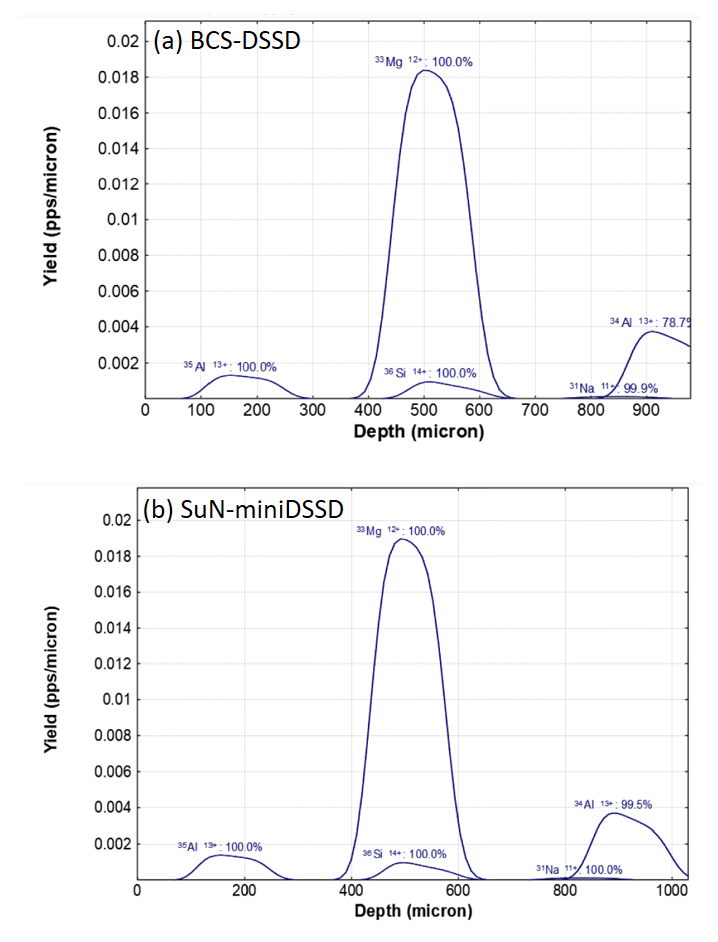}
    \caption{\large LISE++ \cite{Tarasov_2008} simulations for the optimized range distribution of $^{33}$Mg$^{12+}$ in the implantation silicon detectors for both parts of the experiment. All the fragments of interest are completely deposited in the center with minimal contaminants. Silicon detectors also have enough thickness to detect the $\beta$-decay electron.}
    \label{fig:lise}
\end{figure}

A beam of $^{33}$Mg fragments was delivered by the National Superconducting Cyclotron Laboratory (NSCL) facility. NSCL (now decommissioned) used the in-flight fragmentation method to produce exotic beams of radioactive nuclei. In this method, an accelerated heavy-ion primary beam undergoes fragmentation in its flight path and produces a cocktail beam with many different particles. The fragment of interest is separated using a fragment separator that is typically comprised of electromagnetic optical elements. By carefully tuning the separator and making appropriate cuts, the beam rate and purity of the fragments delivered to user experiments can be optimized. Figure \ref{fig:nscl} shows the different components of the NSCL beamlines.  

For this experiment, a completely stripped primary beam of $^{48}$Ca$^{20+}$ was accelerated to 140 MeV/u with the help of the K500 and K1200 coupled cyclotrons. This beam was then impinged on a $^{9}$Be fragmentation target with a thickness of 846 mg/cm$^2$. Fragmentation of $^{48}$Ca produced $^{33}$Mg along with many other isotopes. The $^{33}$Mg was separated using the A1900 fragment separator \cite{Morrissey2003,Stolz2005}. The A1900 fragment separator consists of four magnetic dipoles that deflect different beam components based on their momentum-to-charge ratio. It also has 24 magnetic quadrupoles for focusing the beam throughout its flight path. Additionally, an aluminum wedge-shaped degrader with a $0^{\circ}$ thickness of 450 mg/cm$^2$ was placed at I2 image position to further separate the beam components by energy loss and subsequent energy dispersion. An eXtended Focal Plane (XFP) scintillator was used for the time of flight measurement for particle identification as discussed in Section \ref{sec:pid}. Finally, another aluminum degrader with variable thickness (based on the angle) was placed just before the experimental end-station to ensure that the fragments stopped in the implantation detectors. All the parameters for beam delivery were optimized for the rate and purity of $^{33}$Mg fragment based on LISE++ \cite{Tarasov_2008} simulations. Figure \ref{fig:lise} shows the range distribution for all fragments implanted in the silicon implantation detectors for both parts of the experiment (refer to Section \ref{sec:detectors} for an overview). $^{33}$Mg$^{12+}$ is deposited at the center of the silicon implantation detectors with minimal contaminants. 

\bigskip
\Large \subsection{Particle Identification}\label{sec:pid}
\bigskip
\large 

Even though the contaminants in the beam delivered to the experimental end station are minimal, they can still contribute to the background and make the data analysis difficult. Therefore, particle identification (PID) is done on an event-by-event basis, and only the events that pass the PID cuts are considered for data analysis. There are a few ways to perform particle identification. In this experiment, the $\delta$E-TOF method was used. For a particle with atomic number $Z$, atomic mass number $A$, and charge $Q$, the energy loss in a thin detector $\delta E$ is proportional to $Z^2$, and the time of flight TOF between any two points which depends on the velocity of the particle is proportional to $A/Q$. Plotting both quantities on a 2-dimensional plot leads to a PID plot with each blob corresponding to a particular nuclide as seen in Figure \ref{fig:pid}. The energy loss was measured in a thin silicon PIN detector just upstream of the SuN detector. The time of flight was measured between the XFP scintillator and the silicon PIN detector. Once the $^{33}$Mg blob on the PID plot is identified, only events that pass the cuts for that blob are used for analysis. 

\begin{figure}
    \centering
    \includegraphics[width=500pt,keepaspectratio]{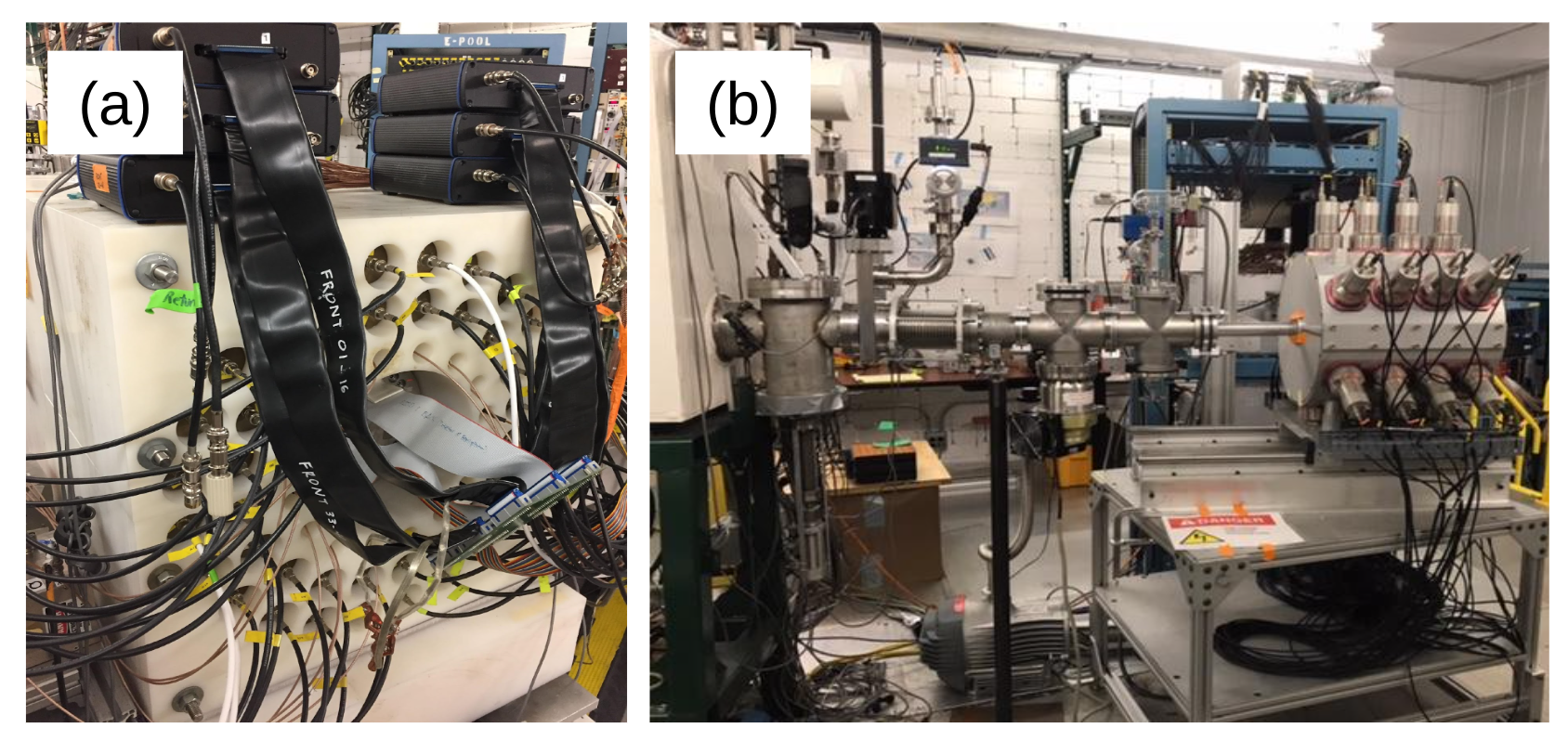}
    \caption{\large Picture of the (a) NERO and (b) SuN detector set-up at the experimental decay station in S2 vault.}
    \label{fig:setup}
\end{figure}

\bigskip
\LARGE \section{Detector Setup}\label{sec:detectors}
\bigskip
\large

The experiment utilized an ensemble of detectors to measure the charged particles ($^{33}$Mg implants, electrons), neutrons, and $\gamma$ rays. Additional detectors were also used for particle identification, to veto light particle background, to veto punchthrough events, etc. The experiment was performed in two parts: 
\begin{enumerate}
    \large \item BCS-NERO part focused on measuring the $\beta$-delayed neutron branching ratio, i.e. the P$_n$ value. It utilized the Beta Counting System (BCS) for detecting the $^{33}$Mg implant and decay electron and the NERO detector for $\beta$-delayed neutrons. 
    \large \item SuN-miniDSSD part focused on measuring the $\beta$-feeding intensities to excited states in $^{33}$Al. It utilized the miniDSSD for $^{33}$Mg implant and decay electron and the SuN detector for $\beta$-delayed $\gamma$ rays.  
\end{enumerate}

\bigskip
\Large \subsection{BCS-NERO}\label{sec:bcsnero}
\bigskip
\large 

\begin{figure}
    \centering
    \includegraphics[width=470pt,keepaspectratio]{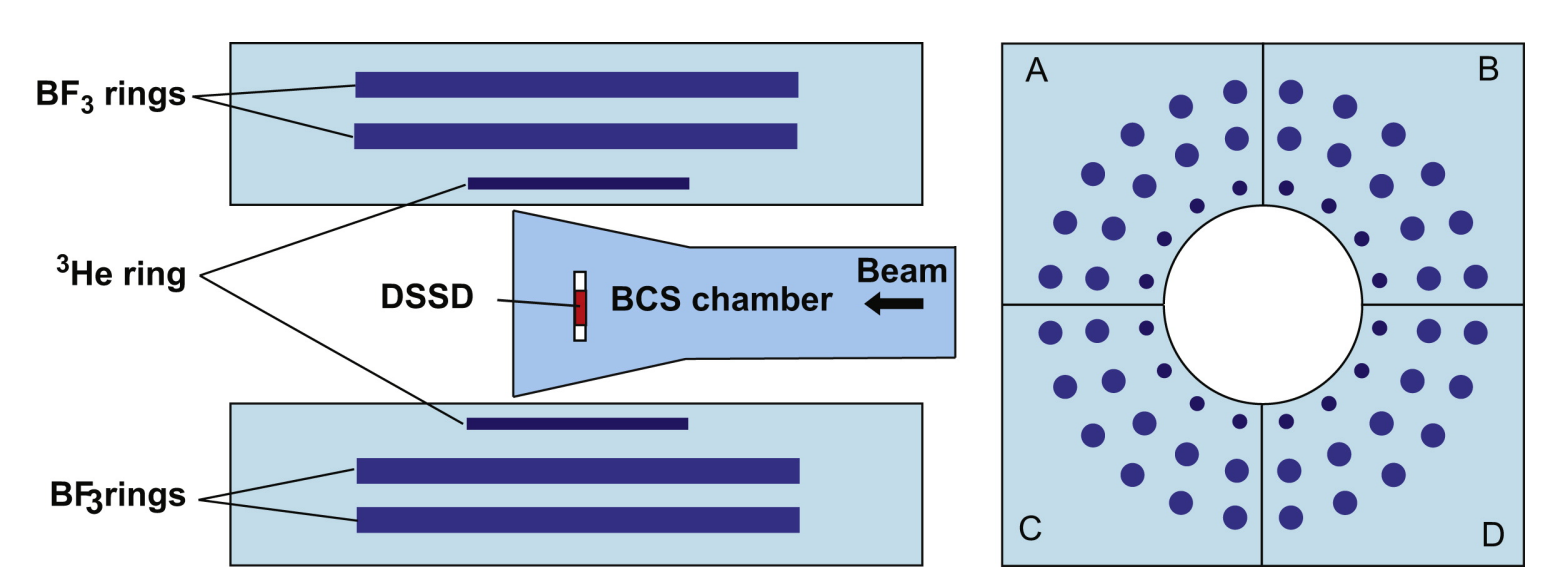}
    \caption{\large Side view and front view of the NERO detector with BCS chamber inside. $^{33}$Mg fragments are implanted onto the DSSD and $\beta$-delayed neutrons are detected by concentric $^{3}$He and BF$_{3}$ rings embedded in a high-density polyethylene (HDPE) matrix. This design is optimized to have a constant neutron detection efficiency over a wide range of neutron energies. Figure reproduced from \cite{Pereira2010}. Refer to Figure \ref{fig:setup}(a) for the actual set-up.}
    \label{fig:nero}
\end{figure}

\begin{table}[]
    \centering
    \begin{tabular}{cccc}
    \toprule
       \large Detector  & \large Thickness ($\mu$m) & \large Bias Voltage (V) & \large Gain \\
    \midrule
      \large PIN1  & \large 525 & \large 40 & \large Low \\
      \large PIN2   & \large 1041 & \large 75 & \large Low \\
      \large PIN3   & \large 488 & \large 50 & \large Low \\
      \large DSSD   & \large 979 & \large 280 & \large Low and High\\
      \large SSSD   & \large 982 & \large -210 & \large High \\
      \large Plastic Scintillator & \large $\sim$25400 & \large -850 & \large High \\
    \bottomrule
    \end{tabular}
    \caption{\large Properties of detectors for particle identification and detecting implantation and decay events in the BCS-NERO part of the experiment. }
    \label{tab:bcs}
\end{table}

BCS or the Beta Counting System is a stack of silicon detectors to detect charged particles. The main detector is a Double-sided Stripped Silicon Detector (DSSD). It is a 979 $\mu$m thick Micron Semiconductor BB1 detector with an active area of 39.90 $\times$ 39.90 mm$^2$. The front side of the DSSD is segmented into 40 vertical strips and the back side is segmented into 40 horizontal strips which together create 1600 pixels of 1 mm$^2$ to localize the interaction vertex of charged particles detected. Each channel is read out by a Multi-Channel Systems CPA-16 dual-gain preamplifier for a total of 160 channel readouts. The 80 low-gain channels are used to detect the heavy ion implants, whereas the 80 high-gain channels are used to detect $\beta$-decay electrons. A coincidence between a signal from a front strip and a back strip of the DSSD was required for a valid signal to be read out. This was achieved using DDAS breakout modules. A combination of OR signals from the 40 front strips AND a combination of OR signals from the 40 back strips was used as an external trigger for the DSSD. 

In addition to the DSSD, a Single Sided Stripped Silicon Detector (SSSD) was placed just behind the DSSD to act as a veto for punchthrough events. In such events, the implant punches through without stopping in the DSSD. These have to be discarded as the decay electron will not be registered in the DSSD and successful implant-decay correlation cannot be done. Silicon PIN detectors are placed upstream of the DSSD for particle identification as discussed in Section \ref{sec:pid}. PIN1 and PIN2 were placed high upstream in the experimental setup. A third detector PIN3 was also placed in the BCS stack to check the transmission from the PINs to the BCS stack. Additionally, the beam may contain light particles that, when passing through the DSSD, will register a signal similar to the decay electrons in the high-gain channels. To detect and discard such events, a plastic scintillator is placed at the end of the stack that is used to distinguish the light particles and electrons. The BCS stack is embedded in the NERO matrix as shown in Figure \ref{fig:nero}. Table \ref{tab:bcs} lists the properties of detectors in the BCS stack along with their bias voltages. 

\begin{table}[]
    \centering
    \begin{tabular}{ccc}
    \toprule
       \large Detector  & \large Type & \large Bias Voltage (V)  \\
    \midrule
      \large Quad A 1--4 & \large $^3$He & \large 1130  \\
      \large Quad A 5--15 & \large BF$_3$ & \large 2620  \\
      \large Quad B 1--4 & \large $^3$He & \large 1130  \\
      \large Quad B 5--15 & \large BF$_3$ & \large 2620 \\
      \large Quad C 1--4 & \large $^3$He & \large 1080 \\
      \large Quad C 5--15 & \large BF$_3$ & \large 2620 \\
      \large Quad D 1--4 & \large $^3$He & \large 1350 \\
      \large Quad D 5--15 & \large BF$_3$ & \large 2620  \\
    \bottomrule
    \end{tabular}
    \caption{\large Properties and types of neutron long counter detectors embedded in the NERO matrix to detect $\beta$-delayed neutrons and estimate the neutron branching ratio.}
    \label{tab:nero}
\end{table}

The Neutron Emission Ratio Observer (NERO) is a neutron long counter system to detect the $\beta$-delayed neutrons \cite{Pereira2010}. It consists of 44 BF$_3$-filled tubes and 16 $^3$He-filled tubes inserted into a High-Density Polyethylene (HDPE) matrix. The total dimensions of the matrix are $60 \times 60 \times 80$ cm$^3$ and the density of the matrix is $\rho = 0.93(1)$ g/cm$^3$. It is divided into 4 quadrants, each housing 11 BF$_3$ tubes and 4 $^3$He tubes as listed in Table \ref{tab:nero}. It also lists the operating bias voltages for these tubes. The matrix has a cylindrical bore of radius 11.4 cm to accommodate the BCS chamber. All the tubes are arranged in three concentric circles with $^3$He tubes occupying the innermost circle as shown in Figure \ref{fig:nero}. The choice of this design is to optimize the efficiency of neutron detection and to have a constant efficiency over a wide range of neutron energies. $^{3}$He tubes have a relatively higher neutron detection efficiency and the fastest neutrons will interact first with the tubes in the innermost rings. The neutrons are moderated and slowed down while interacting with the tubes in the outer rings. 

$\beta$-delayed neutrons are moderated to thermal energies in the HPDE matrix with moderation times of $\sim 200 \mu$s. HDPE is an ideal choice due to the high density of protons since protons require the least amount of scattering events to thermalize the neutrons. The neutrons induce $^3$He(n,p)$^3$H reaction in the $^3$He tubes and $^{10}$B(n,$\alpha$)$^7$Li reaction in the BF$_3$ tubes. The charged particles from these reactions ionize the gases in the tubes and the electron cascades are detected due to the applied voltages across the tubes, indicating a neutron detection. The energy information of the neutron is completely lost in this method of detection. The detector has been designed for constant efficiencies up to $\sim 2$ MeV, so as long as neutron energies are below that, the neutron energy need not be known. NERO achieves a neutron detection efficiency of $\sim$ 35\%.

\bigskip
\Large \subsection{SuN-miniDSSD}\label{sec:sunminidssd}
\bigskip
\large 

\begin{figure}
    \centering
    \includegraphics[width=480pt,keepaspectratio]{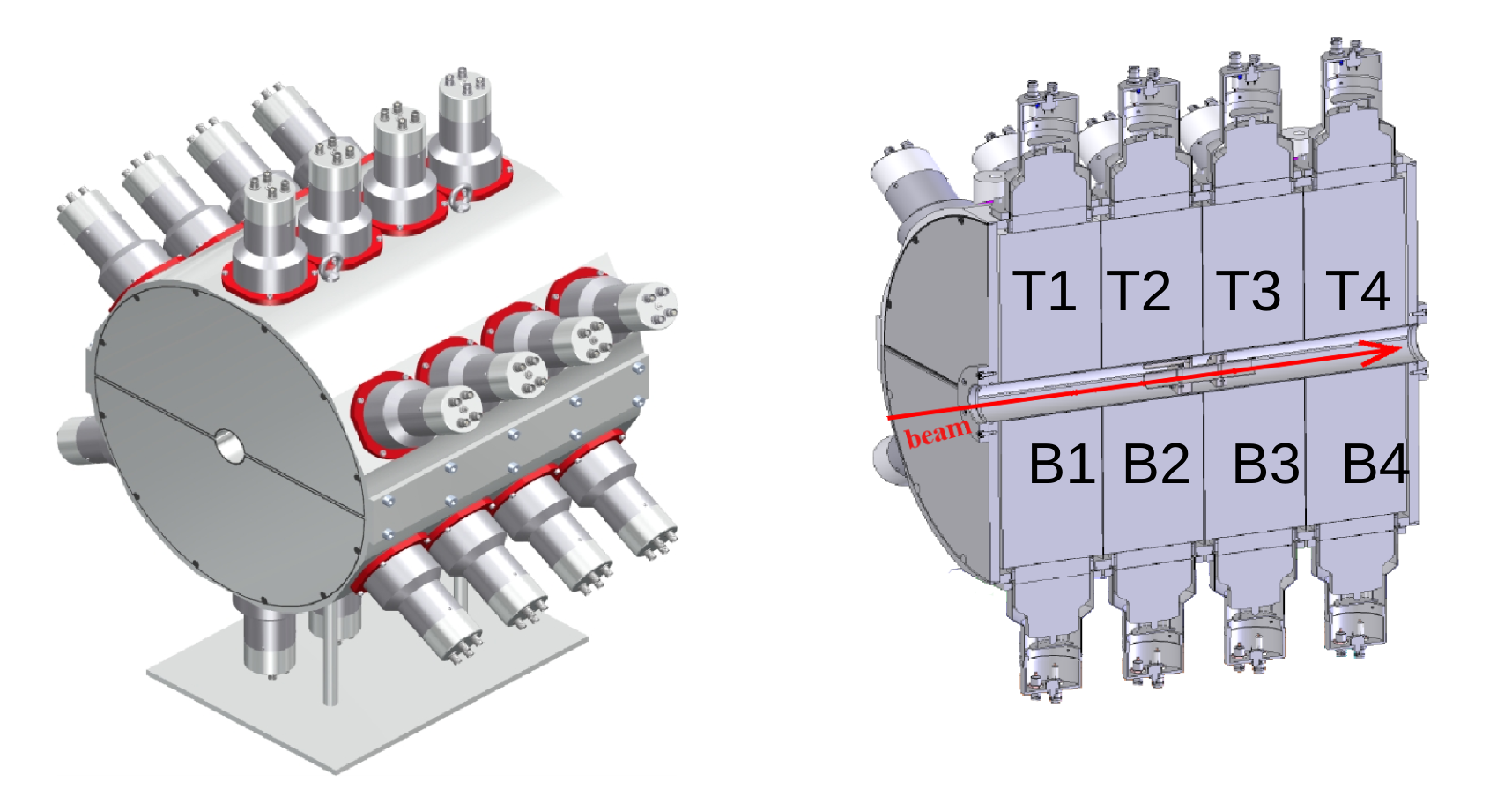}
    \caption{\large Schematic of the SuN detector (left) along with its cross-sectional view (right). SuN comprises of NaI(Tl) scintillator crystals and has a total of 8 segments (T1-T4 on the top, B1-B4 on the bottom) with three photomultiplier tubes (PMTs) for each segment. Figure adapted from \cite{Simon2013}. Refer to Figure \ref{fig:setup}(b) for the actual set-up.}
    \label{fig:sun}
\end{figure}

The second part of the experiment was to detect the $\beta$-feeding intensities to excited states in $^{33}$Al by detecting the $\beta$-delayed $\gamma$ rays in the SuN detector. A miniDSSD was used to detect the implants and decay electrons instead of the BCS. This was because the cylindrical bore in SuN to allow the beam to pass through was much smaller than in NERO. The miniDSSD is a Micron Semiconductor BB8 detector that is 1030 $\mu$m thick and has an active area of 20.0 $\times$ 20.0 mm$^2$. Both sides of the miniDSSD are orthogonally segmented into 16 strips of 1.25 mm pitch, forming a total of 256 pixels. It is also read out by a Multi-Channel Systems CPA-16 dual-gain preamplifier. The low-gain channels detect $^{33}$Mg implants and the high-gain channels detect electrons. Coincidence between a signal from a front strip and a back strip was also required to read out a valid signal from miniDSSD. This was again achieved with an external trigger using DDAS breakout modules, similar to the BCS-DSSD. Punchthrough events to be vetoed were identified with a Silicon surface barrier detector placed downstream of the miniDSSD. It was also used to veto events from light particles in the beam that mimic decay events in the miniDSSD. Particle identification detectors are the same as used in the BCS-NERO part of the experiment, except no PIN3 is required. Table \ref{tab:minidssd} lists the properties of different detectors along with their bias voltages for this part of the experiment.  

\begin{table}[]
    \centering
    \begin{tabular}{cccc}
    \toprule
       \large Detector  & \large Thickness ($\mu$m) & \large Bias Voltage (V) & \large Gain \\
    \midrule
      \large PIN1  & \large 525 & \large 40 & \large Low \\
      \large PIN2   & \large 1041 & \large 75 & \large Low \\
      \large miniDSSD   & \large 1030 & \large 110 & \large Low and High\\
      \large Si Surface barrier   & \large 500 & \large 60 & \large High \\
    \bottomrule
    \end{tabular}
    \caption{\large Properties of detectors for particle identification and detecting implantation and decay events in the SuN-miniDSSD part of the experiment. }
    \label{tab:minidssd}
\end{table}

Summing NaI or the SuN detector is a Total Absorption Spectrometer composed of NaI(Tl) scintillator crystals. These have high intrinsic efficiency in detecting $\gamma$ rays. Moreover, owing to an almost 4$\pi$ solid angle coverage of SuN, the geometric efficiency is also high. Combined, this leads to a very high $\gamma$ detection efficiency which is a primary requirement for Total Absorption Spectroscopy. NaI is a commonly used scintillator material due to the ease of producing large monocrystalline structures, increasing the interaction volume and hence, the efficiency. It is doped with thallium (Tl) to introduce states between the conduction and valence bands in pure NaI. These intermediate states produce photons whose wavelengths overlap well with the response of photomultiplier tubes. 

SuN consists of a total of 8 half-annular segments, 4 at the top (T1-T4) and 4 at the bottom (B1-B4) as shown in Figure \ref{fig:sun}. It forms a cylindrical barrel 16 inch in length and diameter. Lead bricks are placed on the upstream side to provide shielding from beam particles hitting multiple components in the beamline. Each segment is read out by three photomultiplier tubes (PMTs) to make sure all the photons produced in the segment are collected. All three PMTs must be fired within a coincidence window of 2000 ns for a signal in the SuN segment to be recorded. Multiple SuN segments can also be fired in a single physics event either due to scattering of the $\gamma$ ray into another segment or because of different $\gamma$ rays in a cascade. In such events, the energies from all the segments are summed and a single count corresponding to the sum is added to the total absorption (TAS) spectrum. Additionally, multiple counts corresponding to energies deposited in a single segment are added to the singles spectrum (also known as the `sum of segments' or the SS spectra). The multiplicity of an event registers the number of SuN segments that fired in an event and ranges from 1 -- 8. Together, all these spectra provide strict constraints for the $\gamma$-ray cascade under study. 

\bigskip
\LARGE \section{Data Acquisition}
\bigskip
\large 

Raw data from the detectors in this experiment are time series data of voltage pulses that encode information about the energy deposited. For digital data acquisition systems (DDAS), the pulse is discretized. This pulse is processed with some sort of a digital filter that extracts energy information before a hit in a particular channel is recorded (along with a timestamp and energy deposited). A combination of hits from different channels based on pre-specified constraints is packaged into an event for further analysis. 

\bigskip
\Large \subsection{Digital Pulse Processing}\label{sec:dsp}
\bigskip
\large 

\begin{figure}
    \centering
    \includegraphics[width=450pt,keepaspectratio]{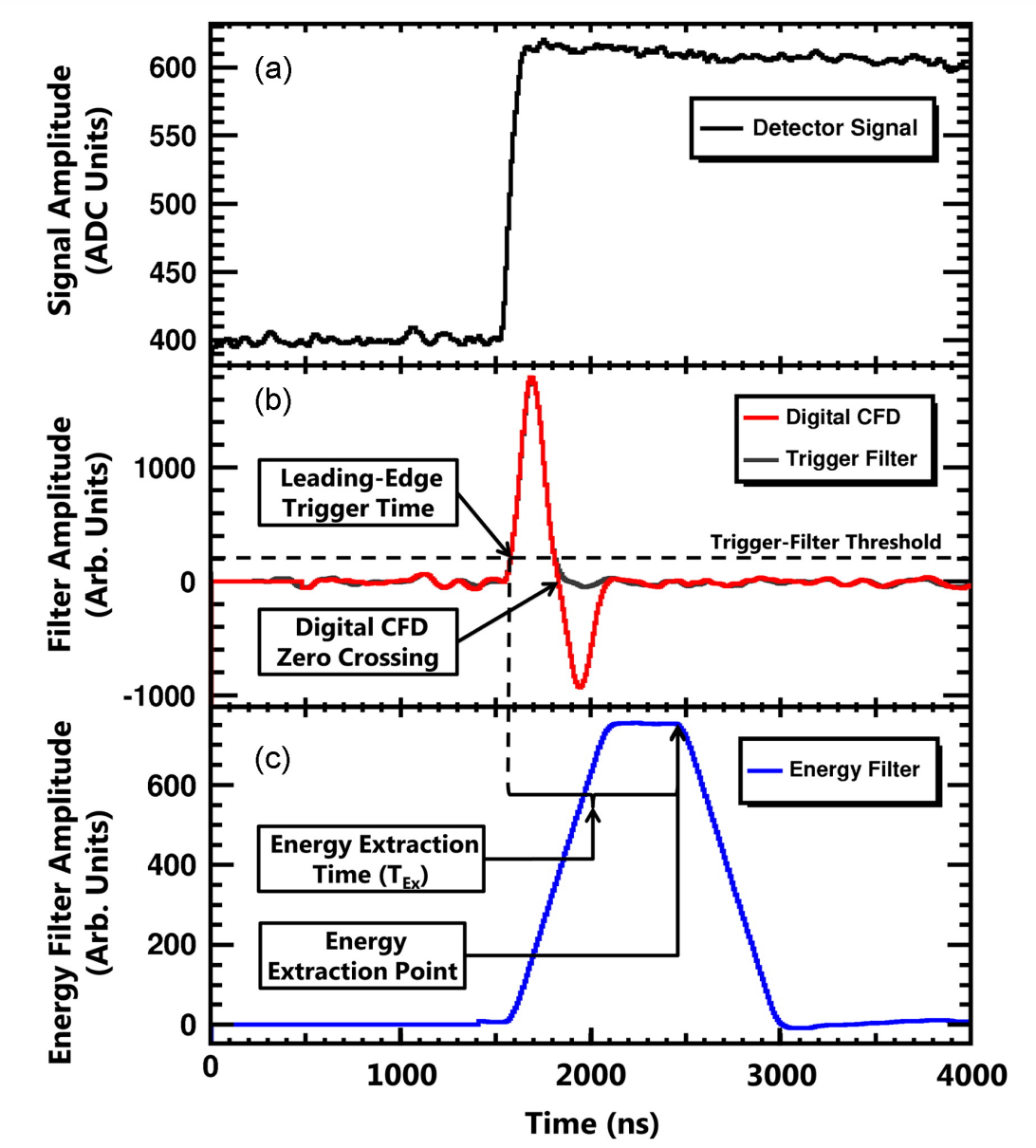}
    \caption{\large Detector signal processing in Pixie-16 modules with the Digital Data Acquisition System (DDAS). (a) shows the raw preamplified signal from the detector, (b) shows the digital Constant Fraction Discriminator (CFD) and trigger-filter response, and (c) shows the energy-filter response. Figure reproduced from \cite{Prokop2014}.}
    \label{fig:processing}
\end{figure}

Signals from all the detectors were processed using the NSCLDAQ digital data acquisition system \cite{Prokop2014}. It is a suite of software pipelines for data processing and online analysis. It utilizes XIA digital gamma finder class of electronics with high frequency (100 -- 250 MHz) Pixie-16 digitizer modules. Each channel is FPGA programmable and can independently record pulse-arrival time and pulse-processed energy. A total of three crates were used to house all the Pixie modules and their clocks were synchronized. (Differences in processing times were determined using a split pulser signal sent to all three crates and were accounted for in the analysis). 

Each input trace is continuously analyzed using a short filter, also known as a trigger filter or fast filter. It is a trapezoidal filter whose purpose is to register the arrival of an incoming pulse. A trapezoidal filter is characterized by two windows of width L ($T_{Peak}$) and G ($T_{Gap}$). The height H of the pulse at point $k$ is calculated as
\begin{equation}
    \large LH_k = \sum_{k-L+1}^{k}H_i - \sum_{2L-G+1}^{k-L-G}H_i
\end{equation}
where $k$ is a signal point and $H_i$ is the value of trace at point $i$. If the height of a pulse passes a preset trigger filter threshold, it is timestamped and the trace then passes through a long filter also known as the slow filter or the energy filter. This is also a trapezoidal filter with longer L and G such that H represents the energy deposition. L, G, and the trigger filter threshold can be set independently for each channel, both for the trigger filter as well as the energy filter. These parameters affect the energy as well as timing resolution and are set as per the following guidelines:
\begin{enumerate}
    \large \item $T_{Gap}$ should be equal to or larger than the rise time of the signal.
    \large \item $T_{Peak}$ for the trigger filter depends on the signal-to-noise ratio. A longer $T_{Peak}$ averages the signal over a longer time which reduces the likelihood of triggering on noise, but also increases the `dead time' of the system. 
    \large \item $T_{Peak}$ for the energy filter can be larger than $T_{Peak}$ for the trigger filter. The energy filter is applied only when the trigger threshold is crossed, so the dead time is less of an issue.
\end{enumerate}
Figure \ref{fig:processing} shows the raw detector signal, and the outputs of the trigger and the energy filter. Additionally, the decay time of the signal $\tau$ is also an important parameter to be set. For very high trigger rates in an experiment, there could be a pileup of pulses. In that case, the decay time gives an idea of how much the previous pulse has decayed so that an accurate baseline subtraction can be done. The decay time also leads to a slope in the plateau which can be corrected to get more energy accuracy. 

\bigskip
\Large \subsection{Event Building}
\bigskip
\large 

\begin{figure}
    \centering
    \includegraphics[width=500pt,keepaspectratio]{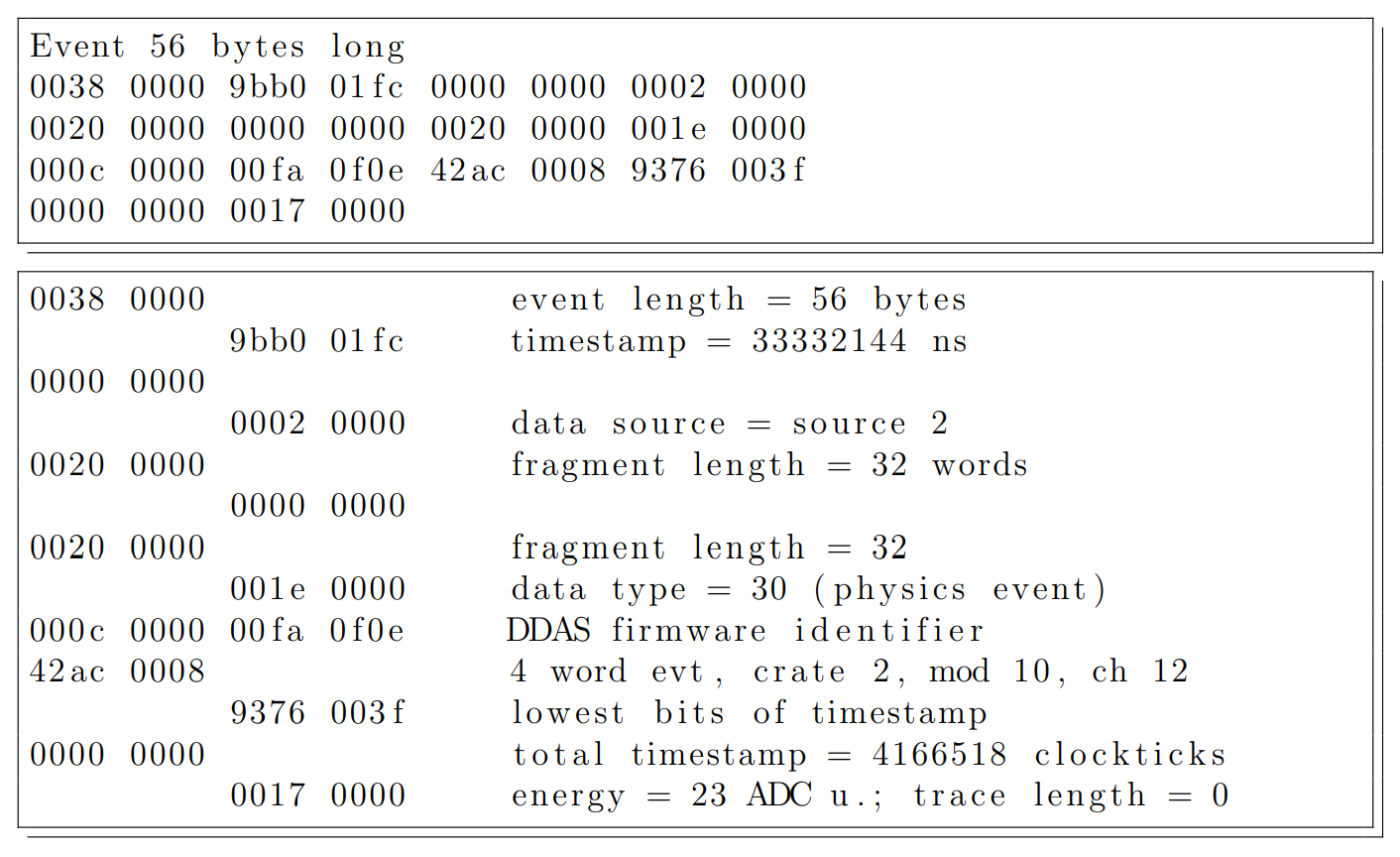}
    \caption{\large Example of an event recorded with the NSCLDAQ/FRIBDAQ. The upper panel shows the event as it is recorded in the binary .evt file. The lower panel shows the same event commented with what each word represents. Figure reproduced from \cite{Ong2018}.}
    \label{fig:event}
\end{figure}

NSCLDAQ is highly suitable for processing and merging data from multiple sources. Each of the three Pixie crates had its individual computer (spdaq) device onboard. Data from each spdaq is fed into a ring buffer called `raw ring'. A data packet in the raw ring contains the crate number, module number, channel number, timestamp, and the energy recorded. The NSCLDAQ online event builder accesses data from the raw rings and creates a `built ring' of merged data. The merging of raw data into events is done in a time-ordered fashion within a user-defined time window. All the detector hits occurring within the event building time window are bundled together into a single event. An event contains a group of triggers that are supposed to arise from the same physics event. For example, an implantation event of $^{33}$Mg will register a hit in the XFP scintillator, PIN detectors, and the low-gain channel of the DSSD (or miniDSSD) whereas a $\beta$-decay event will only register a hit in the high gain channel of the DSSD (or miniDSSD). Depending on the event correlation window, the $\beta$-delayed neutrons and $\gamma$s from the NERO and SuN detector, respectively, can also be included in the $\beta$-decay event. An example built event is shown in Figure \ref{fig:event} in the way it is stored in a binary `.evt' file. It contains information on the length of the event which designates the number of detector channels fired in that event, followed by the contents of raw data for each hit. 

The event building window is selected as per the requirements of the particular experiment. For the BCS-NERO part of the experiment, the event building window was chosen to be 200 $\mu$s to allow for the inclusion of $\beta$-delayed neutrons in the decay event following the moderation of neutrons in the NERO matrix before detection. On the other hand, the SuN-miniDSSD part of the experiment had the event building window set to just 2000 ns to avoid the high rate of background $\gamma$s to be included in the built events. The event building time window can also influence the rate of beam particles that can be accepted by the experimental set-up. A primary consideration is that successive physics events should be separated by at least the event building time window apart from other instrumentation constraints specific to an experiment. 

\huge \chapter{Data Analysis}\label{chp:analysis}
\large

This chapter focuses on the experimental data analysis techniques. It covers how the data recorded from various detectors is mapped onto the nuclear quantities measured. It includes setting up the data acquisition and signal processing parameters, detector thresholds, efficiency and energy calibrations, correlating implant events with decay events, and simulations of detectors. The experiment was performed in two parts - BCS-NERO part and the SuN-miniDSSD part. Analysis of the datasets from each part is presented in separate sections.  

\bigskip
\LARGE \section{BCS - NERO Dataset}
\bigskip
\large 

This part of the experiment aimed to measure the $\beta$-delayed neutron branching ratio or the P$_n$ value. The set-up included XFP scintillator and the PIN detectors, BCS stack with DSSD, SSSD, and plastic scintillator, and the NERO matrix with $^3$He and BF$_3$ long counters as discussed in Section \ref{sec:bcsnero}. 

\bigskip
\Large \subsection{Digital Signal Processing}
\bigskip
\large 

\begin{table}[]
    \centering
    \begin{tabular}{cccccc}
    \toprule
       \multicolumn{1}{c}{\large Detector}  & \multicolumn{3}{c}{\large Trigger Filter} & \multicolumn{2}{c}{\large Energy Filter} \\
    \cmidrule(lr){2-4} \cmidrule(lr){5-6}
       \large Channel & \large $T_{Peak}$ ($\mu$s) & \large $T_{Gap}$ ($\mu$s) & \large Threshold & \large $T_{Peak}$ ($\mu$s) & \large $T_{Peak}$ ($\mu$s) \\
    \midrule
      \large PIN1  & \large 5.12 & \large 3.2 & \large $\sim$6400 & \large 9.52 & \large 0.64 \\
      \large PIN2   & \large 5.12 & \large 3.2 & \large $\sim$6400 & \large 9.52 & \large 0.64 \\
      \large PIN1-XFP Time of flight  & \large 0.48 & \large 0.08 & \large $\sim$600 & \large 1.28 & \large 1.28 \\
       \large PIN3   & \large 5.12 & \large 3.2 & \large $\sim$6400 & \large 9.52 & \large 0.64 \\
      \large DSSD (low gain)   & \large 1.34 & \large 1.2 & \large $\sim$2000 & \large 2.3 & \large 0.24 \\
      \large DSSD (high gain) & \large 6.16 & \large 4.0 & \large $\sim$1400 & \large 9.2 & \large 0.64 \\
      \large SSSD & \large 3.04 & \large 2.4 & \large $\sim$4000 & \large 7.76 & \large 0.48 \\
      \large Plastic Scintillator  & \large 0.8 & \large 0.24 & \large $\sim$1000 & \large 0.8 & \large 0 \\
      \midrule
      \large NERO $^3$He tubes  & \large 2.0 & \large 0.8 & \large $\sim$1000 & \large 4.0 & \large 2.0 \\
      \large NERO BF$_3$ tubes   & \large 2.0 & \large 0.8 & \large $\sim$800 & \large 4.0 & \large 1.04 \\
    \bottomrule
    \end{tabular}
    \caption{\large Digital Signal Processing Parameters optimized for each detector channel for the BCS-NERO part of the experiment. For detectors with multiple channels, each channel was optimized separately and only the average value is shown here.}
    \label{tab:bcsneropar}
\end{table}

The digital signal processing parameters were set for each detector channel individually, by analyzing the signal trace with an oscilloscope. The time windows, as described in Section \ref{sec:dsp}, were chosen to maximize the energy resolutions. Trigger thresholds were chosen to go as low in energy as possible without hitting the low-energy noise peak. Table \ref{tab:bcsneropar} lists the parameters for all the detectors in this setup. Parameters for PIN detectors, high gain channels of the DSSD, and the SSSD were set before the experiment using a $^{241}$Am source that emits $\alpha$ particles with an energy of 5.48 MeV. The parameters for the low gain channel of the DSSD and the plastic scintillator were set online with the beam during the experiment.

\bigskip
\Large \subsection{NERO Thresholds And Calibrations}

\bigskip
\Large \subsubsection{NERO Thresholds}
\bigskip
\large 

\begin{figure}
    \centering
    \includegraphics[width=370pt,keepaspectratio]{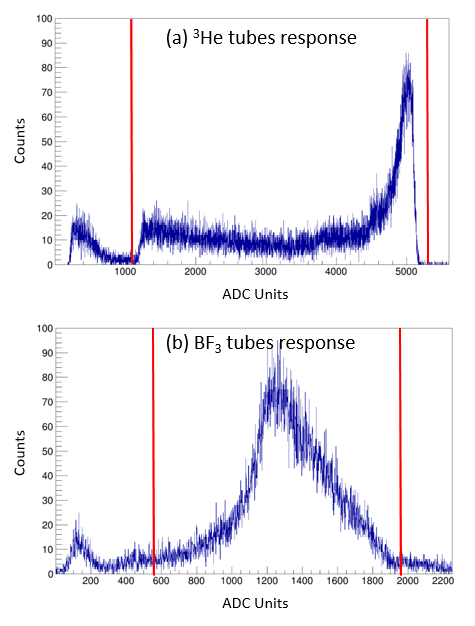}
    \caption{\large Sample neutron response spectra for (a)$^3$He and (b)BF$_3$ long counters in the NERO matrix. The low energy noise peaks are visible in both spectra. Red lines designate the upper and lower energy thresholds. Only the events in between are considered valid neutron events.}
    \label{fig:neroresponse}
\end{figure}

Figures \ref{fig:neroresponse}(a) and \ref{fig:neroresponse}(b) show the neutron response for a sample $^3$He tube and a BF$_3$ tube, respectively from a $^{252}$Cf source. The lower and upper thresholds designated by the red lines define the region of true neutron events. These thresholds are implemented in the software and are different from the trigger thresholds listed in Table \ref{tab:bcsneropar}. These are also determined individually for all 60 tubes in the NERO matrix. Integrating the region between these limits gives the total number of neutrons detected in a single counter. Summing the contributions from all counters gives the total number of neutrons detected by NERO. 

\bigskip
\Large \subsubsection{NERO Efficiency Calibrations}
\bigskip
\large 

\begin{table}[]
    \centering
    \begin{tabular}{cccc}
    \toprule
       \large Isotope  & \large Amount (atom \%) & \large Amount (atom \%) & \large Neutron Activity (s$^{-1}$) \\
       \large & \large  December 1, 1989 & \large September 4, 2020 & \large \\
    \midrule
      \large $^{249}$Cf  & \large 7.32 & \large 6.89 & \large 0 \\
      \large $^{250}$Cf   & \large 13.11 & \large 2.57 & \large 36.42 \\
      \large $^{251}$Cf  & \large 4.55 & \large 4.44 & \large 0 \\
      \large $^{252}$Cf   & \large 75.02 & \large 0.02 & \large 70.85 \\
      \large $^{245}$Cm  & \large 0 & \large 0.43 & \large 0 \\
      \large $^{246}$Cm   & \large 0 & \large 10.52 & \large 0.12 \\
      \large $^{247}$Cm  & \large 0 & \large 0.11 & \large 0 \\
      \large $^{248}$Cm   & \large 0 & \large 75.02 & \large 3.85 \\
    \midrule
      \large Total & \large & \large & \large 111.24 \\
    \bottomrule
    \end{tabular}
    \caption{\large Isotopic composition and the corresponding neutron activity of a $^{252}$Cf source (Z7153) used for NERO efficiency calibrations. The amounts on December 1, 1989 are supplied by the manufacturer of the source and the amounts on September 4, 2020 (the day of efficiency calibrations) is calculated.}
    \label{tab:cf252}
\end{table}

Determining the NERO efficiency is crucial for estimating the total number of neutrons emitted during the experiment. It is benchmarked using a $^{252}$Cf source with known activity. $^{252}$Cf has a half-life of 2.645 years and produces neutrons via spontaneous fission. It has a fission branching ratio of 3.102\% and produces 0.116 neutrons per decay. The remaining $\alpha$ decays to $^{248}$Cm. The composition of the Z7153 source as a mixture of Cf isotopes on December 1, 1989 is listed in Table \ref{tab:cf252} as supplied by the manufacturer. The amounts of each isotope along with their $\alpha$ daughters on September 4, 2020 i.e. the day of the calibration are also listed in another column. This translated to a neutron activity of 111.24 neutrons per second for the $^{252}$Cf source on the day of efficiency calibration. 

The rate of neutron background in the room is measured to be 2.0914 neutrons per second. This is subtracted from the total rate observed during the calibration run to get the rate of detected neutron from the $^{252}$Cf source. Dividing this by the neutron activity of the source gives the NERO efficiency. Table \ref{tab:neroeff} lists the contributions from each quadrant separately. The obtained value for the efficiency of 31.3(4)\% is in agreement with the previously benchmarked value of 31.7(2)\% \cite{Pereira2010} for the $^{252}$Cf source. Subsequently, the same scaling factor as in \cite{Pereira2010} was used to get the efficiency for low-energy neutrons which is relevant for estimating the P$_n$ value. The scaled NERO efficiency used for the P$_n$ value calculations is 36.5(50)\%.  

\begin{table}[]
    \centering
    \begin{tabular}{ccccc}
    \toprule
       \large NERO  & \large Room Background & \large $^3$He tubes (1--4) & \large BF$_3$ tubes (5--15) & \large Total \\
       \large Quadrant & \large  (neutrons/s) & \large  efficiency (\%) & \large efficiency (\%) & \large efficiency (\%) \\
    \midrule
      \large A  & \large 0.6832 & \large 2.63 & \large 4.73 & \large 7.36\\
      \large B   & \large 0.5619 & \large 3.04 & \large 5.09 & \large 8.13\\
      \large C  & \large 0.4526 & \large 3.14 & \large 4.81 & \large 7.95\\
      \large D   & \large 0.3937 & \large 2.96 & \large 4.89 & \large 7.85\\
    \midrule
      \large Total & \large 2.0914 & \large & \large & \large 31.3(4) \\
    \midrule
      \large Scaled & \large & \large & \large & \large \textbf{36.5(50)} \\
    \bottomrule
    \end{tabular}
    \caption{\large Neutron detection efficiencies for each quadrant and the total NERO efficiency. The estimated room background has to be subtracted from the total neutron counts.}
    \label{tab:neroeff}
\end{table}

\bigskip
\Large \subsection{BCS Thresholds And Calibrations}

\bigskip
\Large \subsubsection{DSSD Thresholds}
\bigskip
\large 

\begin{figure}
    \centering
    \includegraphics[width=450pt,keepaspectratio]{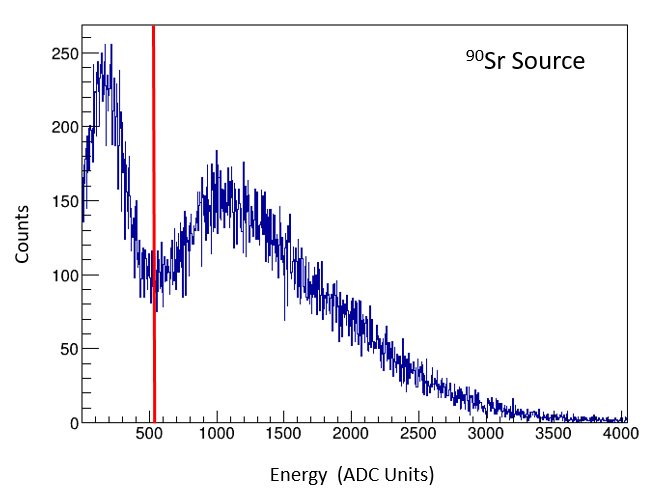}
    \caption{\large Energy spectrum for a sample BCS DSSD strip from a $^{90}$Sr calibration run. This is to set the low energy thresholds (red line) for the high gain channel of the DSSD. The threshold is set individually for each strip at the dip between the noise peak and the low energy electrons peak. This allows for the highest signal-to-noise ratio. }
    \label{fig:bcs90sr}
\end{figure}

The electrons from $\beta$-decay form a continuous energy spectrum from zero energy up to the Q-value. To identify as many decay events as possible, it is important to successfully detect electrons with low energies. However, the electronic noise takes over below a certain limit and it becomes impossible to distinguish the low-energy electrons from the noise. The low energy thresholds are determined using the $\beta$ electrons spectrum from a $^{90}$Sr source. Thresholds are set at the dip between the noise peak and the electrons peak as shown by the red line in Figure \ref{fig:bcs90sr}. These are determined for each detector channel and implemented in the analysis software.

\bigskip
\Large \subsubsection{DSSD Energy Calibrations}
\bigskip
\large 

\begin{figure}
    \centering
    \includegraphics[width=450pt,keepaspectratio]{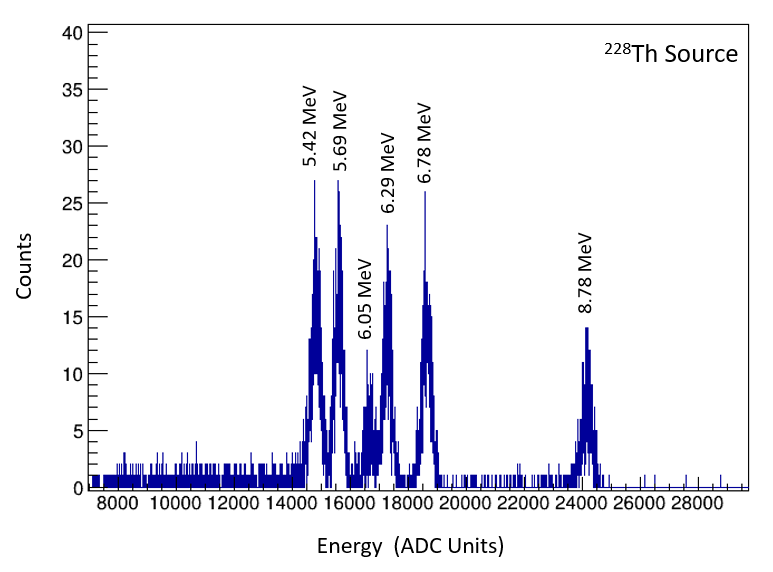}
    \caption{\large Energy spectrum for a sample BCS DSSD strip from a $^{228}$Th energy calibration run. Six characteristic $\alpha$-peaks from the decay chain of $^{228}$Th are shown in the figure. The centroids for Gaussian fits to the peaks were used for a linear energy calibration.}
    \label{fig:bcs228th}
\end{figure}

Each strip of the DSSD in the high gain channel is calibrated using a $^{228}$Th source. The complete decay chain of $^{228}$Th leads to $\alpha$ particle emission with six distinct energies of 5.42 MeV, 5.69 MeV, 6.05 MeV, 6.29 MeV, 6.78 MeV, and 8.78 MeV. Each peak can be identified as shown in Figure \ref{fig:bcs228th}. All the peaks are fit to a Gaussian and the resulting centroids are linearly mapped to the corresponding alpha energies. Residuals for each strip from the linear fit were below 20 keV. This calibration for the high-gain channels can be roughly translated to the low-gain channels by adjusting for the appropriate gain factor. It is useful to check the energy of the incoming beam which can be confirmed with the LISE++ simulations as discussed in Section \ref{sec:beamdelivery}.  

\begin{figure}
    \centering
    \includegraphics[width=450pt,keepaspectratio]{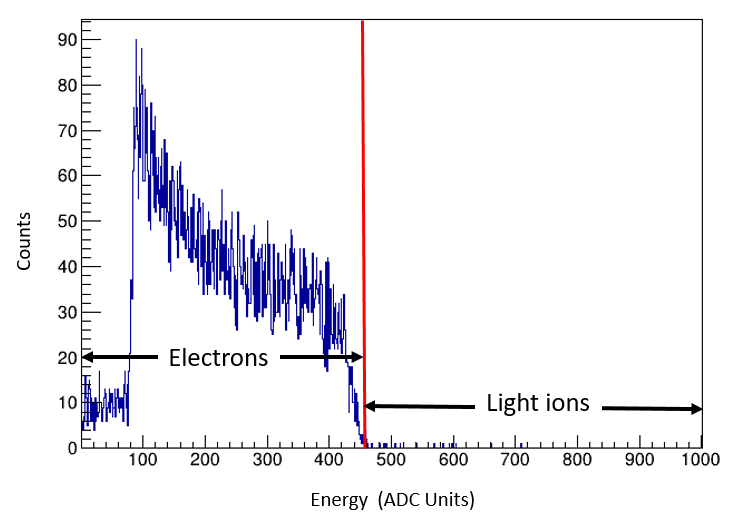}
    \caption{\large Plastic scintillator spectrum for a sample run with $^{33}$Mg beam. It is placed downstream of the BCS stack to veto light particle events that can mimic decay events in the DSSD. The red line shows the discriminator threshold between electrons and light particles. The $^{33}$Mg beam doesn't contain any light particles.}
    \label{fig:scint}
\end{figure}

\bigskip
\Large \subsubsection{Plastic Scintillator Threshold}
\bigskip
\large 

A plastic scintillator was placed downstream of the BCS stack to distinguish between light particles in the beam that mimic decay events in the high-gain channel of the DSSD. However, these light particles as a part of the beam have large forward momentum and deposit relatively higher energy in the plastic scintillator. These high-energy events in the scintillator can then be vetoed. Figure \ref{fig:scint} shows the energy spectrum for the plastic scintillator for a single run with $^{33}$Mg beam. The red line shows the threshold beyond which all the events are vetoed. This threshold is also implemented in the analysis software. It can be seen that the $^{33}$Mg beam had very high purity and negligible light particle contaminants.

\begin{figure}
    \centering
    \includegraphics[width=450pt,keepaspectratio]{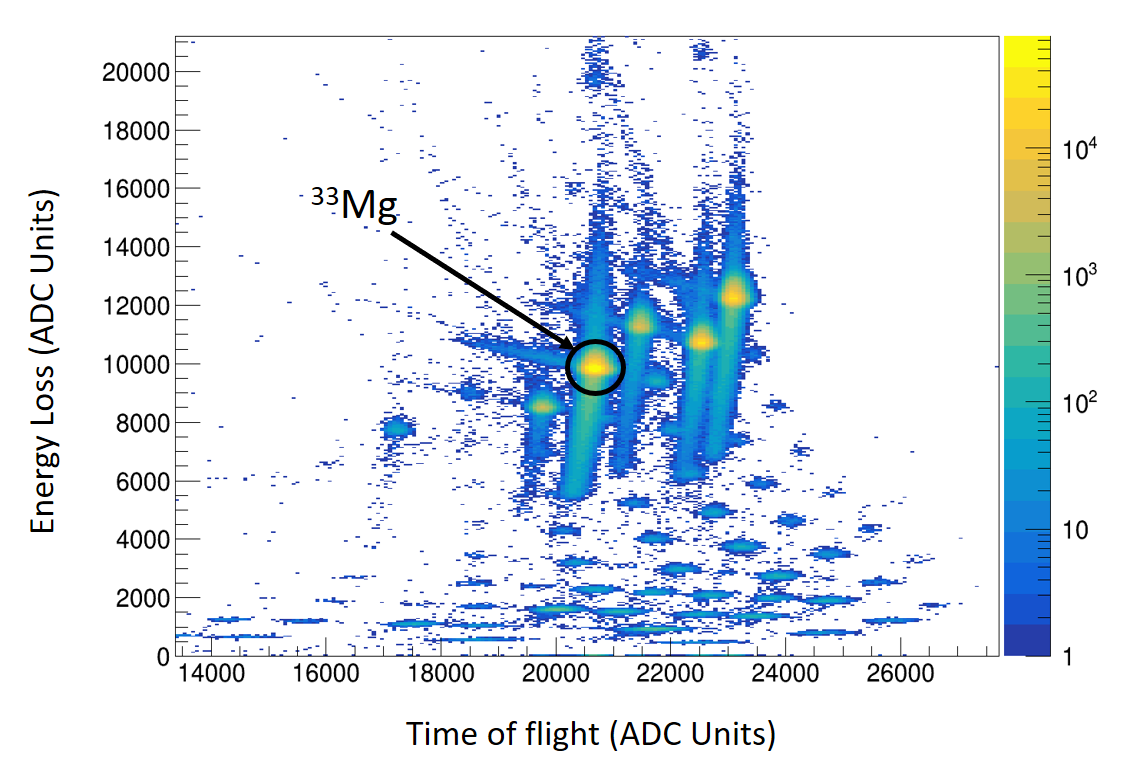}
    \caption{\large Particle Identification (PID) plot for the $\delta$E-TOF method as described in Section \ref{sec:pid}. Energy loss is from the PIN1 detector and the time of flight is between the XFP scintillator and the PIN1 detector. The A1900 fragment separator settings were optimized for the $^{33}$Mg fragment to be delivered to the experimental end-station with the least contaminants. Only the events in the $^{33}$Mg blob were used for further analysis. }
    \label{fig:pid}
\end{figure}

\bigskip
\Large \subsection{Particle Identification}\label{sec:pidbcsnero}
\bigskip
\large

\begin{figure}
    \centering
    \includegraphics[width=430pt,keepaspectratio]{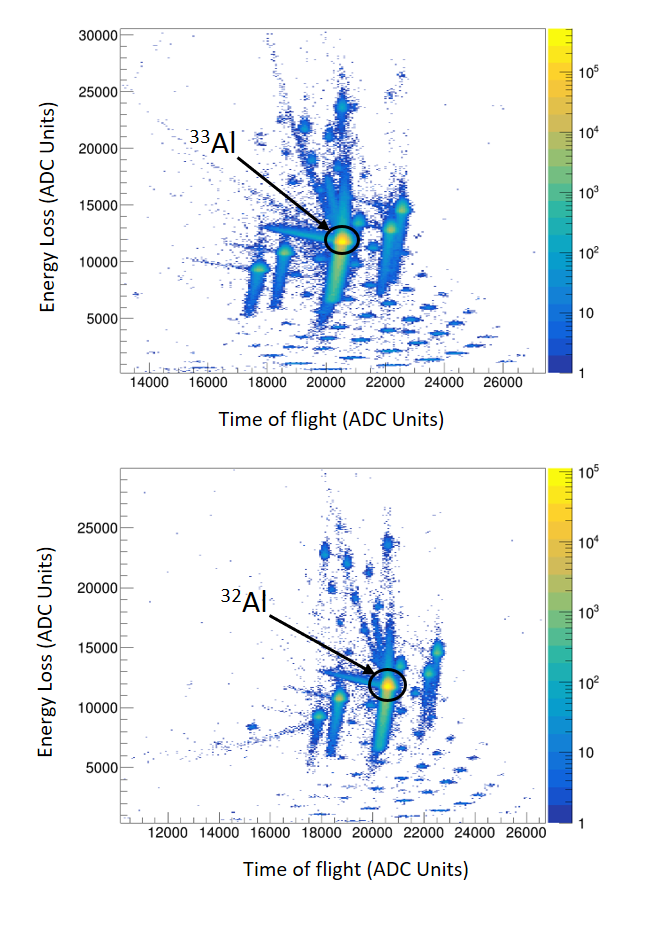}
    \caption{\large PID plots for the $^{33,32}$Al beams delivered to the experimental end-station to estimate the background from $\beta$-daughter and $\beta$-delayed neutron daughter, respectively. Refer to the caption of Figure \ref{fig:pid} for description. }
    \label{fig:pidbg}
\end{figure}

Figure \ref{fig:pid} shows the Particle Identification (PID) plot for the $^{33}$Mg beam using the $\delta$E-TOF method as discussed in Section \ref{sec:pid}. The energy loss $\delta$E is measured in the PIN1 detector whereas the time of flight TOF is measured using a Time-to-Amplitude Converter (TAC) between the XFP scintillator at I2 position in the A1900 fragment separator and the PIN1 detector. The central blob corresponds to the $^{33}$Mg fragments. A tight square gate is implemented around this blob and only the events that fall within that square gate are considered as valid $^{33}$Mg implants. No further corrections to the PID were required since the relatively tight gate had enough statistics for the analysis. 

Any $\beta$-delayed spectroscopy measurements of $^{33}$Mg will always have background contributions from the further decay of the $^{33}$Al daughter and $^{32}$Al $\beta$-delayed neutron daughter. As a result, the complete experiment was also performed using $^{33,32}$Al beams. This allowed background estimation on the same experimental setup to cancel out systematic errors, if any. Figures \ref{fig:pidbg}(a) and \ref{fig:pidbg}(b) show the PID plots for $^{33}$Al and $^{32}$Al beams, respectively. 

\bigskip
\Large \subsection{Implant and Decay Correlations}\label{sec:correlation}
\bigskip
\large

Measuring the $\beta$-decay half-life of an isotope requires measuring the time difference between the implantation of a particle of an isotope and the detection of the electron from its $\beta$-decay. This is achieved by classifying the physics events recorded by the data acquisition system into implantation events and decay events and subsequently correlating them in the analysis software. A physics event can be classified into an implantation event upon satisfying the following conditions:
\begin{enumerate}
    \large \item Energies in PIN1, PIN2, and PIN3 are greater than zero. This ensures the event occurred due to a beam particle.
    \large \item At least one of each of the front and back strips in the low gain channel of the DSSD registers an energy greater than zero. This allows for spatial localization of the implant on the DSSD. When multiple strips fire, the strip with the highest energy deposition (usually the one in the middle of the group of strips fired) is chosen as the position of the implant. 
    \large \item No energy deposition in the SSSD downstream of the DSSD. This is to ensure that the incoming beam particle was implanted in the DSSD and did not just deposit some energy while punching through. 
    \large \item No energy deposition in the plastic scintillator.
    \large \item Additional conditions can be placed on the PID to select implantations of a particular isotope. 
\end{enumerate}
To be classified as a decay event, a physics event needs to satisfy the following conditions:
\begin{enumerate}
    \large \item No energy deposition on either of the PIN detectors. This ensures that no beam particle was involved in the event.
    \large \item No signal in the low gain channels of the DSSD. The decay electrons can have a maximum energy deposition of only a few MeV, which is below the thresholds of the low gain channels. 
    \large \item Energies in at least one of each of the front and back strips high-gain channel of the DSSD to be greater than the low energy thresholds applied in the software. This allows for spatial localization of the decay event and spatial correlations with the implant event during analysis. 
    \large \item Energy deposition of either zero or less than the light-ion threshold in the plastic scintillator. This ensures that the physics event involves an electron. 
\end{enumerate}

Once a list of implantation and decay events is obtained, they are spatially and temporally correlated. Spatial correlation requires that the implantation and decay events occur within a pre-defined correlation field, i.e. the distance between the pixels for these events on the DSSD should be less than the correlation window. A correlation window of 3 $\times$ 3 was found to be optimal for the BCS DSSD. Temporal correlation requires that the time between the implantation and decay events is less than a pre-defined correlation time. The correlation time window should at least be a few times the half-life of the implanted particle. It was chosen to be 1000 ms for all isotopes in this experiment. All pairs of implants and decay events that satisfy both temporal and spatial correlations are considered for further analysis. It should be noted that a single decay event can be correlated to multiple implantation events and vice versa. These multiple correlations include true correlations such as daughter or granddaughter decays as well as random correlations from the background. The rate of random correlations can be estimated with the reverse time correlator as described in Section \ref{sec:reverse}.

\bigskip
\Large \subsubsection{Reverse Time Correlations}\label{sec:reverse}
\bigskip
\large 

The correlation logic described in Section \ref{sec:correlation} is run forward in time, i.e. the decay events happen after implantation events. As discussed earlier, this includes true as well as random correlations. A similar correlation logic can also be run backward in time where the decay events occur before implantation events. Since this is physically not possible, the reverse time correlator will include only the random correlations. This allows for the estimation of the rate of random correlations which constitute the background. 

\bigskip
\Large \subsubsection{Half-life Measurement}
\bigskip
\large 

\begin{figure}
    \centering
    \includegraphics[width=300pt,keepaspectratio]{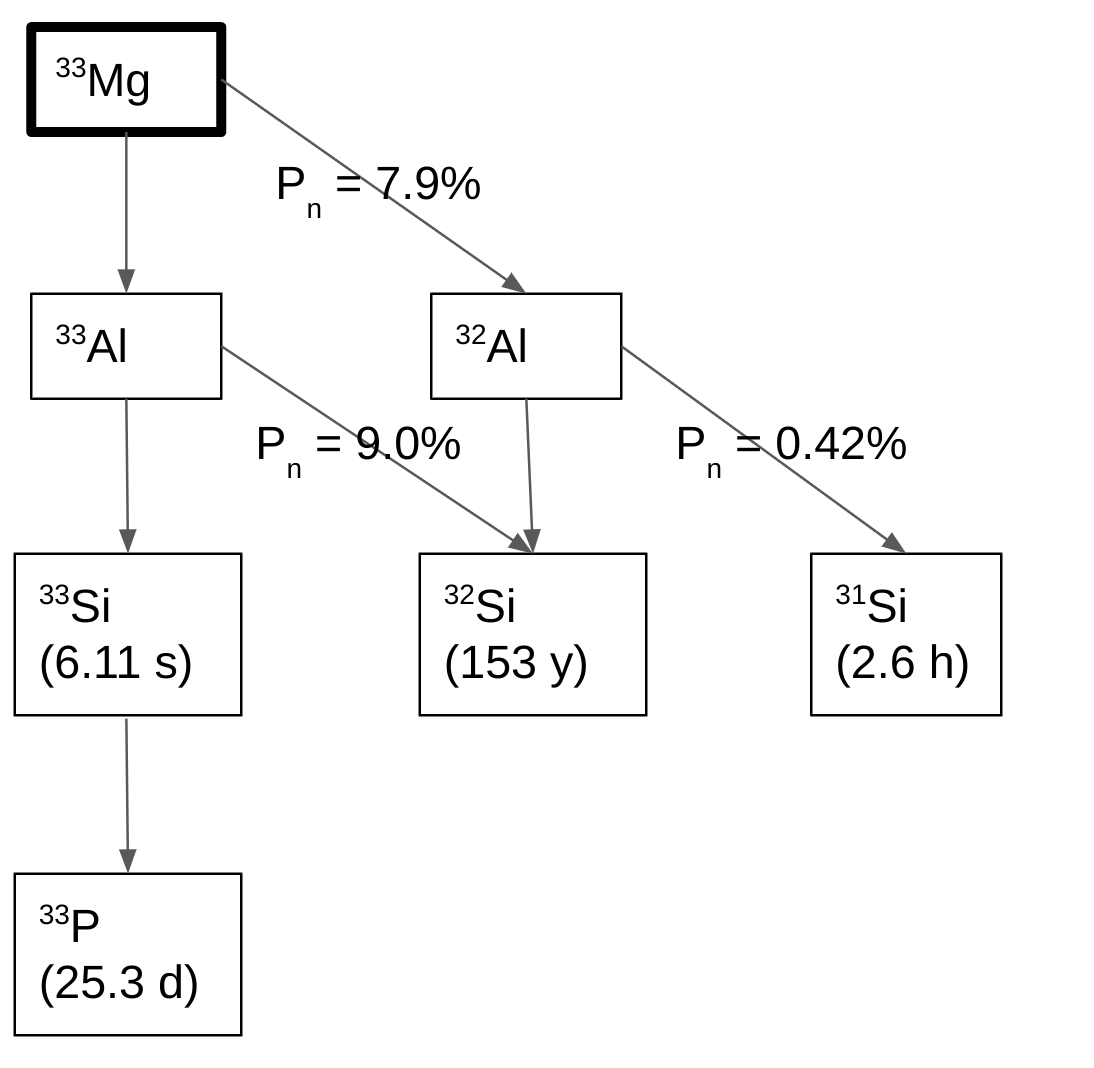}
    \caption{\large The complete decay chain of $^{33}$Mg with contributions from the decay of daughter ($^{33}$Al), granddaughter ($^{33}$Si), and $\beta$-delayed neutron daughter ($^{32}$Al). All other isotopes ($^{33}$P, $^{32}$Si, $^{31}$Si) are considered `pseudo-stable' for the correlation time window of 1000 ms.}
    \label{fig:decay}
\end{figure}

Once all the implant-decay correlation pairs are identified, a histogram of the time difference between the implantation and decay events is generated, also known as the decay curve. This curve resembles the probability distribution of the time an implanted isotope survives before undergoing $\beta$-decay. Due to nuclear decay being a first-order kinetic transition, the decay curve for a single nuclide parent decay is exponentially decreasing. This decay curve can be fit to Equation \ref{eqn:nuclear}, to extract the decay constant, which can be converted into half-life using Equation \ref{eqn:halflife}. In practice, however, the decay curve has multiple components from the decay of daughter, granddaughter, $\beta$-delayed neutron daughter, etc. when working with exotic nuclei far from stable isotopes. As a result, the decay curve is fit with a modified version of Equation \ref{eqn:bateman}. The modifications include accounting for $\beta$-delayed neutron branchings and background from random correlations. The half-lives of all other isotopes in the decay chain should either be known, or a multi-parameter fit can be performed. 

Figure \ref{fig:decay} shows the decay chain for $^{33}$Mg along with $\beta$-delayed neutron branchings. The decay curve for $^{33}$Mg will include contributions from the decays of $^{33}$Al, $^{32}$Al, and $^{33}$Si. All other isotopes are considered 'pseudo stable' for the 1000 ms correlation time window considered here. The half-lives of $^{33}$Al and $^{32}$Al were measured using the same experimental setup and technique with beams of $^{33}$Al and $^{32}$Al, respectively. The half-life for $^{33}$Si was acquired from the Nudat database by the National Nuclear Data Center (NNDC) \cite{Goosman1973}. The P$_n$ values for $^{33}$Mg, $^{33}$Al, and $^{32}$Al were estimated from the NERO data as described in Section \ref{sec:pn}. With these inputs, the decay curve for $^{33}$Mg implants was fit to Equation \ref{eqn:bateman}, and the half-life was extracted. 

\bigskip
\Large \subsection{Calculating P$_n$ Value}\label{sec:pn}
\bigskip
\large

The event-building time window for defining a physics event during the NERO-BCS part of the experiment is set to 200 $\mu$s. This allows for the $\beta$-delayed neutrons detected in NERO to be a part of the decay event. The total number of neutrons in the correlated decay events can then be used to calculate the P$_n$ value. The P$_n$ value is calculated as:
\begin{equation}
    \large P_n = \frac{N_n - N_d}{\epsilon_n N_{\beta}}
\end{equation}
where N$_n$ is the number of neutrons in decay events from forward time correlated decay curve minus the number of neutrons in decay events from reverse time correlated decay curve, N$_d$ is the number of $\beta$-delayed neutrons expected from the decay of all other isotopes in the decay chain except the parent isotope, N$_{\beta}$ is the number of forward time correlated decay events minus the number of reverse time correlated decay events, and $\epsilon_n$ is the NERO neutron detection efficiency.

In terms of $\beta$ and neutron correlations, we need true $\beta$ particles to be correlated with true neutrons. True here means particles from a valid $\beta$-delayed neutron decay. However, there are three major sources of background for $\beta$-n correlations. These are accounted for in the P$_n$ value calculations as follows:
\begin{enumerate}
    \large \item Random $\beta$ correlating with a true neutron. This is accounted for when subtracting the reverse time correlator components in N$_{\beta}$.
    \large \item True $\beta$ correlating with a random neutron. The rate of random neutrons is about 2.1 neutrons per second. So the probability of a random neutron correlating with a true $\beta$ within a 200 $\mu$s event building time window is tiny and negligible. 
    \large \item Random $\beta$ correlating with a random neutron. This factor is also tiny but accounted for via the reverse time correlator components subtraction.
\end{enumerate}

The P$_n$ values for $^{33}$Al and $^{32}$Al were calculated with the respective particle beams implanted in the BCS-DSSD. The N$_d$ for these isotopes is 0 since none of their daughters undergo $\beta$-delayed neutron decay. Using these experimental measurements, the P$_n$ value for $^{33}$Mg was determined with an iterative procedure. The number of neutrons from the daughters' decay N$_d$ depends not only on the daughters' P$_n$ value but also on the parent P$_n$ value that is being determined. N$_d$ is assumed to be 0 at the beginning and then solved recursively until it converges.

\bigskip
\LARGE \section{SuN - miniDSSD Dataset}
\bigskip
\large 

This part of the experiment aimed to measure the $\beta$ feeding intensities into excited states in $^{33}$Al by detecting the $\beta$-delayed $\gamma$ rays. The set-up included XFP scintillator and the PIN detectors for particle identification, miniDSSD and silicon veto detector for the implants and electrons, and the SuN detector for the $\gamma$ rays as discussed in Section \ref{sec:sunminidssd}.

\bigskip
\Large \subsection{Digital Signal Processing}
\bigskip
\large 

\begin{table}[]
    \centering
    \begin{tabular}{cccccc}
    \toprule
       \multicolumn{1}{c}{\large Detector}  & \multicolumn{3}{c}{\large Trigger Filter} & \multicolumn{2}{c}{\large Energy Filter} \\
    \cmidrule(lr){2-4} \cmidrule(lr){5-6}
       \large Channel & \large $T_{Peak}$ ($\mu$s) & \large $T_{Gap}$ ($\mu$s) & \large Threshold & \large $T_{Peak}$ ($\mu$s) & \large $T_{Gap}$ ($\mu$s) \\
    \midrule
     \large PIN1  & \large 5.12 & \large 3.2 & \large $\sim$11200 & \large 9.52 & \large 0.64 \\
      \large PIN2   & \large 5.12 & \large 3.2 & \large $\sim$6400 & \large 9.52 & \large 0.64 \\
      \large PIN1-XFP Time of flight  & \large 0.48 & \large 0.08 & \large $\sim$600 & \large 1.28 & \large 1.28 \\
      \large miniDSSD (low gain)   & \large 0.96 & \large 0.512 & \large $\sim$2000 & \large 4.928 & \large 3.2 \\
      \large miniDSSD (high gain) & \large 0.96 & \large 0.512 & \large $\sim$1500 & \large 4.928 & \large 3.2 \\
      \large Si Surface Barrier Veto & \large 1.664 & \large 0.64 & \large $\sim$3000 & \large 4.096 & \large 2.56 \\
    \midrule
      \large SuN PMTs  & \large 0.16 & \large 0.048 & \large $\sim$20 & \large 0.48 & \large 0.16 \\
    \bottomrule
    \end{tabular}
    \caption{\large Digital Signal Processing Parameters optimized for each detector channel for the SuN-miniDSSD part of the experiment. For detectors with multiple channels, each channel was optimized separately and only the average value is shown here.}
    \label{tab:sunpar}
\end{table}

Digital signal processing parameters were set for each channel individually in a similar manner as done for the BCS-NERO part. The signal trace was analyzed on an oscilloscope and the timing and threshold parameters were chosen to reduce noise and maximize energy resolution. Table \ref{tab:sunpar} lists the parameters for all detectors in this setup.

\bigskip
\Large \subsection{miniDSSD Thresholds And Calibrations}
\bigskip
\large 

\begin{figure}
    \centering
    \includegraphics[width=450pt,keepaspectratio]{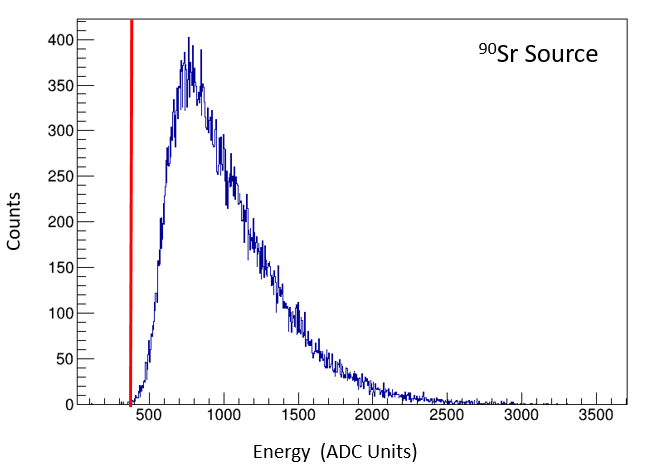}
    \caption{\large Example energy spectrum for a sample miniDSSD strip from a $^{90}$Sr calibration run. The red line designates low energy thresholds for the high gain channel of the miniDSSD. Refer to the caption of Figure \ref{fig:bcs90sr} for more details. }
    \label{fig:minidssd90sr}
\end{figure}

\begin{figure}
    \centering
    \includegraphics[width=450pt,keepaspectratio]{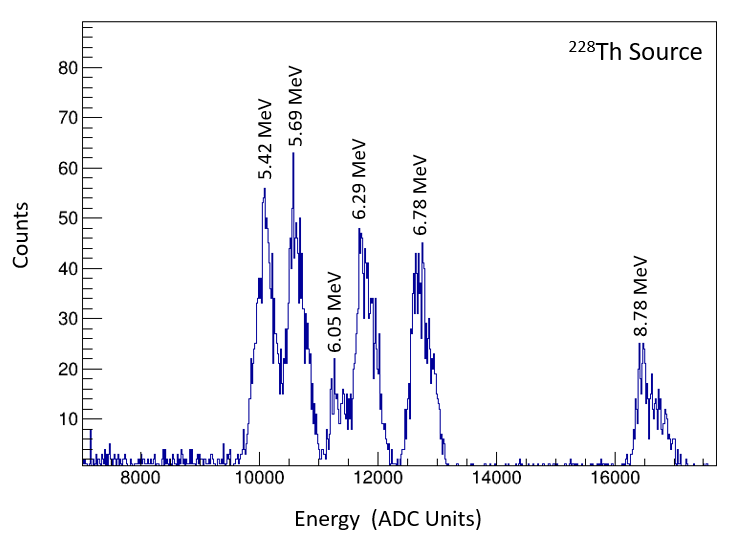}
    \caption{\large Example energy spectrum for a sample miniDSSD strip from a $^{228}$Th energy calibration run. Each peak corresponds to an $\alpha$-energy from the ${228}$Th decay chain as shown in the figure. The high gain channels for each strip were linearly calibrated with these peaks.}
    \label{fig:minidssd228th}
\end{figure}

The detector thresholds and energy calibrations for each strip of the miniDSSD were determined similarly to the BCS DSSD using a $^{90}$Sr source and a $^{228}$Th source, respectively. Figure \ref{fig:minidssd90sr} shows the energy spectrum from a sample miniDSSD strip for a $^{90}$Sr source. The low energy noise peak is barely visible and the thresholds designated by the red line were set such that the entire electron peak is included in the analysis. Figure \ref{fig:minidssd228th} also shows the energy spectrum of a $^{228}$Th source for one of the miniDSSD strips. All six $\alpha$ peaks can be seen and centroids from their Gaussian fits were used for a linear energy calibration. Thresholds and calibrations were determined individually for all 32 strips (16 on the front and 16 on the back) of the miniDSSD. 

\bigskip
\Large \subsection{SuN Gain Matching and Energy Calibrations}
\bigskip
\large

\begin{figure}
    \centering
    \includegraphics[width=450pt,keepaspectratio]{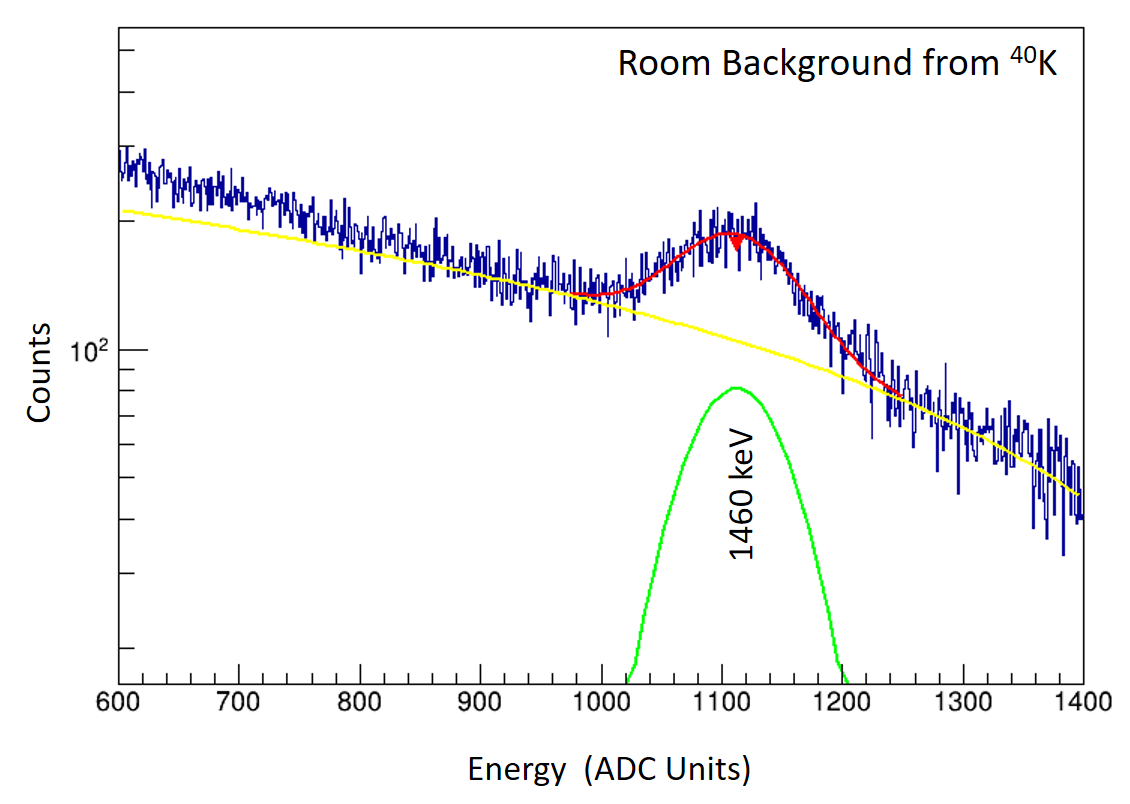}
    \caption{\large Room background spectrum for a sample SuN PMT for gain matching. The PMTs have to be gain-matched since the energies of 3 PMTs are summed to get the energy deposition in a SuN segment. This is done using the 1460 keV $\gamma$ ray from the room background that originates from the decay of $^{40}$K found in trace amounts in the concrete present in the walls of the room. }
    \label{fig:sunbg}
\end{figure}

\begin{figure}
    \centering
    \includegraphics[width=400pt,keepaspectratio]{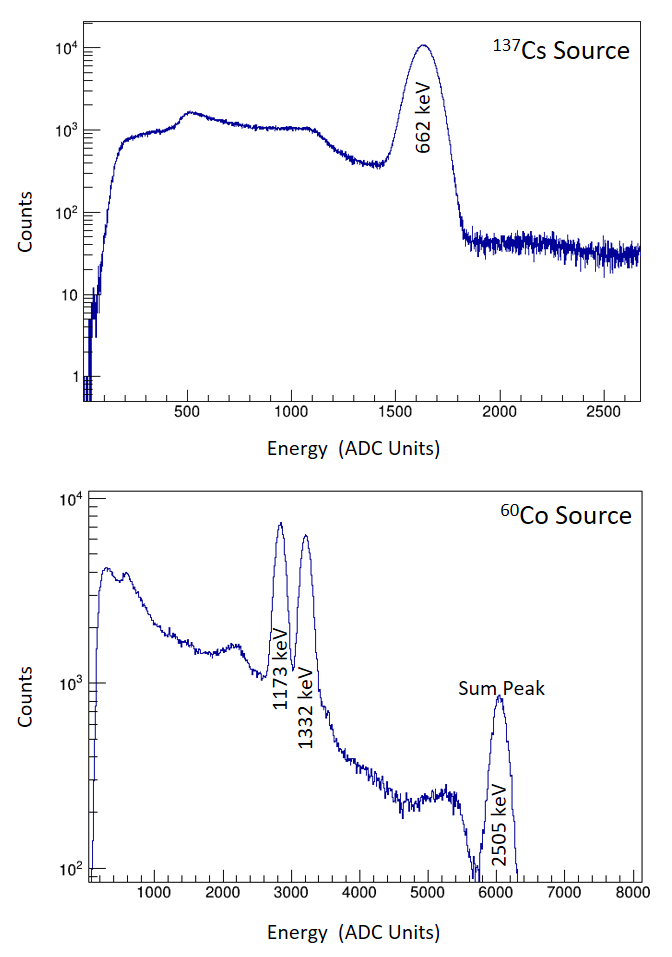}
    \caption{\large Example energy spectrum of a sample SuN Segment (with gain-matched PMTs) from $^{137}$Cs and $^{60}$Co calibration runs. The photopeaks are labeled in the figure corresponding to the $\gamma$ energy. $^{60}$Co calibration run also shows the sum peak.}
    \label{fig:suncalib}
\end{figure}

The sun detector consists of 8 NaI crystals, or segments (4 at the top and 4 at the bottom). Light generated in each segment due to interaction with the incoming $\gamma$ ray is collected by three photomultiplier tubes (PMTs) as discussed in Section \ref{sec:sunminidssd}. These PMTs have to be gain-matched before adding their energies to get the total energy deposited in each SuN segment. This is achieved with a 1460 keV $\gamma$ ray that exists as a room background from the decay of $^{40}$K. Radioactive $^{40}$K is found in trace amounts in the concrete present in the walls of the experimental room. Figure \ref{fig:sunbg} shows the 1460 keV line in one of the SuN PMT. A Gaussian peak is fit onto a decaying background. Gain-matching factors for each PMT are chosen such that the centroids from the Gaussian fits of all PMTs match up. Once the PMTs are gain-matched, the energy deposition in each SuN segment can be determined. 

All of the 8 SuN segments still have to be calibrated to establish a relation between channel numbers and $\gamma$ energy. Four $\gamma$ lines - 59.5 keV from $^{241}$Am source, 662 keV from $^{137}$Cs source, and 1173 keV and 1332 keV from $^{60}$Co were used for SuN energy calibrations. Figure \ref{fig:suncalib} shows three of the four $\gamma$ lines from $^{137}$Cs and $^{60}$Co in one of the SuN segments. In addition, a known 4730 keV $\gamma$ line from the $\beta$-decay for $^{33}$Mg was used as a reference in the higher energy region. A quadratic calibration was performed with a very small quadratic factor of the order of $10^{-7}$. All the residuals were found to be less than 10 keV.

\bigskip
\Large \subsection{Particle Identification}
\bigskip
\large

Particle identification for this part of the experiment was similar to the NERO-BCS part as the beam delivery was the same and only the downstream experimental setup was changed. Beams of $^{33}$Mg for the experiment and $^{33}$Al and $^{32}$Al for the background estimation were delivered to the SuN-miniDSSD experimental setup. Refer to Section \ref{sec:pid} and Section \ref{sec:pidbcsnero} for more details on particle identification. 

\bigskip
\Large \subsection{Implant and Decay Correlations}
\bigskip
\large

Similar to the BCS DSSD, implantation and decay events have to be correlated in the miniDSSD too. The $\beta$-delayed $\gamma$ rays detected in SuN are expected to be a part of the decay event. Since the SuN PMTs have a very fast timing response, the physics event-building time window in the data acquisition system can be set to 2000 ns (compared to 200 $\mu$s for the NERO-BCS part required for neutron moderation). The shorter window also allows for background suppression as the SuN PMTs have relatively high count rates from cosmic ray events. 

Since the experimental setup is different, the conditions for a physics event to be classified as an implantation event and a decay event are also different. To be classified as an implantation event, a physics event must satisfy the following conditions:

\begin{enumerate}
    \large \item Energies in PIN1 and PIN2 are greater than zero. This ensures implantation from a beam particle.
    \large \item At least one of each of the front and back strips in the low gain channel of the miniDSSD registers an energy greater than zero. If multiple strips fire, the strip with the highest energy deposition is chosen as the position of the implant. 
    \large \item No energy deposition in the Si veto detector downstream of the miniDSSD. This condition helps discard punchthrough events as they shouldn't be correlated with decay events. 
\end{enumerate}
To be classified as a decay event, the following conditions must be satisfied:
\begin{enumerate}
    \large \item No energy deposition on either PIN1 or PIN2. 
    \large \item No signal in the low gain channels of the DSSD.  
    \large \item Energies in at least one of each of the front and back strips high-gain channel of the DSSD to be greater than the low energy thresholds applied in the software. 
    \large \item No energy deposition in the Si veto detector.  
\end{enumerate}

Implantation and decay events are temporally and spatially correlated using the same logic as that for the BCS DSSD. However, since the miniDSSD has smaller dimensions and only 16 strips on each side, a spatial correlation window of $1 \times 1$ was optimal, i.e. the implant and decay had to be in the same pixel to be successfully correlated. Along with a forward time run, the correlator was run in reverse time to get an estimation of the background from random correlations. 

\bigskip
\Large \subsection{$\beta$-delayed $\gamma$ Spectra}
\bigskip
\large

The SuN detector as a whole has a very high efficiency for detecting $\gamma$ rays. However, a $\gamma$ ray can deposit partial energies in multiple segments due to scattering across segments. Multiple $\gamma$ rays in a cascade from the same physics event can also be detected in different segments of SuN. As a result, two types of spectra are created with the data from SuN. A total absorption spectrum (TAS) represents the sum of total energy deposited in all segments in a single physics event. A singles spectrum, also known as Sum of Segments (SS), represents the energy deposited in a single segment. Spectra from all eight segments are added together to get the SS. Note that this summation is not on an event-by-event basis like the TAS but it is the summation of complete spectra. Additionally, the multiplicity spectrum (Mul) represents the number of segments fired in a single physics event. Together, these spectra constrain the complete $\gamma$ ray cascade from a physics event. 

$\gamma$ rays from only the decay events that are successfully correlated with implants form the SuN $\gamma$ spectra (TAS, SS, and Mul). However, these include background from random implant-decay correlations. To account for this, similar $\gamma$ spectra from correlated decay events with the reverse time correlator are subtracted from the forward correlated $\gamma$ spectra. This also removes the contribution from random $\gamma$ rays correlating with the decay events. Reverse correlator subtracted spectra still contain contributions from the decay of daughter ($^{33}$Al) and $\beta$-delayed neutron daughter ($^{32}$Al) nuclei. As a result, similar spectra are created with the $^{33,32}$Al beam implantations and are proportionately subtracted, depending on the number of their decays within the correlation time window as estimated with the Bateman Equation \ref{eqn:bateman}(modified to account for neutron branchings). This step also accounts for any granddaughter, neutron granddaughter, etc. decay as they will also be present in the daughter and neutron daughter decay, and be subtracted off. The resulting spectra are the experimental $\beta$-delayed $\gamma$ spectra for just the parent decay ($^{33}$Mg). The $\gamma$ rays in this spectra will correspond to excited states in $^{33}$Al and $^{32}$Al as shown in Figure \ref{fig:expt}. 

\bigskip
\LARGE \section{\texttt{GEANT4} Simulations of SuN}\label{sec:geant4}
\bigskip
\large

\begin{figure}
    \centering
    \includegraphics[width=500pt,keepaspectratio]{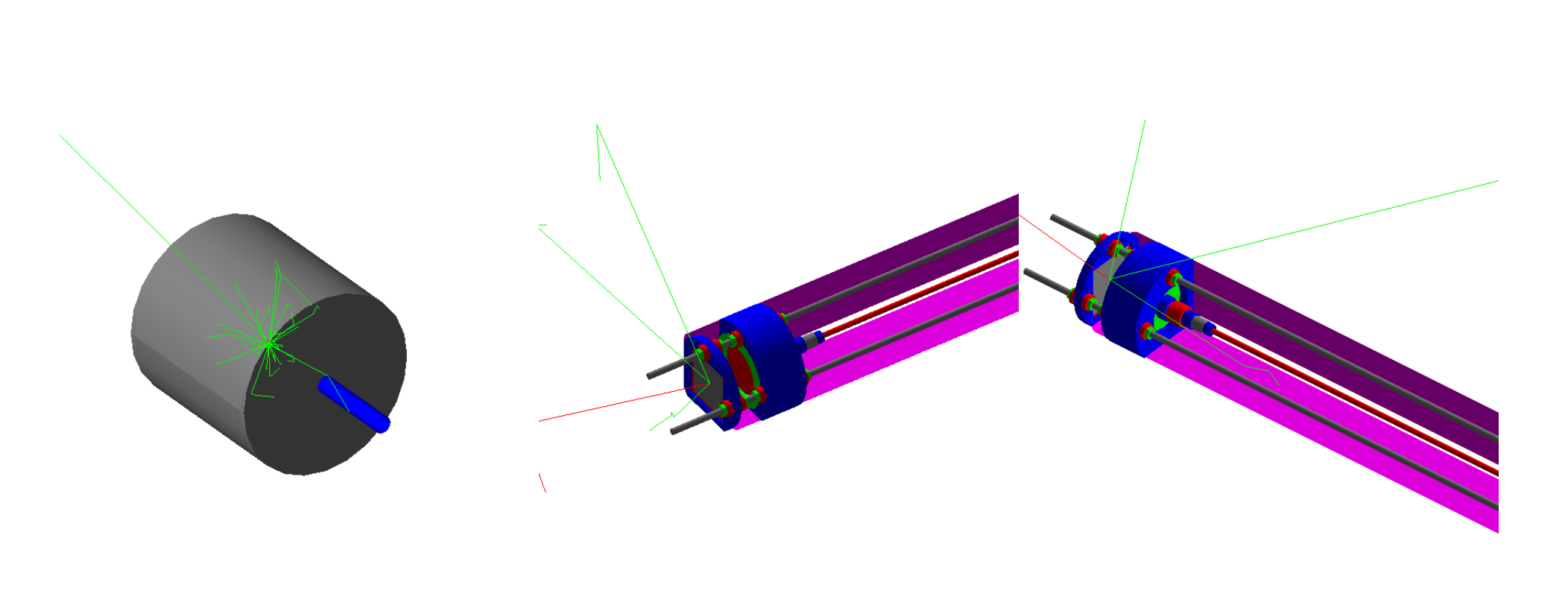}
    \caption{\large \texttt{GEANT4} visualization of SuN detector in the left panel and the miniDSSD implantation detector in the middle and the right panel. Green lines are for visualizing the $\gamma$ ray tracks whereas the red lines correspond to electron tracks. The green track outside the SuN detector volume represents partial energy deposition in the detector. Figure reproduced from \cite{Dombos2018}.}
    \label{fig:geant}
\end{figure}

$\gamma$ rays interact with the detector material in three principal ways: 
\begin{enumerate}
    \large \item Photoelectric absorption where the complete $\gamma$ energy is deposited in the detector material. 
    \large \item Compton Scattering where the $\gamma$ ray scatters off an electron and deposits partial energy in the detector material. 
    \large \item Pair production where the incoming $\gamma$ ray produces an electron-positron pair. The positron can further annihilate with an electron in the detector material and produce two 511 keV $\gamma$ rays. 
\end{enumerate}
The probability of each type of interaction depends on the incoming $\gamma$ ray energy. As a result, the SuN spectra corresponding to multiple $\gamma$ cascades from excited levels in the daughter nucleus ($^{33}$Al) cannot be read simply as counting the number of $\gamma$ rays with a particular energy. Instead, the detector response is simulated for each possible cascade, and a linear combination of such simulated spectra reproduces the experimental spectra. The weight of each simulated cascade at a particular excitation energy represents the $\beta$ feeding intensity at that excitation energy. 

The SuN detector response is simulated with the \texttt{GEANT4} package \cite{Agostinelli2003}. It is a versatile tool to model the interaction of any type of radiation with matter. It provides great flexibility for choosing detector materials and geometries, characterizing and tracking emitted radiation, recording total energy deposited in pre-defined volumes, etc. Previous works \cite{Simon2013,Quinn2015,Dombos2018} have explored the process of simulating SuN detector response in \texttt{GEANT4} in great detail and have successfully validated the SuN detector efficiency and resolution. 

\begin{figure}
    \centering
    \includegraphics[width=450pt,keepaspectratio]{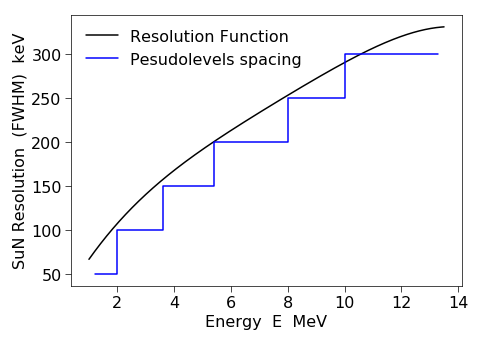}
    \caption{\large SuN photopeak energy resolution in keV calculated as a function of the incident $\gamma$ ray energy. The resolution function used is from \cite{Dombos2018}. Also shown here is the pesudolevels spacing used for \texttt{RAINIER} and \texttt{GEANT4} simulations. In these simulations, each pseudolevel represents the contribution from all the energy levels within that energy bin. As long as the spacing is smaller than SuN resolution, the resulting simulated spectra from any level and its corresponding pesudolevel should be nearly identical.}
    \label{fig:sunresolution}
\end{figure}

A collection of simulated spectra (TAS, SS, and Mul) for the feeding of the daughter nucleus at a single excitation energy is referred to as a template. All the excited levels in the daughter nucleus will have their own templates. At low energies, templates are created for discrete energy levels depending on the $\gamma$ branching cascades from that level. These can either be known from previous measurements or determined from the experimental dataset using a combination of TAS and SS as discussed in Section \ref{sec:structure}. At higher energies, especially above the neutron separation energy, the level density becomes high and it is impractical to create a template for all levels. In that region, referred to as the quasi-continuum, a statistical approach is used. Pseudolevels are placed at finite intervals that represent the contribution from all levels in that interval. The spacing between pesudolevels should be less than the SuN energy resolution so that templates from an energy level and its corresponding pesudolevel are nearly identical. Figure \ref{fig:sunresolution} shows the SuN energy resolution function and the pseudolevel spacings used in this work. The spacings used are 50 keV up to 2 MeV, 100 keV up to 3.9 MeV, 150 keV up to 5.4 MeV, 200 keV up to 8 MeV, 250 keV up to 10 MeV, and 300 keV up to 14.5 MeV. The $\gamma$ branchings from pseudolevels are calculated using \texttt{RAINIER} as described in Section \ref{sec:rainier}. 

\bigskip
\Large \subsection{Particle simulations in SuN}
\bigskip
\large

\begin{figure}
    \centering
    \includegraphics[width=450pt,keepaspectratio]{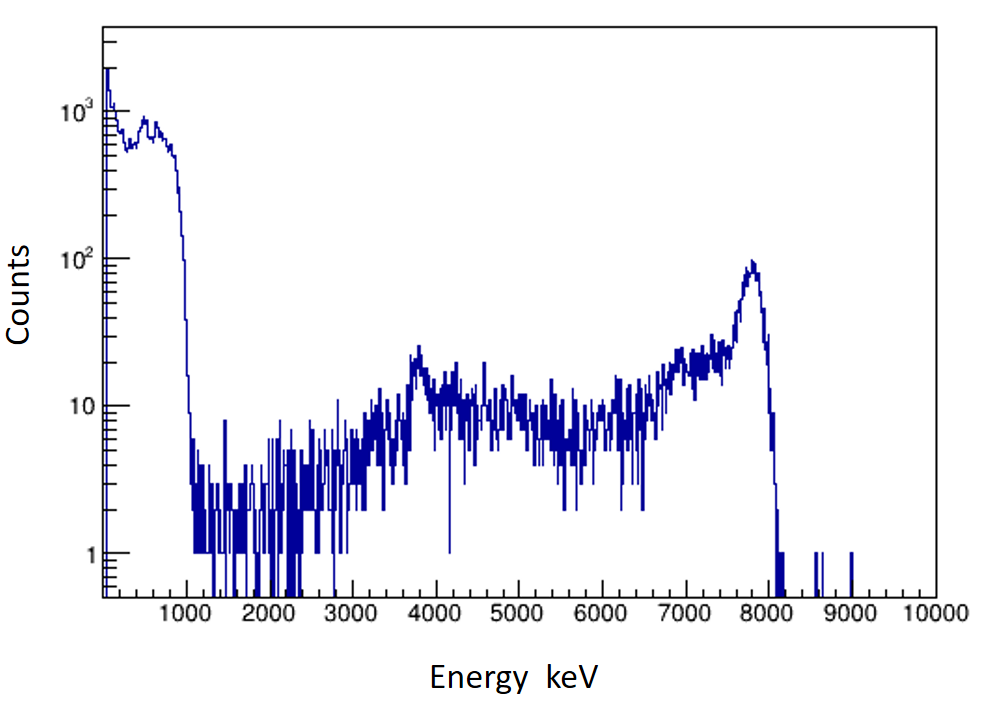}
    \caption{\large \texttt{GEANT4} simulated SuN Total Absorption Spectrum for monoenergetic 1 MeV neutrons emitted isotropically from the center of the detector. The inelastic scattering peaks on $^{127}$I and $^{23}$Na and the 7826 keV neutron capture peak on $^{127}$I (Q-value = 6826 keV) can be identified. }
    \label{fig:sunneutrons}
\end{figure}

While the SuN detector is primarily designed to detect $\gamma$ rays, other particles produced in a decay event such as an electron, a neutron, or a neutrino can interact with the SuN detector and change the observed spectra. As a result, these interactions have to be accounted for in the \texttt{GEANT4} simulated spectra. The neutrinos have a negligible chance of any interactions and can be ignored. The electrons and neutrons interactions, however, deposit additional energy in SuN and shift the observed spectra. The low-energy electrons up to $\sim$400 keV are stopped in the miniDSSD and electrons up to energies of $\sim$1 MeV are stopped in the beam pipe. However, the Q-value of the $^{33}$Mg $\beta$-decay is 13.477 MeV, and electron energies can go as high as that. Moreover, higher-energy electrons will lose less energy in the miniDSSD and the beam pipe and deposit an even higher fraction of their total energy in SuN, shifting the resulting spectra considerably. This is accounted for in \texttt{GEANT4} simulations of SuN. 

The neutron interaction with SuN is a bit more complex. In addition to depositing their kinetic energy in the detector through elastic scattering, neutrons can also undergo inelastic scattering via $^{23}$Na(n,n'$\gamma$)$^{23}$Na or $^{127}$I(n,n'$\gamma$)$^{127}$I as well as neutron capture on $^{127}$I(n,$\gamma$)$^{128}$I. Figure \ref{fig:sunneutrons} shows the simulated total absorption spectrum for monoenergetic 1 MeV neutrons. The 58 keV peak corresponds to $\gamma$ rays from inelastic scattering on $^{127}$I and the 7.8 MeV peak is due to $\gamma$ rays from neutron capture on $^{127}$I. 7.8 MeV corresponds to the Q-value of $^{127}$I(n,$\gamma$)$^{128}$I (6.826 MeV) plus the 1 MeV neutron energy. Although SuN detector does detect neutrons, the efficiency is not high and the neutron capture peak is submerged in the $\gamma$ rays from the quasi-continuum. Overall, the neutron response has to be accounted for in the analysis but it is not well-benchmarked and hence, NERO data is used to constrain the $\beta$-delayed neutron branching. 

\bigskip
\LARGE \section{\texttt{RAINIER} Simulations}\label{sec:rainier}
\bigskip
\large

\texttt{RAINIER} \cite{Kirsch2018} is a software tool used for a probabilistic determination of the $\gamma$-ray branchings from a pseudolevel in the quasi-continuum while accounting for level, width, and cascade fluctuations. It takes user-defined energy, spin, and parity of the psesudolevel as inputs and creates $\gamma$ branching cascades. To do this, it requires a prescription for level density (LD) and $\gamma$ strength function (GSF). These are discussed in the following sections. 

\bigskip
\Large \subsection{Level Density Parameters}\label{sec:ld}
\bigskip
\large

\begin{figure}
    \centering
    \includegraphics[width=350pt,keepaspectratio]{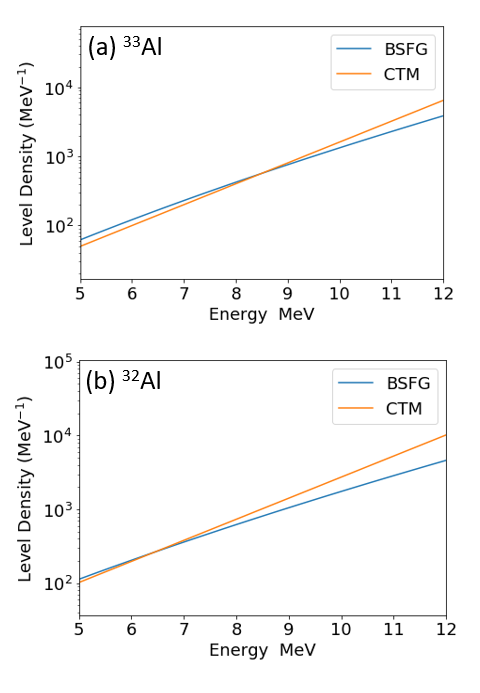}
    \caption{\large Comparison of level densities per MeV calculated using the Back Shifted Fermi Gas (BSFG) model and the Constant Temperature Model (CTM) as described in Section \ref{sec:ld} for both $^{33}$Al and $^{32}$Al. }
    \label{fig:ld}
\end{figure}

\begin{table}[]
    \centering
    \begin{tabular}{ccccc}
    \toprule
       \multicolumn{1}{c}{\large}  & \multicolumn{2}{c}{\large BSFG} & \multicolumn{2}{c}{\large CTM} \\
    \cmidrule(lr){2-3} \cmidrule(lr){4-5}
       \large & \large $a$ (MeV$^{-1}$) & \large $E_1$ (MeV) & \large $T$ (MeV) & \large$E_0$ (MeV) \\
    \midrule
      \large $^{33}$Al & \large 4.96 & \large -0.51 & \large 1.44 & \large -1.13  \\
      \large $^{32}$Al & \large 4.59 & \large -2.05 & \large 1.52 & \large -2.67 \\
    \bottomrule
    \end{tabular}
    \caption{\large Level Density parameters for the Back Shifted Fermi Gas (BSFG) model and the Constant Temperature Model (CTM) calculated as described in Section \ref{sec:ld} for both $^{33}$Al and $^{32}$Al.}
    \label{tab:ld}
\end{table}

No experimental data exists for the level densities in $^{33,32}$Al. As a result, an empirical prescription from \cite{Egidy2005,vonEgidy2009} was used to determine the level density prescriptions. It is based on extrapolations from nuclei where experimental data exists. Two models were used for the description of level density - the Back-Shifted Fermi Gas model (BSFG) and the Constant Temperature Model (CTM). The level density as a function of excitation energy for a nucleus with $Z$ protons and $N$ neutrons for BSFG is given by 
\begin{equation}\label{eqn:bsfg}
    \large \rho(E) = \frac{e^{2\sqrt{a(E-E_1)}}}{12\sqrt{2}\sigma a^{1/4}(E-E_1)^{5/4}}
\end{equation}
where $a$ is the level density parameter, $E_1$ is the energy backshift, and $\sigma$ is the spin cut-off parameter. The level density for CTM is given by 
\begin{equation}
    \large \rho(E) = \frac{1}{T}e^{(E-E_0)/T}
\end{equation}
where $T$ is the temperature, and $E_0$ is the energy backshift. All the parameters in both models are calculated as follows. 
\begin{subequations}
\begin{gather}
    \large a = (0.199 + 0.0096S')A^{0.869} \\
    \large E_1 = -0.381 + 0.5P_a' \\
    \large \sigma^2 = 0.391A^{0.675}(E-0.5P_a')^{0.312} \\
    \large T = A^{-2/3}/(0.0597 + 0.00198S') \\
    \large E_0 = -1.004 + 0.5P_a'
\end{gather}
\end{subequations}
where $A$ is the atomic mass number, $S'$ is the modified shell correction factor and $P_a'$ is the deutron pairing energy. $S'$ is related to $P_a'$ as 
\begin{equation}
    \large S' = S + 0.5P_a'
\end{equation}
where S is the shell correction factor. It is the difference between the experimental value for the mass excess and the theoretical prediction by the liquid drop model that doesn't account for shell effects. 
\begin{equation}
    \large S = M_{exp} - M_{LD}
\end{equation}
The mass excess for the liquid drop model is related to the liquid drop binding energy as 
\begin{equation}
    \large M_{LD} = N*M_n + Z*M_p - BE_{LD}
\end{equation}
which is given by 
\begin{equation}
    \large -\frac{BE_{LD}}{A} = a_{vol} + a_{sf}A^{-1/3} + \frac{3e^2}{5r_0}Z^2A^{-4/3} + (a_{sym} + a_{ss}A^{-1/3}) \bigg ( \frac{N-Z}{A} \bigg )^2 
\end{equation}
The constants for the liquid drop model are - 
$a_{vol}$ = -15.65 MeV,
$a_{sf}$ = 17.63 MeV,
$a_{sym}$ = 27.72 MeV,
$a_{ss}$ = -25.60 MeV, and
$r_0$ = 1.233 fm. Finally, the deutron pairing energy can be written in terms of mass excess as
\begin{equation}
    \large P_a' (A,Z) =  \frac{1}{2} \big[ M(A+2,Z+1) - 2M(A,Z) + M(A-2,Z-1) \big ]
\end{equation}
Table \ref{tab:ld} lists the calculated level density parameters that are provided in the \texttt{RAINIER} input file and Figure \ref{fig:ld} shows the calculated level densities for BSFG and CTM for both $^{33}$Al and $^{32}$Al. 

\bigskip
\Large \subsection{$\gamma$-ray Strength Function Parameters}\label{sec:gsf}
\bigskip
\large

\begin{figure}
    \centering
    \includegraphics[width=350pt,keepaspectratio]{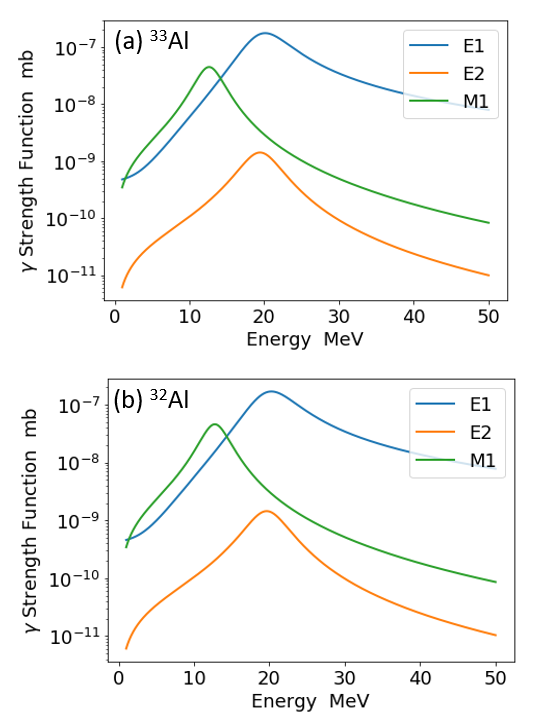}
    \caption{\large $\gamma$-ray strength function in mb for the E1, M1, and E2 transitions calculated as described in Section \ref{sec:gsf} for both $^{33}$Al and $^{32}$Al. E1 is calculated using Generalized Lorentzian functional form whereas E2 and M1 are calculated using Standard Lorentzian functional form.}
    \label{fig:gsf}
\end{figure}

\begin{table}[]
    \centering
    \begin{tabular}{ccccccc}
    \toprule
       \multicolumn{1}{c}{\large}  & \multicolumn{3}{c}{\large $^{33}$Al} & \multicolumn{3}{c}{\large $^{32}$Al} \\
    \cmidrule(lr){2-4} \cmidrule(lr){5-7}
       \large & \large Energy E & \large Width $\Gamma$ & \large Strength $\sigma$ & \large Energy E & \large Width $\Gamma$ & \large Strength $\sigma$ \\
       \large & \large (MeV) & \large (MeV) & \large (mb) & \large (MeV) & \large (MeV) & \large (mb) \\
    \midrule
      \large E1 & \large 21.23 & \large 8.90 & \large 40.57 & \large 21.39 & \large 9.03 & \large 39.19  \\
      \large M1 & \large 12.78 & \large 4.00 & \large 6.59 & \large 12.91 & \large 4.00 & \large 6.81 \\
      \large E2 & \large 19.64 & \large 5.71 & \large 0.53 & \large 19.84 & \large 5.73 & \large 0.55 \\
    \bottomrule
    \end{tabular}
    \caption{\large Energy, width, and strength $\gamma$-ray strength function parameters for the E1, M1, and E2 transitions calculated as described in Section \ref{sec:gsf} for both $^{33}$Al and $^{32}$Al.}
    \label{tab:gsf}
\end{table}

The $\gamma$ strength functions (GSF) for different transitions were calculated similarly to the prescription used in \texttt{TALYS} \cite{Koning2023}. GSF for type $X$ (E or M) and multipole ($\ell$) is modeled as a Standard Lorentzian given by 
\begin{equation}\label{eqn:gsf}
    \large f_{X\ell} (E_{\gamma}) = K_{X\ell}\frac{\sigma_{X\ell}E_{\gamma}\Gamma_{X\ell}^2}{(E_{\gamma}^2-E_{X\ell}^2)^2 + E_{\gamma}^2\Gamma_{X\ell}^2}
\end{equation}
where 
\begin{equation}
    \large K_{X\ell} = \frac{1}{(2\ell+1)\pi^2\hbar^2c^2}
\end{equation}
and $E_{X\ell}$ is the energy of the resonance, $\Gamma_{X\ell}$ is the width of the resonance, and $\sigma_{X\ell}$ is the strength of the resonance. For a nucleus with mass number A and atomic number Z, the parameters for E1 transition are given by
\begin{subequations}
\begin{gather}
    \large E_{E1} = 31.2A^{-1/3} \hspace{3pt} + \hspace{3pt} 20.6A^{-1/6} \hspace{5pt} \text{MeV} \\ \Gamma_{E1} = 0.026E_{E1}^{1.91} \hspace{5pt} \text{MeV} \\
    \sigma_{E1} = 1.2 \times 120NZ/(A\pi\Gamma_{E1}) \hspace{5pt} \text{mb}
\end{gather}
\end{subequations}
The parameters for E2 transition are given by
\begin{subequations}
\begin{gather}
    \large E_{E2} = 63A^{-1/3} \hspace{5pt} \text{MeV} \\ 
    \Gamma_{E2} = 6.11 - 0.012A \hspace{5pt} \text{MeV} \\ 
    \sigma_{E2} = 0.00014Z^2E_{E2}/(A^{-1/3}\Gamma_{E2}) \hspace{5pt} \text{mb}
\end{gather}
\end{subequations}
For the M1 transition, the energy and width are given as
\begin{subequations}
\begin{gather}
    \large E_{M1} = 41A^{-1/3} \hspace{5pt} MeV \\ \Gamma_{M1} = 4.0 \hspace{5pt} MeV
\end{gather}
The strength for M1 is back-calculated using Equation \ref{eqn:gsf} from the value of GSF at 7 MeV given by
\begin{gather}
    \large f_{M1} \text{(7 MeV)} = 1.58 \times 10^{-9}A^{0.47} 
\end{gather}
\end{subequations}

These three are the only types of electromagnetic transitions considered for the GSF. E2 and M1 are calculated using the Standard Lorentzian functional form described in Equation \ref{eqn:gsf} whereas E1 uses the Generalized Lorentzian functional form \cite{Kopecky1990} given by
\begin{equation}
    \large f_{E1}(E_{\gamma},T) = K_{E1} \bigg [ \frac{E_{\gamma}\tilde{\Gamma}_{E1}(E_{\gamma})}{(E_{\gamma}^2+E_{E1}^2)^2 + E_{\gamma}^2\tilde{\Gamma}_{E1}(E_{\gamma})^2} + \frac{0.7\Gamma_{E1}4\pi^2T^2}{E_{E1}^5}   \bigg]\sigma_{E1}\Gamma_{E1}
\end{equation}
where the energy-dependent damping width $\tilde{\Gamma}_{E1}(E_{\gamma})$ is given by
\begin{equation}
    \large \tilde{\Gamma}_{E1}(E_{\gamma}) = \Gamma_{E1}\frac{E_{\gamma}^2+4\pi^2T^2}{E_{E1}^2}
\end{equation}
and T is the nuclear temperature given by 
\begin{equation}
    \large T = \sqrt{\frac{E-\Delta-E_{\gamma}}{a}}
\end{equation}
where $E$ is total excitation energy, $\Delta$ is the pairing correction, and $a$ is the level density parameter. Table \ref{tab:gsf} lists the calculated GSF parameters for E1, M1, and E2 transitions for both $^{33,32}$Al. Figure \ref{fig:gsf} shows the actual GSF as a function of $\gamma$ energy. It can be confirmed that E1 is the strongest transition followed by M1, followed by E2.

\huge \chapter{Experimental Results}\label{chp:results}
\large 

This chapter lists the results from the experimental measurement of the ground state to ground state $\beta$-decay transition strength of $^{33}$Mg for both, the BCS-NERO and the SuN-miniDSSD parts of the experiment. It includes the half-life measurements, $\beta$-delayed neutron branching ratio measurement, and the $\beta$ feeding intensities into the excited levels of $^{33}$Al. The corresponding nuclear log-ft values and Gamow-Teller transition strengths are also listed. It also includes a section on new levels proposed in $^{33}$Al based on the experimental total absorption spectrum (TAS). Finally, the results from implementing two different nuclear level density models for treating the quasi-continuum in $^{33,32}$Al are discussed.  

\bigskip
\LARGE \section{Half-life Measurements}
\bigskip
\large

\begin{table}[]
    \centering
    \begin{tabular}{cccc}
    \toprule
        \large  & \large NNDC Reported & \large BCS-NERO Dataset  & \large SuN-miniDSSD Dataset \\
       \large Isotope & \large T$_{1/2}$ (ms) & \large T$_{1/2}$ (ms) & \large T$_{1/2}$ (ms) \\
    \midrule
      \large $^{33}$Mg  & \large 89.4(10) & \large 92.0(4) & \large 93.2(6)  \\
      \large $^{33}$Al & \large 41.7(2) & \large 42.0(2) & \large 42.6(3) \\
      \large $^{32}$Al & \large 31.9(8) & \large 31.7(1)  & \large 31.5(1)  \\
    \bottomrule
    \end{tabular}
    \caption{\large Experimentally measured half-lives for $^{33}$Mg, $^{33,32}$Al in this work compared to the half-lives reported by the National Nuclear Data Center (NNDC) based on previous measurements.}
    \label{tab:halflives}
\end{table}

\begin{figure}
    \centering
    \includegraphics[width=450pt,keepaspectratio]{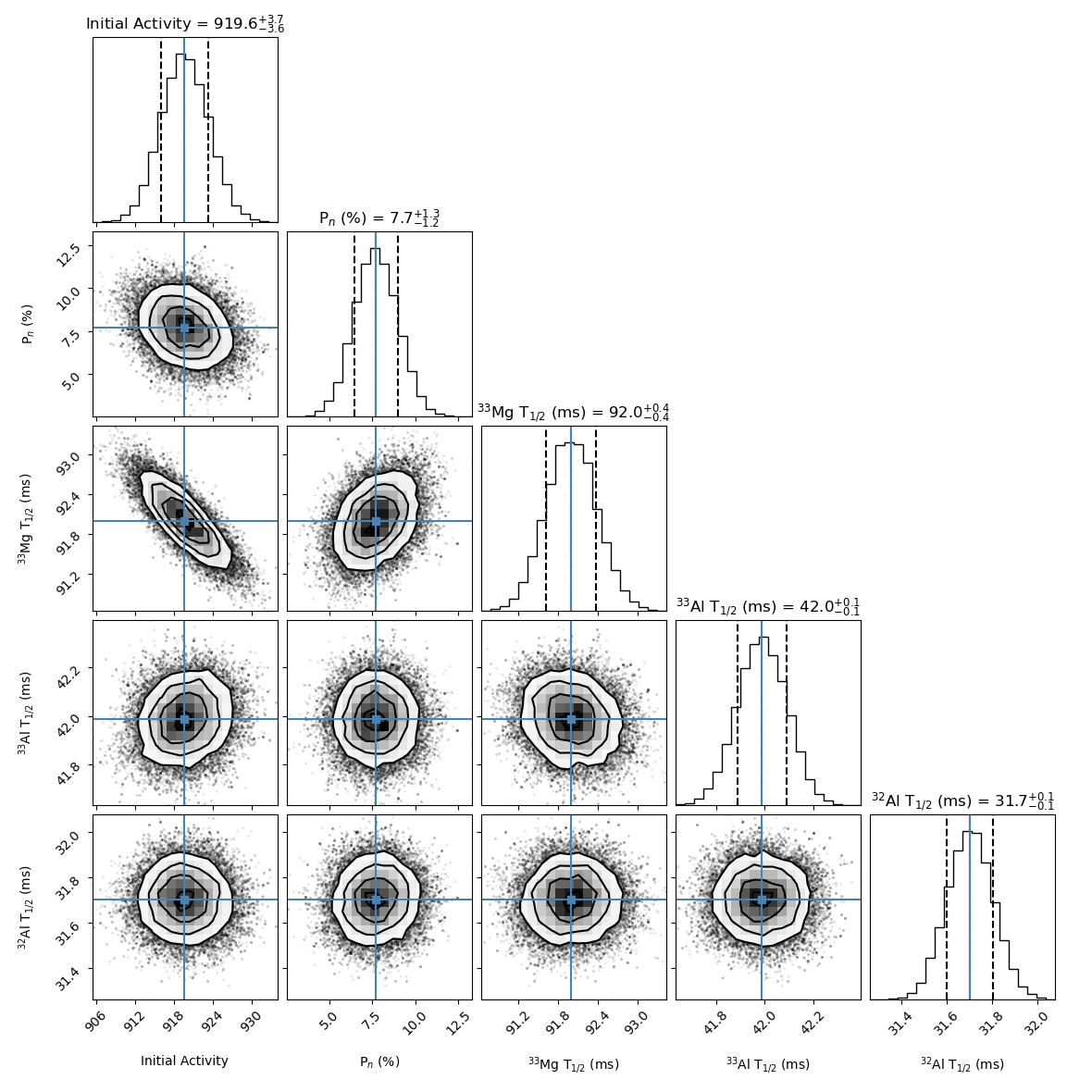}
    \caption{\large Corner plot for posterior distributions from the $^{33}$Mg decay curve fit for the BCS-NERO dataset. Since the posteriors are nearly Gaussian, the median is reported to be the parameter fit whereas the 16$^\text{th}$ and the 84$^\text{th}$ percentile constitute the 1-$\sigma$ uncertainty. }
    \label{fig:decay33mgnero}
\end{figure}

\begin{figure}
    \centering
    \includegraphics[width=450pt,keepaspectratio]{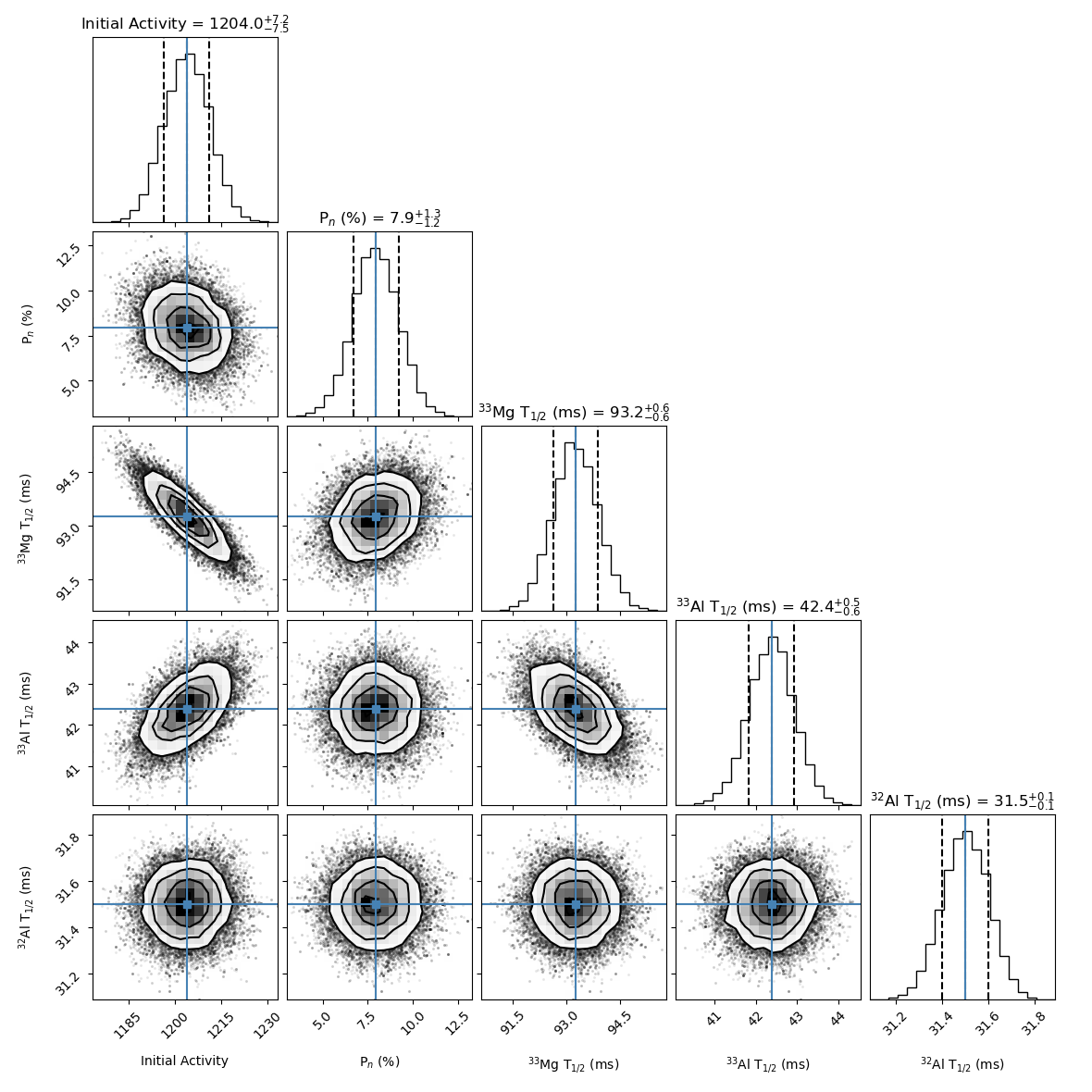}
    \caption{\large Corner plot for posterior distributions from the $^{33}$Mg decay curve fit for the SuN-miniDSSD dataset. Since the posteriors are nearly Gaussian, the median is reported to be the parameter fit whereas the 16$^\text{th}$ and the 84$^\text{th}$ percentile constitute the 1-$\sigma$ uncertainty. }
    \label{fig:decay33mgsun}
\end{figure}

\begin{figure}
    \centering
    \includegraphics[width=450pt,keepaspectratio]{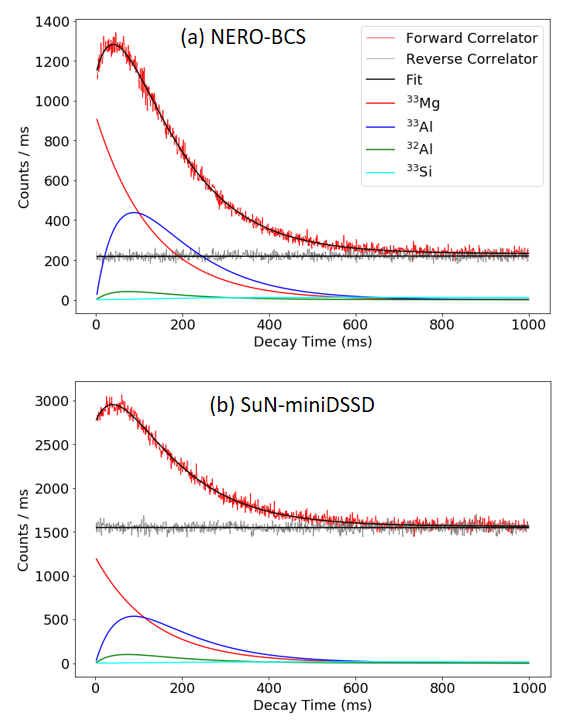}
    \caption{\large $^{33}$Mg decay curve fits for the (a)NERO-BCS and the (b)SuN-miniDSSD datasets. Half-lives are reported in Table \ref{tab:halflives}. }
    \label{fig:hf33mg}
\end{figure}

\begin{figure}
    \centering
    \includegraphics[width=450pt,keepaspectratio]{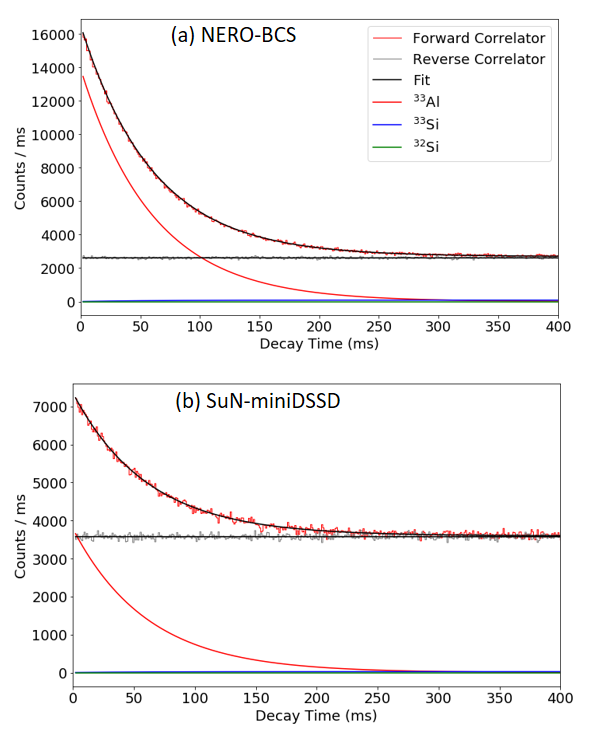}
    \caption{\large $^{33}$Al decay curve fits for the (a)NERO-BCS and the (b)SuN-miniDSSD datasets. Half-lives are reported in Table \ref{tab:halflives}.}
    \label{fig:hf33al}
\end{figure}

\begin{figure}
    \centering
    \includegraphics[width=450pt,keepaspectratio]{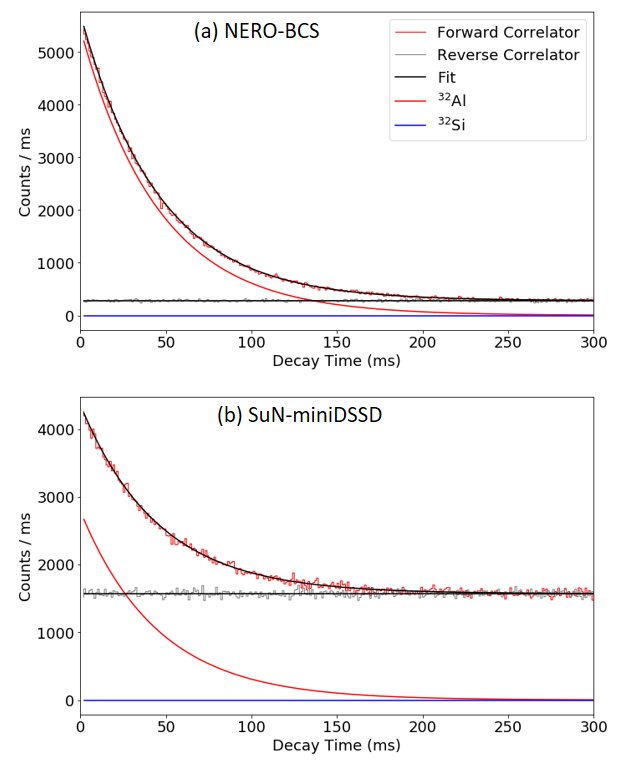}
    \caption{\large $^{32}$Al decay curve fits for the (a)NERO-BCS and the (b)SuN-miniDSSD datasets. Half-lives are reported in Table \ref{tab:halflives}. }
    \label{fig:hf32al}
\end{figure}

Following the procedure outlined in Section \ref{sec:correlation}, the decay curve for $^{33}$Mg implants is generated for a correlation time window of 1000 ms. A spatial correlation window of $3 \times 3$ is used for the BCS DSSD and $1 \times 1$ for the miniDSSD. Figure \ref{fig:hf33mg} shows the decay curves for both. The characteristic shape of the decay curves with a hump at early times is a feature of Bateman Equation \ref{eqn:bateman} and can be attributed to the fact that the daughter $^{33}$Al has a shorter half-life than the parent $^{33}$Mg. The SuN-miniDSSD dataset has a high rate of background since the miniDSSD is small so the incoming beam is focused on a smaller region, which increases the chances of random coincidences. The decay curves for $^{33}$Mg contain decays from the following components:
\begin{enumerate}
    \large \item Decay of parent $^{33}$Mg.
    \large \item Decay of daughter $^{33}$Al.
    \large \item Decay of neutron daughter $^{32}$Al. 
    \large \item Decay of granddaughter $^{33}$Si. 
    \large \item Flat background from random implant-decay correlations. 
\end{enumerate}
The half-lives of $^{33,32}$Al were experimentally determined using the same setup and technique (See Figures \ref{fig:hf33al} and \ref{fig:hf32al}). The P$_n$ value for $^{33}$Mg was determined from the NERO dataset as reported in Section \ref{sec:pnresults}. The rate of random background coincidences was estimated from the time reverse correlated decay curve over the same correlation time window. With all these components included in the Bateman Equation \ref{eqn:bateman} model (modified to allow for the neutron branchings), a multi-parameter Bayesian fit was performed to fit the model to the experimental decay curve. Posterior distributions of the fit parameters for the BCS-NERO dataset and the SuN-miniDSSD dataset can be seen in the corner plots in Figures \ref{fig:decay33mgnero} and \ref{fig:decay33mgsun}, respectively. Gaussian priors were chosen for the P$_n$ value and the half-lives of $^{33,32}$Al based on experimental measurements in this work. Uniform non-informative priors were chosen for the half-life of $^{33}$Mg and the initial activity (normalization factor). Half-life measurements for $^{33}$Mg from the fits from both the datasets are reported in Table \ref{tab:halflives}. These are in agreement with each other but slightly larger than the $^{33}$Mg half-life reported on NNDC based on previous measurements \cite{Tripathi2008}. 

A similar procedure was also performed for $^{33,32}$Al beam implantations. The decay curves and the fits for these are shown in Figure \ref{fig:hf33al} for $^{33}$Al and Figure \ref{fig:hf32al} for $^{32}$Al. The decay curve for $^{33}$Al had contributions from the parent ($^{33}$Al), daughter ($^{33}$Si), and neutron daughter ($^{32}$Si) decay, whereas the decay curve for $^{32}$Al had contributions from just the parent ($^{32}$Al) and the daughter ($^{32}$Si) decay. Half-life measurements for both these isotopes for both datasets are also reported in Table \ref{tab:halflives} along with those reported by the NNDC based on previous measurements \cite{Tripathi2008,Morton2002,Han2017}. The half-life measurements for $^{33}$Al and $^{32}$Al are consistent with previous measurements but the measured half-lives for $^{33}$Mg from both the datasets are higher than the previous measurement \cite{Tripathi2008} by more than 2 standard deviations.   

\bigskip
\LARGE \section{P$_n$ Values}\label{sec:pnresults}
\bigskip
\large

\begin{table}[]
    \centering
    \begin{tabular}{ccc}
    \toprule
       \large &\large NNDC Reported & \large NERO Dataset  \\
       \large Isotope &\large P$_n$ (\%) & \large P$_n$ (\%)  \\
    \midrule
      \large $^{33}$Mg & \large 14(2) & \large 7.9(13)\%  \\
      \large $^{33}$Al & \large 8.5(7) & \large 9.0(13)\%  \\
      \large $^{32}$Al & \large 0.7(5) & \large 0.42(9)\%  \\
    \bottomrule
    \end{tabular}
    \caption{\large Experimentally measured P$_n$ values for $^{33}$Mg, $^{33,32}$Al with the BCS-NERO dataset. Measurements for $^{33,32}$Al listed here were also used in the determination of the P$_n$ value for $^{33}$Mg. }
    \label{tab:pn}
\end{table}

The experimental $\beta$-delayed neutron branching ratio (P$_n$ value) is determined by counting the number of neutrons in the decay events correlated with implants. Section \ref{sec:pn} outlines the procedure to determine the P$_n$ value while accounting for various sources of background. Table \ref{tab:pn} lists the experimentally determined P$_n$ values for $^{33}$Mg, and $^{33,32}$Al along with those reported by NNDC. The P$_n$ value measurements for $^{33,32}$Al are consistent with the literature values reported by the National Nuclear Data Center (NNDC). The P$_n$ value measurement of 7.9(13)\% for $^{33}$Mg is significantly lower than the literature value of 14(2) \% reported by NNDC \cite{Angelique2006}. 

The major contribution to the reported experimental uncertainty comes from the NERO efficiency uncertainty. The NERO efficiency is fairly constant up to neutron energies of $\sim$ 2 MeV. The uncertainty on NERO efficiency, (with the assumption a flat efficiency) increases if higher energy neutrons are expected \cite{Pereira2010}. Although the energies of the $\beta$-delayed neutrons can range anywhere from 0 to Q$_{\beta}$ - S$_n$ $\approx$ 8 MeV, the expected neutron energies as per the neutron branchings reported in \cite{Angelique2006} are less than 2.5 MeV, so a flat efficiency can be used. Nonetheless, the NERO efficiency uncertainty still contributes the most to the uncertainty on the P$_n$ value measurement.

\bigskip
\LARGE \section{Nuclear Structure of $^{33}$Al}\label{sec:structure}
\bigskip
\large 


\begin{figure}
    \centering
    \includegraphics[width=470pt,keepaspectratio]{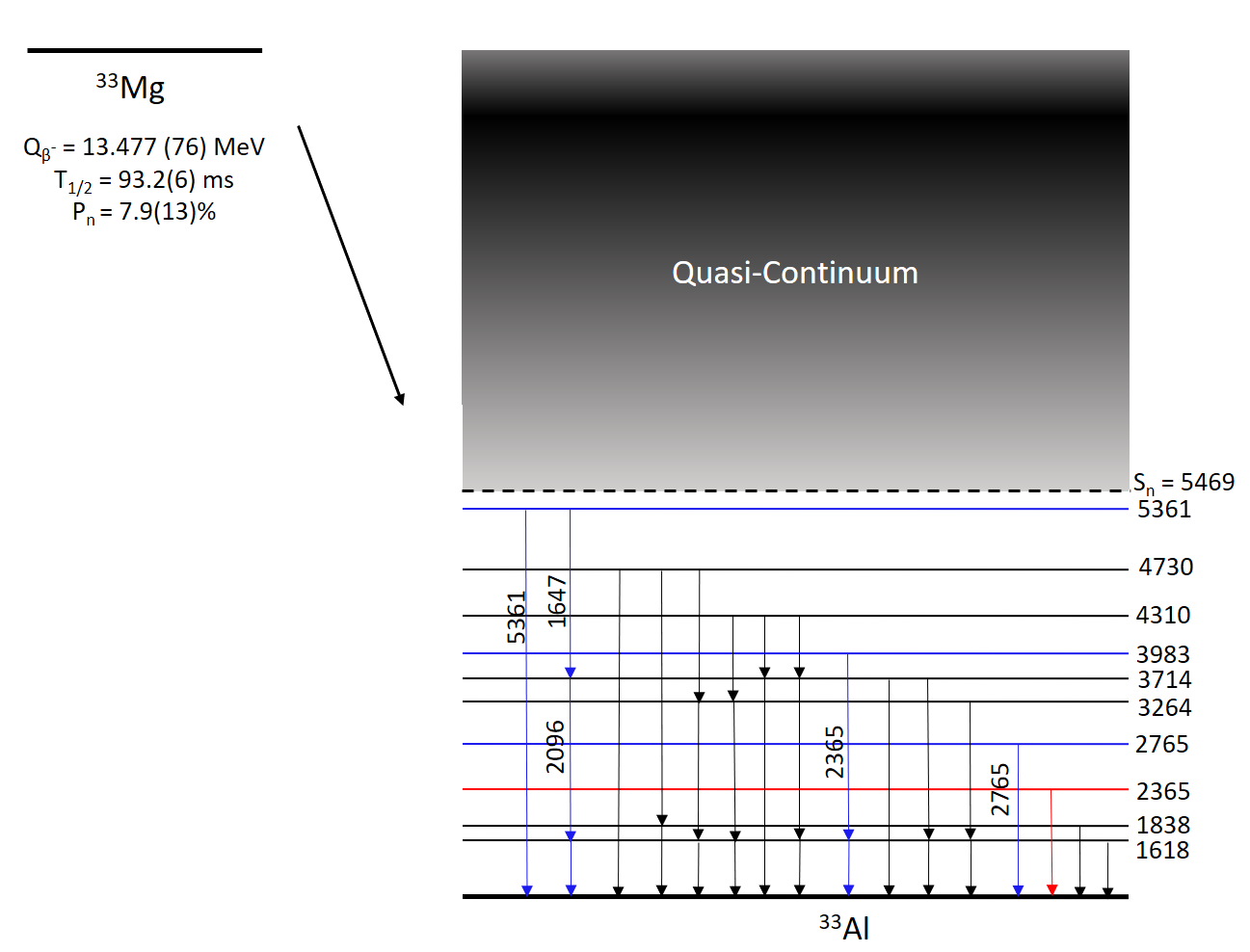}
    \caption{\large Nuclear scheme for $^{33}$Al from the $\beta$-decay of $^{33}$Mg constructed in this work. This is compared to the current level scheme reported by NNDC based on two experimental results by Ang\'{e}lique et al. and Tripathi et al. \cite{Angelique2006,Tripathi2008}. Red levels and $\gamma$-rays are deleted from the current level scheme and blue levels and $\gamma$-rays are new additions to the current level scheme based on this work. The black levels and $\gamma$-rays from the current level scheme were left unchanged. }
    \label{fig:newlevels}
\end{figure}

\begin{figure}
    \centering
    \includegraphics[width=450pt,keepaspectratio]{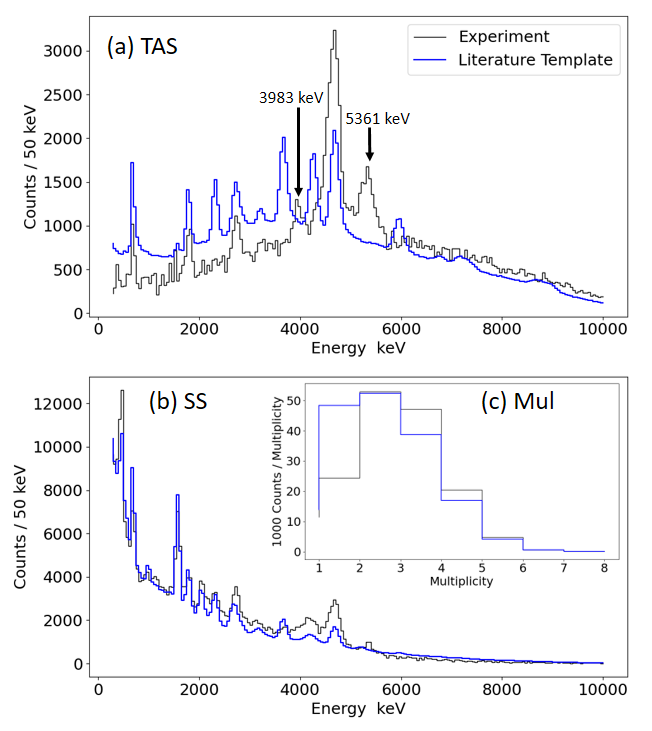}
    \caption{\large Comparison of the $\beta$-delayed $\gamma$ experimental template from the $\beta$-decay of $^{33}$Mg consisting of the Total Absorption Spectrum (TAS), Singles Spectrum (SS), and the multiplicity (Mul) from the SuN-miniDSSD dataset with the literature template obtained from \texttt{GEANT4} simulations of the SuN detector based on $\beta$-feeding intensities reported by the National Nuclear Data Center (NNDC). Two new levels identified in $^{33}$Al are highlighted in the TAS.}
    \label{fig:literature}
\end{figure}

The nuclear structure of $^{33}$Al as reported by the NNDC is based on two experimental results \cite{Angelique2006,Tripathi2008} from the $\beta$-decay of $^{33}$Mg with high-resolution $\gamma$-spectroscopy. These works also report the $^{33}$Mg $\beta$ feeding intensities into excited states in $^{33}$Al. Based on this information, a $\gamma$ template consisting of total absorption spectrum (TAS), singles spectrum (SS), and multiplicity (Mul) is generated with \texttt{GEANT4} simulations of the SuN detector and compared with the experimental template obtained with the SuN-miniDSSD dataset. Figure \ref{fig:literature} shows the comparison. The experimental TAS has a few peaks missing and also has two new peaks previously unreported. SS matches well at low energies of up to $\sim$ 2.5 MeV but not quite at higher energies. The literature template has way more events with a single multiplicity than seen in the experimental spectrum. This is the impact of the Pandemonium effect. Based on this information, a few changes were made to the level scheme in $^{33}$Al before extracting the experimental $\beta$-feeding intensities from the SuN-miniDSSD dataset. Figure \ref{fig:newlevels} show a summary of these changes.  

\begin{figure}
    \centering
    \includegraphics[width=350pt,keepaspectratio]{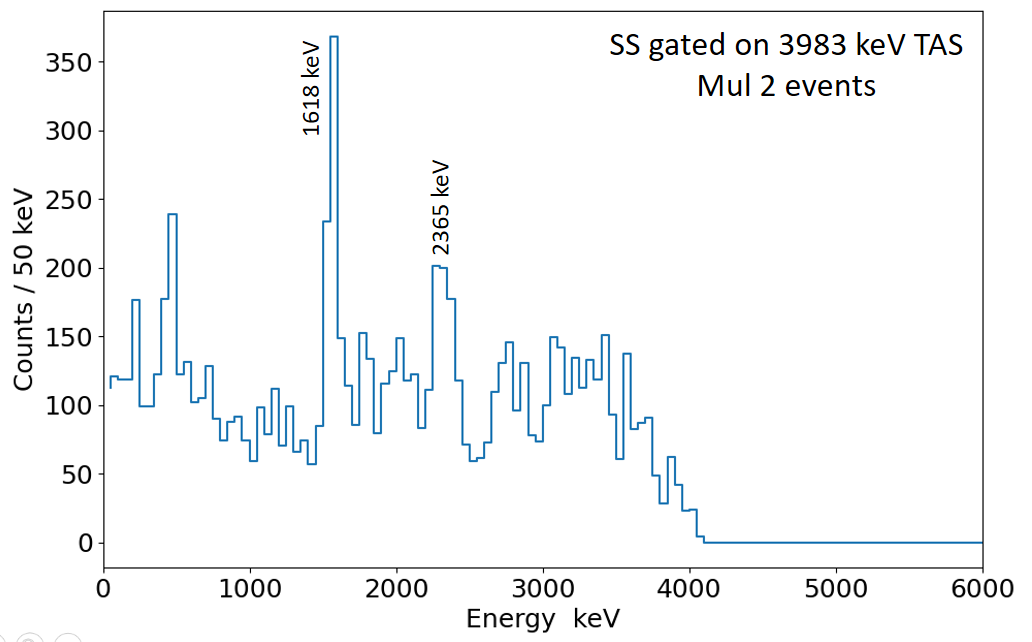}
    \caption{\large Singles Spectrum (SS) gated on the 3983 keV peak in the Total Absorption Spectrum and multiplicity (Mul) 2 events for the $^{33}$Mg $\beta$-delayed $\gamma$ rays from the SuN-miniDSSD dataset. Decay of the new proposed 3983 keV level via the 1618 keV level in $^{33}$Al can be confirmed.}
    \label{fig:gs3983}
\end{figure}

\bigskip
\Large \subsection{2365 and 3983 keV states}
\bigskip
\large 

NNDC reports a 7\% $\beta$-feeding intensity into the 2365 keV level in $^{33}$Al. However, the 2365 keV peak was found missing in the experimental TAS. Instead, a new peak at 3983 keV was found in the experimental TAS. Plotting the SS gated on the new 3983 keV peak reveals a 2365 keV $\gamma$ from this level. Further restricting it to multiplicity 2 events confirms a $\gamma$ cascade consisting of 2365 keV and 1618 keV from the 3983 keV level. Figure \ref{fig:gs3983} shows the SS gated on 3983 keV TAS peak and Mul 2. The proposed cascade can be easily identified. In conclusion, all the feeding previously assigned to the 2365 keV level should actually be to the 3983 keV level.  

\bigskip
\Large \subsection{2765 keV state}
\bigskip
\large 

\begin{figure}
    \centering
    \includegraphics[width=350pt,keepaspectratio]{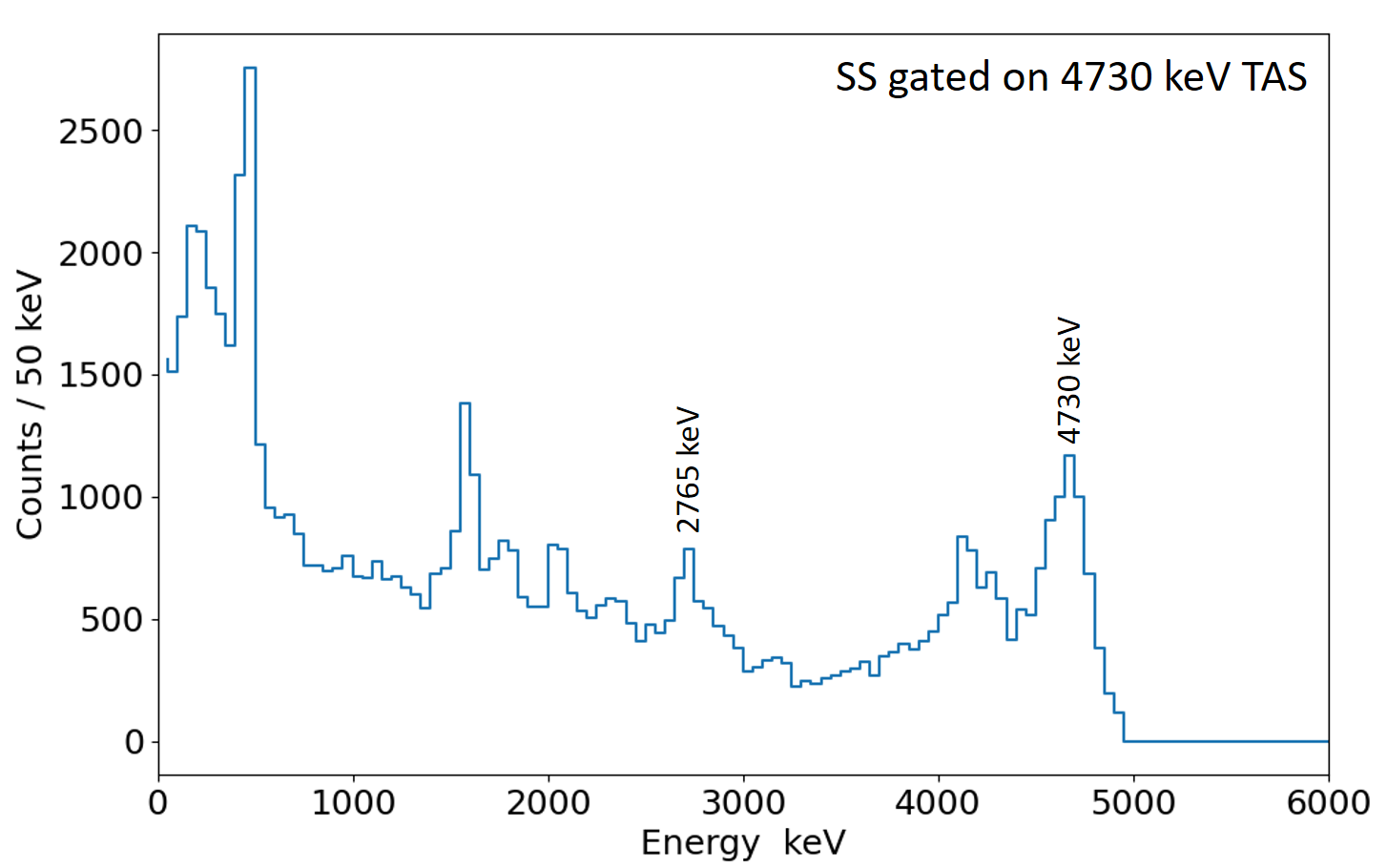}
    \caption{\large Singles Spectrum (SS) gated on the 4730 keV peak in the Total Absorption Spectrum for the $^{33}$Mg $\beta$-delayed $\gamma$ rays from the SuN-miniDSSD dataset. Decay via the 2765 keV level in $^{33}$Al can be confirmed.}
    \label{fig:gs4730}
\end{figure}

Ang\'{e}lique et al. \cite{Angelique2006} found a 2761 keV $\gamma$ ray from the $\beta$-decay of $^{33}$Mg which they attributed to an excited level in $^{33}$Al. Tripathi et al. \cite{Tripathi2008} also report a 2765 keV $\gamma$ ray which they attribute completely to an excited state in $^{32}$Al. As a result, the decay scheme from NNDC doesn't include a 2765 keV excited state for $^{33}$Al. A 2765 keV $\gamma$ peak was also found in the experimental TAS. However, the 2765 keV was also found in the SS gated on the 4730 keV TAS peak which is a known excited state in $^{33}$Al. Figure \ref{fig:gs4730} shows the SS gated on the 4730 keV TAS peak and the 2765 keV peak can be clearly identified. Additionally, assigning 100\% of the 2765 keV peak intensity to an excited level in $^{32}$Al would conflict with the significantly lower P$_n$ value measurement from the BCS-NERO dataset. As a result, the 2765 keV excited level was included in both $^{33}$Al and $^{32}$Al for the analysis. This is also reflected in the relatively larger $\beta$-feeding uncertainty to this level in $^{33}$Al. 

\bigskip
\Large \subsection{5361 keV state}
\bigskip
\large

\begin{figure}
    \centering
    \includegraphics[width=350pt,keepaspectratio]{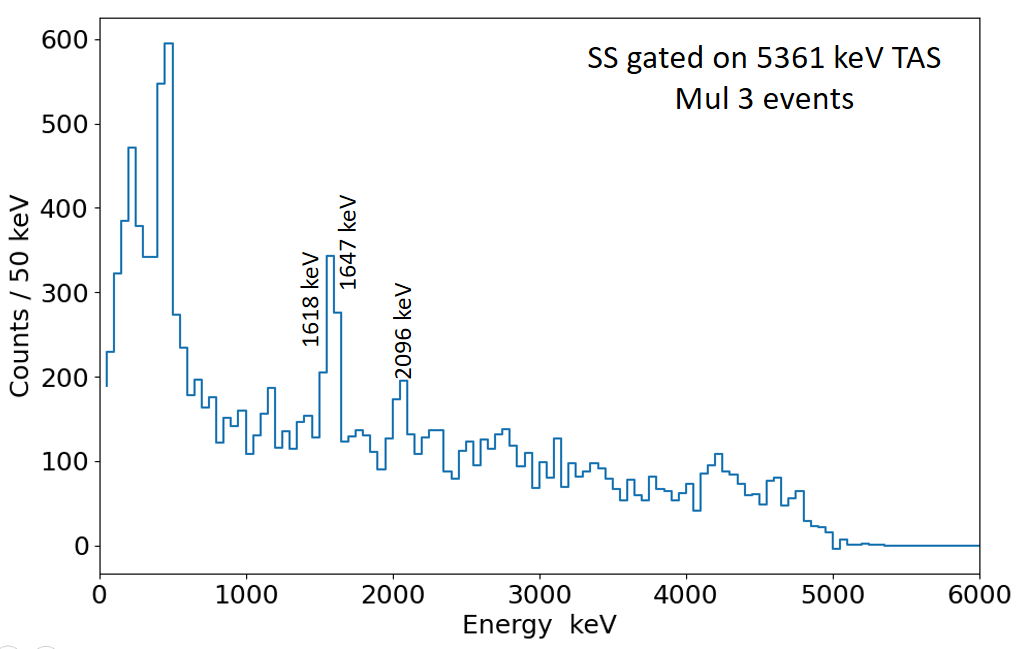}
    \caption{\large Singles Spectrum (SS) gated on the 5361 keV peak in the Total Absorption Spectrum and multiplicity (Mul) 3 events for the $^{33}$Mg $\beta$-delayed $\gamma$ rays from the SuN-miniDSSD dataset. Decay of the new proposed 5361 keV level via the 3714 keV level and the 1618 keV level in $^{33}$Al can be confirmed.}
    \label{fig:gs5361}
\end{figure}

A new peak at 5361 keV was found in the experimental TAS beyond the last reported excited state of 4730 keV in $^{33}$Al. It was also found in the SS confirming a direct decay to the ground state. In addition, a 3 $\gamma$ cascade decay was also found via the 3714 kev level and the 1618 keV level. Figure \ref{fig:gs5361} shows the experimental SS gated on the 5361 keV TAS peak and Mul 3. A 1647 keV $\gamma$ ray to the 3714 keV level, followed by a 2096 keV $\gamma$ ray to the 1618 keV level, and a 1618 keV $\gamma$ ray to the ground state can be identified. 

\bigskip
\LARGE \section{$\beta$ Feeding Intensities}\label{sec:fit}
\bigskip
\large 

\begin{figure}
    \centering
    \includegraphics[width=450pt,keepaspectratio]{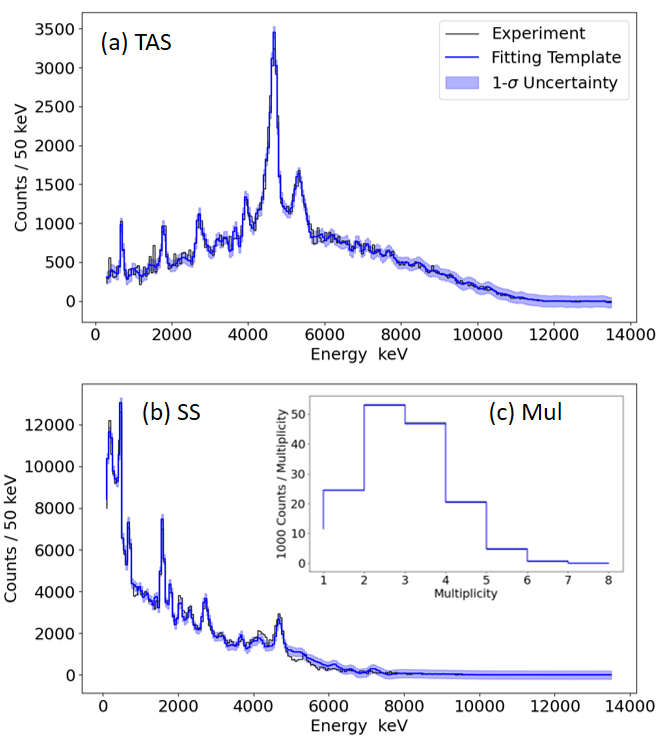}
    \caption{\large Experimental $\beta$-delayed $\gamma$ template (TAS, SS, and Mul) for the $\beta$-decay of $^{33}$Mg from the SuN-miniDSSD dataset (black) fit with templates from the \texttt{GEANT4} simulations of SuN (blue) using Bayesian Ridge Regression as described in Section \ref{sec:fit}. The light blue band around the fit shows the 1-$\sigma$ uncertainty on the posterior distribution of the fit parameters propagated to the simulated template. }
    \label{fig:fitting}
\end{figure}

\begin{table}[]
    \centering
    \begin{tabular}{cccc}
    \toprule
       \large State (keV) in $^{33}$Al  & \large Ang\'{e}lique et al. \cite{Angelique2006} & \large Tripathi et al. \cite{Tripathi2008} & \large This work \\
    \midrule
      \large 1618 & \large 21.3(2) & \large -5(3) & \large - \\
      \large 1838   & \large 4.8(2) & \large 6(2) & \large 3.7(4) \\
      \large 2096   & \large 8.6(7) & \large - & \large - \\
      \large 2365  & \large - & \large 7(2) & \large - \\
      \large 2765   & \large 8.4(8) & \large -  & \large 7.1(11) \\
      \large 3264   & \large - & \large 2(2) & \large 0.7(1) \\
      \large 3714  & \large -  & \large 12(3)  & \large 1.3(3) \\
      \large 3983  & \large -  & \large - & \large 6.9(1) \\
      \large 4310  & \large -  & \large 13(2) & \large 2.5(2) \\
      \large 4730  & \large 18(2)   & \large 15(5) & \large 28.9(4) \\
      \large 5361  & \large - & \large - & \large 10.7(2) \\
      \large  Quasi-continuum  & \large - & \large - & \large 29.6(15) \\
      \large  P$_n$  & \large 14(2) & \large - & \large 7.9(13) \\
    \midrule
      \large Ground state  & \large -  & \large 37(8) & \large 0.7(24) \\
    \bottomrule
    \end{tabular}
    \caption{\large Experimental measurements of the $\beta$-feeding intensities (\%) in this work for all excited states in $^{33}$Al from the $\beta$-decay of $^{33}$Mg compared to two previous measurements \cite{Angelique2006,Tripathi2008}. Quasi-continuum refers to the contribution for all levels above the neutron separation energy. P$_n$ refers to the $\beta$-delayed neutron branch. Subtracting everything from unity yields the ground-state branch.}
    \label{tab:feeding}
\end{table}

The experimental $\beta$-delayed $\gamma$ template (TAS, SS, and Mul) for the $\beta$-decay of $^{33}$Mg from the SuN-miniDSSD dataset was fit as a linear combination of simulated templates generated from \texttt{GEANT4} simulations of SuN. Simulated templates were generated for feeding into all the discrete energy levels in $^{33}$Al and all pseudolevels in the quasi-continuum ranging from the neutron separation energy up to the $\beta$-decay Q-value. The spacing between pesudolevels was determined from the SuN energy resolution functions as discussed in Section \ref{sec:geant4}. The $\gamma$ decay scheme for each pseudolevel was determined using \texttt{RAINIER} simulations for two nuclear level density models as described in Section \ref{sec:rainier}. Additional templates were generated for $\beta$-delayed neutron decay from each pesudolevel in the quasi-continuum of $^{33}$Al to discrete levels in $^{32}$Al. The neutron energy $E_n$ for all these templates was determined as 
\begin{equation}
    \large E_n = E_l - S_n - E_x
\end{equation}
where $E_l$ is the energy of the pesudolevel in $^{33}$Al, $S_n$ is the neutron separation energy of $^{33}$Al, and $E_x$ is the excitation energy of discrete level in $^{32}$Al. All the simulated templates were fit to the experimental template using Bayesian Ridge Regression \cite{Tipping2001} which is a probabilistic model of linear regression. Due to its Bayesian nature, it is found to be more robust to ill-posed problems. The main advantage of this algorithm is that the function predictions are treated as a random variable by adding a Gaussian error term with mean 0 and variance $\lambda^2$ to the deterministic predictions. $\lambda$ is also treated as a free parameter and inferred from the data. The parameters of the fit are normalized to sum up to 100 which then constitute the experimental $\beta$-feeding intensities. Since the posterior distributions are nearly Gaussian, 1-$\sigma$ uncertainties on the posteriors are scaled with the same normalization factor to report the experimental uncertainties. Figure \ref{fig:fitting} shows the fit obtained for simultaneously fitting the TAS, SS, and Mul and Table \ref{tab:feeding} lists the corresponding $\beta$-feeding intensities. A few levels from previous works \cite{Angelique2006,Tripathi2008} are not considered here since there wasn't a TAS peak in the experimental template and the corresponding $\gamma$ rays were identified as a part of the higher multiplicity cascade. A significant fraction of the $\beta$-feeding intensity is found to be in the 3 {--} 5 MeV excitation energy region in $^{33}$Al which is consistent with previous measurements. However, the actual intensities are very different. Table \ref{tab:feeding} compares the $\beta$-feeding intensities from this work with previous measurements \cite{Angelique2006, Tripathi2008}.

\bigskip
\Large \subsection{Nuclear log-ft Values}
\bigskip
\large 

\begin{table}[]
    \centering
    \begin{tabular}{ccc}
    \toprule
       \large State (keV) in $^{33}$Al  & \large Feeding Intensity (\%) & \large log-ft \\
    \midrule
     \large Ground state  & \large 0.7(24)  & \large 7.0$_{-0.7}^{+\infty}$ \\
      \large 1838   & \large 3.7(4) & \large 6.24(5) \\
      \large 2765   & \large 7.1(11) & \large 5.48(7)  \\
      \large 3264   & \large 0.7(1) & \large 6.38(7) \\
      \large 3714  & \large 1.3(3)  & \large 6.02(11)  \\
      \large 3983  & \large 6.9(1)  & \large 5.24(2)  \\
      \large 4310  & \large 2.5(2)  & \large 5.61(4)  \\
      \large 4730  & \large 28.9(4)   & \large 4.45(2)  \\
      \large 5361  & \large 10.7(2) & \large 4.73(2)  \\
    \bottomrule
    \end{tabular}
    \caption{\large Nuclear log-ft values corresponding to the $\beta$-feeding intensities measured in this work for all excited states in $^{33}$Al from the $\beta$-decay of $^{33}$Mg.}
    \label{tab:logft}
\end{table}

The nuclear log-ft values from the $\beta$-decay of $^{33}$Mg were calculated for all the discrete energy levels including the ground state in $^{33}$Al using the LogFT web tool provided by the National Nuclear Data Center (NNDC). These are listed in Table \ref{tab:logft} alongside the $\beta$-feeding intensities for each level. Since the $\beta$-feeding intensity for the ground state is very small, only a $1-\sigma$ lower limit is provided for the log-ft value for the ground state, the upper limit for the log-ft value being zero feeding intensity and infinite log-ft value. 

\bigskip
\Large \subsection{Gamow-Teller Transition Strengths}
\bigskip
\large 

\begin{figure}
    \centering
    \includegraphics[width=450pt,keepaspectratio]{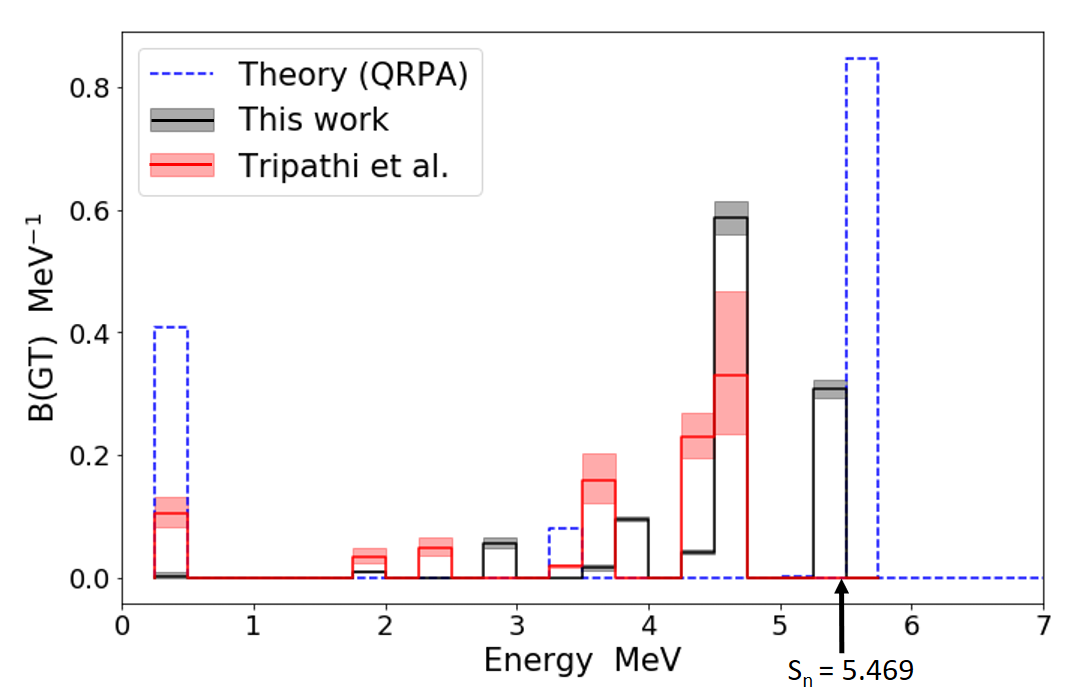}
    \caption{\large Comparison of the experimentally determined Gamow-Teller Transition strengths B(GT) for the $\beta$-decay of $^{33}$Mg with Quasi-Random Phase Approximation (QRPA) calculations. Experimental B(GT) values can only be reliably determined up to the neutron separation energy S$_n$ = 5469 keV. }
    \label{fig:bgt}
\end{figure}

The Gamow-Teller transition strength B(GT) is defined as the square of the nuclear matrix element $|M_{fi}^2|$ connecting the initial and the final states in the parent and daughter nuclei. It is related to the nuclear log-ft value according to Equation \ref{eqn:ft}. B(GT) was calculated for all states in $^{33}$Al from the experimentally determined log-ft values. These are plotted in Figure \ref{fig:bgt} as a function of the excitation energy in $^{33}$Al up to the neutron separation energy. 

The experimental B(GT) values are also compared to the B(GT) values calculated with the Quasi-Random Phase Approximation (QRPA) \cite{Mller1990} using a folded-Yukawa potential, a Lipkin-Nogami pairing model, and including a residual Gamow-Teller interaction. A spheroidal deformation of $\epsilon_2$ = 0.110, $\epsilon_4$ = -0.040, and $\epsilon_6$ = -0.040 determined by the minimum of the nuclear potential energy surface defined by FRLDM macroscopic-microscopic model \cite{Mller1995} was used in this calculation. Figure \ref{fig:bgt} also shows the theoretically calculated B(GT) values using QRPA.

\bigskip
\LARGE \section{Level Density Model Comparison}
\bigskip
\large 

\begin{figure}
    \centering
    \includegraphics[width=300pt,keepaspectratio]{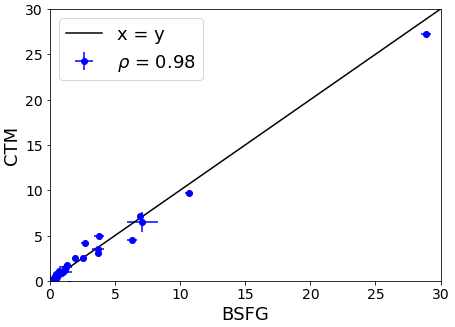}
    \caption{\large Scatter plot for comparison of the experimentally determined $\beta$-feeding intensities into excited levels (and pseudolevels in quasi-continuum) in $^{33}$Al using different level density models to treat the quasi-continuum in $^{33}$Al. Both BSFG and CTM models lead to very similar experimental results and a correlation coefficient of 0.98 is found between the $\beta$-feeding intensities obtained by both the models.}
    \label{fig:ldcomp}
\end{figure}

\begin{figure}
    \centering
    \includegraphics[width=300pt,keepaspectratio]{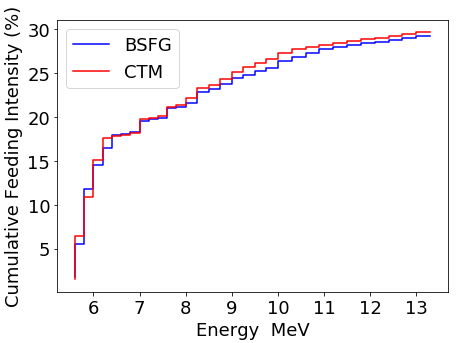}
    \caption{\large Experimental results for cumulative $\beta$-feeding intensities in the quasi-continuum for both the level density models. Quasi-continuum ranges from the neutron separation energy (5.469 MeV) to the Q-value (13.477 MeV). Both models lead to similar feeding in the quasi-continuum. }
    \label{fig:cumfeed}
\end{figure}

\begin{figure}
    \centering
    \includegraphics[width=350pt,keepaspectratio]{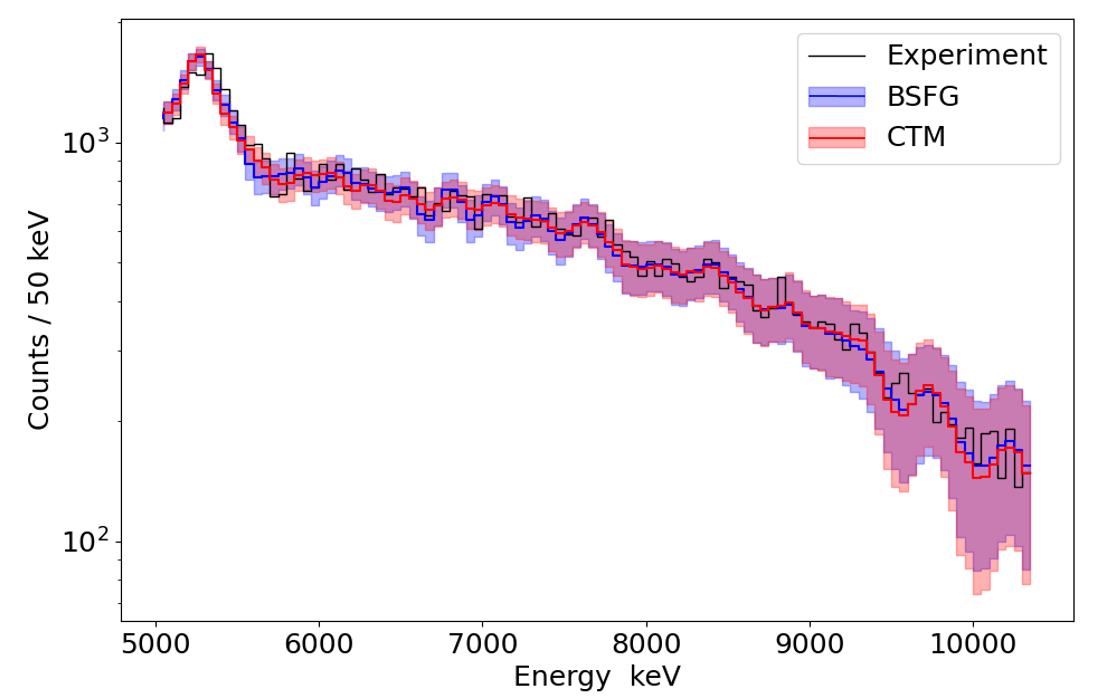}
    \caption{\large Comparison of simulated TAS using different level density models with the experimental TAS on the log scale. Colored bands represent 1-$\sigma$ uncertainties on the posteriors of fit parameters propagated to the simulated spectra. Both the fits are nearly identical.}
    \label{fig:ldtas}
\end{figure}

\texttt{RAINIER} and \texttt{GEANT4} simulations of the pesudolevels templates in the quasi-continuum in $^{33}$Al require a prescription for the nuclear level density as a function of the excitation energy. Two models of the nuclear level densities were considered in this work - the Back-Shifted Fermi Gas model (BSFG) and the Constant Temperature Model (CTM) as described in Section \ref{sec:ld}. The experimental results for the $\beta$-feeding intensities were insensitive to the nuclear level density model chosen. This is evident from Figure \ref{fig:ldcomp} that compares the experimental results from using different level density models. A correlation coefficient of 0.98 is found between the weights of fit templates using BSFG and CTM models. Figure \ref{fig:cumfeed} shows the cumulative $\beta$-feeding intensities in the quasi-continuum for both the models and Figure \ref{fig:ldtas} shows the fit to the experimental TAS at higher energies (on a log scale) for both the models. Both the figures confirm the non-dependence of the experimental results on the choice of nuclear level density model. Finally, based on the marginally better fit to the data, experimental results obtained using BSFG model for the nuclear level density are reported. 

\huge \chapter{Quantified Nuclear Mass Model}\label{chp:massmodel}
\bigskip
\large

Nuclear masses are an important piece of data to accurately quantify the strength of Urca cooling in accreting neutron star crusts. As stated in Section \ref{sec:nuclear-data}, Urca cooling luminosity sensitively depends on the electron-capture threshold, or the Q-value, to the fifth power, which in turn depends on the nuclear mass. Of the 7000 nuclei predicted to exist in nature \cite{Erler2012,Agbemava2014}, less than 3000 have their masses experimentally measured to date \cite{Wang2021}. Even after considerable experimental efforts ongoing worldwide, many of the exotic nuclei would still be out of reach in the foreseeable future. Moreover, since the nuclei undergoing Urca cooling are typically neutron-rich, away from the valley of stability on the nuclear chart, an even smaller fraction of them will have existing experimental mass measurements. As a result, we have to rely on theoretical predictions which have their own unique set of challenges. A goal of this dissertation is to build a global nuclear mass model with quantified uncertainties and covariances and that will be the focus of this chapter. 

\bigskip
\LARGE \section{Current status}
\bigskip
\large

\begin{figure}
    \centering
    \includegraphics[width=450 pt,keepaspectratio]{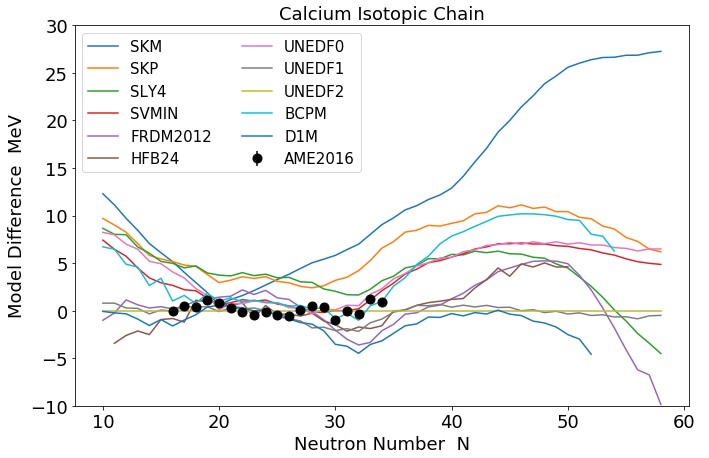}
    \caption{\large Binding Energy predictions for the Calcium isotopic chain for several theoretical mass models considered here (See Section \ref{sec:theorymasses}). The binding energies are normalized to UNEDF2 for plotting. Note the increased divergence as one moves away from the experimentally measured region.}
    \label{fig:divergence}
\end{figure}

Nuclear reaction network calculations generally rely on the choice of a particular theoretical mass model. For example, \texttt{Xnet} (refer Section \ref{sec:intro-reactions}) uses nuclear mass predictions from the Finite Range Droplet Model (FRDM2012) \cite{Mller2016}. A common metric used for making this choice is the root-mean-squared standard deviation from the experimental dataset. However, theoretical mass models with many parameters are prone to overfitting and may not generalize well beyond the experimental dataset. They also lack quantified uncertainties for their predictions. Figure \ref{fig:divergence} shows the binding energy predictions for the calcium isotopic chain for several theoretical mass models (normalized to a random model) along with experimental measurements. It can be seen that the spread in their predictions is least where experimental masses are measured and quickly diverges beyond that, both in the neutron-rich and neutron-deficient regions. This is why generalizable predictions with honest uncertainty quantification are very important. With the rise of Machine Learning (ML) in recent years, especially physics-informed ML, several attempts have been made to tackle this problem. They try to either optimize the parameters of theoretical models \cite{Bayram2017,Bertsch2017,Kejzlar2020}, learn mass residuals \cite{Yuan2016,Utama2016,Utama2017,Utama2018,Zhang2017,Neufcourt2018,Neufcourt2020proton,Niu2018,Rodrguez2019,BaosRodrguez2019,Gao2021,Wu2021}, or even directly predict the masses \cite{Athanassopoulos2004,lovell2022nuclear}. More phenomenological approaches that directly predict masses from neighboring measurements are also possible \cite{Vladimirova2021}. (See recent reviews \cite{Bedaque2021}, \cite{Boehnlein2022} for more information). All of these highlight the importance of incorporating domain knowledge in this exercise which can be done in very different ways. Each approach, however, has a trade-off between complexity, extrapolation accuracy, and computational costs. The model presented in the following sections uses Gaussian Process Regression for reduced complexity and improves extrapolation accuracy using Bayesian Model Averaging over the predictions of many theoretical approaches to utilize their collective wisdom.
 
\bigskip
\Large \subsection{Residuals Modeling}
\bigskip
\large

\begin{figure}
    \centering
    \includegraphics[width=470 pt,keepaspectratio]{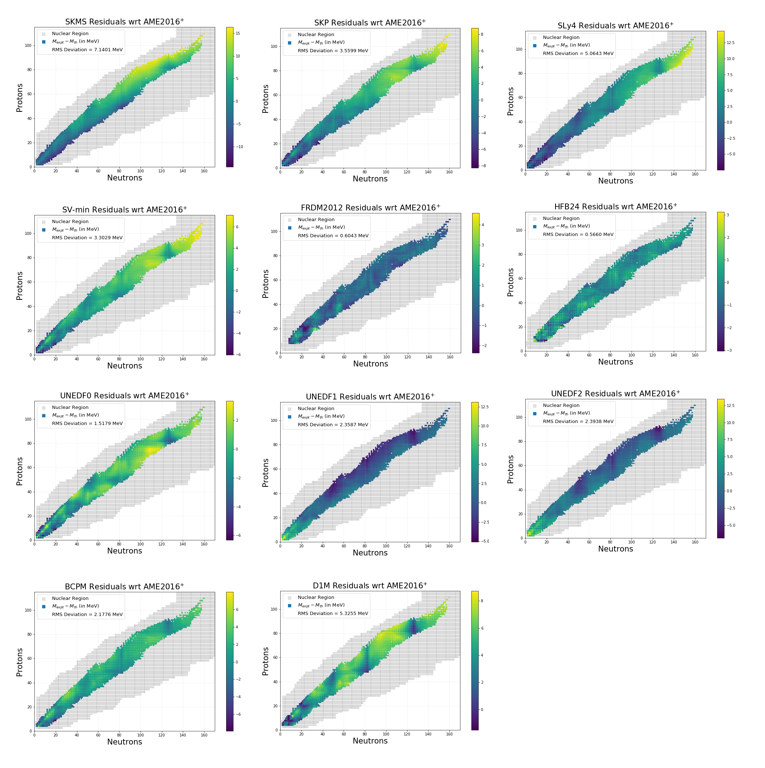}
    \caption{\large Nuclear Binding Energy Residuals, i.e. the difference between experimental measurements and theoretical predictions for all theoretical mass models considered here (See Section \ref{sec:theorymasses}). The residuals have a characteristic structure that provides clues for potential shortcomings of the underlying theoretical model.}
    \label{fig:residuals}
\end{figure}

Any nuclear mass model that has to be used in nuclear reaction network calculations or astrophysical simulations should have the following characteristics.

\begin{enumerate}
    \large \item It should be based on controlled theoretical formalism such that it uses quantified nuclear properties as inputs.
    \large \item It should generalize beyond the region of known nuclear masses and should cover the entire nuclear region. 
    \large \item Along with nuclear masses, it should also be consistent with other nuclear observables, for example, separation energies, Q-values, shell gaps, etc. 
    \large \item It should have a measure of confidence in its predictions so that proper uncertainty quantification can be done. 
\end{enumerate}

Keeping this in mind, data-driven Machine Learning (ML) methods, with their ability to capture the topology of nuclear mass surfaces, are a great tool of choice. However, the nuclear force is quite complex, and its exact parameterization is unknown. And training a complex ML model requires a lot of data. Since we only have experimental mass measurements for less than 3000 nuclei, it puts a restriction on the complexity of ML model and makes it difficult to directly predict accurate nuclear masses. This is where physics-informed ML models can help solve this issue. 

A popular way to incorporate known physics in the modeling process is to model the residuals, i.e. the difference between experimental mass measurements and theoretical mass predictions. Figure \ref{fig:residuals} plots the residuals for a suite of physics-based mass models. The residuals are non-zero since each model is based on a different set of assumptions and approximations, and has a different parameterization. The models are also calibrated on different subsets of experimental data. Some models perform better on the neutron-rich side whereas others perform better on the proton-rich side. There is, however, a characteristic structure of these residuals that shows where the physics-based models lack. Modeling and extrapolating just the residuals and adding them back to base theoretical model predictions gives us a way to build physics-informed data-driven mass models. These have the desired characteristics and outperform currently available mass models. A set of such models can be further combined to form a single mass model that combines the predictive power of each underlying model. This is done via Bayesian Model Averaging (BMA).

\bigskip
\LARGE \section{Binding Energy Datasets}
\bigskip
\large

\begin{figure}
    \centering
    \includegraphics[width=450 pt,keepaspectratio]{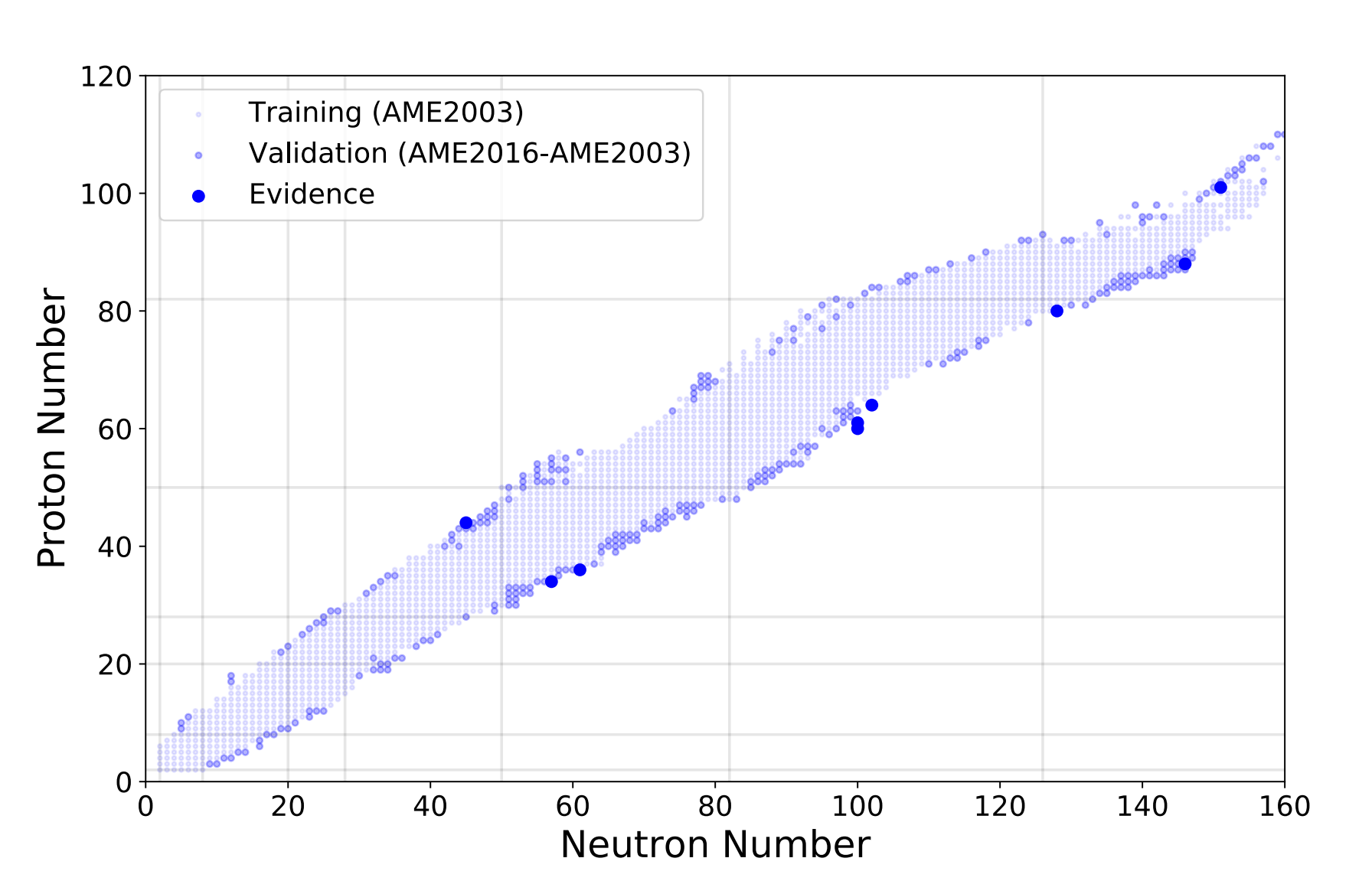}
    \caption{\large Nuclear chart showing the various nuclei in training set (AME2003), validation set (AME2016-AME2003) and evidence set for BMA ($^{89}$Ru, $^{91}$Se, $^{97}$Kr, $^{160}$Nd, $^{161}$Pm, $^{166}$Gd, $^{208}$Hg, $^{234}$Ra, $^{252}$Md). The choice of these 9 evidence nuclei is intended to cover all mass regions as well as both the neutron-rich and the proton-rich side of the training set.}
    \label{fig:dataset}
\end{figure}

The nuclear binding energy is directly related to nuclear masses as
\begin{equation}
    \large - \text{BE (Z, N)} = [Zm_p + Nm_n - \text{M (Z, N)}]c^2,
\end{equation}
where BE (Z,N) is the binding energy of a nucleus with Z protons and N neutrons, $m_p$ is the mass of the proton, $m_n$ is the mass of the neutron, and M (Z,N) is the mass of the nucleus. It is the energy associated with the strength of the binding of the nucleons. The higher the binding, the lower is the mass, and vice versa. Due to this one-on-one correspondence, binding energy is used as a proxy for nuclear masses. Both experimental data and theoretical predictions are required for the modeling of residuals.

\bigskip
\Large \subsection{Experimental Data}
\bigskip
\large

The experimental values are the total binding energies taken from the Atomic Mass Evaluation datasets, AME2003 \cite{Audi2003} and AME2016 \cite{Wang2017}. Only the experimentally measured values are considered and the empirical predictions referred to as trends of mass surfaces (TMS) are ignored. Additionally, a new dataset called NEW2018 was also created by augmenting the AME2016 dataset. This included some newer mass measurements \cite{deRoubin2017,Welker2017,Vilen2018,Leistenschneider2018,Michimasa2018,Orford2018,Ito2018} after the compilation of AME2016. A few masses published in KROLL2020 \cite{Kroll2020} beyond the NEW2018 dataset were also used as evidence for BMA. Whenever there is a conflict of mass measurement for a particular isotope, the most recent value is used.

AME2003 was used for initial training followed by AME2016-AME2003 (isotopes with their experimental mass measurements compiled between 2003 and 2016) for validation. Rather than selecting a random subset of training data for validation, this method of choosing the validation set seeks to gauge the extrapolating power of these models and reduce overfitting with the limited amount of data in possession. After fixing the hyperparameters, the predictions with the model trained on AME2003 were also used to calculate the BMA weights, although the evidence nuclei were selected from NEW2018 and KROLL2020. This was to ensure that a comprehensive set of evidence nuclei is available that are not included in the training. Figure \ref{fig:dataset} shows the nuclear chart with nuclei contained in training, validation, and evidence sets. The final predictions are made using the NEW2018 dataset and the BMA weights from the previous step. 

AME2020 \cite{Wang2021} has also been recently published. However, it would have little impact on the final global mass model at the moment. Future AME publications should still be used to update this model periodically.
\bigskip

\Large \subsection{Theoretical Mass Predictions}\label{sec:theorymasses}
\bigskip
\large

A total of 11 theoretical mass models were considered for modeling residuals, and to subsequently perform BMA. Seven of these models were based on Skyrme Energy Density Functionals (EDFs) for the nuclear Density Functional Theory (DFT): SkM* \cite{Bartel1982}, SkP \cite{Dobaczewski1984}, SLy4 \cite{Chabanat1995}, SV-min \cite{Klpfel2009}, UNEDF0 \cite{Kortelainen2010}, UNEDF1 \cite{Kortelainen2012}, UNEDF2 \cite{Kortelainen2014}. Two models were based on Gogny EDFs: BCPM \cite{Baldo2013} and D1M \cite{Goriely2009}. Two additional models were also considered that are generally used currently for astrophysics simulations: the microscopic-macroscopic FRDM2012 model \cite{Mller2016} and the Skyrme-HFB model HFB-24 \cite{Goriely2013}. The deviations for these models from experimental values, measured as the root mean square over all the isotopes range from a few MeV for the DFT-based models to around 600 keV for the FRDM2012 model. Following statistical treatment as described in Section \ref{sec:gp}, and adding back the residuals, this RMS deviation is reduced to less than 200 keV for all mass models. 

\bigskip
\LARGE \section{Statistical Methods}

\bigskip
\Large \subsection{Gaussian Process Regression}\label{sec:gp}
\bigskip
\large

The residuals between experimental and theoretical binding energies are modeled as a Gaussian Process following \cite{Neufcourt2018}. Let $x_i$ be the feature vector describing the nucleus and $y_i$ be the corresponding residual to be modeled. The residual $y_i = y_i^{exp} - y^{th}(x_i)$ dataset is constructed for each nuclear mass model. The mapping from $x_i$ to $y_i$ can be represented as:
\begin{equation}
    y_i = f(x_i,\theta) + \sigma \epsilon_i \,,
\end{equation}
where $\epsilon_i$ is a standard Gaussian variable and $\sigma$ is the noise scale factor. If $f$ is deterministic, this term accounts for the stochastic nature of the experimental data. Here, the function $f(x_i,\theta)$ itself is stochastic and is modeled as a Gaussian Process:
\begin{equation}
f(x_i,\theta) \sim \mathcal{GP} (\mu, k_{\eta,\rho} (x,x')) \,,
\end{equation}
with mean $\mu$ and covariance kernel $k$. The covariance kernel is chosen to be the squared exponential kernel such that
\begin{equation}
k_{\eta,\rho} (x, x') 
:= \eta^{2} e^{\sum_{i}-\frac{(x_{i}-x_{i}')^2}{\rho_{x_i}}} \,,
\end{equation}
where $i$ goes over the features. $\eta$ is the variance and $\rho_{x_i}$ corresponds to the lengthscale for each feature. Thus, the parameter vector $\theta$ comprises of $\{\mu,\eta,\rho_{x_i}\}$. Additionally, a white noise kernel with a noise level of 0.0235 MeV was added to the squared exponential kernel to account for the experimental errors. This noise level was chosen based on the average experimental uncertainty and was held constant considering its small value. Effectively, it allows the GP a leeway to pass within 0.0235 MeV of the training points to get the best fit. The feature vector is comprised of $\{Z,N\}$ i.e. the proton number and the neutron number, respectively. Two additional features, promiscuity and pairing factor were also considered to aid the learning process as described in Section \ref{sec:feature}. As a Bayesian model, we draw samples for the parameters $\theta$ from their posterior distribution given by the Bayes equation
\begin{equation}
p(y|\theta):= \frac{p(\theta|y) \pi(\theta)}{\int p(\theta|y) \pi(\theta)d\theta} \,,
\end{equation}
where $p(\theta|y)$ is the statistical model's likelihood and $\pi(\theta)$ the prior on its parameters. Gaussian likelihood with an inverse-gamma function for priors was used for modeling. Samples were drawn from the posterior using Markov Chain Monte Carlo (MCMC) after the stationary state was reached. Finally, the GP predictions are added back to the theoretical masses to get the statistically corrected mass models and the GP uncertainties represent the uncertainties for the statistically corrected models. These models can now be averaged using BMA as discussed in the next section. 

\bigskip
\Large \subsection{Bayesian Model Averaging} \label{sec:bma}
\bigskip
\large

The predictions of statistically corrected mass models are combined using Bayesian Model Averaging (BMA). The BMA predictions are an average of the predictions of the individual model, weighted by the posterior probability for that model to be the hypothetical 'true' model given priors and data. These can be written as:
\begin{equation}
    w_k = p(\mathcal{M}_k | y) = \frac{p(y | \mathcal{M}_k)\pi(\mathcal{M}_k)}%
	{\sum_{\ell=1}^p p(y | \mathcal{M}_\ell) \pi(\mathcal{M}_\ell)} \,,
\end{equation}
where $\pi(\mathcal{M}_k)$ is the prior weight of model $\mathcal{M}_k$, 
and the evidence $p(y |\mathcal{M}_k)$ is obtained by integrating the likelihood equation over the parameter space. For the GP emulators, this gives 
\begin{equation}
p(y | \mathcal{M}_k) = \int p(y|\theta_k, \mathcal{M}_k) \pi(\theta_k, \mathcal{M}_k) d\theta_k \,.
\end{equation}

A uniform, non-informative prior distribution is used for the weights. 
The data $y$ are the experimental mass measurements of a subset of 
``evidence'' nuclei. 
As a result, our BMA weights are also sensitive to the choice of the evidence nuclei. Their choice in our study and the resulting weights are discussed in Section \ref{sec:bmaweights}. 

Once the weights are calculated, 
the final BMA predictions and uncertainties can be obtained as:
\begin{subequations}\label{eqn:bma}
\begin{gather}
y(x) = \sum_k w_k y^{(k)}(x) \,; \\
\sigma_y^2(x) = \sum_k w_k (y^{(k)}(x) - y(x))^2 + \sum_k w_k \sigma_{y_k}^2(x) \,,
\end{gather}
\end{subequations}
where $y_k$ are the individual model predictions, 
$\sigma(y_k)$ are the model uncertainties 
and $w_k$ are the model weights \cite{kejzlar2019bayesian}. The two terms in the equation for uncertainties correspond to the uncertainty on model choice and the uncertainty on individual model parameters, respectively. The first term therefore represents a part of the uncertainty that is not accounted for when using predictions derived from a single model. This highlights the role of BMA in proper uncertainty quantification. 

\bigskip
\LARGE \section{Model Aspects} \label{sec:modelaspects} 

\bigskip
\Large \subsection{Sample Convergence}
\bigskip
\large 

\begin{figure}
    \centering
    \includegraphics[width=400 pt,keepaspectratio]{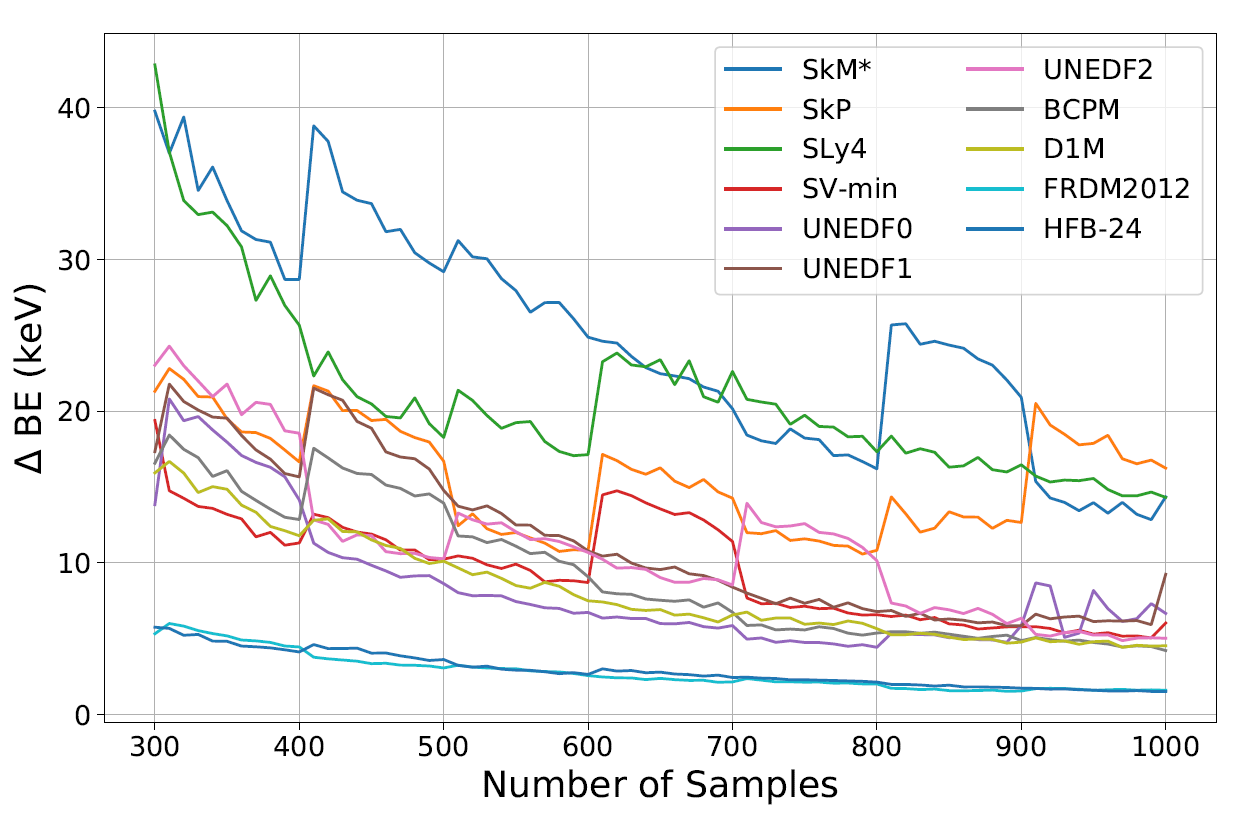}
    \caption{\large The absolute change in Binding Energy residuals ($\Delta BE$) predicted by the GP emulator with 10 additional samples drawn as a function of the number of samples. This is a proxy for MCMC convergence as ($\Delta BE \hspace{1pt} \rightarrow \hspace{1pt} 0$) means additional samples do not significantly change the predictions.}
    \label{fig:convergence}
\end{figure}

For most practical applications of Bayesian analysis, the exact integrals of the posterior distributions for making final predictions are often intractable and we have to resort to sampling methods. As a result, Markov Chain Monte Carlo MCMC sampling methods were used to sample the GP parameters from their posterior distributions. A natural dilemma that arises is the trade-off between computational costs and the convergence of the final predictions. Figure \ref{fig:convergence} shows $\Delta BE$ i.e. the change in the binding energies when 10 additional samples are drawn. (This is after the initial 50,000 samples burnt in for the MCMC to stabilize). The quantity plotted is the rms of the difference over all nuclei. Initially, we see large changes as we draw more samples which nicely converge as the number of samples becomes large. Occasional jumps in $\Delta BE$ correspond to the exploration of new parameter space. We decided to stop at 1000 samples since the change in BE at this stage was about 20 keV for all models which is very small for all practical aspects. 

\bigskip
\Large \subsection{Feature Selection}\label{sec:feature}
\bigskip
\large 

\begin{table}[]
    \centering
    \begin{tabular}{c|c|cccccc}
    \toprule
       \large Model  & \large Raw & \large GP\_Zero & GP\_Con & GP\_Dis & GP++\_Zero & \textbf{GP++\_Con} & GP++\_Dis \\
    \midrule
 \large SkM* & \large 7.74 & \large 3.59 & \large 3.81 & \large 3.24 & \large 1.62 & \large 1.74 & \large 1.59 \\
\large  SkP & \large 3.84 & \large 2.11 & \large 2.04 & \large 1.33 & \large 1.06 & \large 1.12 & \large 1.00 \\
\large  SLy4 & \large 6.01 & \large 3.47 & \large 2.76 & \large 1.34 & \large 1.47 & \large 1.24 & \large 1.17 \\
\large  SV-min & \large 3.37 & \large 1.88 & \large 1.49 & \large 0.97 & \large 1.02 & \large 0.97 & \large 1.84 \\
\large  UNEDF0 & \large 1.73 & \large 1.09 & \large 1.09 & \large 0.89 & \large 0.70 & \large 0.73 & \large 0.82 \\
\large  UNEDF1 & \large 2.58 & \large 1.51 & \large 1.46 & \large 1.31 & \large 0.84 & \large 0.79 & \large 0.75 \\
\large  UNEDF2 & \large 2.51 & \large 1.51 & \large 1.43 & \large 1.24 & \large 0.84 & \large 1.01 & \large 1.22 \\
\large  BCPM & \large 2.13 & \large 1.27 & \large 1.25 & \large 1.07 & \large 0.90 & \large 1.05 & \large 0.90 \\
\large  D1M & \large 5.40 & \large 2.74 & \large 1.01 & \large 0.99 & \large 1.19 & \large 0.74 & \large 1.03 \\
\large  FRDM2012 & \large 0.79 & \large 0.62 & \large 0.63 & \large 0.57 & \large 0.65 & \large 0.54 & \large 0.70 \\
\large  HFB-24 & \large 0.70 & \large 0.66 & \large 0.67 & \large 0.72 & \large 0.67 & \large 0.67 & \large 0.69 \\ 
    \midrule
\large Average & \large 3.35 & \large 1.86 & \large 1.60 & \large 1.24 & \large 1.00 & \textbf{\large 0.96} & \large 1.06 \\
    \bottomrule
    \end{tabular}
    \caption{\large Root-mean-squared (RMS) deviations of binding energies (in MeV) 
	over the AME2016-AME2003 validation set,
	for each nuclear mass model, with and without (raw)
	the GP corrections for variants
	GP\_Zero, GP\_Constant (GP\_Con), GP\_Discrete (GP\_Dis), GP++\_Zero, GP++\_Constant (++\_Con), GP\_++\_Discrete (++\_Dis)
	as described in Section \ref{sec:modelaspects}. 
	The average values in the last row 
	correspond to a simple average over all the nuclear mass models for each variant. Based on this measure, the GP++\_Constant is the best-performing variant overall.}
    \label{tab:residuals}
\end{table}

A unique nucleus can be identified with just the atomic number Z and the number of neutrons N. However, in the spirit of physics-informed ML,  additional aspects of the nuclear force can be incorporated that can aid in the learning process. Nuclei exhibit shell structure where a certain number of nucleons corresponding to closed shells are more tightly bound \cite{Mayer1948,Mayer1949}. They also exhibit pairing effects since nuclei with an even number of nucleons have slightly lower energies than nuclei with an odd number of nucleons. As a result, two additional features were constructed - the promiscuity factor and the pairing factor. The promiscuity factor $p$, accounted for shell effects and is defined as 
\begin{equation}
    \large p = \frac{\nu_{p}\nu_{n}}{\nu_{p}+\nu_{n}}
\end{equation} where $\nu_{p}$ and $\nu_{n}$ are the differences between the actual nucleon
numbers $Z$ and $N$ and the nearest magic number (20, 28, 40, 50, 82 and 126). The pairing factor 
\begin{equation}
   \large \delta = \frac{[(-1)^{Z}+(-1)^{N}]}{2}
\end{equation}
accounts for pairing effects. Models with these two additional features are designated as GP++ in the model validation step (see Table \ref{tab:residuals}). The best-performing model was found to be the one where these two additional features were used in the training. 

\bigskip
\Large \subsection{Choice of GP Mean}
\bigskip
\large 

There are no hyperparameters to be tuned, given the Bayesian nature of the analysis. The model predictions essentially integrate over all possible hyperparameters weighted by their posterior probabilities. However, there is a choice in terms of parameterizing the mean of GP. The mean of GP is usually set to 0 (GP\_Zero), if no significant bias is expected in the training data. This wasn't guaranteed for some of the base theoretical models considered since they were calibrated on different subsets of experimental data. To allow for this, GP\_Constant (GP\_Con) models had a constant mean for the GP throughout the nuclear chart. Additionally, the GP\_Discrete (GP\_Dis) model had a constant mean value for each subset of nuclei with the same open shell, i.e. all nuclei between two proton and neutron magic numbers. Combined with the two feature variants, we get a total of 6 models as shown in Table \ref{tab:residuals}. 

After training on the AME2003 dataset, all six GP variants were validated against the 281 nuclei in the AME2016-AME2003 dataset, containing all the nuclei that had their experimental masses 
compiled between 2003 and 2016. Table \ref{tab:residuals} shows the root mean squared error over the validation set for each variant as well as for each theoretical mass model. 
All the GP variants perform better than the raw models, with some raw models showing more improvements than others. More phenomenological models like HFB24 or FRDM2012 have already captured most of the information contained in the experimental data in the raw models themselves. 
The ++ variants outperform their counterparts, showing definite correlations along $p$ and $\delta$, highlighting the importance of shell effects and pairing effects for predicting the masses of nuclei. The GP\_++\_Constant performs better than GP\_++\_Zero, as already noted before \cite{Neufcourt2018,Neufcourt2020proton,Neufcourt2020limits}. The most complex GP\_++\_Discrete variant, surprisingly doesn't perform as well as the GP\_++\_Constant variant. This may be because of redundancy where the additional features and the discrete mean both try to capture the shell effects, thereby complicating the training. 
Based on these observations, the GP\_++\_Constant was chosen for all further analysis owing to its accuracy and simplified implementation. 

\bigskip
\Large \subsection{BMA Weights}\label{sec:bmaweights}
\bigskip
\large 

\begin{table}[]
    \centering
    \begin{tabular}{cc}
    \toprule
       \large Corrected Model  & \large BMA Weight \\
    \midrule
      \large  SKM*  & \large 0.07  \\
      \large  SkP  & \large 0.13  \\
      \large SLy4   & \large 0.08   \\
      \large SV-min   & \large 0.17 \\
      \large UNEDF0   & \large 0.12 \\
      \large  UNEDF1  & \large 0.11  \\
      \large  UNEDF2  & \large 0.03  \\
      \large  BCPM  & \large 0.13   \\
      \large  D1M  & \large 0.10 \\
      \large  FRDM2012  & \large 0.01 \\
      \large  HFB-24  & \large 0.05 \\
    \bottomrule
    \end{tabular}
    \caption{\large BMA model weights (rounded to two digits past decimal) for statistically corrected mass models (see Section \ref{sec:bma} based on evidence nuclei as described in Section \ref{sec:bmaweights}. Note that these weights are for statistically corrected models as per Section \ref{sec:gp} and not the raw theoretical models.}
    \label{tab:bmaweights}
\end{table}

Calculations of the model weights require a choice of evidence nuclei. In the classical BMA literature \cite{KASS1995}, the evidence can be a subset of the training set itself. However, the predictions of GP models are very close to the training data. As a result, the evidence set is chosen to be comprised of nuclei outside the training or the validation set. The weights then represent the true extrapolative power of the models. A total of 9 nuclei chosen in the evidence set are $^{89}$Ru, $^{91}$Se, $^{97}$Kr, $^{160}$Nd, $^{161}$Pm, $^{166}$Gd, $^{208}$Hg, $^{234}$Ra, $^{252}$Md. These are chosen to cover all the regions in the nuclear chart (light, medium, and heavy mass regions) on both the neutron and the proton-rich sides. Figure \ref{fig:dataset} shows the intended balanced distribution of evidence nuclei throughout the nuclear chart and Table \ref{tab:bmaweights} gives the corresponding weights for each model. 
Models based on Syrme EDF like SV-min, SKP, and the BCPM model based on Gogny EDF have large weights as compared to the phenomenological model like HFB-24. This shows that the statistically corrected DFT models can better extrapolate nuclear masses than the corrected phenomenological models. Changing the set of evidence nuclei can change the absolute weights for each model. Our choice here was motivated by the need for a global mass model. Moreover, it serves as an example for BMA analysis. Future studies can customize their evidence set depending on the applications one is interested in. For example, see \cite{Neufcourt2020proton,Hamaker2021}.

The restricted choice of 9 evidence nuclei was limited in part by the availability of additional data and in part by the computational cost of computing high-dimensional evidence. As more mass measurements become available in the future, the evidence can be approximated more strongly for faster computations and increased stability, for instance via the Laplace Approximation \cite{KASS1995} where the posteriors are matched to a Gaussian distribution. For the given evidence set here, the weights calculated by sampling the posterior were very similar to the weights calculated by Laplace approximation. Thus, we can include more evidence nuclei in future analysis by using appropriate approximations. 

\bigskip
\LARGE \section{Final Mass Model}
\bigskip
\large

\begin{figure}
    \centering
    \includegraphics[width=400 pt,keepaspectratio]{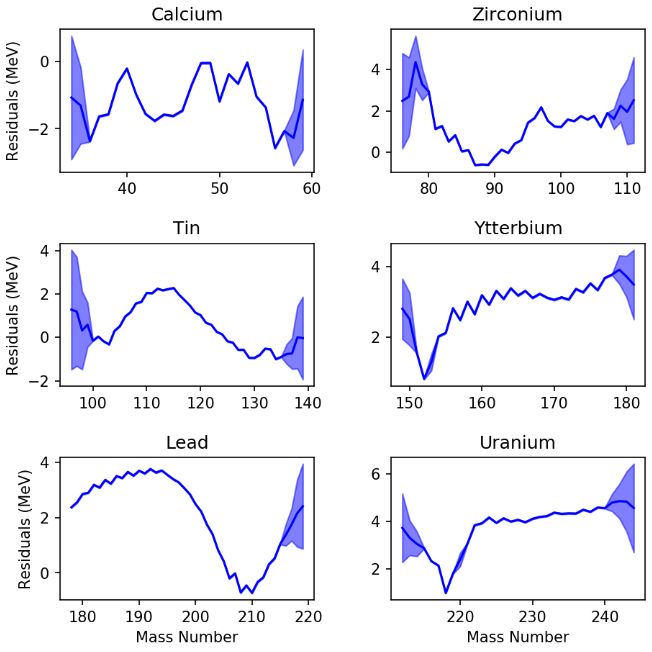}
    \caption{\large BMA residuals for six isotopic chains. Uncertainties become larger as one moves away from experimentally accessed regions. Ytterbium and Uranium mass chains also feature the need for interpolations along with extrapolations.}
    \label{fig:extrapolations}
\end{figure}

\bigskip
\Large \subsection{Binding Energy Predictions}
\bigskip

\begin{figure}
    \centering
    \includegraphics[width=400 pt,keepaspectratio]{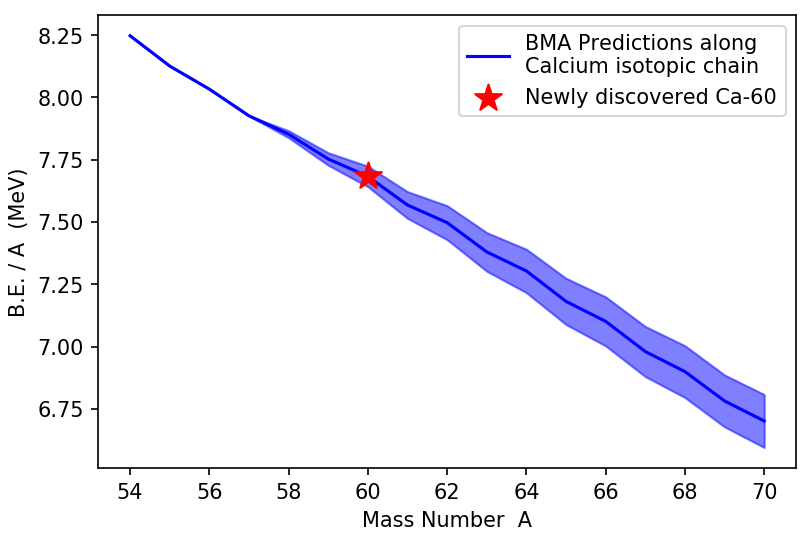}
    \caption{\large The trend in BMA predictions of binding energy per nucleon for the calcium isotopic chain. $^{60}$Ca is the most exotic calcium isotope to be seen in the laboratory so far \cite{Neufcourt2019} but its mass hasn't been experimentally measured yet. The red star shows BMA prediction along with uncertainty.}
    \label{fig:calcium}
\end{figure}

\large
The residuals predicted by the GP++\_Constant model are weighted as per Equation \ref{eqn:bma} for BMA. Figure \ref{fig:extrapolations} shows the residuals for 6 different isotopic chains. It also shows the need for interpolations, for example, in Ytterbium and Uranium isotopic chains, along with extrapolations which is the primary goal of this model. These residuals are then added back to base theoretical model predictions (also weighted as per BMA) to get the final mass predictions. Figure \ref{fig:calcium} shows how the final trend in total binding energy per nucleon is extrapolated for the calcium isotopic chain. The uncertainties are modest near the experimentally accessed region and grow upon moving further. This is natural since the model is least confident in its predictions in the most exotic regions. $^{60}$Ca is the most exotic isotope made in the laboratory recently \cite{Neufcourt2019} whose mass is yet to be measured. BMA predicts its mass along with uncertainty as shown in Figure \ref{fig:calcium}.

\bigskip
\Large \subsection{Correlations} \label{sec:correlations}
\bigskip
\large 

\begin{figure}
    \centering
    \includegraphics[width=450 pt,keepaspectratio]{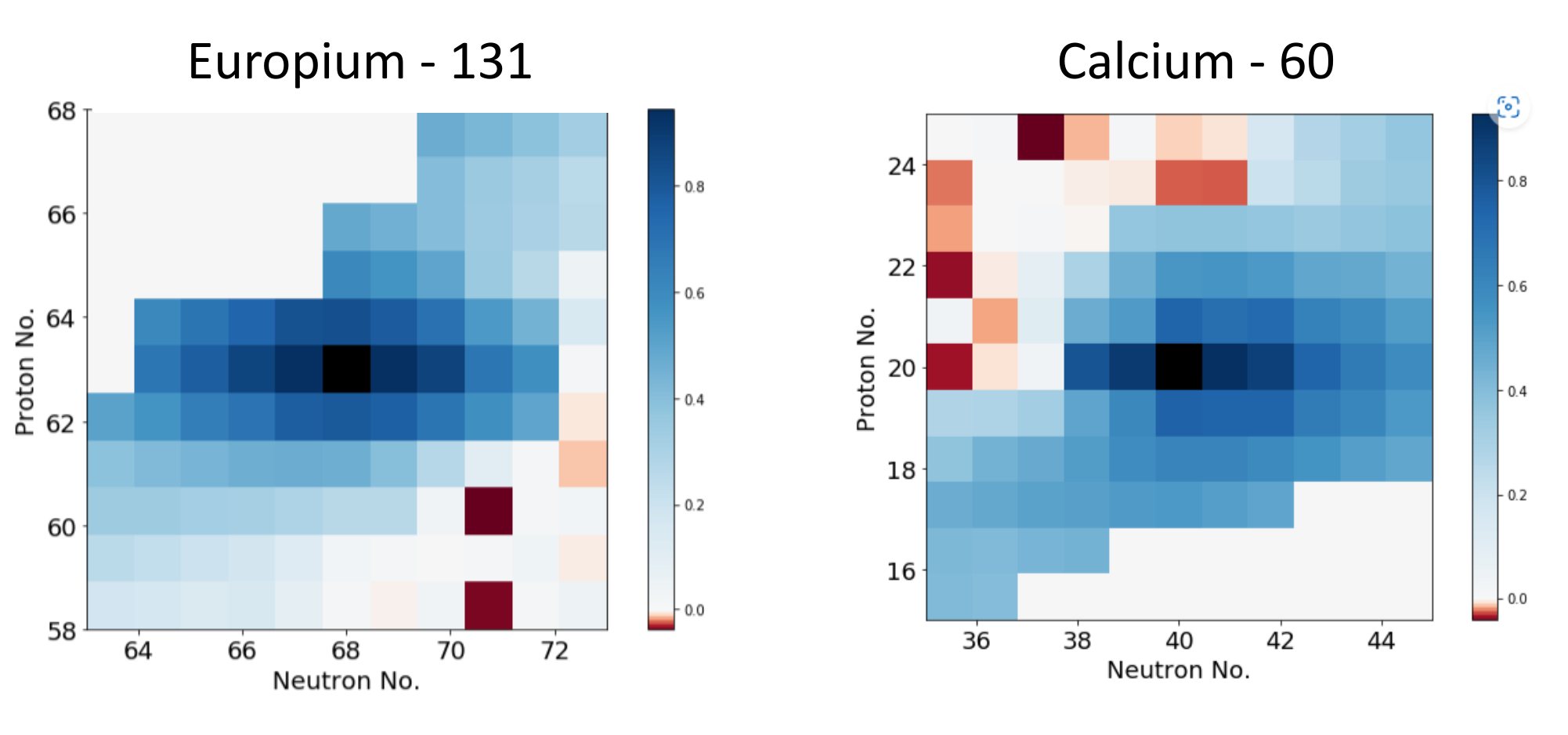}
    \caption{\large Heatmap of correlation coefficients of BMA predictions of the mass of a nucleus with its neighboring nuclei. The black square in the center corresponds to a complete correlation of the mass of a nuclei with itself. BMA predictions have high positive correlations on both sides of the valley of stability, and along both the proton number as well as the neutron number axes. }
    \label{fig:correlations}
\end{figure}

Neighboring nuclei in the samples of mass tables from posterior distributions show plenty of correlations along the proton number as well as the neutron number axes. Figure \ref{fig:correlations} shows the correlation coefficients with neighboring nuclei for $^{131}$Eu on the proton-rich side and $^{60}$Ca on the neutron-rich side of the valley of stability. The mass prediction of any nucleus is perfectly correlated with itself, which is shown by the black square in the center corresponding to a correlation coefficient of unity. The correlation is strong on an average of up to 3 units in the proton number axis and up to 5 units in the neutron number axis. 

\bigskip
\Large \subsection{Quantified Uncertainties with Covariances}\label{sec:ecp}
\bigskip
\large

\begin{figure}
    \centering
    \includegraphics[width=450 pt,keepaspectratio]{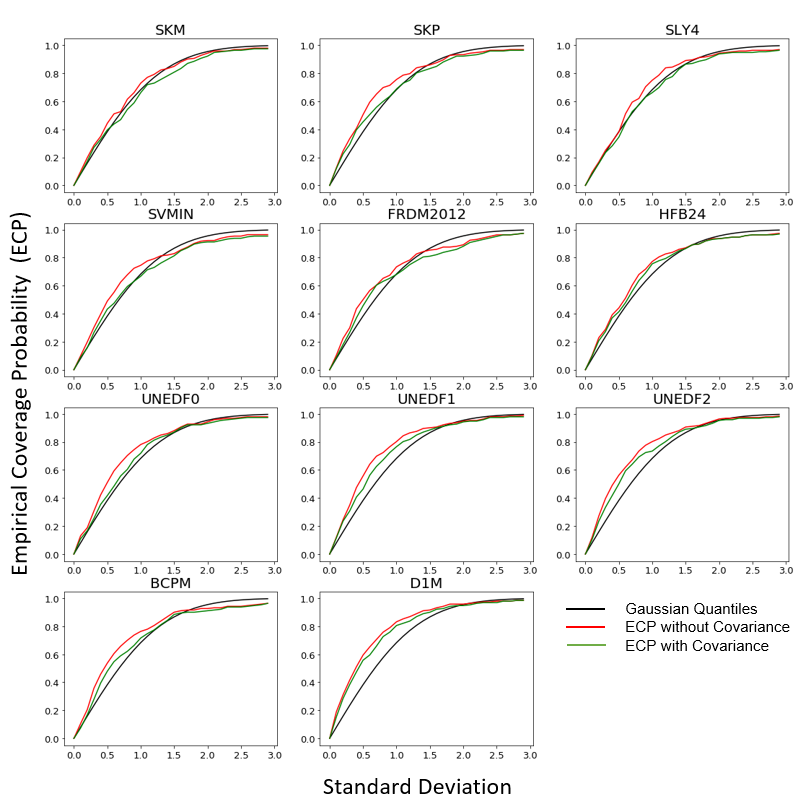}
    \caption{\large Empirical Coverage Probabilities (ECP) for two-neutron separation energies derived from BMA mass predictions (see Section \ref{sec:ecp}). The red line doesn't account for the covariance term in uncertainty calculation whereas the green line accounts for it and as a result, showcases a more honest uncertainty quantification (closer to the black line). }
    \label{fig:ecp}
\end{figure}

BMA doesn't just predict quantified uncertainties, but also the full covariance matrix. As seen in Section \ref{sec:correlations}, masses of neighboring nuclei are correlated. This face should be exploited when sampling nuclear masses, for example, for a sensitivity study. Moreover, masses are seldomly used directly in nuclear astrophysics simulations. They are converted into mass filters like separation energies, Q-values, etc. before being utilized. It then becomes crucial to use the covariances to properly propagate uncertainties to mass filters. Let's say we want to calculate the two-neutron separation energies $(S_{2n})$ from BMA masses. It can be calculated as - 
\begin{equation}
    \large S_{2n} (Z,N) = BE(Z,N) - BE(Z,N-2)
\end{equation}
where BE is the binding energy, Z is the proton number and N is the neutron number. The uncertainties without covariances can be calculated as:
\begin{equation}\label{eqn:unc}
    \large \delta(S_{2n}) = \sqrt{\sigma_{(Z,N)}^2 + \sigma_{(Z,N-2)}^2}
\end{equation}
This corresponds to the red curves in Figure \ref{fig:ecp}. It shows the empirical coverage probabilities (ECP), i.e. the fraction of validation data that lies within a given standard deviation from the mean prediction. For the normal uncertainties reported, approximately 68\% of validation data should fall within 1-$\sigma$, 95\% within 2-$\sigma$, and so on. A perfectly honest uncertainty quantification is shown by the black lines that correspond to Gaussian quantiles. Rewriting Equation \ref{eqn:unc} with the covariance term, we get
\begin{equation}\label{eqn:unc}
    \large \delta(S_{2n}) = \sqrt{\sigma_{(Z,N)}^2 + \sigma_{(Z,N-2)}^2 - 2 \sigma_{(Z,N)(Z,N-2)}}
\end{equation}
where the last term represents the covariance between nuclei (Z,N) and (Z,N-2). Plotting ECPs with covariances gives the green curves in Figure \ref{fig:ecp}. It can be seen that the green curves are closer to the black curves than the red curves. While not necessarily dishonest, the red curves are overly pessimistic in their predictions. This highlights the role of covariances in proper uncertainty quantification and for calculating mass filters that are used in nuclear astrophysics calculations.

\huge \chapter{Impact on Astrophysical Models}\label{chp:impact}
\large 

This chapter discusses the astrophysical impact of updated nuclear data from this work on the strength of Urca cooling in accreting neutron stars. The \texttt{Xnet} nuclear reaction network used for such calculations is introduced, followed by the changes in Urca cooling due to the revised ground-state to ground-state $\beta$-decay transition strength in $^{33}$Mg $\rightarrow $ $^{33}$Al. The changes in the profile of Urca cooling due to the updated BMA mass model (Chapter \ref{chp:massmodel}) are also discussed. 

\bigskip
\LARGE \section{\texttt{Xnet} Nuclear Reaction Network}
\bigskip
\large

\texttt{Xnet} is a customized nuclear reaction network for simulating the steady-state composition in accreting neutron star crusts. It follows the compositional changes in an accreted fluid element as it is pushed deeper into the neutron star by matter from ongoing accretion. It is a comprehensive reaction network that allows for all possible nuclear reactions that can take place in the crust including electron capture, $\beta$ decay, neutron capture/emission, thermonuclear and pycnonuclear fusion, and neutron transfer reactions as discussed in Section \ref{sec:intro-reactions}. It also tracks the composition of the fluid element and the amount of nuclear energy generation as a function of the depth. This gives the steady-state composition of the crust, and the nuclear heating profile in the crust, respectively. It is, therefore, a great tool to study the effects of changing nuclear properties on the composition and heat deposition in the accreting neutron star crust.  

The amount of Urca cooling in the outer crust is the most sensitive to weak interaction rates and nuclear masses. The $\beta$-decay and electron capture rates due to weak interactions depend on external conditions like temperature, density, and electron Fermi energy. As a result, only the transition strengths that depend on the initial and final nuclear wavefunctions are stored in the \texttt{Xnet} database. All the weak reaction rates are calculated on the fly for each timestep using a fast phase space approximation \cite{Gupta2007}. Experimental data for the transition strengths is used wherever available whereas theoretically calculated transition strengths with the Quasi-Random Phase Approximation (QRPA) \cite{Mller1990} are used otherwise. Nuclear masses from the latest Atomic Mass Evaluation (AME) database \cite{Wang2017} are used if available along with the Finite Range Droplet Model (FRDM) mass predictions \cite{Mller2016} where experimental data is not available. 

\bigskip
\LARGE \section{$^{33}$Mg - $^{33}$Al Transition}
\bigskip
\large

\begin{table}[]
    \centering
    \begin{tabular}{ccc}
    \toprule
       \large $^{33}$Mg$_{g.s.} \rightarrow \hspace{3pt} ^{33}$Al$_{g.s.}$ & \large  Tripathi et al. \cite{Tripathi2008}  & \large This work \\
    \midrule
      \large Feeding Intensity (\%)  & \large 37(8) & \large 0.7(24)  \\
      \large log-ft value   & \large 5.2(1) & \large 7.0$_{-0.7}^{+\infty}$ \\
      \large Intrinsic Urca Cooling Luminosity (L$_{34}$)  & \large 3847.3  & \large 61.0  \\
    \bottomrule
    \end{tabular}
    \caption{\large Comparison of the updated parameters for $^{33}$Mg$_{g.s.} \rightarrow \hspace{3pt} ^{33}$Al$_{g.s.}$ transition from this work with previous measurements. A two orders of magnitude reduction in the Intrinsic Urca Cooling Luminosity is found.}
    \label{tab:comparison}
\end{table}

\begin{figure}
    \centering
    \includegraphics[width=450pt,keepaspectratio]{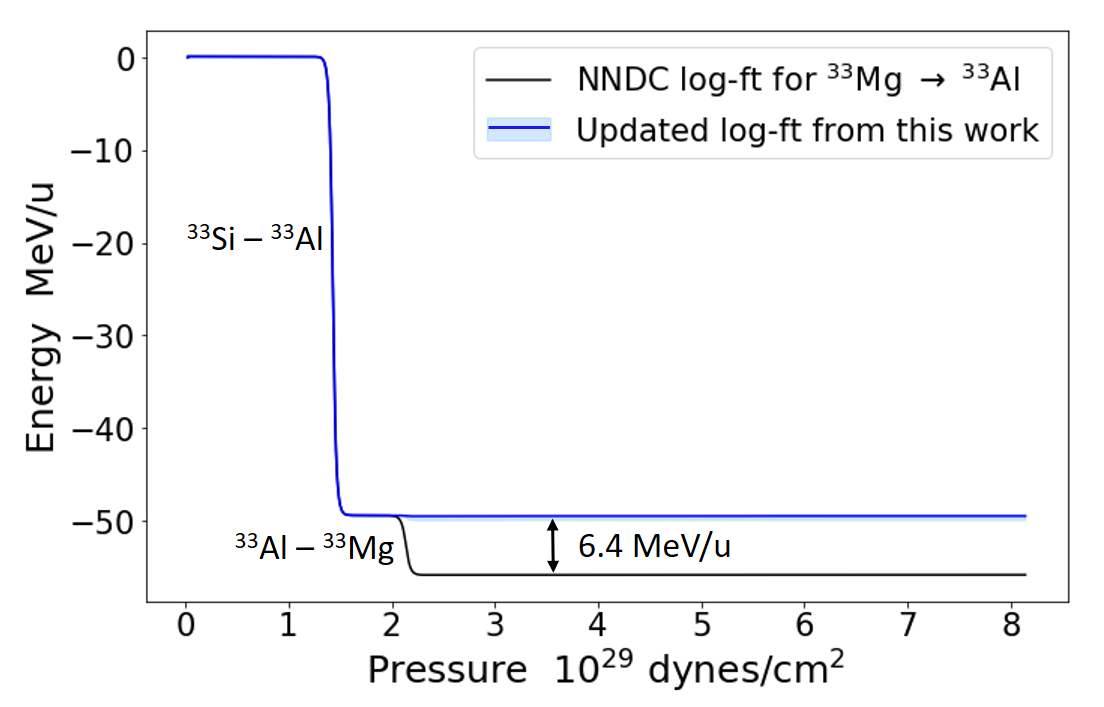}
    \caption{\large Total integrated energy deposited by nuclear reactions as a function of pressure in the accreting neutron star crust (a proxy for depth) as calculated by \texttt{Xnet} for a pure \emph{A} = 33 mass chain. Updated log-ft values for $^{33}$Mg$ \hspace{3pt} \rightarrow \hspace{3pt} ^{33}$Al transition from this work leads to a 6.4 MeV/u reduction in Urca cooling.} 
    \label{fig:mg33cooling}
\end{figure}

The transition strengths for the $\beta$-decay of $^{33}$Mg $\rightarrow$ $^{33}$Al had been previously measured by Tripathi et al.\cite{Tripathi2008}, which are included in the National Nuclear Data Center (NNDC) database. These experimental values are used in \texttt{Xnet} for the crust cooling calculations by default. However, the ground state to ground state transition strength is found to be significantly lower in this work. Table \ref{tab:comparison} compares the feeding intensities, nuclear log-ft values, and the corresponding Urca cooling luminosity for previous and current measurements. The lower feeding intensity to the ground state, and the corresponding larger log-ft value translates into a two orders of magnitude reduction for the intrinsic Urca luminosity. The updated log-ft value and the transition strengths from this work were used in \texttt{Xnet} for refining the crust cooling calculations. \texttt{Xnet} was run for a pure \emph{A} = 33 mass chain with and without the updated experimental results from this work. Figure \ref{fig:mg33cooling} shows the integrated nuclear energy deposition in the crust as a function of the pressure, which is a proxy for depth in the neutron star crust. Decreases in energy depositions signify negative net energy generation and thus, the presence of Urca cooling. The reduced ground state to ground state transition strength for the $\beta$-decay of $^{33}$Mg $\rightarrow$ $^{33}$Al from this work leads to a 6.4 MeV/u reduction in Urca cooling as shown in Figure \ref{fig:mg33cooling}. It should be noted that these simplistic calculations are for the maximum Urca cooling possible as the entire crust is composed of just mass \emph{A} = 33 nuclei. In realistic crusts, the mass fraction of \emph{A} = 33 nuclei will be folded into the calculations, which is still significant for the crust composed of X-ray burst ashes.

\bigskip
\LARGE \section{Nuclear Masses}
\bigskip
\large

\begin{figure}
    \centering
    \includegraphics[width=450pt,keepaspectratio]{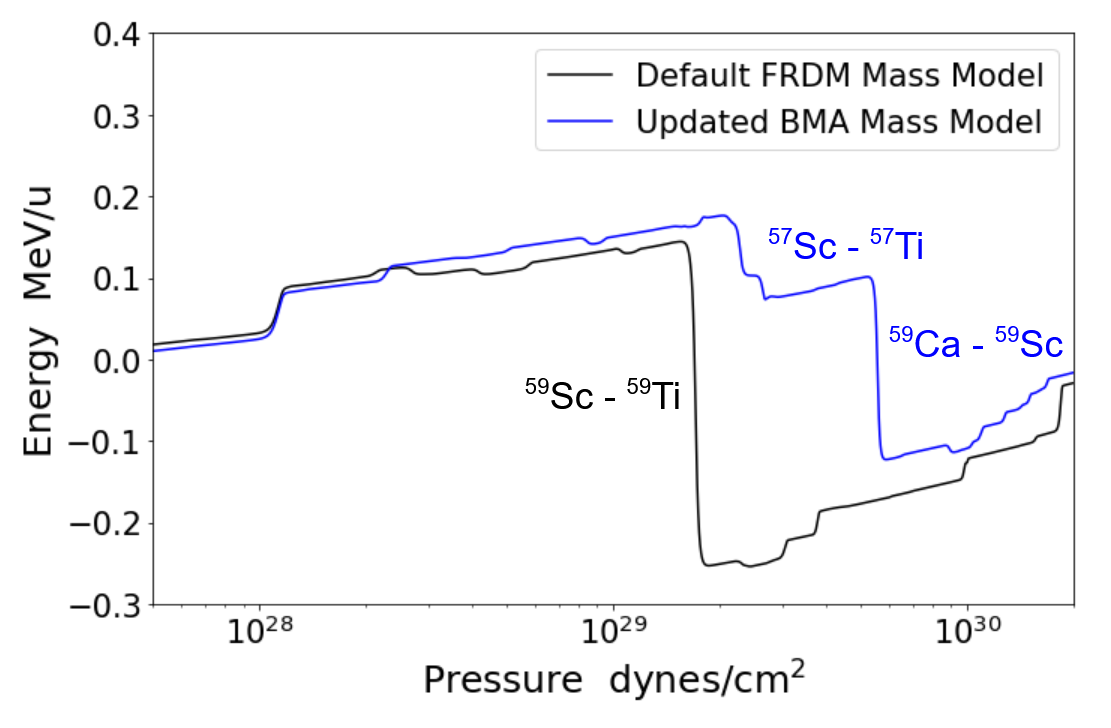}
    \caption{\large Total integrated energy deposited by nuclear reactions as a function of pressure in the accreting neutron star crust (a proxy for depth) as calculated by \texttt{Xnet} for crust composed of superburst ashes. Updating the nuclear masses in \texttt{Xnet} from the BMA model leads to completely different Urca cooling pairs in the crust and the overall location of Urca cooling is shifted deeper in the crust.}
    \label{fig:massescooling}
\end{figure}

Nuclear masses in \texttt{Xnet} were updated from the default FRDM model to the BMA mass model developed in this work. Changing nuclear masses affects multiple nuclear reaction rates. This can change the trajectory of the entire network due to a complex non-linear dependence of the abundances on the reaction rates. Changes in electron capture and $\beta$-decay Q-value change the depth and strength of electron capture heating as well as Urca cooling. Changes in neutron separation energies change the sequence of neutron release and the transition from the outer crust to the inner crust. They also change the neutron transfer reaction rates. This changes the abundance distribution per mass chain, and again impacts electron capture heating and Urca cooling. \texttt{Xnet} calculations were performed for a crust composed of superburst ashes \cite{Keek2012} with the default FRDM mass model as well as the updated BMA mass model. Figure \ref{fig:massescooling} shows the comparison of nuclear energy deposition as a function of the pressure or the depth in the neutron star crust for both cases. Dips in energy generation correspond to Urca cooling. Updating the nuclear masses from the BMA mass model changes the relevant Urca cooling pairs. $^{59}$Ca {--} $^{59}$Sc is the most important Urca pair for mass \emph{A} = 59 chain instead of the $^{59}$Sc {--} $^{59}$Ti Urca pair with the FRDM mass model. The $^{57}$Sc {--} $^{57}$Ti Urca pair in the mass \emph{A} = 57 chain becomes relevant which was not a significant Urca cooling chain with the FRDM mass model. Thus, changing nuclear masses leads to a very different Urca cooling profile. 

Although the BMA masses lead to a slightly lower Urca cooling overall, it happens at deeper locations. The depth of Urca cooling has important implications for the thermal structure of the neutron star crust. Since Urca cooling is extremely sensitive to temperature $(\sim \text{T}^5)$ \cite{Shapiro1983}, it thermally decouples the regions above and below an Urca cooling layer. If a particular heat source is below an Urca cooling layer, it will increase the rate of Urca cooling, and most of the energy deposited will be radiated away by the neutrinos. This puts a strong constraint on the location of heat sources in the neutron star crust. A previous study \cite{Meisel2020} explored the effect of nuclear masses on Urca cooling where new experimental mass measurements led to a cooler crust. Results in this work also confirm this crust cooling sensitivity to nuclear masses. As a result, it is very important to use an accurate nuclear mass model for crust cooling calculations.

\huge \chapter{Conclusions and Future Directions}\label{chp:conclusions}
\large

The experimental results presented in this work confirm the forbidden nature of the $\beta$-decay transition from the ground state in $^{33}$Mg to the ground state in $^{33}$Al. This can be concluded from the small 0.7(24)\% branching measured to the ground state corresponding to a nuclear log-ft value of 7.0$_{-0.7}^{+\infty}$. This result was achieved using a combination of high-efficiency $\gamma$ detectors for performing Total Absorption Spectroscopy and a high-efficiency neutron detector to constrain the $\beta$-delayed neutron branching ratio. The choice of this experimental setup helped mitigate the systematic errors arising from the Pandemonium effect that could lead to an overestimation of the low-energy branchings. Based on the $\beta$-decay selection rules, this result is in agreement with a negative parity assigned to the ground state in $^{33}$Mg \cite{Datta2016,Bazin2021}. The impact of this measurement on the Urca cooling in accreting neutron star crusts was also investigated. For a crust composed completely of mass \emph{A} = 33 mass chain, the new ground state to ground state transition strength of $^{33}$Mg $\longleftrightarrow \hspace{2pt} ^{33}$Al reduces the Strength of Urca cooling from 55.8 MeV/u to 49.4 MeV/u by 6.4 MeV/u. Previous results from the $\beta$-decay of $^{61}$V \cite{Ong2020} and the ongoing analysis from the $\beta$-decay of $^{55}$Ca seem to indicate a trend that both QRPA and experimentally determined Urca cooling rates seem to be over-estimated. 

In addition to the ground state transition, the $\beta$-feeding intensities to all the excited states in $^{33}$Al were also measured. The strongest feeding of 28.9(4)\% was found to the 4730 keV excited state in $^{33}$Al, corresponding to a nuclear log-ft value of 4.45(2). A total of 29.6\% feeding was found to the quasi-continuum region above the neutron separation energy. The $\beta$-delayed neutron branching ratio was measured to be 7.9(13)\%. Additionally, based on the analysis of $\gamma$ ray energy spectra, two new excited states at 3983 keV and 5361 keV were proposed with $\beta$-feeding intensities of 6.9(1)\% and 10.7(2)\% corresponding to log-ft values of 5.24(2) and 4.73(2), respectively. The 3983 keV state is found to decay via the 1618 keV state emitting a 2365 keV $\gamma$ ray in the cascade. This intensity was wrongly attributed to a 2365 keV state in previous measurements and the Total Absorption Spectroscopy technique helped identify the new level. 

A new technique for modeling nuclear mass residuals was developed using Bayesian Gaussian Process Regression leading to statistically corrected theoretical mass models. Further, predictions from multiple statistically corrected models were combined using Bayesian Model Averaging (BMA) to yield a global nuclear mass model with quantified uncertainties. The BMA mass model agrees with all experimental mass measurements while also retaining the power to extrapolate the nuclear mass surface to more exotic nuclei with quantified uncertainties. Accreting neutron star crust calculation with the updated BMA mass model for a crust composed of superburst ashes leads to a slightly reduced Urca cooling but deeper in the crust compared to the default calculations with the FRDM mass model. This sensitivity highlights the need to use accurate nuclear mass models in such calculations. Previous works \cite{Meisel2020} also report the correlations between nuclear masses and the strength of Urca cooling. 

\bigskip
\LARGE \section{Future Directions for Nuclear Experiments}
\bigskip
\large

The experimental results presented in this work highlight the efficacy of the Total Absorption Spectroscopy to measure the ground state to ground state $\beta$-decay transition strengths in exotic neutron-rich nuclei. Even for nuclei with experimentally measured log-ft values, the Pandemonium effect may have systematically skewed the results and there seems to be a trend of QRPA and previous experiments to overestimate Urca cooling. As a result, it is important to perform more experimental measurements for all relevant Urca pairs with the technique used in this work. With the Facility for Rare Isotope Beams (FRIB) coming online in the last couple of years, more exotic Urca cooling candidates will be within experimental reach in the near future. Detector systems like the FRIB Decay Station initiator (FDSi) and the FRIB Decay Station (FDS) in the future with their modular ability to perform high-resolution $\gamma$-ray spectroscopy with DEGA, total absorption spectroscopy with MTAS, and neutron spectroscopy with VANDLE and NEXT in the same experiment provide the best platform for such studies. Multiple scientific results from FDSi experiments have already been published \cite{Crawford2022,Gray2023,Lubna2023} and a $\beta$-decay experiment of multiple isotopes in the region around $^{55}$Ca motivated by Urca cooling in superbursting systems is currently being analyzed. In addition to $\beta$-decay measurements, the results presented in this work also highlight the importance of nuclear mass measurements given the sensitivity of Urca cooling to nuclear masses. All such efforts will help improve our understanding and properly quantifying the role of Urca cooling in accreting neutron star crusts.  

\bigskip
\LARGE \section{Future Directions for Modeling Neutron Star Crusts}
\bigskip
\large

The immediate future direction is to calculate the impact of the experimental results presented in this work for more realistic crust compositions, including X-ray burst ashes. \texttt{Xnet} is a tool of choice for such accreting neutron star crust calculations relying on a lot of nuclear data. One avenue for improvement is to increase the quality of nuclear data in its database and regularly update it with new experimental results. Another avenue for improvement would be on the computational side to speed up the calculations. The availability of a mass model like the BMA with quantified uncertainties and covariances allows for the full propagation of mass uncertainties on the crust calculations. However, to do this with an MCMC-type sampling process, the reaction network would need to run much faster than it currently does. 

Recently suggested alternate approach \cite{Gusakov2020,Gusakov2021,Shchechilin2021} with the diffusion of free neutrons in the inner crust would violate the hydrostatic equilibrium condition and the assumption of a constant red-shifted neutron chemical potential. This would lead to a very different Equation of State (EOS) that would transform the inner crust into homogeneous nuclear matter. As a result, no pycnonuclear fusion reactions can take place and the heat deposited would be much lower than current estimates. Although the reaction pathways for this process have not yet been identified, \texttt{Xnet} with the appropriate modifications can be used to identify such reaction pathways and improve the accreting neutron star crust models. 

\bigskip
\LARGE \section{Future Directions for Modeling Nuclear Masses}
\bigskip
\large

The BMA nuclear mass model developed in this work extrapolates the nuclear mass surface beyond the experimentally measured region on the nuclear chart. The correlation length for mass predictions is found to be few nucleons in the neutron number axis as well as the proton number axis. As a result, with newer experimental nuclear mass measurements, the BMA model needs to be updated periodically to ensure most of the experimental trend is still captured. More mass measurement will also lead to a larger evidence set that can be used to refine BMA weights. Another promising avenue for further progress would be toward Local Bayesian Model Mixing (LBMM) \cite{Kejzlar2023} where the weights of participating models can vary over the nuclear chart as opposed to being constant like in BMA. It is a kind of hierarchical model where the parameters of the functions for model mixing weights are learned from the data. This allows for the development of niche theoretical models that can be mixed locally.

Nuclear mass residuals modeled with the Gaussian Process emulator contain information about the shortcomings of a theoretical model. This information can be used to develop high-fidelity theoretical models and drive progress in nuclear theory. Several frameworks combining data science and machine learning with the domain knowledge in nuclear physics have been explored in the past and are actively being explored for many nuclear observables. The Bayesian Gaussian Process Regression and Bayesian Model Averaging techniques explored in this work for predicting nuclear masses are excellent tools that can also be applied to other nuclear quantities.

%
\chapter*{\centering \LARGE \vfill APPENDICES \vfill }

\begin{appendices}
\huge \chapter{List of my contributions}

\LARGE \section{Publications}

\Large \subsection{Neutron star crust modeling}
\bigskip

\begin{enumerate}
    \large \item  ``Impact of Pycnonuclear Fusion Uncertainties on the Cooling of Accreting Neutron Star Crusts."\\ 
    \href{https://iopscience.iop.org/article/10.3847/1538-4357/acebc4}{\underline{R. Jain} \emph{et al.}, The Astrophysics Journal \textbf{955}, 51 (2023).}
    \begin{itemize}
        \large \item Performed and analyzed the sensitivity studies calculations.
        \large \item Wrote the first draft for the paper. 
    \end{itemize}

    \large \item ``The impact of neutron transfer on heating and cooling of accreted neutron star crusts."\\
    \href{https://iopscience.iop.org/article/10.3847/1538-4357/ac4271}{ H. Schatz \emph{et al.}, The Astrophysical Journal \textbf{925}, 205 (2022).}
     \begin{itemize}
        \large \item Performed reaction network calculations with neutron transfer reactions included for multiple crust compositions.
        \large \item Discussed and analyzed the outputs of these calculations. 
    \end{itemize}

    \large \item  ``Properties of the Shallow Heat Source in KS 1731{--}260."\\ 
    {\underline{R. Jain} \emph{et al.}, (in prep).}
    \begin{itemize}
        \large \item Performed Bayesian fitting for the properties of the shallow heat source in KS 1731{--}260 for various crust compositions.
        \large \item Wrote the first draft for the paper. 
    \end{itemize}

     \large \item  ``Effects of nuclear pasta impurity on the cooling of KS 1731{--}260."\\ 
    {\underline{R. Jain} \emph{et al.}, (in prep).}
    \begin{itemize}
        \large \item Inferred the properties of the shallow heat source in KS 1731{--}260 with and without nuclear pasta impurity and analyzed the results.  
    \end{itemize}

    \large \item  ``Effect of nuclear masses on Urca Cooling in accreting neutron star crusts."\\ 
    {\underline{R. Jain}, S. Lalit \emph{et al.}, (in prep).}
    \begin{itemize}
        \large \item Provided nuclear masses for the Bayesian Model Averaging mass model.
        \large \item Analyzed the effects of changing mass models on numerous Urca pairs.
        \large \item Contributed to the writing of the paper. 
    \end{itemize}

\end{enumerate}

\bigskip
\Large \subsection{Nuclear Physics Experiments}
\bigskip

\begin{enumerate}
     \large \item  ``$\beta$-decay of $^{33}$Mg with Total Absorption Spectroscopy."\\ 
    {\underline{R. Jain} \emph{et al.}, (in prep).}
    \begin{itemize}
        \large \item Set up and performed the experiment, and analyzed the experimental data as my Ph.D. thesis experiment.
        \large \item Wrote the first draft for the paper. 
    \end{itemize}
    
    \large \item  ``Proton shell gaps in N=28 nuclei from the first complete spectroscopy study with FRIB Decay Station Initiator."\\ \href{https://journals.aps.org/prl/abstract/10.1103/PhysRevLett.132.152503}{ I. Cox \emph{et al.}, Physical Review Letters \textbf{132}, 152503 (2024).}

    \large \item ``$\beta$-decay of neutron-rich $^{45}$Cl located at the magic number N = 28."\\
    \href{https://journals.aps.org/prc/abstract/10.1103/PhysRevC.108.024312}{ S. Bhattacharya \emph{et al.}, Physical Review C \textbf{108}, 024312 (2023).}

    \large \item ``$\beta$-decay of $^{36}$Mg and $^{36}$Al: Identification of $\beta$-decaying isomer in $^{36}$Al."\\
    \href{https://journals.aps.org/prc/abstract/10.1103/PhysRevC.108.014329}{R. S. Lubna \emph{et al.}, Physical Review C \textbf{108}, 014329 (2023).}

    \large \item ``Microsecond Isomer at the N = 20 Island of Shape Inversion Observed at FRIB."\\
    \href{https://journals.aps.org/prl/abstract/10.1103/PhysRevLett.130.242501}{ T. J. Gray \emph{et al.}, Physical Review Letters \textbf{130}, 242501 (2023).}

    \large \item ``Crossing N = 28 Toward the Neutron Drip Line: First Measurement of Half-Lives at FRIB."\\
    \href{https://journals.aps.org/prl/abstract/10.1103/PhysRevLett.129.212501}{H. L. Crawford \emph{et al.}, Physical Review Letters \textbf{129}, 212501 (2022).}

    \large \item ``Evidence of a Near-Threshold Resonance in $^{11}{\mathrm B}$ Relevant to the $\beta$-Delayed Proton Emission of $^{11}{\mathrm Be}$."\\
    \href{https://journals.aps.org/prl/abstract/10.1103/PhysRevLett.129.012501}{ Y. Ayyad \emph{et al.}, Physical Review Letters \textbf{129}, 012501 (2022).}

    \large \item ``$^{57}$Zn $\beta$-delayed proton emission establishes the $^{56}$Ni rp-process waiting point bypass."\\
    \href{https://www.sciencedirect.com/science/article/pii/S0370269322001939?via\%3Dihub}{M. Saxena \emph{et al.}, Physics Letters B \textbf{829}, 137059 (2022).}

    \large \item ``Online Bayesian optimization for a recoil mass separator."\\
    \href{https://journals.aps.org/prab/abstract/10.1103/PhysRevAccelBeams.25.044601}{ S. A. Miskovich \emph{et al.}, Physical Review Accelerators and Beams \textbf{25}, 044601 (2022).}

    \large \item ``First direct measurement of $^{59}$Cu(p,$\alpha$)$^{56}$Ni: A step towards constraining the Ni-Cu cycle in the cosmos."\\
    \href{https://doi.org/10.1103/PhysRevC.104.L042801}{ J. S. Randhawa \emph{et al.}, Physical Review C Letters \textbf{104}, L042801 (2021).}    
\end{enumerate}

\bigskip
\Large \subsection{Nuclear mass models}
\bigskip

\begin{enumerate}
    \large \item  ``Quantified nuclear mass model using Bayesian Model Averaging."\\ 
    {\underline{R. Jain}. L. Neufcourt, S. Giuliani, W. Nazarewicz (in prep).}
    \begin{itemize}
        \large \item Performed the Bayesian analysis of nuclear mass models.
        \large \item Analyzed the results and wrote the first draft of the paper. 
    \end{itemize}

    \large \item ``Precision mass measurement of lightweight self-conjugate nucleus $^{80}$Zr."\\
    \href{https://www.nature.com/articles/s41567-021-01395-w}{ A. Hamaker, E. Leistenschneider,  \underline{R. Jain} \emph{et al.}, Nature Physics \textbf{17}, 1408-1412 (2021).}
    \begin{itemize}
        \large \item Performed Bayesian analysis.
        \large \item Contributed to the writing of the methods section. 
    \end{itemize}
    
\end{enumerate}

\bigskip
\Large \subsection{Others}
\bigskip

\begin{enumerate}
   \large \item ``Horizons: nuclear astrophysics in the 2020s and beyond."\\
    \href{https://iopscience.iop.org/article/10.1088/1361-6471/ac8890}{ H. Schatz \emph{et al.}, Journal of Physics G: Nuclear and Particle Physics \textbf{49}, 110502 (2022).}
    \begin{itemize}
        \large \item Organizer of the JINA Horizons Meeting Junior Workshop that lead to this white paper.  
        \large \item Contributed to the writing of Section 6 - Diversity in nuclear astrophysics and Section 7 - Career development: perspectives of early career researchers. 
    \end{itemize}

\end{enumerate}

\bigskip
\LARGE \section{Presentations}

\Large \subsection{Invited Talks}
\bigskip

\begin{enumerate}

    \large \item ``Nuclear Physics of Accreting Neutron Stars.''\\
    \emph{Postdoctoral Candidate Seminar, Lawrence Livermore National Laboratory.}\\
    November 13, 2023, Virtual Seminar.
    
    \large \item ``Heating and Cooling of Accreting Neutron Star Crusts.''\\
    \emph{Cyclotron Colloquium, Texas A\&M University.}\\
    October 12, 2023, College Station, Texas.

    \large \item ``Quantified Nuclear Mass Model for r-process simulations using Bayesian Machine Learning Techniques''\\ \emph{The $1^{st}$ IReNA-Ukakuren Joint Workshop ``Advancing Professional Development in Nuclear Astrophysics and Beyond".}\\ August 28 - September 1, 2023, Mitaka, Tokyo, Japan.

    \large \item ``Weak Interactions for Modeling Accreting Neutron Star Crusts''\\
    \emph{IReNA Workshop on Weak Interactions in Nuclear Astrophysics.}\\
    July 10-12, 2023, East Lansing, Michigan.

    \large \item ``Nuclear Experiments for Astrophysics: An Overview''\\
    \emph{JINA-CEE Frontiers in Nuclear Astrophysics - Junior Researchers Workshop.}\\
    May 23-27, 2022, South Bend, Indiana.
\end{enumerate}

\bigskip
\Large \subsection{Contributed Talks}
\bigskip

\begin{enumerate}
    \large \item ``The role of \emph{A} = 33 mass chain in Urca Cooling of Accreting Neutron Star Crusts.''\\
    \emph{American Physical Society Division of Nuclear Physics Meeting.}\\
    November 26 - December 1, 2023, Waikolao Village, Hawaii.

    \large \item ``Nuclear Physics of Accreting Neutron Stars.''\\\emph{$17^{\text{th}}$ International Symposium on Nuclei in the Cosmos (NIC XVII).}\\
    September 17-22, 2023, Daejeon, South Korea.

    \large \item ``Quantified Nuclear Mass Model for Nuclear Astrophysics Simulations.''\\
    \emph{CeNAM Frontiers in Nuclear Astrophysics.}\\
    May 21-26, 2023, East Lansing, Michigan.
 
    \large \item ``Inferring properties of a mysterious shallow heat source in accreting neutron star crusts.''\\
    \emph{$18^{th}$ Russbach School on Nuclear Astrophysics.}\\
    March 12-18, 2023, Russbach, Austria.

    \large \item ``Can Machines Learn using Gaussian Distributions?''\\ 
    \emph{Physics Graduate Organization (PGO) Seminar, Dept. of Physics \& Astronomy, Michigan State University.}\\
    February 6, 2023, East Lansing, Michigan.

    \large \item ``$\beta$-decay of $^{33}$Mg using Total Absorption Spectroscopy.''\\
    \emph{American Physical Society Division of Nuclear Physics Meeting.}\\
    October 27-30, 2022, New Orleans, Louisiana.

    \large \item ``$\beta$-decay in Neutron Star Crusts.''\\
    \emph{Nuclear Physics in Astrophysics - X Conference.}\\
    September 4-9, 2022, CERN, Switzerland.

    \large \item ``$\beta$-decay Experiments for Urca Cooling in Neutron Star Crusts.''\\
    \emph{JINA-INT Workshop on Neutron Star Cooling.}\\
    February 14-18, 2022, INT, University of Washington, Seattle.

    \large \item ``X-rays from the Space.''\\
    \emph{AGEP Student Success Conference.}\\
    November 5-6, 2021, Michigan State University, East Lansing, Michigan.

    \large \item ``Quantified Nuclear Mass Model for Nuclear Astrophysics Simulations.''\\
    \emph{American Physical Society Division of Nuclear Physics Meeting.}\\
    October 10-14, 2021, Virtual Meeting.

    \large \item ``Pycnonuclear Fusion and the Shallow Heat Source in Accreting Neutron Star Crusts.''\\
    \emph{American Physical Society Division of Nuclear Physics Meeting.}\\
    October 29 - November 1, 2020, Virtual Meeting.

    \large \item ``Extrapolating Nuclear Masses using Bayesian Gaussian Process Regression.''\\
    \emph{A.I. for Nuclear Physics Workshop.}\\
    March 4-6, 2020, Jefferson Lab, Newport News, Virginia.

    \large \item ``Sensitivity Studies of Fusion Reactions in the Crusts of Accreting Neutron Stars''\\
    \emph{The 10th European Summer School on Experimental Nuclear Astrophysics.}\\
    June 16-23, 2019, LNS, Catania, Italy.
\end{enumerate}





  
\end{appendices}
%
%
%
%

\chapter*{\centering \LARGE \vfill BIBLIOGRAPHY \vfill }
\bibliography{references}

\begin{thebibliography}{100}

\bibitem{Angelique2006}
J.~C. Angelique.
\newblock Spectroscopy near the n=20 shell closure: beta-n decay studies of 33mg and 35al.
\newblock In {\em AIP Conference Proceedings}. AIP, 2006.

\bibitem{Tripathi2008}
Vandana Tripathi, S.~L. Tabor, P.~F. Mantica, Y.~Utsuno, P.~Bender, J.~Cook, C.~R. Hoffman, Sangjin Lee, T.~Otsuka, J.~Pereira, M.~Perry, K.~Pepper, J.~S. Pinter, J.~Stoker, A.~Volya, and D.~Weisshaar.
\newblock Intruder configurations in 33mg.
\newblock {\em Physical Review Letters}, 101(14), October 2008.

\bibitem{Brown2009Mapping-Crustal}
Edward~F. Brown and Andrew Cumming.
\newblock Mapping crustal heating with the cooling lightcurves of quasi-persistent transients.
\newblock {\em The Astrophysics Journal}, 698:1020, 2009.

\bibitem{Wang2021}
Meng Wang, W.J. Huang, F.G. Kondev, G.~Audi, and S.~Naimi.
\newblock The {AME} 2020 atomic mass evaluation ({II}). {T}ables, graphs and references.
\newblock {\em Chin. Phys. C}, 45(3):030003, March 2021.

\bibitem{Hardy1977}
J.C. Hardy, L.C. Carraz, B.~Jonson, and P.G. Hansen.
\newblock The essential decay of pandemonium: A demonstration of errors in complex beta-decay schemes.
\newblock {\em Physics Letters B}, 71(2):307–310, November 1977.

\bibitem{Aguado2012}
María Esther~Estévez Aguado.
\newblock Tas measurements for neutrino physics and nuclear structure: study of the beta decays of 150er, 152,156yb and 188,190,192pb.
\newblock {\em Ph. D. Thesis}, 2012.

\bibitem{Tarasov_2008}
O.B. Tarasov and D.~Bazin.
\newblock Lise++: Radioactive beam production with in-flight separators.
\newblock {\em Nuclear Instruments and Methods in Physics Research Section B: Beam Interactions with Materials and Atoms}, 266(19–20):4657–4664, October 2008.

\bibitem{Pereira2010}
J.~Pereira, P.~Hosmer, G.~Lorusso, P.~Santi, A.~Couture, J.~Daly, M.~Del~Santo, T.~Elliot, J.~Görres, C.~Herlitzius, K.-L. Kratz, L.O. Lamm, H.Y. Lee, F.~Montes, M.~Ouellette, E.~Pellegrini, P.~Reeder, H.~Schatz, F.~Schertz, L.~Schnorrenberger, K.~Smith, E.~Stech, E.~Strandberg, C.~Ugalde, M.~Wiescher, and A.~Wöhr.
\newblock The neutron long counter nero for studies of neutron emission in the r-process.
\newblock {\em Nuclear Instruments and Methods in Physics Research Section A: Accelerators, Spectrometers, Detectors and Associated Equipment}, 618(1–3):275–283, June 2010.

\bibitem{Simon2013}
A.~Simon, S.J. Quinn, A.~Spyrou, A.~Battaglia, I.~Beskin, A.~Best, B.~Bucher, M.~Couder, P.A. DeYoung, X.~Fang, J.~Görres, A.~Kontos, Q.~Li, S.N. Liddick, A.~Long, S.~Lyons, K.~Padmanabhan, J.~Peace, A.~Roberts, D.~Robertson, K.~Smith, M.K. Smith, E.~Stech, B.~Stefanek, W.P. Tan, X.D. Tang, and M.~Wiescher.
\newblock Sun: Summing nai(tl) gamma-ray detector for capture reaction measurements.
\newblock {\em Nuclear Instruments and Methods in Physics Research Section A: Accelerators, Spectrometers, Detectors and Associated Equipment}, 703:16–21, March 2013.

\bibitem{Prokop2014}
C.J. Prokop, S.N. Liddick, B.L. Abromeit, A.T. Chemey, N.R. Larson, S.~Suchyta, and J.R. Tompkins.
\newblock Digital data acquisition system implementation at the national superconducting cyclotron laboratory.
\newblock {\em Nuclear Instruments and Methods in Physics Research Section A: Accelerators, Spectrometers, Detectors and Associated Equipment}, 741:163–168, March 2014.

\bibitem{Ong2018}
Wei~Jia Ong.
\newblock Quantifying the urca cooling impact of mass 61 nuclei in x-ray bursting systems.
\newblock {\em Ph. D. Thesis}, 2018.

\bibitem{Dombos2018}
Alexander~Connor Dombos.
\newblock beta-decay total absorption spectroscopy around a = 100-110 relevant to nuclear structure and the astrophysical r process.
\newblock {\em Ph. D. Thesis}, 2018.

\bibitem{Neufcourt2019}
L{\'{e}}o Neufcourt, Yuchen Cao, Witold Nazarewicz, Erik Olsen, and Frederi Viens.
\newblock Neutron drip line in the ca region from bayesian model averaging.
\newblock {\em Phys. Rev. Lett.}, 122(6):062502, February 2019.

\bibitem{Bell1969}
Jocelyn Bell~Burnell.
\newblock The measurement of radio source diameters using a diffraction method.
\newblock 1969.

\bibitem{ozel2016}
Feryal \"{O}zel and Paulo Freire.
\newblock Masses, radii, and the equation of state of neutron stars.
\newblock {\em Annual Review of Astronomy and Astrophysics}, 54(1):401–440, September 2016.

\bibitem{Burrows2021}
A.~Burrows and D.~Vartanyan.
\newblock Core-collapse supernova explosion theory.
\newblock {\em Nature}, 589(7840):29–39, January 2021.

\bibitem{Lattimer2016}
James~M. Lattimer and Madappa Prakash.
\newblock The equation of state of hot, dense matter and neutron stars.
\newblock {\em Physics Reports}, 621:127–164, March 2016.

\bibitem{Baym2018}
Gordon Baym, Tetsuo Hatsuda, Toru Kojo, Philip~D Powell, Yifan Song, and Tatsuyuki Takatsuka.
\newblock From hadrons to quarks in neutron stars: a review.
\newblock {\em Reports on Progress in Physics}, 81(5):056902, March 2018.

\bibitem{Oertel2017}
M.~Oertel, M.~Hempel, T.~Kl\"{a}hn, and S.~Typel.
\newblock Equations of state for supernovae and compact stars.
\newblock {\em Reviews of Modern Physics}, 89(1), March 2017.

\bibitem{Bailes2021}
M.~Bailes, B.~K. Berger, P.~R. Brady, M.~Branchesi, K.~Danzmann, M.~Evans, K.~Holley-Bockelmann, B.~R. Iyer, T.~Kajita, S.~Katsanevas, M.~Kramer, A.~Lazzarini, L.~Lehner, G.~Losurdo, H.~L\"{u}ck, D.~E. McClelland, M.~A. McLaughlin, M.~Punturo, S.~Ransom, S.~Raychaudhury, D.~H. Reitze, F.~Ricci, S.~Rowan, Y.~Saito, G.~H. Sanders, B.~S. Sathyaprakash, B.~F. Schutz, A.~Sesana, H.~Shinkai, X.~Siemens, D.~H. Shoemaker, J.~Thorpe, J.~F.~J. van~den Brand, and S.~Vitale.
\newblock Gravitational-wave physics and astronomy in the 2020s and 2030s.
\newblock {\em Nature Reviews Physics}, 3(5):344–366, April 2021.

\bibitem{Abbott2017}
B.~P. Abbott and Abbott et~al.
\newblock Gw170817: Observation of gravitational waves from a binary neutron star inspiral.
\newblock {\em Physical Review Letters}, 119(16), October 2017.

\bibitem{Lattimer1989}
James~M. Lattimer and A.~Yahil.
\newblock Analysis of the neutrino events from supernova 1987a.
\newblock {\em The Astrophysical Journal}, 340:426, May 1989.

\bibitem{Pian2017}
E.~Pian, P.~D’Avanzo, S.~Benetti, M.~Branchesi, E.~Brocato, S.~Campana, E.~Cappellaro, S.~Covino, V.~D’Elia, J.~P.~U. Fynbo, F.~Getman, G.~Ghirlanda, G.~Ghisellini, A.~Grado, G.~Greco, J.~Hjorth, C.~Kouveliotou, A.~Levan, L.~Limatola, D.~Malesani, P.~A. Mazzali, A.~Melandri, P.~Møller, L.~Nicastro, E.~Palazzi, S.~Piranomonte, A.~Rossi, O.~S. Salafia, J.~Selsing, G.~Stratta, M.~Tanaka, N.~R. Tanvir, L.~Tomasella, D.~Watson, S.~Yang, L.~Amati, L.~A. Antonelli, S.~Ascenzi, M.~G. Bernardini, M.~Boër, F.~Bufano, A.~Bulgarelli, M.~Capaccioli, P.~Casella, A.~J. Castro-Tirado, E.~Chassande-Mottin, R.~Ciolfi, C.~M. Copperwheat, M.~Dadina, G.~De~Cesare, A.~Di~Paola, Y.~Z. Fan, B.~Gendre, G.~Giuffrida, A.~Giunta, L.~K. Hunt, G.~L. Israel, Z.-P. Jin, M.~M. Kasliwal, S.~Klose, M.~Lisi, F.~Longo, E.~Maiorano, M.~Mapelli, N.~Masetti, L.~Nava, B.~Patricelli, D.~Perley, A.~Pescalli, T.~Piran, A.~Possenti, L.~Pulone, M.~Razzano, R.~Salvaterra, P.~Schipani, M.~Spera, A.~Stamerra, L.~Stella, G.~Tagliaferri, V.~Testa,
  E.~Troja, M.~Turatto, S.~D. Vergani, and D.~Vergani.
\newblock Spectroscopic identification of r-process nucleosynthesis in a double neutron-star merger.
\newblock {\em Nature}, 551(7678):67–70, October 2017.

\bibitem{Watson2019}
Darach Watson, Camilla~J. Hansen, Jonatan Selsing, Andreas Koch, Daniele~B. Malesani, Anja~C. Andersen, Johan P.~U. Fynbo, Almudena Arcones, Andreas Bauswein, Stefano Covino, Aniello Grado, Kasper~E. Heintz, Leslie Hunt, Chryssa Kouveliotou, Giorgos Leloudas, Andrew~J. Levan, Paolo Mazzali, and Elena Pian.
\newblock Identification of strontium in the merger of two neutron stars.
\newblock {\em Nature}, 574(7779):497–500, October 2019.

\bibitem{Schatz2022horizons}
H~Schatz, A~D Becerril~Reyes, A~Best, E~F Brown, K~Chatziioannou, K~A Chipps, C~M Deibel, R~Ezzeddine, D~K Galloway, C~J Hansen, F~Herwig, A~P Ji, M~Lugaro, Z~Meisel, D~Norman, J~S Read, L~F Roberts, A~Spyrou, I~Tews, F~X Timmes, C~Travaglio, N~Vassh, C~Abia, P~Adsley, S~Agarwal, M~Aliotta, W~Aoki, A~Arcones, A~Aryan, A~Bandyopadhyay, A~Banu, D~W Bardayan, J~Barnes, A~Bauswein, T~C Beers, J~Bishop, T~Boztepe, B~C\^oté, M~E Caplan, A~E Champagne, J~A Clark, M~Couder, A~Couture, S~E de~Mink, S~Debnath, R~J deBoer, J~den Hartogh, P~Denissenkov, V~Dexheimer, I~Dillmann, J~E Escher, M~A Famiano, R~Farmer, R~Fisher, C~Fr\"{o}hlich, A~Frebel, C~Fryer, G~Fuller, A~K Ganguly, S~Ghosh, B~K Gibson, T~Gorda, K~N Gourgouliatos, V~Graber, M~Gupta, W~C Haxton, A~Heger, W~R Hix, W~C~G Ho, E~M Holmbeck, A~A Hood, S~Huth, G~Imbriani, R~G Izzard, R~Jain, H~Jayatissa, Z~Johnston, T~Kajino, A~Kankainen, G~G Kiss, A~Kwiatkowski, M~La~Cognata, A~M Laird, L~Lamia, P~Landry, E~Laplace, K~D Launey, D~Leahy, G~Leckenby, A~Lennarz,
  B~Longfellow, A~E Lovell, W~G Lynch, S~M Lyons, K~Maeda, E~Masha, C~Matei, J~Merc, B~Messer, F~Montes, A~Mukherjee, M~R Mumpower, D~Neto, B~Nevins, W~G Newton, L~Q Nguyen, K~Nishikawa, N~Nishimura, F~M Nunes, E~O’Connor, B~W O’Shea, W-J Ong, S~D Pain, M~A Pajkos, M~Pignatari, R~G Pizzone, V~M Placco, T~Plewa, B~Pritychenko, A~Psaltis, D~Puentes, Y-Z Qian, D~Radice, D~Rapagnani, B~M Rebeiro, R~Reifarth, A~L Richard, N~Rijal, I~U Roederer, J~S Rojo, J~S K, Y~Saito, A~Schwenk, M~L Sergi, R~S Sidhu, A~Simon, T~Sivarani, Á~Skúladóttir, M~S Smith, A~Spiridon, T~M Sprouse, S~Starrfield, A~W Steiner, F~Strieder, I~Sultana, R~Surman, T~Sz\"{u}cs, A~Tawfik, F~Thielemann, L~Trache, R~Trappitsch, M~B Tsang, A~Tumino, S~Upadhyayula, J~O Valle~Martínez, M~Van~der Swaelmen, C~Viscasillas~Vázquez, A~Watts, B~Wehmeyer, M~Wiescher, C~Wrede, J~Yoon, R~G~T Zegers, M~A Zermane, and M~Zingale.
\newblock Horizons: nuclear astrophysics in the 2020s and beyond.
\newblock {\em Journal of Physics G: Nuclear and Particle Physics}, 49(11):110502, November 2022.

\bibitem{Duchene2013-ez}
Gaspard Duch{\^e}ne and Adam Kraus.
\newblock Stellar multiplicity.
\newblock {\em Annu. Rev. Astron. Astrophys.}, 51(1):269--310, August 2013.

\bibitem{Charles2011}
P.~{Charles}.
\newblock {LMXBs: An Overview}.
\newblock In L.~{Schmidtobreick}, M.~R. {Schreiber}, and C.~{Tappert}, editors, {\em Evolution of Compact Binaries}, volume 447 of {\em Astronomical Society of the Pacific Conference Series}, page~19, September 2011.

\bibitem{wijnands:ks1731}
R.~Wijnands, Jon~M. Miller, Paul~J. Groot, Craig Markwardt, Walter H.~G. Lewin, and Michiel {van der Klis}.
\newblock Chandra observation of the long-duration x-ray transient ks 1731-260 in quiescence.
\newblock {\em The Astrophysics Journal Letters}, 560:L159, 2001.

\bibitem{wijnands.ea:xmm_1731}
R.~{Wijnands}, M.~{Guainazzi}, M.~{van der Klis}, and M.~{M{\'e}ndez}.
\newblock {XMM-Newton Observations of the Neutron Star X-Ray Transient KS 1731-260 in Quiescence}.
\newblock {\em The Astrophysics Journal Letters}, 573:L45--L49, 2002.

\bibitem{Cackett2006Cooling-of-the-}
Edward~M. Cackett, Rudy Wijnands, Maunel Linares, Jon~M. Miller, Jeroen Homan, and Walter H.~G. Lewin.
\newblock Cooling of the quasi-persistent neutron star x-ray transients ks1731-260 and mxb 1659-29.
\newblock {\em Monthly Notices of the Royal Astronomical Society}, 372:479, 2006.

\bibitem{Cackett2010Continued-Cooli}
E.~M. {Cackett}, E.~F. {Brown}, A.~{Cumming}, N.~{Degenaar}, J.~M. {Miller}, and R.~{Wijnands}.
\newblock {Continued Cooling of the Crust in the Neutron Star Low-mass X-ray Binary KS 1731-260}.
\newblock {\em The Astrophysics Journal Letters}, 722:L137--L141, October 2010.

\bibitem{Wijnands2017}
Rudy Wijnands, Nathalie Degenaar, and Dany Page.
\newblock Cooling of accretion-heated neutron stars.
\newblock {\em Journal of Astrophysics and Astronomy}, 38(3), September 2017.

\bibitem{Parikh2019}
A.~S. {Parikh}, R.~{Wijnands}, L.~S. {Ootes}, D.~{Page}, N.~{Degenaar}, A.~{Bahramian}, E.~F. {Brown}, E.~M. {Cackett}, A.~{Cumming}, C.~{Heinke}, J.~{Homan}, A.~{Rouco Escorial}, and M.~J.~P. {Wijngaarden}.
\newblock {Consistent accretion-induced heating of the neutron-star crust in MXB 1659-29 during two different outbursts}.
\newblock {\em Astronomy \& Astrophysics}, 624:A84, April 2019.

\bibitem{Degenaar2019}
N.~{Degenaar}, L.~S. {Ootes}, D.~{Page}, R.~{Wijnands}, A.~S. {Parikh}, J.~{Homan}, E.~M. {Cackett}, J.~M. {Miller}, D.~{Altamirano}, and M.~{Linares}.
\newblock {Crust cooling of the neutron star in Aql X-1: different depth and magnitude of shallow heating during similar accretion outbursts}.
\newblock {\em Monthly Notices of the Royal Astronomical Society}, 488(4):4477--4486, October 2019.

\bibitem{Rutledge2002}
Robert~E. {Rutledge}, Lars {Bildsten}, Edward~F. {Brown}, George~G. {Pavlov}, Vyacheslav~E. {Zavlin}, and Greg {Ushomirsky}.
\newblock {Crustal Emission and the Quiescent Spectrum of the Neutron Star in KS 1731-260}.
\newblock {\em The Astrophysics Journal}, 580(1):413--422, November 2002.

\bibitem{Cackett2006}
Edward~M. {Cackett}, Rudy {Wijnands}, Manuel {Linares}, Jon~M. {Miller}, Jeroen {Homan}, and Walter H.~G. {Lewin}.
\newblock {Cooling of the quasi-persistent neutron star X-ray transients KS 1731-260 and MXB 1659-29}.
\newblock {\em Monthly Notices of the Royal Astronomical Society}, 372(1):479--488, October 2006.

\bibitem{Shternin2007}
P.~S. {Shternin}, D.~G. {Yakovlev}, P.~{Haensel}, and A.~Y. {Potekhin}.
\newblock {Neutron star cooling after deep crustal heating in the X-ray transient KS 1731-260}.
\newblock {\em Monthly Notices of the Royal Astronomical Society}, 382(1):L43--L47, November 2007.

\bibitem{Horowitz2015}
C.~J. {Horowitz}, D.~K. {Berry}, C.~M. {Briggs}, M.~E. {Caplan}, A.~{Cumming}, and A.~S. {Schneider}.
\newblock {Disordered Nuclear Pasta, Magnetic Field Decay, and Crust Cooling in Neutron Stars}.
\newblock {\em Physical Review Letter}, 114(3):031102, January 2015.

\bibitem{Deibel2017}
Alex {Deibel}, Andrew {Cumming}, Edward~F. {Brown}, and Sanjay {Reddy}.
\newblock {Late-time Cooling of Neutron Star Transients and the Physics of the Inner Crust}.
\newblock {\em The Astrophysics Journal}, 839(2):95, April 2017.

\bibitem{Cyburt2016}
R.~H. {Cyburt}, A.~M. {Amthor}, A.~{Heger}, E.~{Johnson}, L.~{Keek}, Z.~{Meisel}, H.~{Schatz}, and K.~{Smith}.
\newblock {Dependence of X-Ray Burst Models on Nuclear Reaction Rates}.
\newblock {\em The Astrophysics Journal}, 830(2):55, October 2016.

\bibitem{Naylor1993}
T.~Naylor and P.~Podsiadlowski.
\newblock How young are the low-mass x-ray binaries? conclusions from a flux-limited sample.
\newblock {\em Monthly Notices of the Royal Astronomical Society}, 262(4):929–935, June 1993.

\bibitem{Haensel1990}
P.~{Haensel} and J.~L. {Zdunik}.
\newblock {Non-equilibrium processes in the crust of an accreting neutron star}.
\newblock {\em Astronomy \& Astrophysics}, 227(2):431--436, January 1990.

\bibitem{Haensel2008}
P.~Haensel and J.~L. Zdunik.
\newblock Models of crustal heating in accreting neutron stars.
\newblock {\em Astronomy {\&} Astrophysics}, 480(2):459--464, January 2008.

\bibitem{Lau2018}
R.~Lau, M.~Beard, S.~S. Gupta, H.~Schatz, A.~V. Afanasjev, E.~F. Brown, A.~Deibel, L.~R. Gasques, G.~W. Hitt, W.~R. Hix, L.~Keek, P.~M\"{o}ller, P.~S. Shternin, A.~W. Steiner, M.~Wiescher, and Y.~Xu.
\newblock Nuclear reactions in the crusts of accreting neutron stars.
\newblock {\em The Astrophysical Journal}, 859(1):62, May 2018.

\bibitem{Randhawa2019}
J.~S. Randhawa, Z.~Meisel, S.~A. Giuliani, H.~Schatz, B.~S. Meyer, K.~Ebinger, A.~A. Hood, and R.~Kanungo.
\newblock Spallation-altered accreted compositions for x-ray bursts: Impact on ignition conditions and burst ashes.
\newblock {\em The Astrophysical Journal}, 887(1):100, December 2019.

\bibitem{Schatz1998}
H.~Schatz, A.~Aprahamian, J.~G\"{o}rres, M.~Wiescher, T.~Rauscher, J.F. Rembges, F.-K. Thielemann, B.~Pfeiffer, P.~M\"{o}ller, K.-L. Kratz, H.~Herndl, B.A. Brown, and H.~Rebel.
\newblock rp-process nucleosynthesis at extreme temperature and density conditions.
\newblock {\em Physics Reports}, 294(4):167–263, February 1998.

\bibitem{Meisel2022Constraining-Ac}
Zach {Meisel}.
\newblock {Constraining Accreted Neutron Star Crust Shallow Heating with the Inferred Depth of Carbon Ignition in X-ray Superbursts}.
\newblock {\em arXiv e-prints}, page arXiv:2208.03347, August 2022.

\bibitem{Page2022A-Hyperburst-in}
Dany {Page}, Jeroen {Homan}, Martin {Nava-Callejas}, Yuri {Cavecchi}, Mikhail~V. {Beznogov}, Nathalie {Degenaar}, Rudy {Wijnands}, and Aastha~S. {Parikh}.
\newblock {A ``Hyperburst'' in the MAXI J0556-332 Neutron Star: Evidence for a New Type of Thermonuclear Explosion}.
\newblock {\em arXiv e-prints}, page arXiv:2202.03962, February 2022.

\bibitem{Schatz2013}
H.~Schatz, S.~Gupta, P.~M\"{o}ller, M.~Beard, E.~F. Brown, A.~T. Deibel, L.~R. Gasques, W.~R. Hix, L.~Keek, R.~Lau, A.~W. Steiner, and M.~Wiescher.
\newblock Strong neutrino cooling by cycles of electron capture and $\upbeta$- decay in neutron star crusts.
\newblock {\em Nature}, 505(7481):62--65, December 2013.

\bibitem{Chugunov2018}
A~I Chugunov.
\newblock Neutron transfer reactions in accreting neutron stars.
\newblock {\em Monthly Notices of the Royal Astronomical Society: Letters}, 483(1):L47–L51, November 2018.

\bibitem{Schatz2022}
H.~Schatz, Z.~Meisel, E.~F. Brown, S.~S. Gupta, G.~W. Hitt, W.~R. Hix, R.~Jain, R.~Lau, P.~M\"{o}ller, W.-J. Ong, P.~S. Shternin, Y.~Xu, and M.~Wiescher.
\newblock The impact of neutron transfer reactions on the heating and cooling of accreted neutron star crusts.
\newblock {\em The Astrophysical Journal}, 925(2):205, February 2022.

\bibitem{Beard2010}
M.~Beard, A.V. Afanasjev, L.C. Chamon, L.R. Gasques, M.~Wiescher, and D.G. Yakovlev.
\newblock Astrophysical s factors for fusion reactions involving c, o, ne, and mg isotopes.
\newblock {\em Atomic Data and Nuclear Data Tables}, 96(5):541--566, September 2010.

\bibitem{Yakovlev2006}
D.~G. Yakovlev, L.~R. Gasques, A.~V. Afanasjev, M.~Beard, and M.~Wiescher.
\newblock Fusion reactions in multicomponent dense matter.
\newblock {\em Phys. Rev. C}, 74:035803, Sep 2006.

\bibitem{Jain2023}
R.~Jain, E.~F. Brown, H.~Schatz, A.~V. Afanasjev, M.~Beard, L.~R. Gasques, S.~S. Gupta, G.~W. Hitt, W.~R. Hix, R.~Lau, P.~M\"{o}ller, W.~J. Ong, M.~Wiescher, and Y.~Xu.
\newblock Impact of pycnonuclear fusion uncertainties on the cooling of accreting neutron star crusts.
\newblock {\em The Astrophysical Journal}, 955(1):51, September 2023.

\bibitem{Gupta2007}
Sanjib {Gupta}, Edward~F. {Brown}, Hendrik {Schatz}, Peter {M{\"o}ller}, and Karl-Ludwig {Kratz}.
\newblock {Heating in the Accreted Neutron Star Ocean: Implications for Superburst Ignition}.
\newblock {\em The Astrophysics Journal}, 662(2):1188--1197, June 2007.

\bibitem{Gusakov2020}
M.~E. {Gusakov} and A.~I. {Chugunov}.
\newblock {Thermodynamically Consistent Equation of State for an Accreted Neutron Star Crust}.
\newblock {\em Physical Review Letter}, 124(19):191101, May 2020.

\bibitem{Gusakov2021}
M.~E. {Gusakov} and A.~I. {Chugunov}.
\newblock {Heat release in accreting neutron stars}.
\newblock {\em Physical Review D}, 103(10):L101301, May 2021.

\bibitem{Shchechilin2021}
N~N Shchechilin, M~E Gusakov, and A~I Chugunov.
\newblock {Deep crustal heating for realistic compositions of thermonuclear ashes}.
\newblock {\em Monthly Notices of the Royal Astronomical Society}, 507(3):3860--3870, 08 2021.

\bibitem{Waterhouse2016}
A.~C. {Waterhouse}, N.~{Degenaar}, R.~{Wijnands}, E.~F. {Brown}, J.~M. {Miller}, D.~{Altamirano}, and M.~{Linares}.
\newblock {Constraining the properties of neutron star crusts with the transient low-mass X-ray binary Aql X-1}.
\newblock {\em Monthly Notices of the Royal Astronomical Society}, 456(4):4001--4014, March 2016.

\bibitem{Turlione2015}
A.~{Turlione}, D.~N. {Aguilera}, and J.~A. {Pons}.
\newblock {Quiescent thermal emission from neutron stars in low-mass X-ray binaries}.
\newblock {\em Astronomy \& Astrophysics}, 577:A5, May 2015.

\bibitem{Degenaar2014Probing-the-Cru}
N.~{Degenaar}, Z.~{Medin}, A.~{Cumming}, R.~{Wijnands}, M.~T. {Wolff}, E.~M. {Cackett}, J.~M. {Miller}, P.~G. {Jonker}, J.~{Homan}, and E.~F. {Brown}.
\newblock {Probing the Crust of the Neutron Star in EXO 0748-676}.
\newblock {\em The Astrophysics Journal}, 791:47, August 2014.

\bibitem{Potekhin2021}
A.~Y. Potekhin and G.~Chabrier.
\newblock Crust structure and thermal evolution of neutron stars in soft x-ray transients.
\newblock {\em Astronomy \& Astrophysics}, 645:A102, January 2021.

\bibitem{Degenaar2015}
N.~{Degenaar}, R.~{Wijnands}, A.~{Bahramian}, G.~R. {Sivakoff}, C.~O. {Heinke}, E.~F. {Brown}, J.~K. {Fridriksson}, J.~{Homan}, E.~M. {Cackett}, A.~{Cumming}, J.~M. {Miller}, D.~{Altamirano}, and D.~{Pooley}.
\newblock {Neutron star crust cooling in the Terzan 5 X-ray transient Swift J174805.3-244637}.
\newblock {\em Monthly Notices of the Royal Astronomical Society}, 451(2):2071--2081, August 2015.

\bibitem{Deibel2015A-Strong-Shallo}
A.~{Deibel}, A.~{Cumming}, E.~F. {Brown}, and D.~{Page}.
\newblock {A Strong Shallow Heat Source in the Accreting Neutron Star MAXI J0556-332}.
\newblock {\em The Astrophysics Journal Letters}, 809:L31, August 2015.

\bibitem{Piro2007Turbulent-Mixin}
A.~L. {Piro} and L.~{Bildsten}.
\newblock Turbulent mixing in the surface layers of accreting neutron stars.
\newblock {\em The Astrophysics Journal}, in press, April 2007.

\bibitem{Horowitz2008}
C.~J. {Horowitz}, H.~{Dussan}, and D.~K. {Berry}.
\newblock {Fusion of neutron-rich oxygen isotopes in the crust of accreting neutron stars}.
\newblock {\em Physical Review C}, 77(4):045807, April 2008.

\bibitem{Chamel2020}
N.~Chamel, A.~F. Fantina, J.~L. Zdunik, and P.~Haensel.
\newblock Experimental constraints on shallow heating in accreting neutron-star crusts.
\newblock {\em Phys. Rev. C}, 102:015804, Jul 2020.

\bibitem{Fattoyev2018}
F.~J. Fattoyev, Edward~F. Brown, Andrew Cumming, Alex Deibel, C.~J. Horowitz, Bao-An Li, and Zidu Lin.
\newblock Deep crustal heating by neutrinos from the surface of accreting neutron stars.
\newblock {\em Phys. Rev. C}, 98:025801, Aug 2018.

\bibitem{sunyaev:transient}
R.~{Sunyaev}.
\newblock Transient x-ray burster ks 1731-260.
\newblock {\em International Astronomical Union Circulars}, 4839:1, aug 1989.

\bibitem{Negreiros2012}
R.~Negreiros, R.~Ruffini, C.~L. Bianco, and J.~A. Rueda.
\newblock Cooling of young neutron stars in grb associated to supernovae.
\newblock {\em Astronomy \&amp; Astrophysics}, 540:A12, March 2012.

\bibitem{Deibel2015}
Alex {Deibel}, Andrew {Cumming}, Edward~F. {Brown}, and Dany {Page}.
\newblock {A Strong Shallow Heat Source in the Accreting Neutron Star MAXI J0556-332}.
\newblock {\em The Astrophysics Journal Letters}, 809(2):L31, August 2015.

\bibitem{Shapiro1983}
Stuart~L. Shapiro and Saul~A. Teukolsky.
\newblock {\em Black Holes, White Dwarfs, and Neutron Stars: The Physics of Compact Objects}.
\newblock Wiley, July 1983.

\bibitem{Krane1989}
Kenneth~S. Krane and William~G. Lynch.
\newblock Introductory nuclear physics.
\newblock {\em Physics Today}, 42(1):78–78, January 1989.

\bibitem{Mller2016}
P.~M\"{o}ller, A.J. Sierk, T.~Ichikawa, and H.~Sagawa.
\newblock Nuclear ground-state masses and deformations: {FRDM}(2012).
\newblock {\em At. Data Nucl. Data Tables}, 109-110:1--204, May 2016.

\bibitem{Goriely2013}
S.~Goriely, N.~Chamel, and J.~M. Pearson.
\newblock Further explorations of {Skyrme-Hartree-Fock-Bogoliubov} mass formulas. {XIII}. the 2012 atomic mass evaluation and the symmetry coefficient.
\newblock {\em Phys. Rev. C}, 88(2):024308, August 2013.

\bibitem{Morton2002}
A.C Morton, P.F Mantica, B.A Brown, A.D Davies, D.E Groh, P.T Hosmer, S.N Liddick, J.I Prisciandaro, H~Schatz, M~Steiner, and A~Stolz.
\newblock Beta decay studies of nuclei near 32mg: Investigating the v(f7/2)–(d3/2) inversion at the n=20 shell closure.
\newblock {\em Physics Letters B}, 544(3–4):274–279, September 2002.

\bibitem{Yordanov2010}
D.~T. Yordanov, K.~Blaum, M.~De~Rydt, M.~Kowalska, R.~Neugart, G.~Neyens, and I.~Hamamoto.
\newblock Comment on “intruder configurations in the 33mg”.
\newblock {\em Physical Review Letters}, 104(12), March 2010.

\bibitem{Yordanov2007}
D.~T. Yordanov, M.~Kowalska, K.~Blaum, M.~De~Rydt, K.~T. Flanagan, P.~Lievens, R.~Neugart, G.~Neyens, and H.~H. Stroke.
\newblock Spin and magnetic moment of 33mg: Evidence for a negative-parity intruder ground state.
\newblock {\em Physical Review Letters}, 99(21), November 2007.

\bibitem{Datta2016}
Ushasi Datta, A.~Rahaman, T.~Aumann, S.~Beceiro-Novo, K.~Boretzky, C.~Caesar, B.~V. Carlson, W.~N. Catford, S.~Chakraborty, M.~Chartier, D.~Cortina-Gil, G.~de~Angelis, P.~Diaz~Fernandez, H.~Emling, O.~Ershova, L.~M. Fraile, H.~Geissel, D.~Gonzalez-Diaz, B.~Jonson, H.~Johansson, N.~Kalantar-Nayestanaki, T.~Kr\"{o}ll, R.~Kr\"{u}cken, J.~Kurcewicz, C.~Langer, T.~Le~Bleis, Y.~Leifels, J.~Marganiec, G.~M\"{u}nzenberg, M.~A. Najafi, T.~Nilsson, C.~Nociforo, V.~Panin, S.~Paschalis, R.~Plag, R.~Reifarth, V.~Ricciardi, D.~Rossi, H.~Scheit, C.~Scheidenberger, H.~Simon, J.~T. Taylor, Y.~Togano, S.~Typel, V.~Volkov, A.~Wagner, F.~Wamers, H.~Weick, M.~Weigand, J.~S. Winfield, D.~Yakorev, and M.~Zoric.
\newblock Direct experimental evidence for a multiparticle-hole ground state configuration of deformed 33mg.
\newblock {\em Physical Review C}, 94(3), September 2016.

\bibitem{Bazin2021}
D.~Bazin, N.~Aoi, H.~Baba, J.~Chen, H.~Crawford, P.~Doornenbal, P.~Fallon, K.~Li, J.~Lee, M.~Matsushita, T.~Motobayashi, H.~Sakurai, H.~Scheit, D.~Steppenbeck, R.~Stroberg, S.~Takeuchi, H.~Wang, K.~Yoneda, and C.~X. Yuan.
\newblock Spectroscopy of 33mg with knockout reactions.
\newblock {\em Physical Review C}, 103(6), June 2021.

\bibitem{Richard2017}
A.~L. Richard, H.~L. Crawford, P.~Fallon, A.~O. Macchiavelli, V.~M. Bader, D.~Bazin, M.~Bowry, C.~M. Campbell, M.~P. Carpenter, R.~M. Clark, M.~Cromaz, A.~Gade, E.~Ideguchi, H.~Iwasaki, M.~D. Jones, C.~Langer, I.~Y. Lee, C.~Loelius, E.~Lunderberg, C.~Morse, J.~Rissanen, M.~Salathe, D.~Smalley, S.~R. Stroberg, D.~Weisshaar, K.~Whitmore, A.~Wiens, S.~J. Williams, K.~Wimmer, and T.~Yamamato.
\newblock Strongly coupled rotational band in 33mg.
\newblock {\em Physical Review C}, 96(1), July 2017.

\bibitem{Langevin1984}
M.~Langevin, C.~Détraz, D.~Guillemaud-Mueller, A.C. Mueller, C.~Thibault, F.~Touchard, and M.~Epherre.
\newblock beta-delayed neutrons from very neutron-rich sodium and magnesium isotopes.
\newblock {\em Nuclear Physics A}, 414(1):151–161, February 1984.

\bibitem{Tripathi2010}
Vandana Tripathi, S.~L. Tabor, P.~F. Mantica, Y.~Utsuno, P.~Bender, J.~Cook, C.~R. Hoffman, Sangjin Lee, T.~Otsuka, J.~Pereira, M.~Perry, K.~Pepper, J.~S. Pinter, J.~Stoker, A.~Volya, and D.~Weisshaar.
\newblock Tripathiet al.reply:.
\newblock {\em Physical Review Letters}, 104(12), March 2010.

\bibitem{Bateman1910}
Bateman Harry.
\newblock The solution of a system of differential equations occurring in the theory of radioactive transformations.
\newblock {\em Proc. Cambridge Philos. Soc}, 15:423--427, 1910.

\bibitem{Cetnar2006}
Jerzy Cetnar.
\newblock General solution of bateman equations for nuclear transmutations.
\newblock {\em Annals of Nuclear Energy}, 33(7):640–645, May 2006.

\bibitem{Morrissey2003}
D.J. Morrissey, B.M. Sherrill, M.~Steiner, A.~Stolz, and I.~Wiedenhoever.
\newblock Commissioning the a1900 projectile fragment separator.
\newblock {\em Nuclear Instruments and Methods in Physics Research Section B: Beam Interactions with Materials and Atoms}, 204:90–96, May 2003.

\bibitem{Stolz2005}
A.~Stolz, T.~Baumann, T.N. Ginter, D.J. Morrissey, M.~Portillo, B.M. Sherrill, M.~Steiner, and J.W. Stetson.
\newblock Production of rare isotope beams with the nscl fragment separator.
\newblock {\em Nuclear Instruments and Methods in Physics Research Section B: Beam Interactions with Materials and Atoms}, 241(1–4):858–861, December 2005.

\bibitem{Goosman1973}
D.~R. Goosman, C.~N. Davids, and D.~E. Alburger.
\newblock Accurate masses and beta-decay schemes for 34p and 33si.
\newblock {\em Physical Review C}, 8(4):1324–1330, October 1973.

\bibitem{Agostinelli2003}
S.~Agostinelli, J.~Allison, K.~Amako, J.~Apostolakis, H.~Araujo, P.~Arce, M.~Asai, D.~Axen, S.~Banerjee, G.~Barrand, F.~Behner, L.~Bellagamba, J.~Boudreau, L.~Broglia, A.~Brunengo, H.~Burkhardt, S.~Chauvie, J.~Chuma, R.~Chytracek, G.~Cooperman, G.~Cosmo, P.~Degtyarenko, A.~Dell’Acqua, G.~Depaola, D.~Dietrich, R.~Enami, A.~Feliciello, C.~Ferguson, H.~Fesefeldt, G.~Folger, F.~Foppiano, A.~Forti, S.~Garelli, S.~Giani, R.~Giannitrapani, D.~Gibin, J.J. Gómez~Cadenas, I.~González, G.~Gracia~Abril, G.~Greeniaus, W.~Greiner, V.~Grichine, A.~Grossheim, S.~Guatelli, P.~Gumplinger, R.~Hamatsu, K.~Hashimoto, H.~Hasui, A.~Heikkinen, A.~Howard, V.~Ivanchenko, A.~Johnson, F.W. Jones, J.~Kallenbach, N.~Kanaya, M.~Kawabata, Y.~Kawabata, M.~Kawaguti, S.~Kelner, P.~Kent, A.~Kimura, T.~Kodama, R.~Kokoulin, M.~Kossov, H.~Kurashige, E.~Lamanna, T.~Lampén, V.~Lara, V.~Lefebure, F.~Lei, M.~Liendl, W.~Lockman, F.~Longo, S.~Magni, M.~Maire, E.~Medernach, K.~Minamimoto, P.~Mora~de Freitas, Y.~Morita, K.~Murakami, M.~Nagamatu,
  R.~Nartallo, P.~Nieminen, T.~Nishimura, K.~Ohtsubo, M.~Okamura, S.~O’Neale, Y.~Oohata, K.~Paech, J.~Perl, A.~Pfeiffer, M.G. Pia, F.~Ranjard, A.~Rybin, S.~Sadilov, E.~Di~Salvo, G.~Santin, T.~Sasaki, N.~Savvas, Y.~Sawada, S.~Scherer, S.~Sei, V.~Sirotenko, D.~Smith, N.~Starkov, H.~Stoecker, J.~Sulkimo, M.~Takahata, S.~Tanaka, E.~Tcherniaev, E.~Safai~Tehrani, M.~Tropeano, P.~Truscott, H.~Uno, L.~Urban, P.~Urban, M.~Verderi, A.~Walkden, W.~Wander, H.~Weber, J.P. Wellisch, T.~Wenaus, D.C. Williams, D.~Wright, T.~Yamada, H.~Yoshida, and D.~Zschiesche.
\newblock Geant4—a simulation toolkit.
\newblock {\em Nuclear Instruments and Methods in Physics Research Section A: Accelerators, Spectrometers, Detectors and Associated Equipment}, 506(3):250–303, July 2003.

\bibitem{Quinn2015}
Stephen~J Quinn.
\newblock Capture cross sections for the astrophysical p process.
\newblock {\em Ph. D. Thesis}, 2015.

\bibitem{Kirsch2018}
L.E. Kirsch and L.A. Bernstein.
\newblock Rainier: A simulation tool for distributions of excited nuclear states and cascade fluctuations.
\newblock {\em Nuclear Instruments and Methods in Physics Research Section A: Accelerators, Spectrometers, Detectors and Associated Equipment}, 892:30–40, June 2018.

\bibitem{Egidy2005}
Till~von Egidy and Dorel Bucurescu.
\newblock Systematics of nuclear level density parameters.
\newblock {\em Physical Review C}, 72(4), October 2005.

\bibitem{vonEgidy2009}
T.~von Egidy and D.~Bucurescu.
\newblock Experimental energy-dependent nuclear spin distributions.
\newblock {\em Physical Review C}, 80(5), November 2009.

\bibitem{Koning2023}
Arjan Koning, Stephane Hilaire, and Stephane Goriely.
\newblock Talys: modeling of nuclear reactions.
\newblock {\em The European Physical Journal A}, 59(6), June 2023.

\bibitem{Kopecky1990}
J.~Kopecky and M.~Uhl.
\newblock Test of gamma-ray strength functions in nuclear reaction model calculations.
\newblock {\em Physical Review C}, 41(5):1941–1955, May 1990.

\bibitem{Han2017}
R.~Han, X.Q. Li, W.G. Jiang, Z.H. Li, H.~Hua, S.Q. Zhang, C.X. Yuan, D.X. Jiang, Y.L. Ye, J.~Li, Z.H. Li, F.R. Xu, Q.B. Chen, J.~Meng, J.S. Wang, C.~Xu, Y.L. Sun, C.G. Wang, H.Y. Wu, C.Y. Niu, C.G. Li, C.~He, W.~Jiang, P.J. Li, H.L. Zang, J.~Feng, S.D. Chen, Q.~Liu, X.C. Chen, H.S. Xu, Z.G. Hu, Y.Y. Yang, P.~Ma, J.B. Ma, S.L. Jin, Z.~Bai, M.R. Huang, Y.J. Zhou, W.H. Ma, Y.~Li, X.H. Zhou, Y.H. Zhang, G.Q. Xiao, and W.L. Zhan.
\newblock Northern boundary of the “island of inversion” and triaxiality in 34 si.
\newblock {\em Physics Letters B}, 772:529–533, September 2017.

\bibitem{Tipping2001}
Michael Tipping.
\newblock Sparse bayesian learning and relevance vector machine.
\newblock {\em J. Mach. Learn. Res.}, 1:211--244, 01 2001.

\bibitem{Mller1990}
Peter M\"{o}ller and Jørgen Randrup.
\newblock New developments in the calculation of beta-strength functions.
\newblock {\em Nuclear Physics A}, 514(1):1–48, July 1990.

\bibitem{Mller1995}
P.~Moller, J.R. Nix, W.D. Myers, and W.J. Swiatecki.
\newblock Nuclear ground-state masses and deformations.
\newblock {\em Atomic Data and Nuclear Data Tables}, 59(2):185–381, March 1995.

\bibitem{Erler2012}
Jochen Erler, Noah Birge, Markus Kortelainen, Witold Nazarewicz, Erik Olsen, Alexander~M. Perhac, and Mario Stoitsov.
\newblock The limits of the nuclear landscape.
\newblock {\em Nat.}, 486(7404):509--512, June 2012.

\bibitem{Agbemava2014}
S.~E. Agbemava, A.~V. Afanasjev, D.~Ray, and P.~Ring.
\newblock Global performance of covariant energy density functionals: Ground state observables of even-even nuclei and the estimate of theoretical uncertainties.
\newblock {\em Phys. Rev. C}, 89(5):054320, May 2014.

\bibitem{Bayram2017}
Tuncay Bayram and Serkan Akkoyun.
\newblock An approach to adjustment of relativistic mean field model parameters.
\newblock {\em {EPJ} Web Conf.}, 146:12033, 2017.

\bibitem{Bertsch2017}
G.{\hspace{0.167em}}F. Bertsch and Derek Bingham.
\newblock Estimating parameter uncertainty in binding-energy models by the frequency-domain bootstrap.
\newblock {\em Phys. Rev. Lett.}, 119(25):252501, December 2017.

\bibitem{Kejzlar2020}
V~Kejzlar, L~Neufcourt, W~Nazarewicz, and P-G Reinhard.
\newblock Statistical aspects of nuclear mass models.
\newblock {\em J. Phys. G}, 47(9):094001, July 2020.

\bibitem{Yuan2016}
Cenxi Yuan.
\newblock Uncertainty decomposition method and its application to the liquid drop model.
\newblock {\em Phys. Rev. C}, 93(3):034310, March 2016.

\bibitem{Utama2016}
R.~Utama, J.~Piekarewicz, and H.~B. Prosper.
\newblock Nuclear mass predictions for the crustal composition of neutron stars: A bayesian neural network approach.
\newblock {\em Phys. Rev. C}, 93(1):014311, January 2016.

\bibitem{Utama2017}
R.~Utama and J.~Piekarewicz.
\newblock Refining mass formulas for astrophysical applications: A bayesian neural network approach.
\newblock {\em Phys. Rev. C}, 96(4):044308, October 2017.

\bibitem{Utama2018}
R.~Utama and J.~Piekarewicz.
\newblock Validating neural-network refinements of nuclear mass models.
\newblock {\em Phys. Rev. C}, 97(1):014306, January 2018.

\bibitem{Zhang2017}
Hai~Fei Zhang, Li~Hao Wang, Jing~Peng Yin, Peng~Hui Chen, and Hong~Fei Zhang.
\newblock Performance of the levenberg{\textendash}marquardt neural network approach in nuclear mass prediction.
\newblock {\em J. Phys. G}, 44(4):045110, March 2017.

\bibitem{Neufcourt2018}
L{\'{e}}o Neufcourt, Yuchen Cao, Witold Nazarewicz, and Frederi Viens.
\newblock Bayesian approach to model-based extrapolation of nuclear observables.
\newblock {\em Phys. Rev. C}, 98(3):034318, September 2018.

\bibitem{Neufcourt2020proton}
L{\'{e}}o Neufcourt, Yuchen Cao, Samuel Giuliani, Witold Nazarewicz, Erik Olsen, and Oleg~B. Tarasov.
\newblock Beyond the proton drip line: Bayesian analysis of proton-emitting nuclei.
\newblock {\em Phys. Rev. C}, 101(1):014319, January 2020.

\bibitem{Niu2018}
Z.M. Niu and H.Z. Liang.
\newblock Nuclear mass predictions based on bayesian neural network approach with pairing and shell effects.
\newblock {\em Phys. Lett. B}, 778:48--53, March 2018.

\bibitem{Rodrguez2019}
Ubaldo~Ba{\~{n}}os Rodr{\'{\i}}guez, Cristofher~Zu{\~{n}}iga Vargas, Marcello Gon{\c{c}}alves, Sergio~Barbosa Duarte, and Fernando Guzm{\'{a}}n.
\newblock Alpha half-lives calculation of superheavy nuclei with $q_{\alpha}$ -value predictions based on the bayesian neural network approach.
\newblock {\em J. Phys. G}, 46(11):115109, October 2019.

\bibitem{BaosRodrguez2019}
Ubaldo~Ba{\~{n}}os Rodr{\'{\i}}guez, Cristofher~Zu{\~{n}}iga Vargas, Marcello Gon{\c{c}}alves, Sergio~Barbosa Duarte, and Fernando Guzm{\'{a}}n.
\newblock Bayesian neural network improvements to nuclear mass formulae and predictions in the {SuperHeavy} elements region.
\newblock {\em Europhys. Lett.}, 127(4):42001, September 2019.

\bibitem{Gao2021}
Ze-Peng Gao, Yong-Jia Wang, Hong-Liang L\"{u}, Qing-Feng Li, Cai-Wan Shen, and Ling Liu.
\newblock Machine learning the nuclear mass.
\newblock {\em Nucl. Sci. Tech.}, 32(10):109, October 2021.

\bibitem{Wu2021}
X.H. Wu, L.H. Guo, and P.W. Zhao.
\newblock Nuclear masses in extended kernel ridge regression with odd-even effects.
\newblock {\em Phys. Lett. B}, 819:136387, August 2021.

\bibitem{Athanassopoulos2004}
S.~Athanassopoulos, E.~Mavrommatis, K.A. Gernoth, and J.W. Clark.
\newblock Nuclear mass systematics using neural networks.
\newblock {\em Nucl. Phys. A}, 743(4):222--235, November 2004.

\bibitem{lovell2022nuclear}
A.~E. Lovell, A.~T. Mohan, T.~M. Sprouse, and M.~R. Mumpower.
\newblock Nuclear masses learned from a probabilistic neural network, 2022.

\bibitem{Vladimirova2021}
Elena~V. Vladimirova, Makar~V. Simonov, and Tatiana~Yu. Tretyakova.
\newblock Phenomenological methods for nuclear mass evaluation at the limits of the nuclear chart.
\newblock In {\em {AYSS}-2020}. {AIP} Publishing, 2021.

\bibitem{Bedaque2021}
Paulo Bedaque, Amber Boehnlein, Mario Cromaz, Markus Diefenthaler, Latifa Elouadrhiri, Tanja Horn, Michelle Kuchera, David Lawrence, Dean Lee, Steven Lidia, Robert McKeown, Wally Melnitchouk, Witold Nazarewicz, Kostas Orginos, Yves Roblin, Michael Scott~Smith, Malachi Schram, and Xin-Nian Wang.
\newblock A.i. for nuclear physics.
\newblock {\em The European Physical Journal A}, 57(3), March 2021.

\bibitem{Boehnlein2022}
Amber Boehnlein, Markus Diefenthaler, Nobuo Sato, Malachi Schram, Veronique Ziegler, Cristiano Fanelli, Morten Hjorth-Jensen, Tanja Horn, Michelle~P. Kuchera, Dean Lee, Witold Nazarewicz, Peter Ostroumov, Kostas Orginos, Alan Poon, Xin-Nian Wang, Alexander Scheinker, Michael~S. Smith, and Long-Gang Pang.
\newblock Colloquium: Machine learning in nuclear physics.
\newblock {\em Reviews of Modern Physics}, 94(3), September 2022.

\bibitem{Audi2003}
G.~Audi, A.H. Wapstra, and C.~Thibault.
\newblock The ame2003 atomic mass evaluation.
\newblock {\em Nucl. Phys. A}, 729(1):337--676, December 2003.

\bibitem{Wang2017}
Meng Wang, G.~Audi, F.~G. Kondev, W.J. Huang, S.~Naimi, and Xing Xu.
\newblock The {AME}2016 atomic mass evaluation ({II}). {T}ables, graphs and references.
\newblock {\em Chin. Phys. C}, 41(3):030003, March 2017.

\bibitem{deRoubin2017}
A.~de~Roubin, D.~Atanasov, K.~Blaum, S.~George, F.~Herfurth, D.~Kisler, M.~Kowalska, S.~Kreim, D.~Lunney, V.~Manea, E.~Minaya Ramirez, M.~Mougeot, D.~Neidherr, M.~Rosenbusch, L.~Schweikhard, A.~Welker, F.~Wienholtz, R.~N. Wolf, and K.~Zuber.
\newblock Nuclear deformation in the a$\approx$100 region: Comparison between new masses and mean-field predictions.
\newblock {\em Phys. Rev. C}, 96(1):014310, July 2017.

\bibitem{Welker2017}
A.~Welker, N.{\hspace{0.167em}}A.{\hspace{0.167em}}S. Althubiti, D.~Atanasov, K.~Blaum, T.{\hspace{0.167em}}E. Cocolios, F.~Herfurth, S.~Kreim, D.~Lunney, V.~Manea, M.~Mougeot, D.~Neidherr, F.~Nowacki, A.~Poves, M.~Rosenbusch, L.~Schweikhard, F.~Wienholtz, R.{\hspace{0.167em}}N. Wolf, and K.~Zuber.
\newblock Binding energy of $^{79}$cu : Probing the structure of the doubly magic $^{78}$ni from only one proton away.
\newblock {\em Phys. Rev. Lett.}, 119(19):192502, November 2017.

\bibitem{Vilen2018}
M.~Vilen, J.{\hspace{0.167em}}M. Kelly, A.~Kankainen, M.~Brodeur, A.~Aprahamian, L.~Canete, T.~Eronen, A.~Jokinen, T.~Kuta, I.{\hspace{0.167em}}D. Moore, M.{\hspace{0.167em}}R. Mumpower, D.{\hspace{0.167em}}A. Nesterenko, H.~Penttil\"{a}, I.~Pohjalainen, W.{\hspace{0.167em}}S. Porter, S.~Rinta-Antila, R.~Surman, A.~Voss, and J.~\"{A}yst\"{o}.
\newblock Precision mass measurements on neutron-rich rare-earth isotopes at {JYFLTRAP}: Reduced neutron pairing and implications for r -process calculations.
\newblock {\em Phys. Rev. Lett.}, 120(26):262701, June 2018.

\bibitem{Leistenschneider2018}
E.~Leistenschneider, M.{\hspace{0.167em}}P. Reiter, S.~Ayet~San Andr{\'{e}}s, B.~Kootte, J.{\hspace{0.167em}}D. Holt, P.~Navr{\'{a}}til, C.~Babcock, C.~Barbieri, B.{\hspace{0.167em}}R. Barquest, J.~Bergmann, J.~Bollig, T.~Brunner, E.~Dunling, A.~Finlay, H.~Geissel, L.~Graham, F.~Greiner, H.~Hergert, C.~Hornung, C.~Jesch, R.~Klawitter, Y.~Lan, D.~Lascar, K.{\hspace{0.167em}}G. Leach, W.~Lippert, J.{\hspace{0.167em}}E. McKay, S.{\hspace{0.167em}}F. Paul, A.~Schwenk, D.~Short, J.~Simonis, V.~Som{\`{a}}, R.~Steinbr\"{u}gge, S.{\hspace{0.167em}}R. Stroberg, R.~Thompson, M.{\hspace{0.167em}}E. Wieser, C.~Will, M.~Yavor, C.~Andreoiu, T.~Dickel, I.~Dillmann, G.~Gwinner, W.{\hspace{0.167em}}R. Pla{\ss}, C.~Scheidenberger, A.{\hspace{0.167em}}A. Kwiatkowski, and J.~Dilling.
\newblock Dawning of the n=32 shell closure seen through precision mass measurements of neutron-rich titanium isotopes.
\newblock {\em Phys. Rev. Lett.}, 120(6):062503, February 2018.

\bibitem{Michimasa2018}
S.~Michimasa, M.~Kobayashi, Y.~Kiyokawa, S.~Ota, D.{\hspace{0.167em}}S. Ahn, H.~Baba, G.{\hspace{0.167em}}P.{\hspace{0.167em}}A. Berg, M.~Dozono, N.~Fukuda, T.~Furuno, E.~Ideguchi, N.~Inabe, T.~Kawabata, S.~Kawase, K.~Kisamori, K.~Kobayashi, T.~Kubo, Y.~Kubota, C.{\hspace{0.167em}}S. Lee, M.~Matsushita, H.~Miya, A.~Mizukami, H.~Nagakura, D.~Nishimura, H.~Oikawa, H.~Sakai, Y.~Shimizu, A.~Stolz, H.~Suzuki, M.~Takaki, H.~Takeda, S.~Takeuchi, H.~Tokieda, T.~Uesaka, K.~Yako, Y.~Yamaguchi, Y.~Yanagisawa, R.~Yokoyama, K.~Yoshida, and S.~Shimoura.
\newblock Magic nature of neutrons in $^{54}$ca : First mass measurements of $^{55{\textendash}57}$ca.
\newblock {\em Phys. Rev. Lett.}, 121(2):022506, July 2018.

\bibitem{Orford2018}
R.~Orford, N.~Vassh, J.{\hspace{0.167em}}A. Clark, G.{\hspace{0.167em}}C. McLaughlin, M.{\hspace{0.167em}}R. Mumpower, G.~Savard, R.~Surman, A.~Aprahamian, F.~Buchinger, M.{\hspace{0.167em}}T. Burkey, D.{\hspace{0.167em}}A. Gorelov, T.{\hspace{0.167em}}Y. Hirsh, J.{\hspace{0.167em}}W. Klimes, G.{\hspace{0.167em}}E. Morgan, A.~Nystrom, and K.{\hspace{0.167em}}S. Sharma.
\newblock Precision mass measurements of neutron-rich neodymium and samarium isotopes and their role in understanding rare-earth peak formation.
\newblock {\em Phys. Rev. Lett.}, 120(26):262702, June 2018.

\bibitem{Ito2018}
Y.~Ito, P.~Schury, M.~Wada, F.~Arai, H.~Haba, Y.~Hirayama, S.~Ishizawa, D.~Kaji, S.~Kimura, H.~Koura, M.~MacCormick, H.~Miyatake, J.{\hspace{0.167em}}Y. Moon, K.~Morimoto, K.~Morita, M.~Mukai, I.~Murray, T.~Niwase, K.~Okada, A.~Ozawa, M.~Rosenbusch, A.~Takamine, T.~Tanaka, Y.{\hspace{0.167em}}X. Watanabe, H.~Wollnik, and S.~Yamaki.
\newblock First direct mass measurements of nuclides around z=100 with a multireflection time-of-flight mass spectrograph.
\newblock {\em Phys. Rev. Lett.}, 120(15):152501, April 2018.

\bibitem{Kroll2020}
Liam Kroll, Balraj Singh, and Alan~A. Chen.
\newblock Compilation of recent atomic mass measurements and deduced quantities.
\newblock {\em At. Data Nucl. Data Tables}, 133-134:101336, May 2020.

\bibitem{Bartel1982}
J.~Bartel, P.~Quentin, M.~Brack, C.~Guet, and H.-B. H{\aa}kansson.
\newblock Towards a better parametrisation of skyrme-like effective forces: A critical study of the {SkM} force.
\newblock {\em Nucl. Phys. A}, 386(1):79--100, September 1982.

\bibitem{Dobaczewski1984}
J.~Dobaczewski, H.~Flocard, and J.~Treiner.
\newblock Hartree-fock-bogolyubov description of nuclei near the neutron-drip line.
\newblock {\em Nucl. Phys. A}, 422(1):103--139, June 1984.

\bibitem{Chabanat1995}
E~Chabanat, P~Bonche, P~Haensel, J~Meyer, and R~Schaeffer.
\newblock New skyrme effective forces for supernovae and neutron rich nuclei.
\newblock {\em Physica Scr.}, T56:231--233, January 1995.

\bibitem{Klpfel2009}
P.~Kl\"{u}pfel, P.-G. Reinhard, T.~J. B\"{u}rvenich, and J.~A. Maruhn.
\newblock Variations on a theme by {Skyrme}: A systematic study of adjustments of model parameters.
\newblock {\em Phys. Rev. C}, 79(3):034310, March 2009.

\bibitem{Kortelainen2010}
M.~Kortelainen, T.~Lesinski, J.~Mor{\'{e}}, W.~Nazarewicz, J.~Sarich, N.~Schunck, M.~V. Stoitsov, and S.~Wild.
\newblock Nuclear energy density optimization.
\newblock {\em Phys. Rev. C}, 82(2):024313, August 2010.

\bibitem{Kortelainen2012}
M.~Kortelainen, J.~McDonnell, W.~Nazarewicz, P.-G. Reinhard, J.~Sarich, N.~Schunck, M.~V. Stoitsov, and S.~M. Wild.
\newblock Nuclear energy density optimization: Large deformations.
\newblock {\em Phys. Rev. C}, 85(2):024304, February 2012.

\bibitem{Kortelainen2014}
M.~Kortelainen, J.~McDonnell, W.~Nazarewicz, E.~Olsen, P.-G. Reinhard, J.~Sarich, N.~Schunck, S.~M. Wild, D.~Davesne, J.~Erler, and A.~Pastore.
\newblock Nuclear energy density optimization: Shell structure.
\newblock {\em Phys. Rev. C}, 89(5):054314, May 2014.

\bibitem{Baldo2013}
M.~Baldo, L.~M. Robledo, P.~Schuck, and X.~Vi{\~{n}}as.
\newblock New kohn-sham density functional based on microscopic nuclear and neutron matter equations of state.
\newblock {\em Phys. Rev. C}, 87(6):064305, June 2013.

\bibitem{Goriely2009}
S.~Goriely, S.~Hilaire, M.~Girod, and S.~P{\'{e}}ru.
\newblock First gogny-hartree-fock-bogoliubov nuclear mass model.
\newblock {\em Phys. Rev. Lett.}, 102(24):242501, June 2009.

\bibitem{kejzlar2019bayesian}
Vojtech Kejzlar, Léo Neufcourt, Taps Maiti, and Frederi Viens.
\newblock Bayesian averaging of computer models with domain discrepancies: a nuclear physics perspective, 2019.

\bibitem{Mayer1948}
Maria~G. Mayer.
\newblock On closed shells in nuclei.
\newblock {\em Physical Review}, 74(3):235–239, August 1948.

\bibitem{Mayer1949}
Maria~Goeppert Mayer.
\newblock On closed shells in nuclei. ii.
\newblock {\em Physical Review}, 75(12):1969–1970, June 1949.

\bibitem{Neufcourt2020limits}
L{\'{e}}o Neufcourt, Yuchen Cao, Samuel~A. Giuliani, Witold Nazarewicz, Erik Olsen, and Oleg~B. Tarasov.
\newblock Quantified limits of the nuclear landscape.
\newblock {\em Phys. Rev. C}, 101(4):044307, April 2020.

\bibitem{KASS1995}
Robert~E. Kass and Adrian~E. Raftery.
\newblock {Bayes Factors}.
\newblock {\em J. Am. Stat. Assoc.}, 90(430):773--795, 1995.

\bibitem{Hamaker2021}
A.~Hamaker, E.~Leistenschneider, R.~Jain, G.~Bollen, S.~A. Giuliani, K.~Lund, W.~Nazarewicz, L.~Neufcourt, C.~R. Nicoloff, D.~Puentes, R.~Ringle, C.~S. Sumithrarachchi, and I.~T. Yandow.
\newblock Precision mass measurement of lightweight self-conjugate nucleus $^{80}$zr.
\newblock {\em Nat. Phys.}, 17(12):1408--1412, November 2021.

\bibitem{Keek2012}
L.~{Keek}, A.~{Heger}, and J.~J.~M. {in't Zand}.
\newblock {Superburst Models for Neutron Stars with Hydrogen- and Helium-rich Atmospheres}.
\newblock {\em The Astrophysics Journal}, 752(2):150, June 2012.

\bibitem{Meisel2020}
Z.~Meisel, S.~George, S.~Ahn, D.~Bazin, B.~A. Brown, J.~Browne, J.~F. Carpino, H.~Chung, R.~H. Cyburt, A.~Estradé, M.~Famiano, A.~Gade, C.~Langer, M.~Matoš, W.~Mittig, F.~Montes, D.~J. Morrissey, J.~Pereira, H.~Schatz, J.~Schatz, M.~Scott, D.~Shapira, K.~Smith, J.~Stevens, W.~Tan, O.~Tarasov, S.~Towers, K.~Wimmer, J.~R. Winkelbauer, J.~Yurkon, and R.~G.~T. Zegers.
\newblock Nuclear mass measurements map the structure of atomic nuclei and accreting neutron stars.
\newblock {\em Physical Review C}, 101(5), May 2020.

\bibitem{Ong2020}
W.-J. Ong, E. F. Brown, J.~Browne, S.~Ahn, K.~Childers, B. P. Crider, A. C. Dombos, S. S. Gupta, G. W. Hitt, C.~Langer, R.~Lewis, S. N. Liddick, S.~Lyons, Z.~Meisel, P.~M\"{o}ller, F.~Montes, F.~Naqvi, J.~Pereira, C.~Prokop, D.~Richman, H.~Schatz, K.~Schmidt, and A.~Spyrou.
\newblock Decay of 61v and its role in cooling accreted neutron star crusts.
\newblock {\em Physical Review Letters}, 125(26), December 2020.

\bibitem{Crawford2022}
H. L. Crawford, V.~Tripathi, J. M. Allmond, B. P. Crider, R.~Grzywacz, S. N. Liddick, A.~Andalib, E.~Argo, C.~Benetti, S.~Bhattacharya, C. M. Campbell, M. P. Carpenter, J.~Chan, A.~Chester, J.~Christie, B. R. Clark, I.~Cox, A. A. Doetsch, J.~Dopfer, J. G. Duarte, P.~Fallon, A.~Frotscher, T.~Gaballah, T. J. Gray, J. T. Harke, J.~Heideman, H.~Heugen, R.~Jain, T. T. King, N.~Kitamura, K.~Kolos, F. G. Kondev, A.~Laminack, B.~Longfellow, R. S. Lubna, S.~Luitel, M.~Madurga, R.~Mahajan, M. J. Mogannam, C.~Morse, S.~Neupane, A.~Nowicki, T. H. Ogunbeku, W.-J. Ong, C.~Porzio, C. J. Prokop, B. C. Rasco, E. K. Ronning, E.~Rubino, T. J. Ruland, K. P. Rykaczewski, L.~Schaedig, D.~Seweryniak, K.~Siegl, M.~Singh, S. L. Tabor, T. L. Tang, T.~Wheeler, J. A. Winger, and Z.~Xu.
\newblock Crossing n = 28 toward the neutron drip line: First measurement of half-lives at frib.
\newblock {\em Physical Review Letters}, 129(21), November 2022.

\bibitem{Gray2023}
T. J. Gray, J. M. Allmond, Z.~Xu, T. T. King, R. S. Lubna, H. L. Crawford, V.~Tripathi, B. P. Crider, R.~Grzywacz, S. N. Liddick, A. O. Macchiavelli, T.~Miyagi, A.~Poves, A.~Andalib, E.~Argo, C.~Benetti, S.~Bhattacharya, C. M. Campbell, M. P. Carpenter, J.~Chan, A.~Chester, J.~Christie, B. R. Clark, I.~Cox, A. A. Doetsch, J.~Dopfer, J. G. Duarte, P.~Fallon, A.~Frotscher, T.~Gaballah, J. T. Harke, J.~Heideman, H.~Huegen, J. D. Holt, R.~Jain, N.~Kitamura, K.~Kolos, F. G. Kondev, A.~Laminack, B.~Longfellow, S.~Luitel, M.~Madurga, R.~Mahajan, M. J. Mogannam, C.~Morse, S.~Neupane, A.~Nowicki, T. H. Ogunbeku, W.-J. Ong, C.~Porzio, C. J. Prokop, B. C. Rasco, E. K. Ronning, E.~Rubino, T. J. Ruland, K. P. Rykaczewski, L.~Schaedig, D.~Seweryniak, K.~Siegl, M.~Singh, A. E. Stuchbery, S. L. Tabor, T. L. Tang, T.~Wheeler, J. A. Winger, and J. L. Wood.
\newblock Microsecond isomer at the n = 20 island of shape inversion observed at frib.
\newblock {\em Physical Review Letters}, 130(24), June 2023.

\bibitem{Lubna2023}
R.~S. Lubna, S.~N. Liddick, T.~H. Ogunbeku, A.~Chester, J.~M. Allmond, Soumik Bhattacharya, C.~M. Campbell, M.~P. Carpenter, K.~L. Childers, P.~Chowdhury, J.~Christie, B.~R. Clark, R.~M. Clark, I.~Cox, H.~L. Crawford, B.~P. Crider, A.~A. Doetsch, P.~Fallon, A.~Frotscher, T.~Gaballah, T.~J. Gray, R.~Grzywacz, J.~T. Harke, A.~C. Hartley, R.~Jain, T.~T. King, N.~Kitamura, K.~Kolos, F.~G. Kondev, E.~Lamere, R.~Lewis, B.~Longfellow, S.~Lyons, S.~Luitel, M.~Madurga, R.~Mahajan, M.~J. Mogannam, C.~Morse, S.~K. Neupane, W.-J. Ong, D.~Perez-Loureiro, C.~Porzio, C.~J. Prokop, A.~L. Richard, E.~K. Ronning, E.~Rubino, K.~Rykaczewski, D.~Seweryniak, K.~Siegl, U.~Silwal, M.~Singh, D.~P. Siwakoti, D.~C. Smith, M.~K. Smith, S.~L. Tabor, T.~L. Tang, Vandana Tripathi, A.~Volya, T.~Wheeler, Y.~Xiao, and Z.~Xu.
\newblock beta decay of 36 mg and 36 al : Identification of a beta -decaying isomer in 36 al.
\newblock {\em Physical Review C}, 108(1), July 2023.

\bibitem{Kejzlar2023}
Vojtech Kejzlar, Léo Neufcourt, and Witold Nazarewicz.
\newblock Local bayesian dirichlet mixing of imperfect models.
\newblock {\em Scientific Reports}, 13(1), November 2023.

\end{thebibliography}
\makebibliographypage

\end{document}